%% file: balland_final_editor.tex
\documentclass[]{aa}

\usepackage[varg]{txfonts}
\usepackage[english]{babel}
\usepackage{graphicx}
\graphicspath{{./figures/}}
\usepackage{supertabular}
\usepackage{longtable}
\usepackage{lscape}
\usepackage[]{natbib}
\bibpunct{(}{)}{;}{a}{}{,}
\usepackage{verbatim}
\bibpunct{(}{)}{;}{a}{}{,}
\usepackage{ulem}

\def\Ia{SN~Ia}
\def\Iae{SNe~Ia}
\def\Ias{SN~Ia$\star$}
\def\Iaes{SNe~Ia$\star$}

\usepackage{twoopt}
\makeatletter

\begin{document}

\title{The ESO's VLT Type Ia supernova spectral set of the final two
  years of SNLS
  \thanks{Based on observations obtained with
    MegaPrime/MegaCam, a joint project of CFHT and CEA/DAPNIA, at the
    Canada-France-Hawaii Telescope (CFHT) which is operated by the
    National Research Council (NRC) of Canada, the Institut National
    des Sciences de l'Univers of the Centre National de la Recherche
    Scientifique (CNRS) of France, and the University of Hawaii. This
    work is based in part on data products produced at TERAPIX and the
    Canadian Astronomy Data Centre as part of the Canada-France-Hawaii
    Telescope Legacy Survey, a collaborative project of NRC and
    CNRS.}\fnmsep
  \thanks{Based on observations obtained with FORS1 and FORS2
    at the Very Large Telescope on Cerro Paranal, operated by the
    European Southern Observatory, Chile (ESO Large Programs
    171.A-0486 and 176.A-0589).}}

\author{C.~Balland\inst{\ref{inst1}}\thanks{Contact author \email{balland@lpnhe.in2p3.fr}} 
\and F.~Cellier-Holzem \inst{\ref{inst1}}
\and C.~Lidman \inst{\ref{inst4},\ref{inst5}}
\and P.~Astier \inst{\ref{inst1}}
\and M.~Betoule \inst{\ref{inst1}}
\and R.~G.~Carlberg \inst{\ref{inst6}}
\and A.~Conley \inst{\ref{inst7}}
\and R.~S.~Ellis \inst{\ref{inst8},\ref{inst9}}
\and J.~Guy \inst{\ref{inst1}}
\and D.~Hardin \inst{\ref{inst1}}
\and I.~M.~Hook \inst{\ref{inst10},\ref{inst11}}
\and D.~A.~Howell \inst{\ref{inst12},\ref{inst13}}
\and R.~Pain \inst{\ref{inst1}}
\and C.~J.~Pritchet \inst{\ref{inst14}}
\and N.~Regnault \inst{\ref{inst1}}
\and M.~Sullivan \inst{\ref{inst15}} 
\and V.~Arsenijevic \inst{\ref{inst15b}}
\and S.~Baumont \inst{\ref{inst1}}
\and P.~El-Hage \inst{\ref{inst1}}
\and S.~Fabbro \inst{\ref{inst14}}
\and D.~Fouchez \inst{\ref{inst17}}
\and A.~Mitra \inst{\ref{inst1}}
\and A.~M\"oller \inst{\ref{inst5},\ref{inst18}}
\and A.~M. Mour\~{a}o \inst{\ref{inst16}}
\and J.~Neveu \inst{\ref{inst19}}
\and M.~Roman \inst{\ref{inst1}}
\and V.~Ruhlmann-Kleider \inst{\ref{inst20}}
}
\institute{Laboratoire de Physique Nucléaire et de Hautes Energies, Sorbonne Universit\'e, CNRS-IN2P3, 4 Place Jussieu, 75005 Paris, France\label{inst1}
\and
Australian Astronomical Observatory, Epping, New South Wales, Australia\label{inst4}
\and
ARC Centre of Excellence for All-sky Astrophysics (CAASTRO), Canberra, Australia\label{inst5}
\and 
Department of Astronomy and Astrophysics, University of Toronto, 50 St. George Street, Toronto ON M5S 3H4, Canada\label{inst6}
\and
Center for Astrophysics and Space Astronomy 389-UCB, University of Colorado, Boulder, CO 80309, USA\label{inst7}
\and
European Southern Observatory (ESO), Karl-Schwarzschild Strasse 2, 85748 Garching, Germany\label{inst8}
\and
Department of Physics and Astronomy, University College London, Gower Street, London, WC1E 6BT, UK\label{inst9}
\and
Department of Physics (Astrophysics), University of Oxford, Denys Wilkinson Building, Keble Road, Oxford OX1 3RH, UK\label{inst10}
\and
Department of Physics, Lancaster University, Lancaster LA1 4YB, United Kingdom\label{inst11}
\and
Las Cumbres Observatory, 6740 Cortona Dr., Suite 102, Goleta, CA 93117\label{inst12}
\and
Department of Physics, University of California, Santa Barbara, Broida Hall, Mail Code 9530, Santa Barbara, CA 93106-9530\label{inst13}
\and
NRC Herzberg Institute for Astrophysics, 5071 West Saanich Road, Victoria V9E 2E7, British Columbia, Canada\label{inst14}
\and
Department of Physics and Astronomy, University of Southampton, Highfield, Southampton, SO17 1SX, UK\label{inst15}
\and
Seven Bridges Genomics Inc., 1 Main St, 5th Floor, Suite 500, Cambridge, MA 02142, USA\label{inst15b}
\and
CENTRA - Centro Multidisciplinar de Astrof\'isica and Dep. F\'isica, Instituto Superior T\'ecnico, Universidade de Lisboa, Portugal\label{inst16}
\and 
Aix Marseille Univ, CNRS/IN2P3, CPPM, Marseille, France\label{inst17}
\and
Research School of Astronomy and Astrophysics, Australian National University, Canberra, Australia\label{inst18}
\and
LAL, Univ. Paris-Sud, CNRS-IN2P3, Université Paris-Saclay, Orsay, France\label{inst19}
\and
CEA, Centre de Saclay, Irfu/SPP, F-91191 Gif-sur-Yvette, France\label{inst20}
}
\date{Received / Accepted}
\abstract {} {We aim to present 70 spectra of 68 new high-redshift type Ia supernovae
  (\Iae) measured at ESO's VLT during the final two years of
  operation (2006-2008) of the Supernova Legacy Survey (SNLS). This new sample
  complements the VLT three year spectral set. Altogether, these two data
  sets form the five year sample of SNLS \Ia~spectra measured at the VLT
  on which the final SNLS cosmological analysis will partly be based.
  In the redshift range considered, this sample is unique in terms of
  homogeneity and number of spectra. We use it to investigate the
  possibility of a spectral evolution of \Iae~populations with
  redshift as well as \Iae~spectral properties as a function of
  lightcurve fit parameters and the mass of the host-galaxy.}
  {Reduction and extraction are based on both IRAF standard tasks and
   our own reduction pipeline. Redshifts are estimated from host-galaxy lines
   whenever possible or alternatively from supernova features. We used
   the spectro-photometric \Ia~model SALT2 combined with a set of
   galaxy templates that model the host-galaxy contamination to assess the type Ia 
   nature of the candidates.}  
   {We identify 68 new \Iae~with
   redshift ranging from $z=0.207$ to $z=0.98$ for an average redshift
   of $z=0.62$. Each spectrum is presented individually along with its
   best-fit SALT2 model. Adding this new sample to the three year VLT
   sample of SNLS, the final dataset contains 209 spectra
   corresponding to 192 \Iae~identified at the VLT. We also publish
   the redshifts of other candidates (host galaxies or other
   transients) whose spectra were obtained at the same time as the
   spectra of live SNe Ia. This list provides a new redshift catalog
   useful for upcoming galaxy surveys.  Using the full VLT
   \Iae~sample, we build composite spectra around maximum light with
   cuts in color,  the lightcurve shape parameter ('stretch'),  host-galaxy mass and redshift.  We find that
   high-$z$ \Iae~are bluer, brighter and have weaker intermediate mass
   element absorption lines than their low-$z$ counterparts at a level
   consistent with what is expected from selection effects. We also
   find a flux excess in the range [3000-3400] \AA~for \Iae~in low
   mass host-galaxies ($M < 10^{10} M_\odot$) or with locally blue $U-V$
   colors, and suggest that the UV flux (or local color) may be used in future cosmological studies as a third standardization parameter in addition to stretch and color.}  {}


\titlerunning{The ESO's VLT \Iae~sample of the last two years of SNLS}

\maketitle

\section{Introduction}
\label{sec:intro}

The use of type Ia supernovae (\Iae) as standardisable candles has led
to the discovery of the acceleration of the universal expansion
\citep{Perlmutter97,Perlmutter99,Riess98b,Schmidt98}.  This
acceleration is usually attributed to a dark energy (DE) component
that contributes  ~70\% of the energy budget of the
Universe. Characterizing the nature of this component by constraining
its equation-of-state parameter $w$ (the DE pressure to energy density
ratio) has become a major goal of observational cosmology. For this
purpose, a combination of various probes has been used. Among those
probes, the measurement of luminosity distances to \Iae~provides the
simplest and most direct way of probing DE at low to intermediate
redshifts \citep{Astier06, Wood-Vasey07, Kowalski08, Sullivan11,
  Suzuki12, Campbell13, Rest14, Betoule14}.

Since the original Supernova Cosmology Project and High-z team
projects, new generations of observational programs have been
developed in order to fill the gaps in the Hubble diagram.  Among
those, the Supernova Legacy Survey (SNLS), with its 427
spectroscopically confirmed \Iae~in the range $0.15 < z < 1.1$, is the
 largest supernova survey at high redshift to date.

SNLS was a five year experiment conducted as part of the Deep Survey of
the Canada-France-Hawaii Telescope Legacy Survey
\citep{Sullivan03}. It is a spectro-photometric program aiming at
discovering and following \Iae~at intermediate to high
redshifts. Conducted from mid-2003 to late 2008, the experiment was
split into two surveys. A photometric program at the
Canada-France-Hawaii Telescope (CFHT) implemented a rolling search
technique that permitted the detection of new \Ia~candidates as well
as the follow-up of their lightcurves in several photometric bands
(mainly MegaCam $g_M$, $r_M$, $i_M$ and $z_M$;
\citealt{Boulade03}). Over the five years of operation, around 1000
\Ia~candidates with well sampled multi-band lightcurves were
obtained (\citealt{Guy10,Bazin11}).  SNLS spectroscopic follow-up
programs have been performed in parallel on 8-10 meter diameter
telescopes to secure the candidate type and redshift. Spectra of about
half of the SNLS \Iae~have been acquired at the ESO Very Large Telescope
(VLT), while the remaining spectra have been obtained at the
Gemini-North and South, Keck I and II telescopes.

\citet{Astier06} published the first SNLS cosmological analysis based
on the first year data set consisting of 71 \Iae~measured from August
2003 to July 2004.  Photometry calibration and luminosity distances to
252 \Iae~measured in the first three years of operation were presented
in \citet{Regnault09} and \citet{Guy10}, respectively. The
corresponding spectra and redshifts were published in
\citet{Howell05,Bronder08, Balland09, Walker11}.  Combining the SNLS
three year \Iae~with low-redshift (z<0.1) \Iae~from the literature \citep{Hicken09,Contreras10}, intermediate
redshift \Iae~from the 1st year of Sloan Digital Sky Survey SDSS-II Supernova Survey \citep{Holtzman08,Kessler09} and a
dozen high-redshift \Iae~from the Hubble Space Telescope (HST) \citep{Riess07},
\citet{Conley11} produced the most advanced supernova Hubble diagram
at the time, with 472 \Iae~in total.  More recently, based on an
 combined flux calibration of SNLS and SDSS \citep{Betoule13},
\citet{Betoule14} performed a joint analysis of SNLS and SDSS-II
supernova sets (the Joint Lightcurve Analysis -- JLA).  Including
external, non-SN datasets in their cosmological analysis, such as the
CMB measurements from the Planck \citep{Planck1513} and WMAP
experiments \citep{Bennett13}, and the Baryon Acoustic Oscillation
(BAO) results \citep{Beutler11, Padmanabhan12, Anderson12},
\citet{Betoule14} obtain a value of the equation of state $w = -1.028\pm 0.055$ (including both systematics and statistical uncertainties, assuming a
flat universe), the most precise measurement to date (see
\citealt{Aubourg15,Alam17} for recent constraints with similar
precision based on BAO scale measurements from the Baryon Oscillation
Spectroscopic Survey (BOSS) experiment).

The present paper focuses on the description and analysis of SNLS
spectra taken at the VLT between August 2006 and September 2008 (the
final two years of SNLS). It complements the analysis of the first
three years SNLS-VLT spectra published in \citet{Balland09}. Confirmed
\Iae~of the new sample having sufficient photometric information will
be included in the SNLS 5yr Hubble diagram \citep{Betoule17}.  Also
published in this paper are the redshifts of other objects (host
galaxies or other transients) whose spectra were obtained at the same
time as the spectra of live SNe Ia through the use of the multi-object
spectroscopic (MOS) mode of the FORS instruments at the VLT.  This
list provides a new redshift catalog useful for upcoming galaxy
surveys.

Spectroscopy is essential not only for securing the type and redshift
of the \Ia~candidates, but also because \Iae~spectra are a rich source
of physical information about their explosion conditions and
composition. Even though the signal-to-noise ($S/N$) ratio of SNLS
spectra is not ideal for detailed analyses, and despite the fact that
they are usually only obtained at a single phase\footnote{We define
  the spectrum phase $\phi$ as the restframe age of the supernova in
  days with respect to the B-band maximum light, divided by stretch.}
(most often around maximum light), they have been used in several
studies that search for empirical correlations beetwen \Iae~peak
luminosity and spectroscopic features in order to reduce the residual
dispersion in the Hubble diagram. \citet{Bronder08} and
\citet{Walker11}, using SNLS spectra and low-redshift spectra compiled
from the literature, show the existence of a significant correlation
between Hubble residuals and the equivalent width of \ion{Si}{ii} ($\sim$ 4130 \AA) and
\ion{Mg}{ii} ($\sim$ 4300 \AA) absorptions, even though not at the level of the
correlations of Hubble residuals with photometric color and stretch
used in the standardization process. In the same spirit, using spectra
of \Iae~at low and higher redshift, \citet{Nordin11} find a
correlation between the SALT2 color and the pseudo-equivalent width of
\ion{Si}{ii}$\lambda 4130$ that could be used to improve cosmological
distance measurements with \Iae~roughly at the level obtained by the usual standardization procedure based on lightcurve shape and color parameters. (We note, however, that this effect is not seen
by \citealt{Chotard11}). More recently, \citet{Milne15} use SNLS-VLT
three year spectra, among others, to explore the spectral region producing
UV/optical color differences seen in \Iae~populations.

Underlying cosmological analyses with \Iae, the hypothesis that
\Iae~properties do not evolve with redshift statistically in a
way that is not corrected for by the usual stretch- and
color-luminosity correction, can be assessed with spectral data.
\citet{Foley08}, using various \Iae~samples from Lick and Keck
observatories and from the ESSENCE program \citep{Wood-Vasey07}, make
a comparison of composite spectra at low and high redshift and show
that once galaxy light contamination is accounted for, the two samples
are remarkably similar. Although several minor localized deviations
between the low and high-redshift spectra are found, the difference
constrains the evolution of SN spectral features to be less than 10\%
in relative flux in the optical rest-frame.  Using the SNLS three year
sample, \citet{Balland09} compute a composite spectrum around maximum
light at $z>0.5$ and compare it with its lower redshift counterpart.
Absorption features due to intermediate mass elements (IME) around
4000 \AA~are found to be shallower in the high redshift spectrum. This
is consistent with the \Iae~observed at higher redshift being more
luminous, hence displaying hotter ejecta with a higher ionization level than at low redshift
\citep{Sullivan09}. Comparing \Iae~spectra obtained with the Hubble
Space Telescope Imaging Spectrograph (HST-STIS) with ground-based
counterparts with similar stretches and phases, \citet{Cooke11} find a
similar trend in the UV region of the spectra. They interpret these
discrepancies as representing compositional variations among their
sample rather than being due to an evolutionary effect.
\citet{Maguire12} improve over the \citet{Cooke11} analysis by using
32 low redshift ($0.001 < z < 0.08$) \Iae~HST-STIS spectra and compare
them to the sample of \citet{Ellis08}. They find that their mean low
redshift near-UV spectrum has a depressed flux compared to its
intermediate redshift counterpart, consistent with evolution of the
near-UV continum at the 3$\sigma$ level. This is in qualitative
agreement with \citet{Foley12} who compare a sample of 17 Keck/SDSS
high-quality spectra at intermediate redshift to a low-redshift sample
with otherwise similar properties: the Keck/SDSS \Iae~have, on
average, extremely similar rest-frame optical spectra, but a UV flux
excess with respect to the low redshift sample.  In this paper, we
reassess the issue of supernova evolution with redshift in the light
of our new sample. We also look for spectral differences arising from
splitting our sample by stretch or host-galaxy mass that go beyond mere
selection effects.

We describe the SNLS photometric survey and spectroscopic programs in
Sect. 2. In Sect. 3, we present the acquisition and reduction of our new
spectral data. We present our redshift estimate and \Ia~identification
procedures in Sect. 4. Results on individual \Ia~spectra are presented in
Sect. 5, as well the redshift catalog of other, non-\Iae.  The average
properties of our \Ia~sample is discussed in Sect. 6. A comparison with
the VLT three year sample is also provided in that section. In Sect. 7, we
build composite spectra below and above $z=0.6$ (the average redshift
of our sample) and discuss their differences in the context of a
possible redshift evolution of \Iae~spectral properties. We also study
spectral differences arising from splitting our sample by stretch or
host-galaxy stellar mass. We discuss our findings and draw our conclusions in
Sects. 8 and 9.

\section{The SNLS experiment}
\label{sec:SNLS}

\subsection{The photometric survey}
\label{subsec:photo}

The SNLS survey is a Stage II Dark Energy experiment
\citep{Albrecht06} aiming at constraining the DE equation of state
parameter at the 10 percent level using several hundreds of precisely
calibrated \Iae~lightcurves sampled around the B-band maximum
luminosity.  Detection and photometric
follow-up are made through an optical imaging survey using the MegaCam
camera on the 3.6m CFHT in Hawaii \citep{Boulade03}. The supernova
candidates are detected thanks to their time-varying luminosity in
relation to a reference image in four 1 square degree fields (D1-D4)
with low Galactic-extinction.  The fields have been observed every 3-5
nights (2-4 days in supernova restframe) during 5 to 6 lunations per
year.  This rolling search technique permits simultaneous observations
of a large number of \Ia~candidates and the construction of
multi-color lightcurves over time.

\subsection{The spectroscopic surveys}
\label{subsec:spectro}

Spectroscopic follow-up is necessary to assess the \Ia~nature of the
candidates and estimate their redshift. Due to the faintness of
distant \Iae~($r_M\sim 24$ at $z\sim 1$), the spectroscopic follow-up
is done with 8-10 meter class telescopes. Two large programs at the
VLT were allocated a total of 500h of observing time during the first
four years of SNLS. A similar amount of time was allocated on the
Gemini North and South telescopes. On the Keck I and II telescopes,
about four nights per semester were allocated to SN followup during
the five years of SNLS.  Candidates were selected for spectroscopy based
on the quality of the first few measured photometric points on their
lightcurves \citep {Sullivan06a}. Spectra of 755 candidates have been
measured, amounting to roughly half of the photometric \Ia~candidates.

High redshift candidates ($z>0.6$) were preferentially observed by the
Gemini telescopes because of the improved sky lines subtraction with
the nod-and-shuffle mode available at these telescopes
\citep{Glazebrook01}. More than 200 candidates were observed on Gemini
from August 2003 to May 2008, 150 of which have been identified as
\Iae~\citep{Howell05, Bronder08, Walker11}.

Lower redshift targets were usually observed at the VLT using the
visual and near UV spectrographs FORS1 and FORS2 \citep{Appenzeller98}.
Roughly 50\% of the SNLS spectra (321 out of 755) were obtained during
the two VLT large programs (European Southern Observatory Large
Programs 171.A-0486 and 176.A-0589) from June 2003 to September
2007. 124 objects, observed between June 2003 and July 2006, have been
identified as \Iae~(either \Ia~ or \Ia$\star$, see our classification
Sect. \ref{subsec:ID}).  \Iae~measured in long-slit spectroscopy mode
(LSS) during this period constitute the VLT three year \Ia~data set from
SNLS \citep{Balland09}. This set is supplemented by 59 \Iae~observed
from August 2006 until September 2007 and published in the present
paper. We also add to the present sample 8 \Iae~ that were measured in
MOS mode during the first 3 years of SNLS but not published in
\citet{Balland09} and one that was identified as a \Ia~ by
\citet{Bazin11} after a new extraction (see Sect. \ref{subsec:snia}). In
total, the sample presented in this paper thus contains 68 \Iae~(see
Sect. \ref{sec:data}) whose redshift ranges from 0.21 to 0.98. Besides
\Iae, 28 SN~II, 9 SN~Ib/Ic and 12 AGN were identified in the full VLT
spectral sample. Finally, the identification was inconclusive for 89
(27\%) candidates.

Several dozen candidates were observed at the Keck telescopes, in
particular objects in the northernmost SNLS field (D3) that was not
observable from the VLT. Around 150 SNLS candidate observations have
been performed at Keck. A subset of these Keck spectra with a
substantially higher $S/N$ than needed purely for typing are presented
in \citet{Ellis08}.

\section{The \Iae~spectral set}
\label{sec:data}

\subsection{The SN~Ia sample}
\label{subsec:snia}

The \Iae~sample (\Ia~ and \Ia$\star$) of the present analysis contains
the \Iae~spectra observed at VLT during the last year and a half of
the SNLS survey, namely from August, 1st 2006 up to the end of the
survey mid-2008.  As mentioned above, we also publish the spectra of
SN~05D1dx, SN~05D1hm, SN~05D1if, SN~05D2le, SN~06D2ag, SN~06D4ba,
SN~06D4bo and SN~06D4bw that were acquired in MOS mode before August,
1st 2006 but were not included in the three year VLT spectral sample
\citep{Balland09} as the extraction pipeline used for that analysis
did not support MOS mode. For completeness, we further add the
spectrum of SN~06D2bo, a Type Ia supernova measured in LSS mode in
February 2006 that was first misclassified\footnote{For this
  supernova, the automatic extraction procedure failed at identifying
  the correct object in the crowded field in which it exploded and a
  neighboring spectrum was extracted instead.} as a non-SN~Ia object,
then reclassified by \citet{Bazin11}.  Altogether, 107 spectra of 104
candidates have been analyzed in the present study, 68 of
which\footnote{A release of the spectral set presented in this paper
  is available at http://supernovae.in2p3.fr/Snls5VltRelease.} are
identified as \Iae~(see Sect. \ref{sec:results}).

\subsection{Data acquisition and reduction}

Up until the beginning of September 2005, all the real-time
spectroscopic follow-up with FORS1 and FORS2 was done in long slit
mode.  By that time, SNLS had been running for slightly more than two
years and was starting to accumulate a significant number of
transients that lacked real-time spectroscopic follow-up.  From
September 2005 onward we used the multi-object spectroscopic (MOS)
mode of FORS1 and FORS2 to target live transients, the host-galaxy of any
other transient that happened to be within the FORS2 field of view and
randomly selected field galaxies (see Sect. \ref{subsec:otherZ}).

The MOS mode consists of 38 movable blades that can be used to make 19
slits anywhere in the FORS focal plane.  The advantage of the MOS mode
over precut masks (the MXU mode) is that it allows one to configure
the focal plane in a very short amount of time.  The MXU masks require
at least one day to manufacture and need to be inserted into the FORS
mask exchange unit during the afternoon, which means that there is
some delay in targeting the live transient.  Furthermore, the MXU mode
is available with FORS2 only.  The disadvantage of the MOS mode is
that the length of the slits are fixed -- the slit lengths vary
between 20\arcsec~(even numbered slits) and 22\arcsec~(odd numbered
slits).  With respect to the MXU mode, this reduces the flexibility
one has in selecting targets, and it usually means that only one
object (generally the SN) is observed in the center of the slit
(length wise).  Some objects are observed close to the slit ends,
which complicates the processing of the data.

We used one of two setups with FORS.  If the SN was likely to be below
$z\sim0.7$, we used the 300V grism with the GG435 order sorting
filter.  If it was likely to be at a higher redshift we used the 300I
grism with the OG590 order sorting filter.  The slit widths were set
to 1\arcsec. For all observations, we used the atmospheric dispersion
corrector (ADC) and set the position angle to pass through the SN and
the center of its host-galaxy. The FORS data were reduced using a mixture of
standard IRAF\footnote{IRAF is distributed by the National Optical
  Astronomy Observatories which are operated by the Association of
  Universities for Research in Astronomy, Inc., under the cooperative
  agreement with the National Science Foundation} tasks and our own
routines that were specifically written to process MOS data from FORS1
and FORS2.  Each spectrum was calibrated in wavelength and flux. For the purpose of flux calibration, we
produced a number of calibration curves from the observation of spectro-photometric standard stars and updated them periodically for a
given setup.  Differential slit losses were partially corrected by the
ADC. Residual losses were taken into account with the recalibration
procedure described in Sect. \ref{subsec:ID}.  No correction of
telluric features was performed.  For the SNe, we also computed an
error spectrum derived from Poisson noise in regions of the 2D
sky-subtracted spectrum that are free of objects.

\section{Spectral analysis}
\label{sec:analysis}

\subsection{Redshift estimates from host-galaxy lines ($z_{host}$)}
\label{subsec:zgal}

Redshifts are estimated from strong spectral features when present and
are not corrected to the heliocentric reference frame.  The most
commonly identified host-galaxy features in the spectra are emission lines
from the [\ion{O}{ii}] unresolved $\lambda\lambda$3727, 3729 doublet,
H$\beta$ $\lambda$4861, the [\ion{O}{iii}] doublet
$\lambda\lambda$4959,5007, H$\alpha$ $\lambda 6563$ and absorption
lines from higher order Balmer transitions, and \ion{Ca}{ii} H\&K
$\lambda\lambda$3934, 3968.  In some spectra, other host-galaxy lines such as
[\ion{Ne}{iii}] $\lambda$3869, [\ion{N}{ii}] $\lambda$6549,
[\ion{N}{ii}] $\lambda$6583 and [\ion{S}{ii}] $\lambda\lambda$6716,
6730 are identified.  To estimate the redshift, we perform a gaussian
fit of each identified host-galaxy emission or absorption line.  Where
possible, we preferentially use well defined host-galaxy features located in
the center of the spectral range.  We assign an error of $\delta z
\sim 0.001$ on the redshift derived from the host-galaxy lines width, typical
of the uncertainty obtained on host-galaxy redshifts \citep{Lidman05, Hook05,
  Howell05, Balland06, Balland07, Baumont08}.  About 75\% of the
redshifts of our sample come from host-galaxy spectral lines, the remaining
25\% come from SN features, because of insufficient or nonexistent
host-galaxy signal in the spectra.
 
\subsection{Redshift estimates from SN features $(z_{SN}$)}
\label{subsec:zsn}

If there is no apparent host-galaxy line, the redshift is estimated from the
supernova features themselves.  First, a rough estimate is visually
inferred from one of the large troughs characteristic of \Ia~ spectra
(e.g., \ion{Ca}{ii} or \ion{Si}{ii} $\lambda$4130).  Then, we perform
a combined fit of the observed lightcurves and spectrum of the object
using SALT2 (\citealt{Guy07}, see Sect. \ref{subsec:ID} below) over a
grid of redshift values regularly spaced around the input value, with
a step of $\Delta z=0.005$.  The redshift value is given by the
best-fit $\chi^2$. As the supernova features have widths larger than
those of host-galaxy lines, we assign an error of $\delta z \sim 0.01$ on the
supernova redshifts and we consider this value as typical of the error
obtained from SN features. Indeed, the distribution of $z_{SN}-z_{host}$
computed on a subset of candidates for which both redshifts are available has a mean of -0.001 and a
standard deviation $\sigma\sim 0.007$. This supports the value of 0.01 adopted as a typical error for $z_{SN}$.
 In 20\% of cases, the best-fit redshift is
further refined by visual inspection of the spectral fit, that is, the
redshift is slightly shifted from the best-fit value to visually
ensure the best overall agreement between the model and the spectrum.
These cases usually correspond to noisy spectra with $S/N$ per
wavelength bin $ \lesssim 1 $ and a high $\gtrsim 40$ \% fraction of
host galaxy contamination.  As an independent check, we also use
SNID\footnote{SuperNova IDentification
  \\ http://www.oamp.fr/people/blondin/software/snid/index.html}
\citep{Blondin07, Blondin11} to secure our estimates.

\subsection{SN identification}
\label{subsec:ID}

To assess the Ia nature of the candidates, we follow the procedure
described in \citet{Baumont08} and extensively used in
\citet{Balland09}. We perform a combined fit of observed lightcurves
and spectra with SALT2 using a $\chi^2$ minization procedure. Telluric
absorptions present in the spectra are masked. A galaxy component
modeling the host-galaxy is added to the SN model to take into account host-galaxy
contamination. The overall fraction of the host-galaxy in the full model is a
free parameter that is adjusted during the fit. Host-galaxy models include
PEGASE2 synthetic templates \citep{Fioc97,Fioc99} for elliptical (E),
lenticular (S0), Sa, Sb, Sc and Sd Hubble types at various ages. We
also consider \citet{Kinney96} templates for the same types, excluding
Sc and Sd. The best-fit host-galaxy model is obtained by interpolation
between two contiguous types in the Kinney template sequence or two
contiguous galaxy ages in the PEGASE2 templates (see
\citealt{Baumont08, Balland09} for more details).  The final supernova
spectrum is obtained by subtracting the best fit host-galaxy model from the
full spectrum. We do not add host-galaxy lines to the PEGASE2 templates and,
consequently, line residuals often contaminate the final SN spectrum.
This does not impact the identification of the candidate type as these
residuals are localized and do not affect the overall spectral shape.

The best-fit SALT2 model is characterized by a SN color $c$ (defined
as a color excess: $c$ is the $B-V$ color of the candidate at maximum
light minus the average color of the SALT2 training sample
supernovae), a lightcurve width parameter $x_1$ linked to the stretch
$s$ (a $s=1$ \Ia~ has $x_1=0$)\footnote{In this paper, we use
  alternatively the $x_1$ parameter and the stretch $s$. The latter is
  used in particular when we build composite spectra for the sake of
  comparison with similar composites published in the litterature.},
the MJD date of B-band maximum light and an overall normalization
parameter $x_0$ \footnote{In SALT2, the SN flux as a function of phase
  $\phi$ and wavelength $\lambda$ is modeled by
  $F(SN,\phi,\lambda)=x_0\times \big[ M_0(\phi,\lambda)+x_1
    M_1(\phi,\lambda)\big]\times \exp (cCL(\lambda))$, where $M_0$ is
  the average spectral sequence and $M_1$ describes the main
  variability among the \Iae~ of the training sample. $CL(\lambda)$ is
  the average color correction law.}.  Imperfect flux calibration of
the candidate spectrum is taken into account by allowing the
photometric model to be recalibrated (twisted) to fit the spectrum
using a recalibration function parametrized as $\exp(\sum \gamma_i
{\tilde \lambda}^i)$.  Here, the \{$\gamma_i$\} are a set of
recalibration coefficients and ${\tilde \lambda}$ is a pivot
wavelength defined as $\tilde \lambda=\lambda/4400-1$ (see
\citealt{Guy07} for further details). The exponential function
enforces positivity to avoid negative flux values. In practice, two
recalibration coefficient (an overall shift and a tilt of the
spectrum) are sufficient to account for imperfect calibration.

The nature of the candidate is then assessed by visual inspection of
the fit results. Following \citet{Balland09}, we classify each
spectrum in one of the following categories :
\begin{itemize}
\item \textbf{\Ia~ :} the object is certainly a \Ia.  A
  candidate falls into this category if at least one defining feature
  of \Iae~ is seen (\ion{Si}{ii} $\lambda 4130$, \ion{Si}{ii} $\lambda
  6150$ or the \ion{S}{ii} W-shaped feature around 5600 \AA) or
  if the spectral fit is good over a large spectral range, the
  spectrum phase is earlier than 5 days past B-band maximum light (to
  avoid possible confusion with type Ib/c, see \citealt{Hook05,
    Howell05}) and no strong recalibration is needed (flux correction
  lower than 20\% over the whole spectral range, see
  \citealt{Baumont08}).

\item \textbf{\Ia$\star$ :} the candidate is likely a \Ia~but
  other types cannot be excluded given the $S/N$ ratio and phase.  The
  spectra that fall into this category do not present clear
  \Ia~features. They typically have a low $S/N$ and/or a phase larger
  than $+5$ days and/or heavy host-galaxy contamination.

\item \textbf{SN? :} a signal is visible on the spectrum. It is
  possibly a supernova, but the type is unclear.

\item \textbf{not \Ia~:}  candidates that fall into this category refer to two different cases: 

  either 1) the spectrum is clearly not the one of a \Ia~
  but rather falls into one of the following possibilities: SN~II,
  SN~Ib, SN~Ic or AGN. As SALT2 is a \Ia~model built on a training
  sample that only contains \Iae, it does not allow for a direct
  identification of these types. The final classification is made by
  eye, but the non-\Ia~ nature of the candidate appears through
  features badly reproduced and/or a stronger than usual
  recalibration. 
  
  or 2) there is not enough signal for a clear identification. This is usually the case for spectra with a heavy host-galaxy contamination (fraction higher than 95\%),  as essentially no supernova
  signal is left after host-galaxy subtraction.

\item \textbf{\Ia-pec:} the spectrum is peculiar and is likely to be
  of the type of SN~1991T/SN~1999aa or SN~1991bg. We found only one
  object (SN~07D1ah, see Sect. \ref{subsec:VLT5}) of a peculiar (SN~1991T
  like) type in the present sample. Peculiar \Iae~ are
  discarded from the sample used for cosmology fits in SNLS.
\end{itemize}

In the identification procedure outlined above, the values of the
photometric parameters are essentially not considered as we want our
classification to be based as much as possible on spectral
features. However, when in doubt, inspection of these parameters can
add valuable information and help for the final decision. For example,
a large value of the shape parameter $x_1$ (typically $|x_1| > 2$) can confirm some
peculiarity seen in the spectrum. A large color parameter $c$ can be
due to a dense dust environment or a very red intrinsic color, but
also be the signature of a SN~Ic.  As in \citet{Balland09}, we have
cross-checked our identifications using the {\tt superfit} template
fitting technique of \citet{Howell05}, and in some cases,
SNID. Putting together all the tools and information at our disposal,
convergence to a final type is obtained in all cases with a low
probability of misclassification.

\section{Results}
\label{sec:results}

Among the 104 \Iae~candidates in the sample presented in this paper
(see Sect. \ref{subsec:snia} above), 51 have been classified as \Ia, 16
as \Ia$\star$, 1 as \Ia-pec, 12 as SN? and 24 as not \Ia.  In this
section, the spectra of the 68 identified \Iae, \Iae$\star$ and
\Ia-pec are presented individually.  In Sect. \ref{subsec:otherZ}, we
present, for completeness, a catalog (redshift and type) of the other
objects (galaxies and other supernova types) observed over the course
of the survey.

\subsection{Observing log}
\label{subsec:obs}

A listing of the objects identified as \Ia, \Ia$\star$ or \Ia-pec is
provided in Table \ref{table:obs}, together with a brief observing
log.  For each spectrum, the coordinates (RA and Dec) of the object,
the UTC date of acquisition and the exposure time are presented in
columns 2, 3, 4 and 5, respectively.  Observing conditions (median
seeing and airmass) are given in columns 6 and 7, and the observer
frame $i_M$-band magnitude at the date of acquisition ({\it a
  posteriori} interpolated from the lightcurve) is listed in column
8. Two SNe have been observed twice: SN~05D1dx and SN~06D1eb, so 70
spectra for 68 objects are presented in Table \ref{table:obs}. All
spectra have been obtained with the MOS mode except for SN 06D1cm, SN
06D1du, SN 06D1ez, SN 06D1fd, SN 06D1ix, SN 06D2bo, SN 06D4gs and SN
07D2aa which were acquired in LSS mode.  All but four spectra have
been measured using the 300V grism + GG435 order sorting filter.

\subsection{Redshifts and types}
\label{subsec:zID}

Redshifts and identifications are presented in Table
\ref{table:result} for each of the 68 \Iae~candidates.  The type (\Ia,
\Ia$\star$ or \Ia-pec) and the redshift are given in columns 2 and 3.
The redshift source (H for host-galaxy; S for SN) is shown in column
4. Eighteen redshifts out of 68 (27\%) are estimated from supernova
features. Column 5 lists the spectrum phase\footnote{The spectrum phase is computed from the best-fit date of maximum light and the MJD date of the spectrum.}, while columns 6 and 7
present, for each spectrum, the best-fit host-galaxy model and the overall
fraction it amounts in the full SN + host-galaxy model. We label the host-galaxy
model with a letter for the Hubble type followed by the age (in Gyrs)
of the corresponding PEGASE2 template. For the Kinney templates, the
two contiguous types between which the best-fit host-galaxy model has been
interpolated is indicated.  In some cases, the best-fit is obtained
when no galaxy component is added to the model. We indicate these
cases with the label NoGalaxy in column 6. An average $S/N$ ratio, <S/N>,
computed in 5 \AA~ bins over the full spectral range, is shown in
column 8.
 
\subsection{Notes on individual spectra}
\label{subsec:indiv}
 
The 70 spectra of the 68 \Iae~are presented individually in Fig.
\ref{fig:Spec05D1dx_1013} - \ref{fig:Spec07D4ei_1725}.  In the
left panel, the extracted spectrum is shown (in gray). It is not
corrected for telluric features.  The SALT2 best-fit model is
overlayed, either with no recalibration (red dashed line) or once
recalibration is applied (red solid line).  The best-fit host-galaxy template
is also plotted (blue solid line). The best-fit host-galaxy model, redshift
and spectrum phase are noted in the figure captions.  In the right
panel, the host-galaxy subtracted spectrum, that is the spectrum minus the blue
solid line of the left panel, is shown (in gray) with the SN model
overlayed as a red solid line.  The latter SN model corresponds to the
full model (red solid curve) minus the host-galaxy model (blue solid curve).
When the best-fit is obtained for a model with no galaxy component
(the NoGalaxy cases), only the extracted spectrum (in gray) and the
best-fit model without (red dashed curve) and with (red solid line)
recalibration are presented.

In almost 90\% of cases, the spectrum is well reproduced by SALT2.
However, we find 8 spectra in our certain \Ia~sample for which the fit
is not fully satisfying, at least in some wavelength regions.  They
mostly correspond to spectra highly contaminated by the host-galaxy signal
for which the host-galaxy subtraction is difficult: SN~06D1dc
($f_{gal}$=77\%), SN~06D1fx (70\%), SN~06D1jz (73\%), SN~06D2hu
(71\%), SN~07D1ad (69\%) and SN~07D4dq (78\%). Two spectra (SN~06D1dl
and SN~07D1ab) show a sharp drop in their flux in the reddest part,
hinting to a potential calibration problem beyond 9000 \AA~(observer
frame). Moreover, in the case of SN~06D1dl, one notes quite narrow features similar to those
found in SN~1991bg-like supernovae, as well as a high velocity
\ion{Ca}{ii} absorption. However, both its stretch and
color are typical of a normal \Ia. More generally, all these spectra
have been unambigusouly identified as \Ia.

Some spectra need quite a strong recalibration.  This is the case for
SN~07D2ct (a very distant \Ia$\star$ at $z=0.94$), or for one spectrum
of SN~06D1eb (a \Ia~at $z=0.704$).  In this latter case, the phase is
more than $5$ days before maximum light, and the strong recalibration
could be at least partly explained by the fact that the SALT2 model is
not as well constrained for early \Iae~spectra because of the paucity
of training data.  In the case of SN~07D2fz, a \Ia~at $z=0.743$ that
needs substantial recalibration, the SALT2 color ($c=-0.15$) is bluer
than the average SNe~Ia color at maximum light ($<c>^{max}=0$). The
spectrum is on the contrary quite red for a \Ia~slightly before
maximum. This mismatch of the spectrum and lightcurve colors explains
the level of recalibration incorporated in the SALT2 fit and is at
least partly due to the lack of constraining lightcurve measurements.

\subsection{Catalog of non \Iae~MOS objects}
\label{subsec:otherZ}

The observations with FORS1 and FORS2 allowed us to target up to 19
objects simultaneously. One of the slits was always placed on the
active supernova. The other 18 were placed on a variety of targets,
which included the host-galaxy galaxies of SNLS supernovae that had faded
from view, variable sources, and randomly selected field galaxies. We
list the redshifts of all these targets in Table \ref{table:MOS}. We
use the ID column to distinguish between host-galaxy galaxies that were
targeted after the transient had faded from view, live transients
(using the labels SNIbc, SNII, SNII? or ?), and random field galaxies.
Most redshifts are derived from two or more clearly identifiable
spectral features. Redshifts based on a single feature, usually
[\ion{O}{ii}], are marked with an asterisk.

Also listed are the supernovae that were not \Iae. These supernovae
were predominantely Type II SNe. Interestingly enough, two of these
supernovae, SN 06D4eu and SN 07D2bv were subsequently identified as
superluminous supernovae in \citet{Howell13}. The redshift of
SN~06D4eu is derived from the host-galaxy and was measured with X-Shooter
\citep{Howell13}, whereas the redshift of SN~07D2bv was measured from
the FORS spectrum.

About a sixth of the SNLS five year SN sample was observed with the MOS
mode of FORS1 and FORS2, leading us to speculate that other
superluminous supernovae were observed spectroscopically with Gemini
and Keck during the five years of the SNLS. Indeed, \citet{Prajs16} recently found a superluminous candidate observed at Keck.

\section{Average properties of the VLT \Iae~samples}
\label{sec:propri}

In this section, we characterize the average properties of the \Ia~
and \Ia$\star$ subsamples that result from the classification decribed
in Sect. \ref{sec:results}. We compare those to the properties of the
same subsamples of the three year analysis \citep{Balland09}. The new \Ia~
and \Ia$\star$ subsamples are fully independent from those of the
three year (no supernova in common and different extraction procedures)
and we expect their average properties to be similar to those of their
three year counterparts. Studying the properties of the new subsamples
thus provides a good way to check that the conclusions drawn from the
three year samples are not biased.

\subsection{Spectro-photometric properties of the new VLT sample}
\label{subsec:VLT5}

The average photometric and spectroscopic parameters of the \Ia~and
\Ia$\star$ new subsamples are shown in Table
\ref{table:VLT5_propri_IaIa?}.  We also compute the average properties
of the combined \Ia~and \Ia$\star$ samples. For each parameter of
interest, we indicate the mean value and associated error.  We further
indicate in parentheses the $1\sigma$ dispersion around the mean
values.

\begin{figure}
	\resizebox{\hsize}{!}{\includegraphics{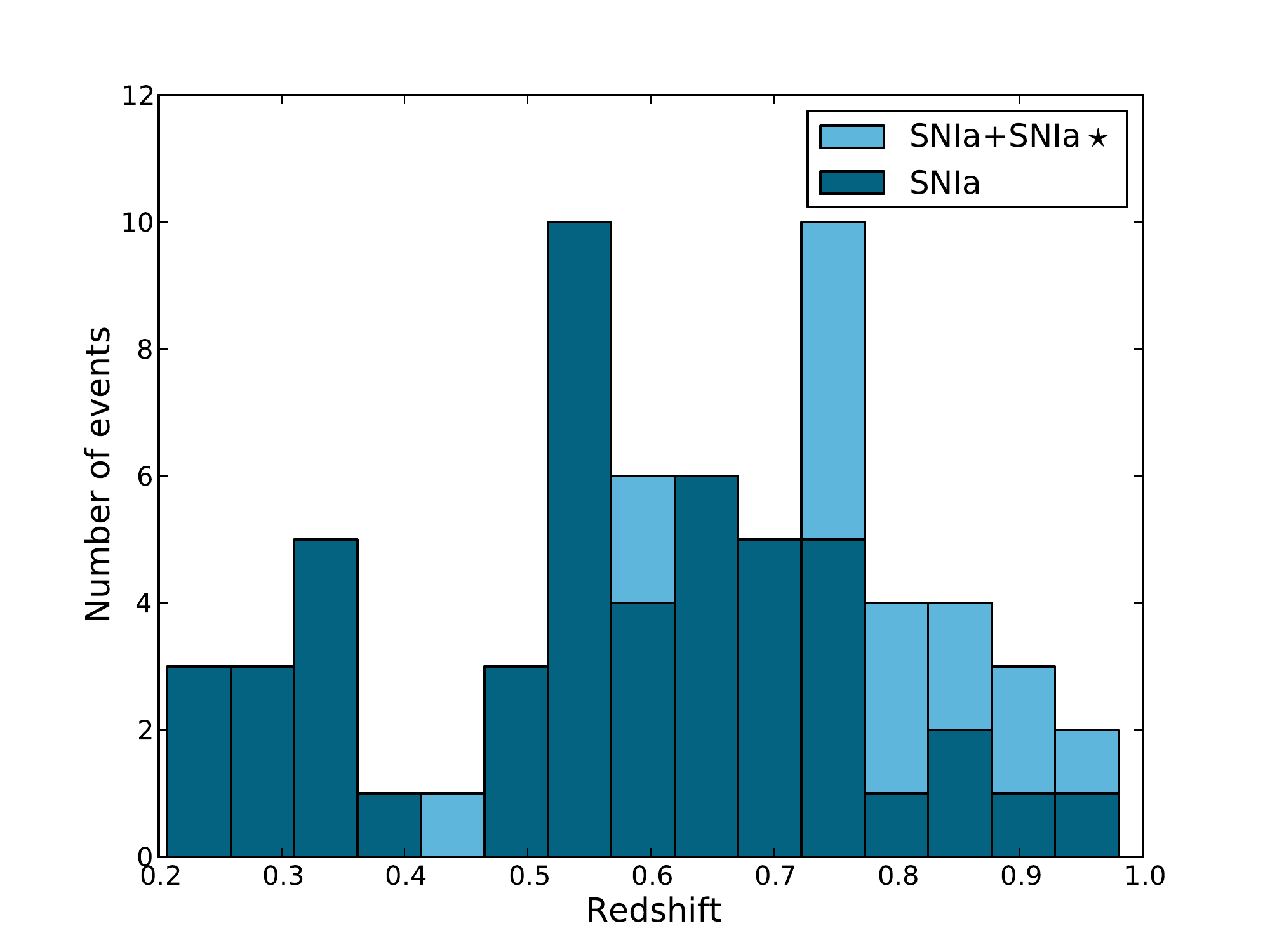}}
		\caption{Redshift distribution of the VLT \Iae~from
                  the last two years of SNLS. \Ia~are shown in dark
                  blue, \Ia~+ \Ia$\star$ in light blue. The mean
                  redshift is $\langle z \rangle_{SNIa} =0.57 \pm
                  0.03$ for the 51 \Ia~and $\langle z
                  \rangle_{SNIa\star} =0.77 \pm 0.03$ for the 16
                  \Ia$\star$}
		\label{fig:redshift}
\end{figure}

\begin{figure}
	\resizebox{\hsize}{!}{\includegraphics{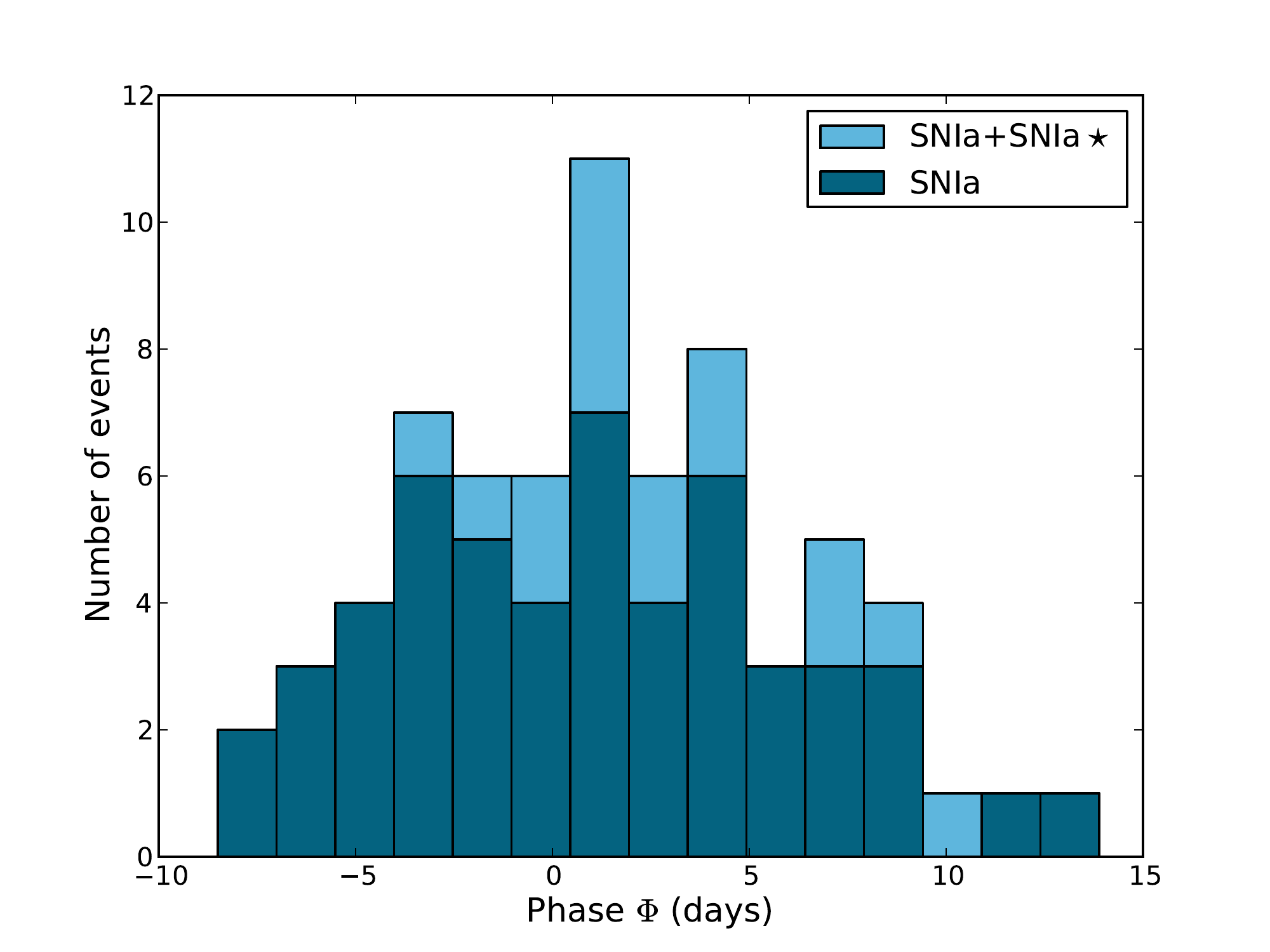}}
		\caption{Phase distribution of the VLT \Iae~from the
                  last two years of SNLS. \Ia~are shown in dark blue,
                  \Ia~+ \Ia$\star$ in light blue. The mean phase is
                  $\langle \Phi \rangle_{SNIa} = 1.0 \pm 0.7$ days for
                  the 51 \Ia~spectra and $\langle \Phi
                  \rangle_{SNIa\star} =2.8 \pm 0.9$ days for the 16
                  \Ia$\star$ spectra}
		\label{fig:phase}
\end{figure}

Figures \ref{fig:redshift} and \ref{fig:phase} present the redshift,
and phase distributions of the two samples.  The average redshift (row
1 of Table \ref{table:VLT5_propri_IaIa?} and Fig. \ref{fig:redshift})
of the \Ia$\star$ subsample ($\langle z \rangle_{SNIa\star} =0.77 \pm
0.03$) is significantly higher than the one of the \Ia~ subsample
($\langle z \rangle_{SNIa} =0.57 \pm 0.03$). This is expected as,
given our observational strategy, spectra of distant supernovae are
noisier and thus more likely to be classified as \Ia$\star$. Indeed,
as can be seen from row 4 of Table \ref{table:VLT5_propri_IaIa?},
\Ia~spectra have a higher $S/N$ per 5 \AA~bin on average ($\langle S/N
\rangle_{SNIa} = 5.5 \pm 0.9$) than \Ia$\star$ spectra ($\langle S/N
\rangle_{SNIa\star} = 1.6\pm 0.2$).

Another important parameter is the spectrum phase.  As explained in
Sect.  \ref{subsec:ID}, spectra observed more than 5 days past
maximum are preferentially classified as a \Ia$\star$ because
confusion with SN~Ib/Ic is possible (although unlikely when lightcurve
information is taken into account along with the spectrum, see end of
Sect. \ref{subsec:ID}).  Row 2 of Table \ref{table:VLT5_propri_IaIa?} and
Fig. \ref{fig:phase} illustrate this tendency : \Ia~ spectra are
marginally at a lower phase ($\langle \Phi \rangle_{SNIa} = 1.0 \pm
0.7$ days) than \Ia$\star$ spectra ($\langle \Phi \rangle_{SNIa\star}
=2.8 \pm 0.9$ days). In particular, we note that there are no \Ia$\star$ with phases lower than -5 days.

In the 5th row of Table~\ref{table:VLT5_propri_IaIa?}, we give the
average values of the $\gamma_1$ recalibration parameter.  $\gamma_1$
is marginally higher for \Ia$\star$ than for \Ia~($\langle \gamma_1
\rangle_{SNIa\star} = 0.66 \pm 0.39$ vs $\langle \gamma_1
\rangle_{SNIa} = 0.42 \pm 0.12$).

As expected, the average host-galaxy fraction (third row of Table
\ref{table:VLT5_propri_IaIa?})  in the best-fit model for \Ia$\star$
spectra is higher ($\langle f_{gal} \rangle_{SNIa\star} = 0.61 \pm
0.04$) than for \Ia~spectra ($\langle f_{gal} \rangle_{SNIa\star} =
0.39 \pm 0.03$).  We note that supernovae with a high host-galaxy
contamination are often high redshift objects, as the reduction in the
galaxy angular size of objects relative to the seeing with redshift
makes it difficult to extract the supernova candidate separately from
the host-galaxy.

The apparent restframe B-band magnitude $m_{B}^{*}$ is given in row 6
of Table \ref{table:VLT5_propri_IaIa?}. Candidates classified as
\Ia$\star$ appear, on average, fainter ($\langle m_{B}^{*}
\rangle_{SNIa\star} = 24.217 \pm 0.086$) than \Ia~($\langle m_{B}^{*}
\rangle_{SNIa} = 23.285 \pm 0.127$), as they are more distant on
average than the \Ia.

We compute a distance corrected apparent magnitude as
$m_{B}^{*\;c} = m_{B}^{*} - 5 log(d_{L}H_{0}c^{-1})$ where $d_{L}$ is
the luminosity distance which depends on redshift and on cosmology
through the parameter set $\{H_{0}, \Omega_{M},
\Omega_{\Lambda}\}$. We adopt the parameter values $\{70,0.27, 0.73\}$
\citep{Conley11}. We find a similar averaged value of $m_{B}^{*\;c}$
for the two subsamples : $\langle m_{B}^{*\;c} \rangle_{SNIa} = 23.965
\pm 0.045$ and $\langle m_{B}^{*\;c} \rangle_{SNIa\star} = 23.981 \pm
0.089$. This is somewhat surprising as we expect intrinsically fainter
objects to be classified as \Ia$\star$ rather than \Ia, and it was
indeed the case in the three year analysis. It is possible that, given the
relatively small number of spectra entering the \Ia$\star$ subsample
(16 objects), and given that other competing effects (later phase or
heavily host-galaxy contaminated spectra) are involved, this tendency is
subdominant in our new sample.

\begin{figure}
	\resizebox{\hsize}{!}{\includegraphics{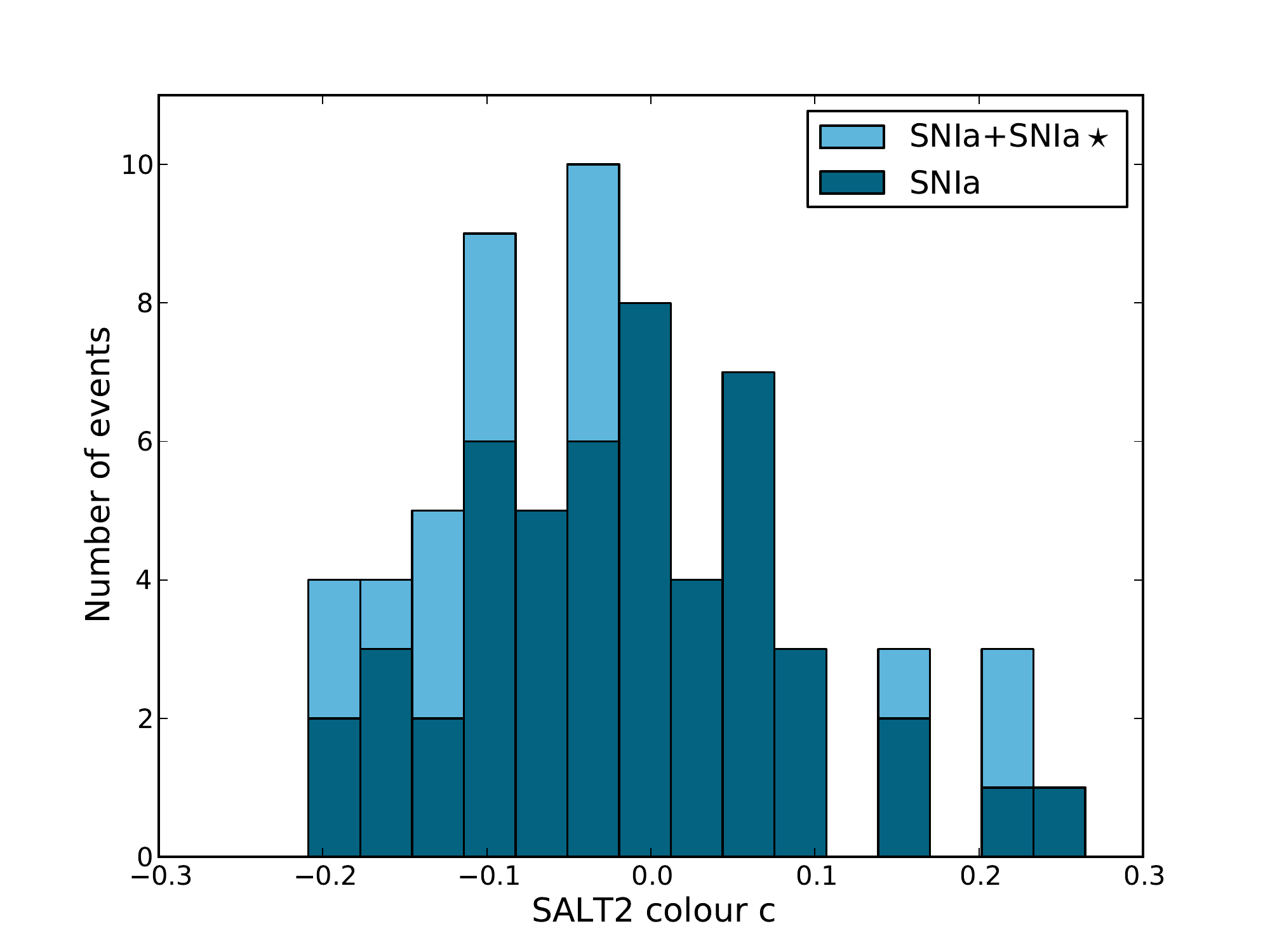}}
		\caption{SALT2 color distribution of the VLT \Iae~from
                  the last two years of SNLS. \Ia~are shown in dark
                  blue, \Ia~+ \Ia$\star$ in light blue. The mean
                  color is $\langle c \rangle_{SNIa} =-0.016 \pm
                  0.014$ for the 51 \Ia~and $\langle c
                  \rangle_{SNIa\star} =-0.045 \pm 0.033$ for the 16
                  \Ia$\star$}
		\label{fig:color}
\end{figure}

\begin{figure}
	\resizebox{\hsize}{!}{\includegraphics{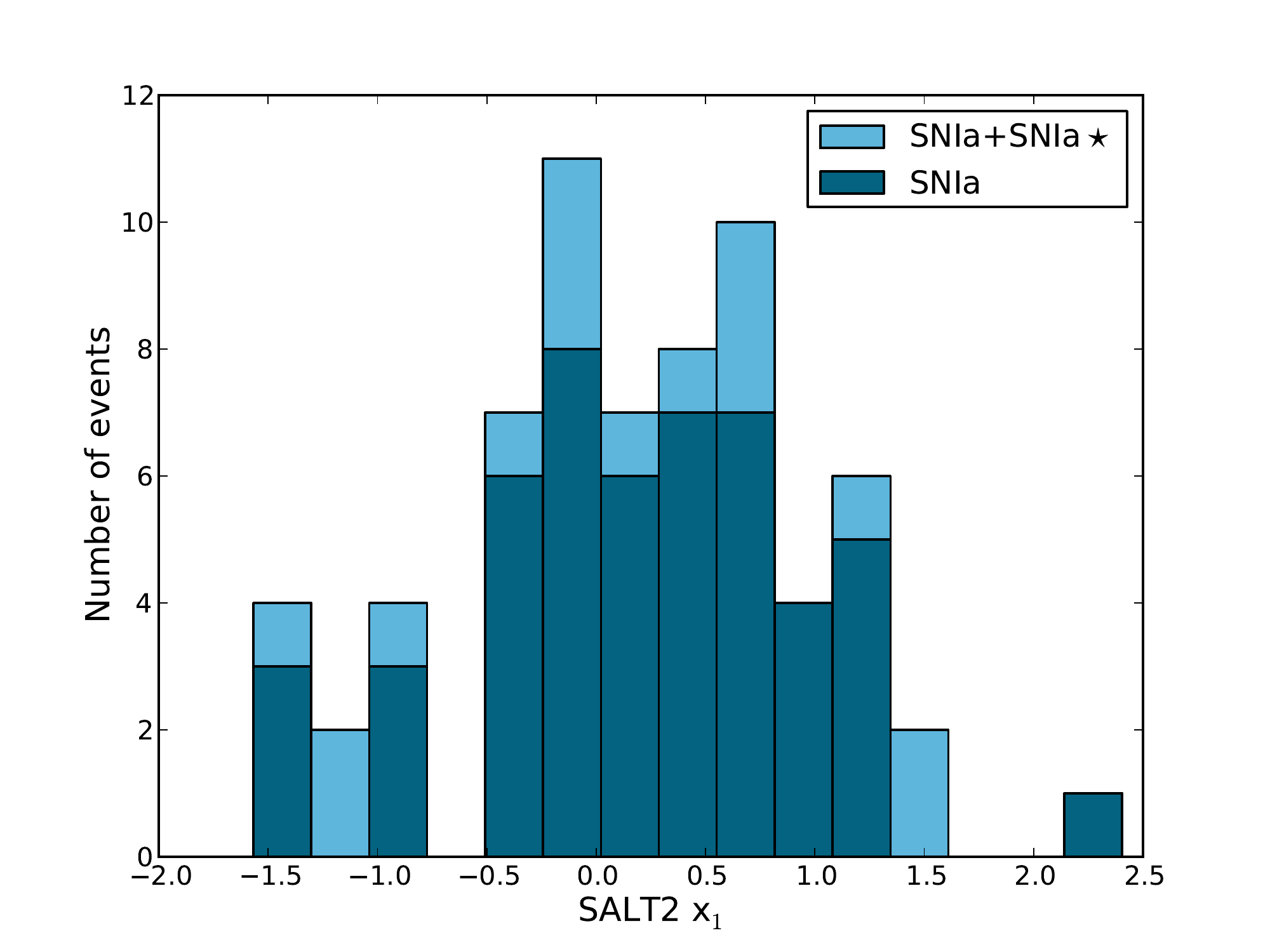}}
		\caption{SALT2 $x_{1}$ distribution of the VLT
                  \Iae~from the last two years of SNLS. \Ia~are shown
                  in dark blue, \Ia~+ \Ia$\star$ in light blue. The
                  mean $x_{1}$ is $\langle x_{1} \rangle_{SNIa} =0.215
                  \pm 0.109$ for the 51 \Ia~and $\langle x_{1}
                  \rangle_{SNIa\star} =0.065 \pm 0.226$ for the 16
                  \Ia$\star$}
		\label{fig:x1}
\end{figure}

The SALT2 color and $x_{1}$ distributions are presented in Figs.
\ref{fig:color} and \ref{fig:x1}.  Typically, a standard
\Ia/\Ia$\star$ has its SALT2 color in the range $-0.2\leq c \leq 0.2$,
and $x_1$ in the range -2 to 2, corresponding to a stretch factor $0.8
\leq s \leq 1.2$.  All our supernovae fall into these ranges, except
seven: SN~06D1kg, SN~07D1bs, SN~07D2bi, and SN~07D4cy, which are
redder ($c>0.2$), SN~07D4ed, which, on the contrary, is bluer
($c<-0.2$), and SN~07D1ah and SN~07D1by, which have high $x_1>2$
values. We discuss each of these SNe Ia in more detail.

\begin{itemize}
\item {\bf SN~06D1kg} (Fig. \ref{fig:Spec06D1kg_1477}): this is a \Ia~at $z=0.32$. Its spectrum at
  $\sim +6$ days is presented in Fig. \ref{fig:Spec06D1kg_1477}. Its
  color ($c=0.265$) is the largest of the sample, while its stretch is
  standard ($s=1.1$). The spectrum is redder than for a standard
  \Ia~at this phase. Explosion occured far from the center of its
  early-type host-galaxy and it is not clear whether its color is due to the
  presence of dust in the line-of-sight (no clear sign of host-galaxy
  \ion{Na}{D} absorption in the spectrum). \ion{Si}{ii} is clearly
  visible both around restframe 4100 \AA~ and 6150 \AA, and \ion{S}{ii}
  is also seen. There is no sign of peculiarity in the spectrum
  besides its reddening and it is hence classified as a \Ia.
\item {\bf SN~07D2bi} (Fig. \ref{fig:Spec07D2bi_1514}): this \Ia~ at $z=0.551$ is the second reddest
  object of our sample ($c=0.233$). Its lightcurve shape is standard
  ($s=0.99$). Its spectrum was taken slightly past maximum light and
  is redder than normal at this phase (with respect to color, it is
  more like a one-week past maximum spectrum than a spectrum at
  maximum light). host-galaxy \ion{Na}{D} absorption falls off the spectral
  range at this redshift and no sign of the presence of dust in the
  line-of-sight of this supernova can be seen.
\item {\bf SN~07D4cy} (Fig. \ref{fig:Spec07D4cy_1694}): as the two \Iae~above, this \Ia$\star$ at
  $z=0.456$ has a red color ($c=0.218$) and a normal stretch
  ($s=0.96$). The spectrum at maximum is quite noisy due to poor
  seeing conditions and the presence of thin cirrus. It is heavily
  host-galaxy contaminated, as the supernova exploded right at the center of
  its host-galaxy (more than 90\% host-galaxy subtraction is necessary to obtain the
  final SN spectrum). The subtracted spectrum is unusally red for a
  Type Ia supernova at maximum light, but this effect could result
  from an imperfect host-galaxy subtraction.
\item {\bf SN~07D1bs} (Fig. \ref{fig:Spec07D1bs_1711}): this \Ia$\star$ at $z=0.617$ has a red color
  ($c=0.207$) and a normal stretch ($s=1.04$). Unlike the cases
  discussed above, its maximum light spectrum is quite standard and is
  not redder than usual at this phase. A high host-galaxy fraction has been
  subtracted as the supernova exploded in the vicinity of the host-galaxy
  center.
\item {\bf SN~07D4ed} (Fig. \ref{fig:Spec07D4ed_1731}): this \Ia~ at $z=0.52$ is slightly bluer than
  usual ($c=-0.209$) but has a normal stretch ($s=1.01$). No host-galaxy is
  visible on the reference image of this supernova. The host-galaxy free -2
  days spectrum shows clear \ion{Si}{ii} $\lambda 4130$. It is bluer
  than standard in the restframe UV. The SALT2 model in the absence of
  recalibration (red dashed curve of Fig. \ref{fig:Spec07D4ed_1731})
  is even bluer and recalibration is necessary to fit the spectrum.
\item {\bf SN~07D1ah} (Fig. \ref{fig:Spec07D1ah_1699}): the stretch and color of this \Ia~ at $z=0.342$
  are $s=1.18$ and $c=-0.03$ respectively. The spectrum has an
  unusually weak \ion{Si}{ii} $\lambda$ 6150 absorption for a \Ia~ at
  maximum light, as well as a pronounced \ion{Ca}{ii} absorption
  trough. Moreover, no \ion{Si}{ii} is seen aroung 4100 \AA. It is
  classified as a \Ia-pec (see Sect. \ref{subsec:ID}).
\item {\bf SN~07D1by} (Fig. \ref{fig:Spec07D1by_1715}): despite its slightly high stretch ($s=1.21$)
  and a mediocre SALT2 fit around restframe 4000 \AA, SN~07D1by has a
  normal color $c=0.07$ and its spectrum appears normal, with a
  clearly visible \ion{Si}{ii} $\lambda 4130$ absorption.
\end{itemize}

We have run the SNID package \citep{Blondin07,Blondin11} on all the
above potentially suspect spectra. The best-fits are always obtained
for a normal \Ia~template, except for SN~07D1ah for which the best-fit
is for the overluminous SN~1991T, confirming our identification.

The mean values of the SALT2 color and $x_1$ parameters are given
respectively in rows 8 and 9 of Table \ref{table:VLT5_propri_IaIa?}.
On average, we find no obvious difference in \Ia~colors ($\langle c
\rangle_{SNIa} =-0.016 \pm 0.014$) with respect to \Ia$\star$ colors
($\langle c \rangle_{SNIa\star} =-0.045 \pm 0.033$). The same
conclusion applies for the $x_1$ parameter (or, equivalently, for the
stretch $s$ computed in row 10 from the formula given in
\citealt{Guy07}) : for \Ia, $\langle x_{1} \rangle_{SNIa} =0.215 \pm
0.109$, while for \Ia$\star$ $\langle x_{1} \rangle_{SNIa\star} =0.065
\pm 0.226$ (again, quoted errors are errors on the mean), or $\langle
s \rangle_{SNIa} =1.001 \pm 0.010$ for \Ia~ and $\langle s
\rangle_{SNIa\star} =0.988 \pm 0.020$ for \Ia$\star$.

Finally, we compute the average properties of the NoGalaxy \Iae~(see
Sect. \ref{subsec:zID}), which we find very similar on average to those of
the full sample. The dispersions of their color and stretch around the
mean are also similar to those of the full sample.

We conclude from this study that the two subsamples have very similar
photometric properties, on average, which suggests that the \Ia$\star$
sample is not contaminated by non-Ia objects with respect to the
\Ia~sample.

\subsection{Assessing the quality of host-galaxy subtraction}
\label{subsec:gal}

As it impacts the quality of the recovered supernova spectrum and its
subsequent identification, host-galaxy subtraction is a challenging step in
the spectral reduction. As explained Sect. \ref{subsec:ID} above, we use
the SALT2 model in order to separate the host-galaxy signal from the
supernova. In this section we {\it a posteriori} assess the quality of
the subtraction by comparing a composite host-galaxy subtracted spectrum
built from supernovae with high contamination to the one made of low
contamination supernovae.

We use only \Ia~spectra (excluding \Ias) and exclude the 8 spectra
discussed in Sect.~\ref{subsec:indiv}. A color cut is applied
($-0.2<c<0.2$) and the phase is chosen in the range $-4 < \Phi < 4$
restframe days to select spectra around maximum light.

\begin{figure}[!ht]\centering
\includegraphics[width=0.8\textwidth]{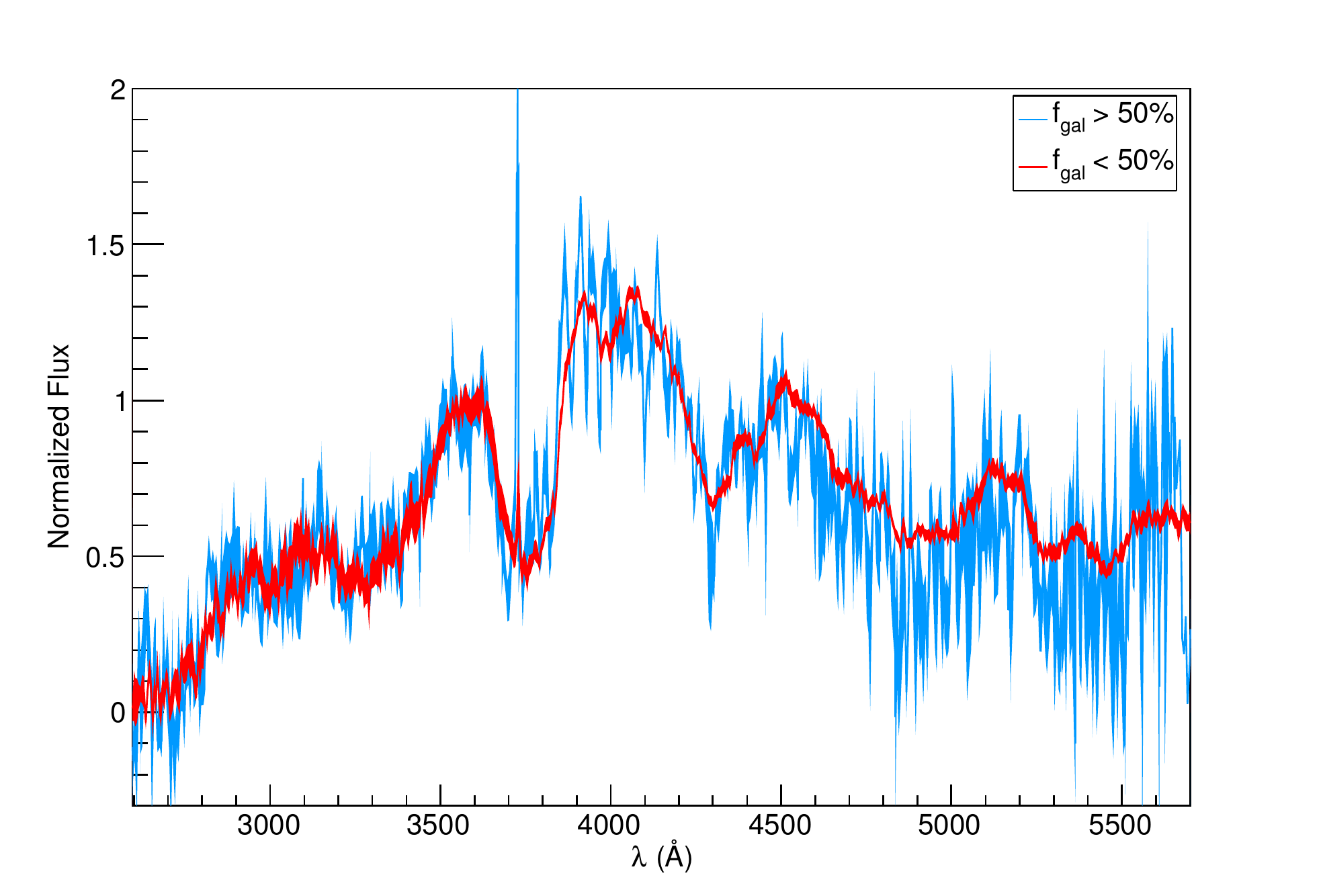}
\caption{Comparison of the composite spectra at maximum light built
  from the VLT \Iae~ of the last two years of SNLS as a function of
  host-galaxy contamination : $f_{gal}<50\%$ (in red) et $f_{gal}\geq50\%$
  (in blue). A $\pm 1\sigma$ range is plotted.}
\label{fig:gal}
\end{figure}

We split the resulting set into two subsamples, based on the value of
the $f_{gal}$ parameter. These contain 17 spectra with $f_{gal}<50\%$
and 7 with $f_{gal}>50\%$, respectively.  Spectra are put into the
restframe and rebinned to 5 \AA. The flux integral is fixed to unity
in the range 4000-4500 \AA~(restframe). The average weighted flux and
its dispersion are then computed in each wavelength bin and we obtain
one average spectrum for each subsample. These two spectra are
overlapped in Fig.~\ref{fig:gal}. A $\pm 1\sigma$ range around each
spectrum is shown.

The two spectra are very similar, showing that the host-galaxy subtraction is
not grossly incorrect for the bulk of our spectra. Residual host-galaxy lines
are visible as we do not attempt to subtract them (we do not model
host-galaxy emission lines in the galaxy model used in SALT2, see Sect.
\ref{subsec:ID}). These are stronger in the higher contaminated
sample, as expected. Absorption line residuals are also visible in the
3500 - 4500 \AA~region and might be due to a lack of spectral
resolution in the host-galaxy model. We note that the average signal-to-noise
ratio is poorer for the $f_{gal}>50\%$ spectrum, due to the small
number of spectra used for this subsample and to the systematic error associated with the subtraction of the host-galaxy signal. We note a slightly
depressed flux of the $f_{gal}>50\%$ spectrum with respect to the
$f_{gal}<50\%$ one beyond restframe 4700 \AA. This might hint toward
a slight tendency to oversubtract the host-galaxy signal at higher
wavelengths, but the poor $S/N$ prevents us from drawing a firm
conclusion. Given that performing a clean host-galaxy subtraction is
notoriously difficult, we conclude that our technique is not grossly
in error.

\subsection{Comparison with the VLT three year data set}
\label{subsec:VLT5vs3}

In this section, we compare the average properties of the new \Ia~+
\Ias~sample with the VLT three year \Ia~+ \Ias~sample.  Results are
presented in Table \ref{table:VLT5_propri_3+5}.  Values for the
redshift, phase, host-galaxy fraction and $S/N$ ratio of the three year sample
are directly taken from \citet{Balland09}, whereas the photometric
parameters (B-band magnitude with or without distance correction,
color, $x_{1}$ and stretch) are re-computed using the updated values
of \citet{Guy10}.

From inspection of Table \ref{table:VLT5_propri_3+5}, it appears that
the two VLT \Ia~+ \Ias~samples have similar photometric properties. It
is also true for the spectroscopic parameters, except for the phase
(row 2) and the host-galaxy fraction (row 3). The phase is indeed marginally
different: the new VLT spectra have been measured $\sim 1.5$ days
earlier, on average, than the VLT three year spectra. This might be due to the fact that, as experience built up along the course of the survey, \Ia~candidates were more efficiently selected in their early phase, allowing spectra to be observed closer to maximum light, on average, than during the first years of the survey. This effect was already noted in \citet{Balland09}.


The higher host-galaxy fraction of the new VLT sample
can be traced back to the change in observing mode, from LSS to MOS,
inducing a different extraction procedure for the two samples.  For
the three year sample, a photometric model of the host-galaxy was built and,
whenever possible, the two components (host-galaxy and SN) were extracted at
the same time\footnote{A separate extraction was performed provided
  the distance between the host-galaxy center and the SN center is larger
  than 2\arcsec \citep{Baumont08}}. For the new sample, the host-galaxy-SN
separation is performed during the identification procedure by adding
a host-galaxy component to the SN model.

The fact that the two independent samples (the three year and the final
two years samples) have similar spectroscopic and photometric
properties shows that they are not biased (or they are biased in the
same way) against the inclusion of peculiar objects or confusion with
other SN types.

\begin{figure}[!ht]\centering
\includegraphics[width=0.8\textwidth]{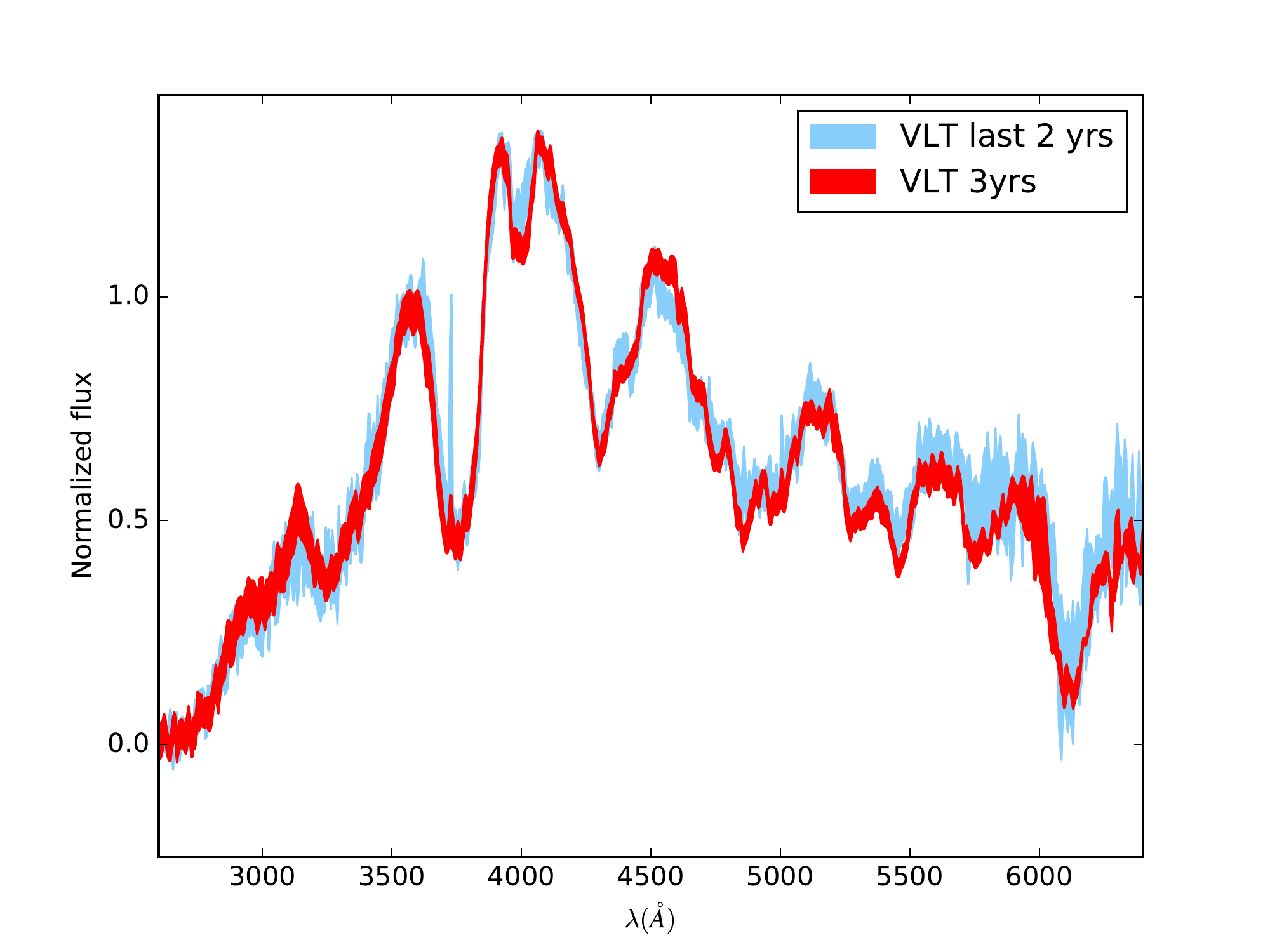}
\caption{Comparison of the composite spectra at maximum light of the
  VLT three year sample (red) and the last two years of SNLS VLT
  sample (blue).}
\label{fig:VLT3vs5}
\end{figure}

To illustrate this similarity, we use 47 VLT three year \Ia~spectra and 24
\Ia~spectra from our new sample around maximum light to build two
average spectra using the procedure described in Sect.
\ref{subsec:gal}. The two spectra are overlapped in
Fig.~\ref{fig:VLT3vs5}. They look remarkably similar over the studied
spectral range. We note a stronger [\ion{O}{ii}] emission in the mean
spectrum of the new sample, as a consequence of not attempting to
subtract host-galaxy lines in the present analysis. Some local differences
are found around restframe 4000 \AA~and 4600 \AA.  The most striking
difference resides in the depth of the absorption feature due to \ion{Si}{ii} $\lambda 5970$
(blueshifted to $5750$ \AA). Indeed, the VLT three year spectrum shows a
deeper absorption than the spectrum of the new sample. Given the
relatively small number of spectra in each subsample, such a
difference could be due to the erroneous inclusion of a non \Ia~object
in one of them. For example, one might argue that the inclusion of a
SN~1991T like supernova in the new sample might weaken the
\ion{Si}{ii} $\lambda 5970$ absorption, but it should also alter the
\ion{Ca}{ii} region, which is not seen. Inclusion of a SN~1999aa like
\Ia~ could smooth the secondary $\lambda 5970$ silicon absorption,
while preserving a strong \ion{Ca}{ii} absorption \citep{Garavini04},
but the \ion{Si}{ii} $\lambda 4100$ would also be affected. Besides,
as discussed above, special care has been taken to eliminate peculiar
events in our samples, and it is unlikely that they are polluted by
such an event.

We thus confirm that the VLT three year and the new samples are consistent
on average.  The two samples can then be combined to build the final
VLT five year spectral set, which contains 192 \Iae~for a total of 209
spectra. This final VLT spectroscopic sample will be merged with the
other \Iae~spectra of the SNLS (from Gemini and Keck telescopes) to
produce the final SNLS spectroscopic sample.  The SNLS five year
cosmological analysis will rely on this sample, once further
photometric cuts are made \citep{Betoule17}.

\section{Properties of SNLS \Iae~from composite spectra}
\label{sec:evo}

In this section, we use the final VLT spectral set described above to
examine how the spectra depend on color, stretch and host-galaxy mass, and to
search for evolution with redshift.  This analysis builds on previous
works \citep{Howell07, Ellis08, Foley08, Sullivan09, Balland09,
  Cooke11, Foley12, Maguire12, Milne15}. To build composite (average)
spectra, we select from the full VLT sample spectra of \Iae~which
belong to the \Ia~category (as opposed to the \Ias~category) with
phase ranging between -4 and +4 days.  This ensures that the analysis
benefits from the highest $S/N$ of the sample. We follow the method
outline in Sect. \ref{subsec:gal} to build the composite spectra. Both SNe with redshift estimated from host-galaxy features and from
template fitting are used.  We first study the effect of color
correction (Sect. \ref{subsec:color}), then split the sample in redshift
(Sect. \ref{subsec:evo}), stretch (Sect. \ref{subsec:stretch}) and host-galaxy mass
(Sect. \ref{subsec:mass}).


\subsection{Color correction}
\label{subsec:color}

We first divide the VLT final sample into two subsamples as a function
of color.  We select 40 spectra with $c<0$ and 31 with $c\geq0$.  The
spectro-photometric properties of these two subsamples are shown in
Table~\ref{table:color}.  The subsamples have similar properties on
average, except for color and distance corrected magnitude.

\begin{figure}[!ht]\centering
\includegraphics[width=0.6\textwidth]{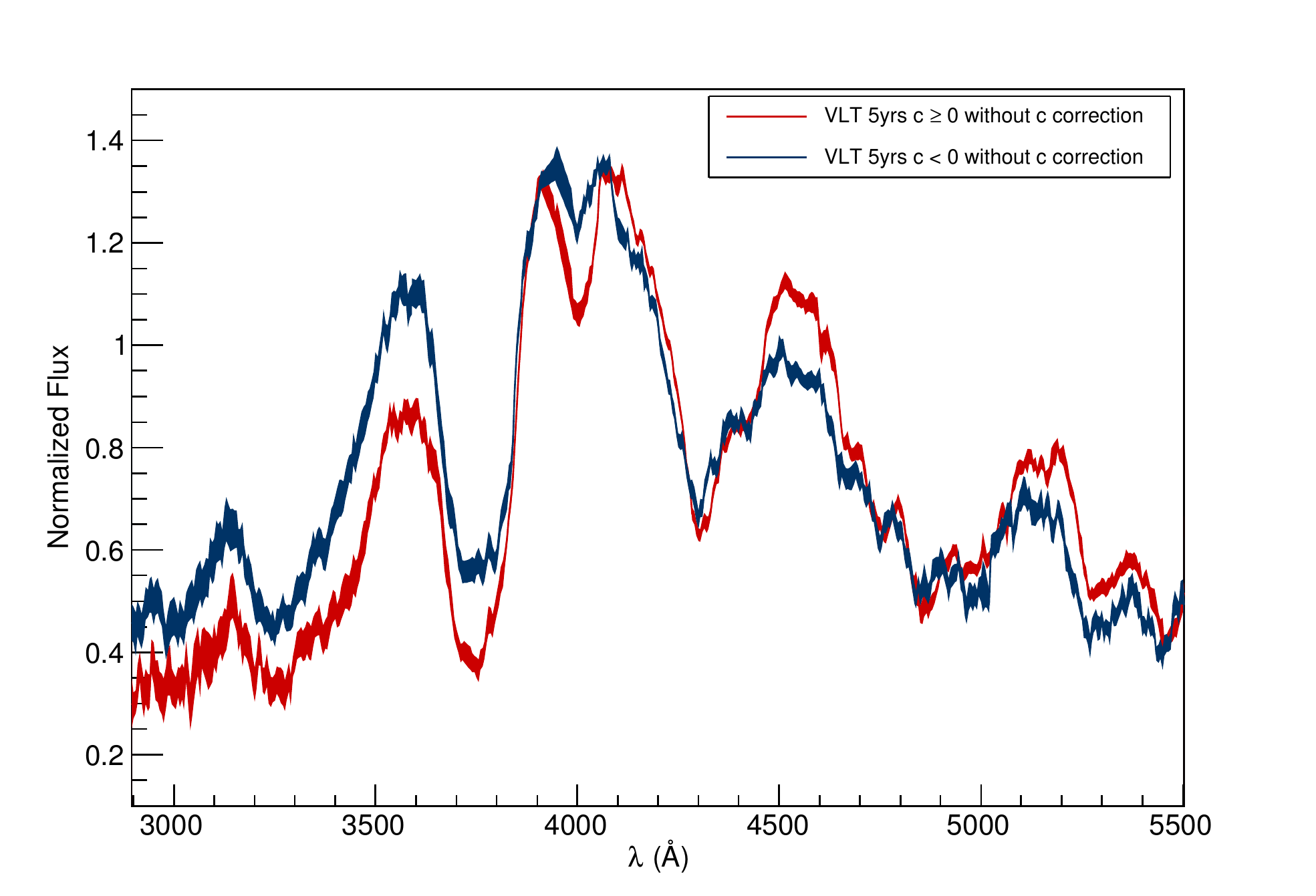}
\\
\includegraphics[width=0.6\textwidth]{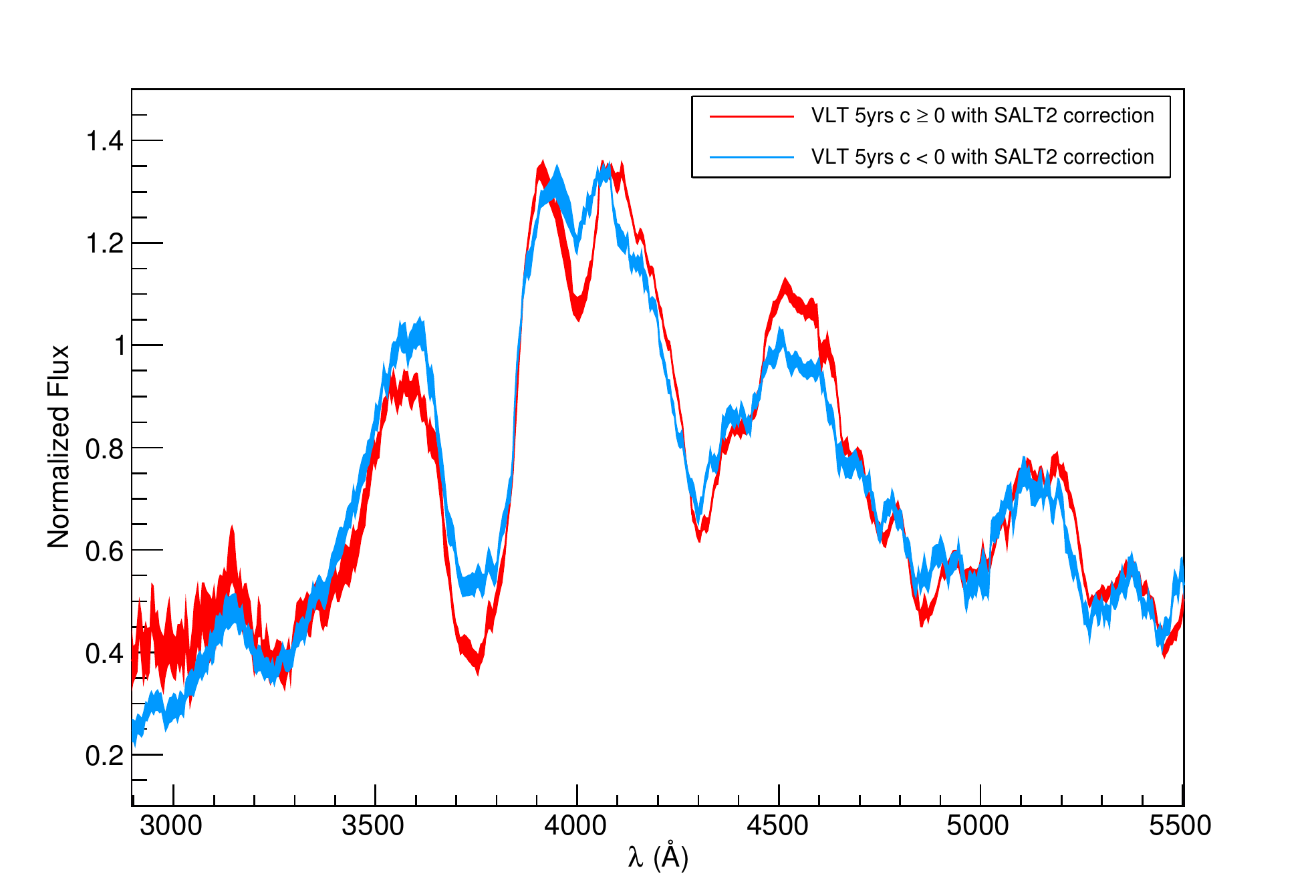}
\\
\includegraphics[width=0.6\textwidth]{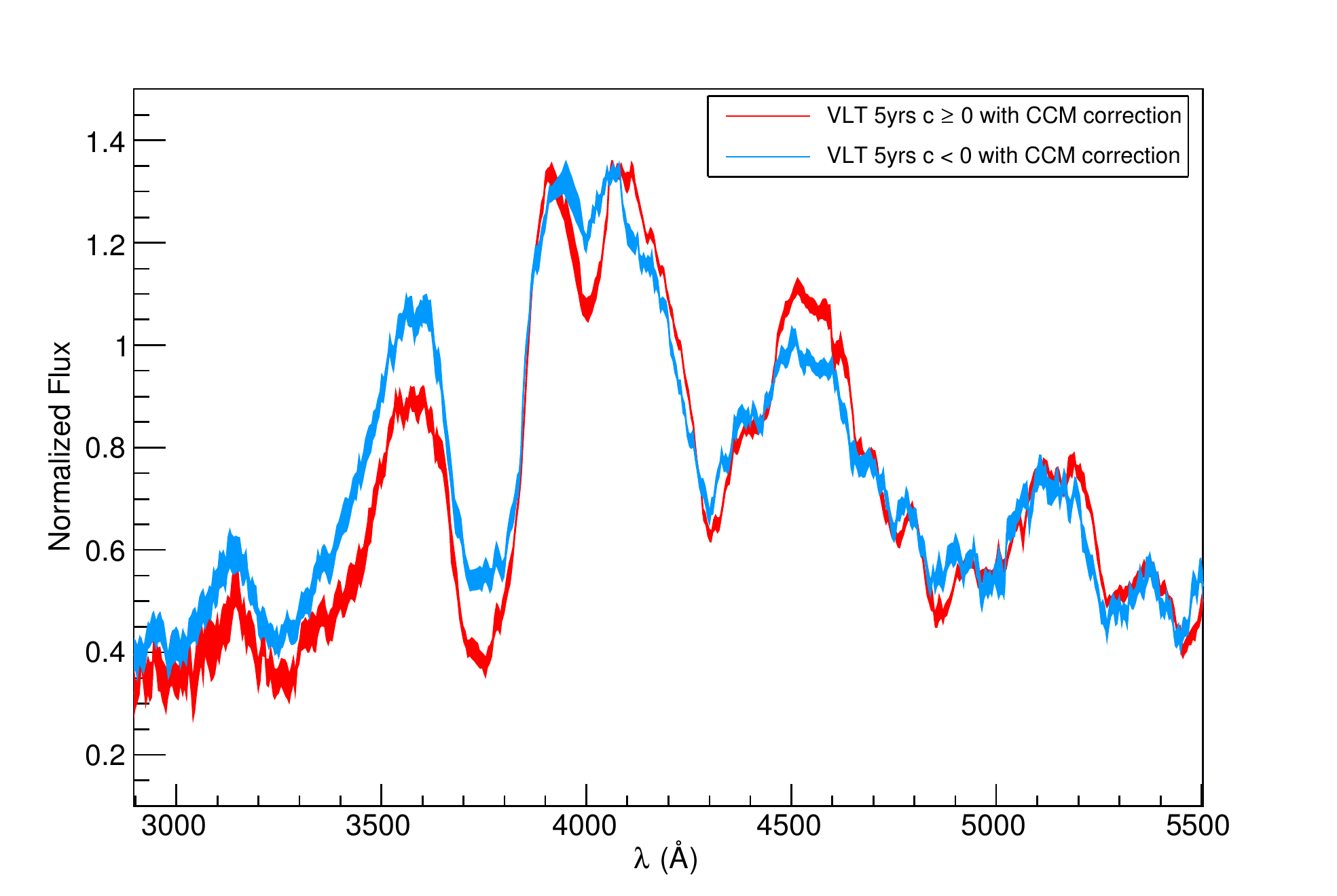}
\caption{Comparison of composite spectra from the full VLT five year
  sample split in color, without color correction (top panel), with
  color correction using the SALT2 color law (middle panel) and with
  color correction using the CCM law (bottom panel). In each panel,
  the blue spectrum is built from $c<0$ \Iae, the red spectrum from
  $c\geq0$ \Iae. No recalibration has been applied to individual
  spectra (see text for details). Residual host-galaxy lines have been
  removed. A $\pm 1\sigma$ range is plotted.}
\label{fig:corcolor}
\end{figure}

We compute a composite spectrum for each subsample before and after
color correction\footnote{In order to estimate the impact of the sole
  color correction, no recalibration is applied to the spectra at this
  stage.}.  When a color correction is applied, either the SALT2 color
law \citep{Guy10} or the CCM color law \citep{Cardelli89} are used.
host-galaxy residual lines ([\ion{O}{ii}] $\lambda\lambda 3727,3729$ and
\ion{Ca}{ii}~H\&K) are removed by interpolating the continuum under
the lines.  The SALT2 color law is derived during the training process
of the SALT2 model \citep{Guy10}. It not only captures the effects of
dust extinction by the host-galaxy interstellar medium, but also at least
partially the intrinsic color scatter among \Iae. In contrast, the CCM
law is a standard dust reddening correction with the total to
selective extinction parameter $R_V=3.1$ \citep{Cardelli89}. After
color correction using the SALT2 color law, the two average spectra
are overlapping nicely over the whole spectral range (see
Fig. \ref{fig:corcolor}), while using the CCM is less efficient,
namely below $\lambda\sim 3700$ \AA. This traces back to differences
in the laws in this spectral region \citep{Guy10}. Local differences
already existing before correction are still present (bluer and
brighter \Iae~have weaker absorption features due to IMEs), as the color correction is
required to be a smooth function of wavelength that cannot correct for
finer scale differences.

Similar comparisons of the effect of color laws on reddening
corrections of \Iae~spectra have been performed by various
authors. \citet{Ellis08}, using a sample of 36 \Iae~ spectra obtained
at Keck within the SNLS, observe the same trend as ours (see their
Fig. 8). Based on a set of HST spectra, \citet{Maguire12} also find
that the combination of the SALT2 law and SALT2 colors derived from
colors of the Sifto lightcurve fitter \citep{Conley08} is more efficient at color correcting
the spectra than the CCM color law with Sifto colors. However, the
matching of the color corrected spectra is better in the latter case,
while the SALT2 correction tends to overcorrect the spectra. We do not
see this trend in our analysis and we choose to color correct our
spectra with the SALT2 law in the following.

\subsection{Redshift evolution}
\label{subsec:evo}

\begin{figure}
	\resizebox{\hsize}{!}{\includegraphics{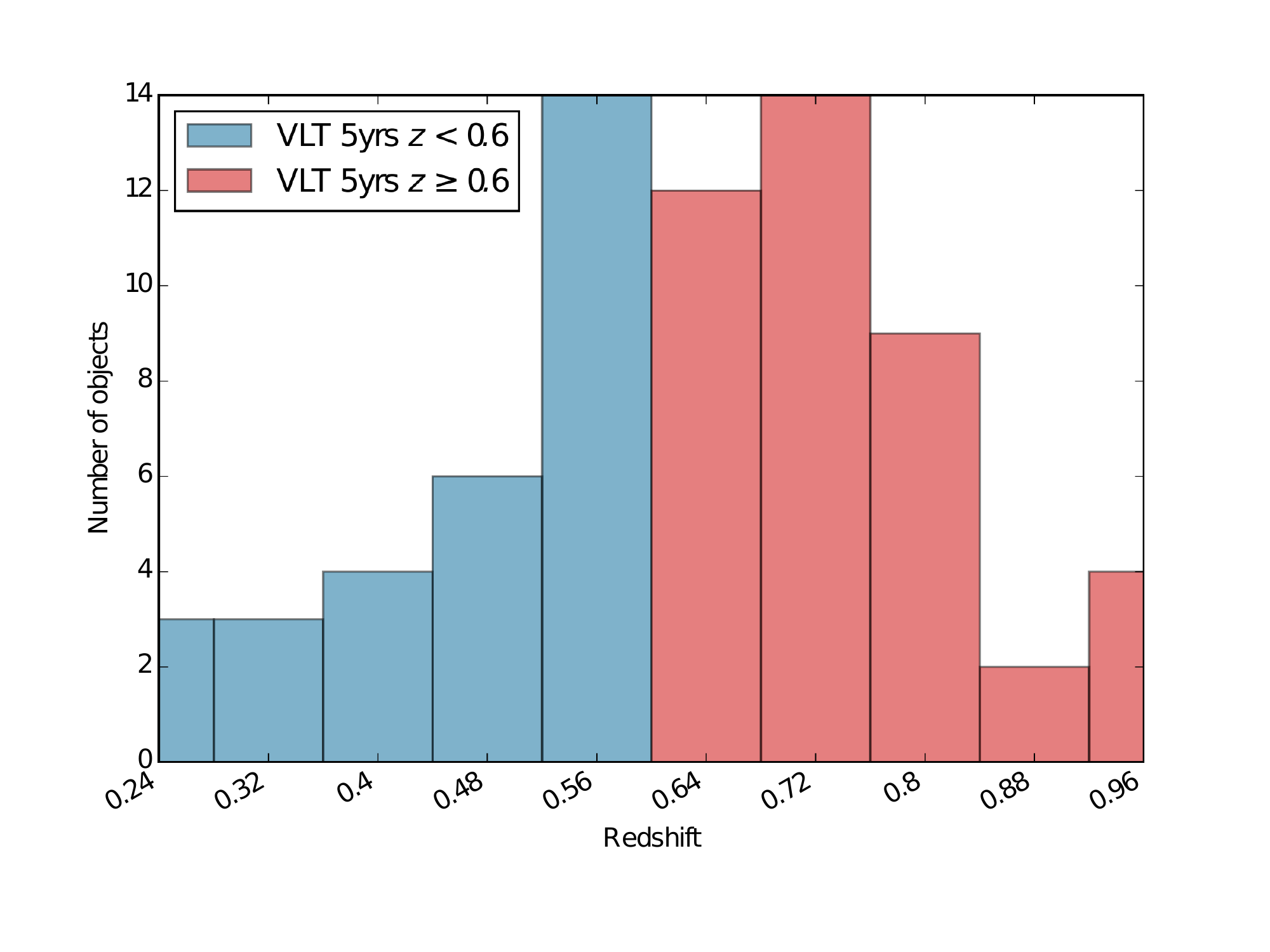}}
		\caption{Redshift distribution of low ($z<0.6$, in
                  blue) and high ($z\geq0.6$, in red) redshift
                  \Iae~subsamples used to build composite spectra. The
                  mean redshift is $\langle z \rangle_{z<0.6} = 0.47
                  \pm 0.02$ for the 30 low-$z$ \Iae~and $\langle z
                  \rangle_{z\geq0.6} =0.73 \pm 0.01$ for the 41
                  high-$z$ \Iae.}
		\label{fig:meanspec_zdistrib}
\end{figure}

\begin{figure}
	\resizebox{\hsize}{!}{\includegraphics{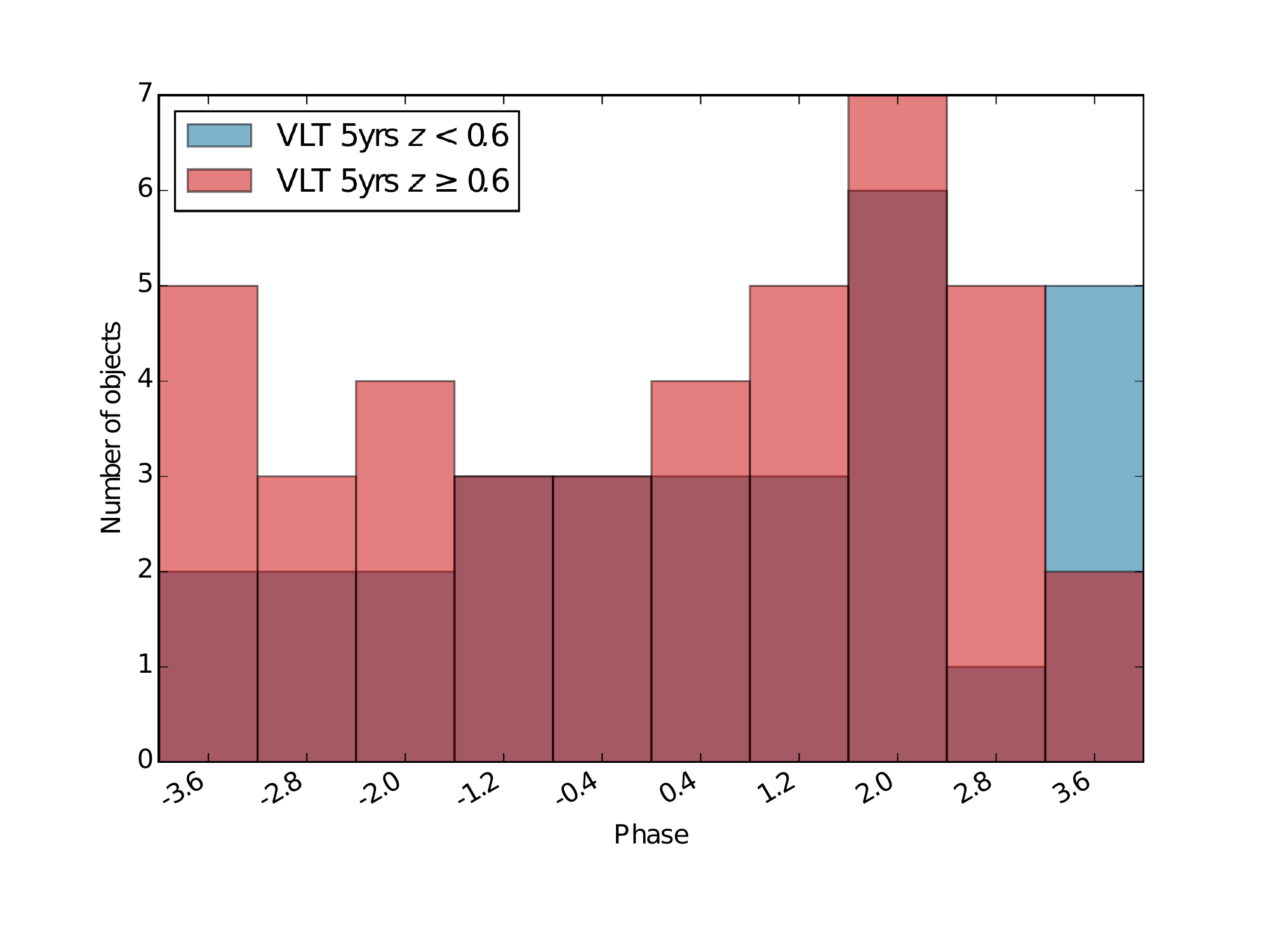}}
		\caption{Phase distribution of low ($z<0.6$, in blue)
                  and high ($z\geq0.6$, in red) redshift
                  \Iae~subsamples used to build composite spectra. The
                  mean phase is $\langle \Phi \rangle_{z<0.6} = 0.5
                  \pm 0.4$ days for the 30 low-$z$ \Iae~and $\langle
                  \Phi \rangle_{z\geq0.6} = 0.1 \pm 0.4$ days for the
                  41 high-$z$ \Iae.}
		\label{fig:meanspec_phasedistrib}
\end{figure}

\begin{figure}
	\resizebox{\hsize}{!}{\includegraphics{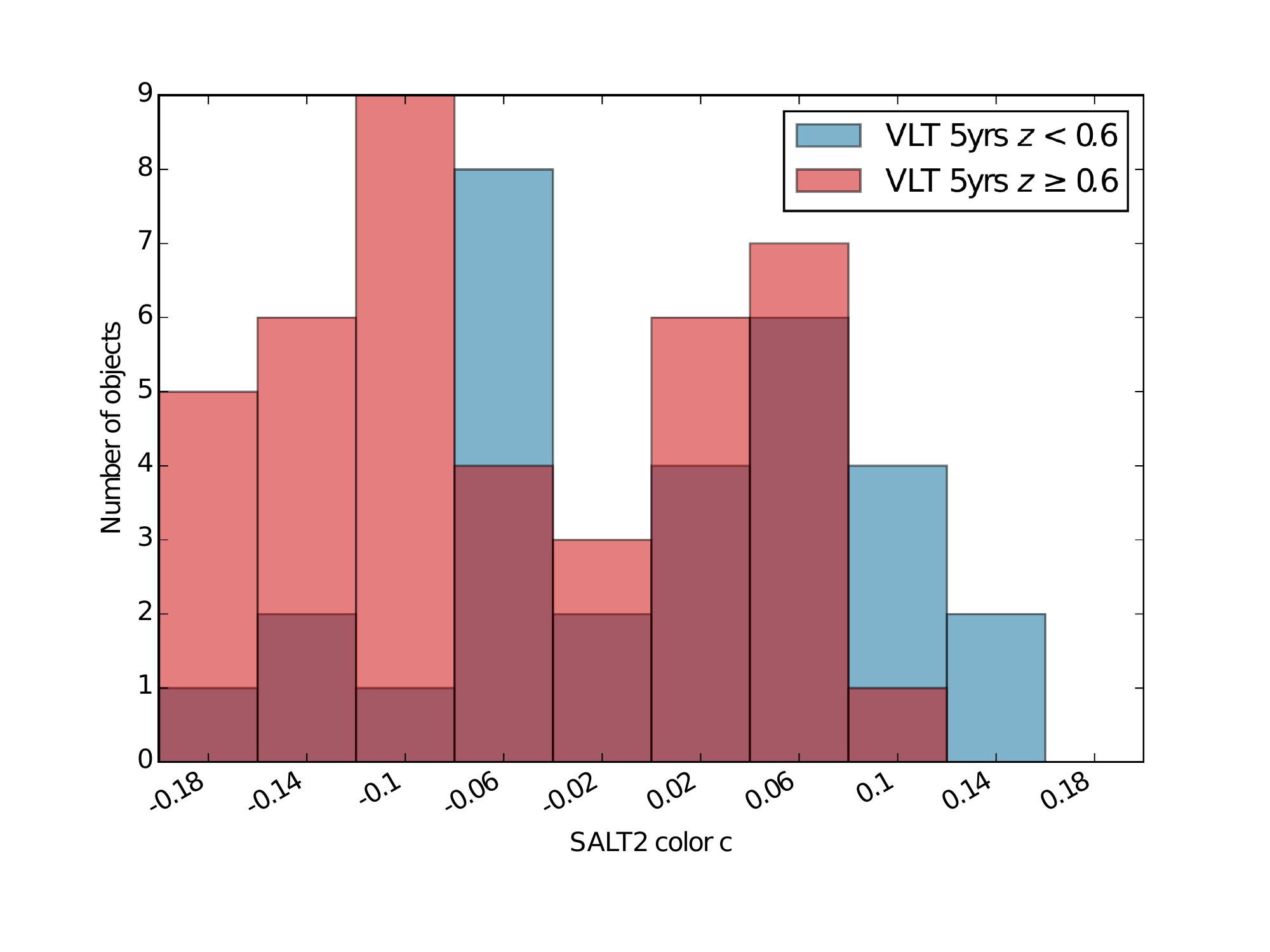}}
		\caption{SALT2 color distribution of low ($z<0.6$, in
                  blue) and high ($z\geq0.6$, in red) redshift
                  \Iae~subsamples used to build composite spectra. The
                  mean color is $\langle c \rangle_{z<0.6} = 0.001 \pm
                  0.015$ for the 30 low-$z$ \Iae~and $\langle c
                  \rangle_{z\geq0.6} =-0.055 \pm 0.014$ for the 41
                  high-$z$ \Iae.}
		\label{fig:meanspec_colordistrib}
\end{figure}

\begin{figure}
	\resizebox{\hsize}{!}{\includegraphics{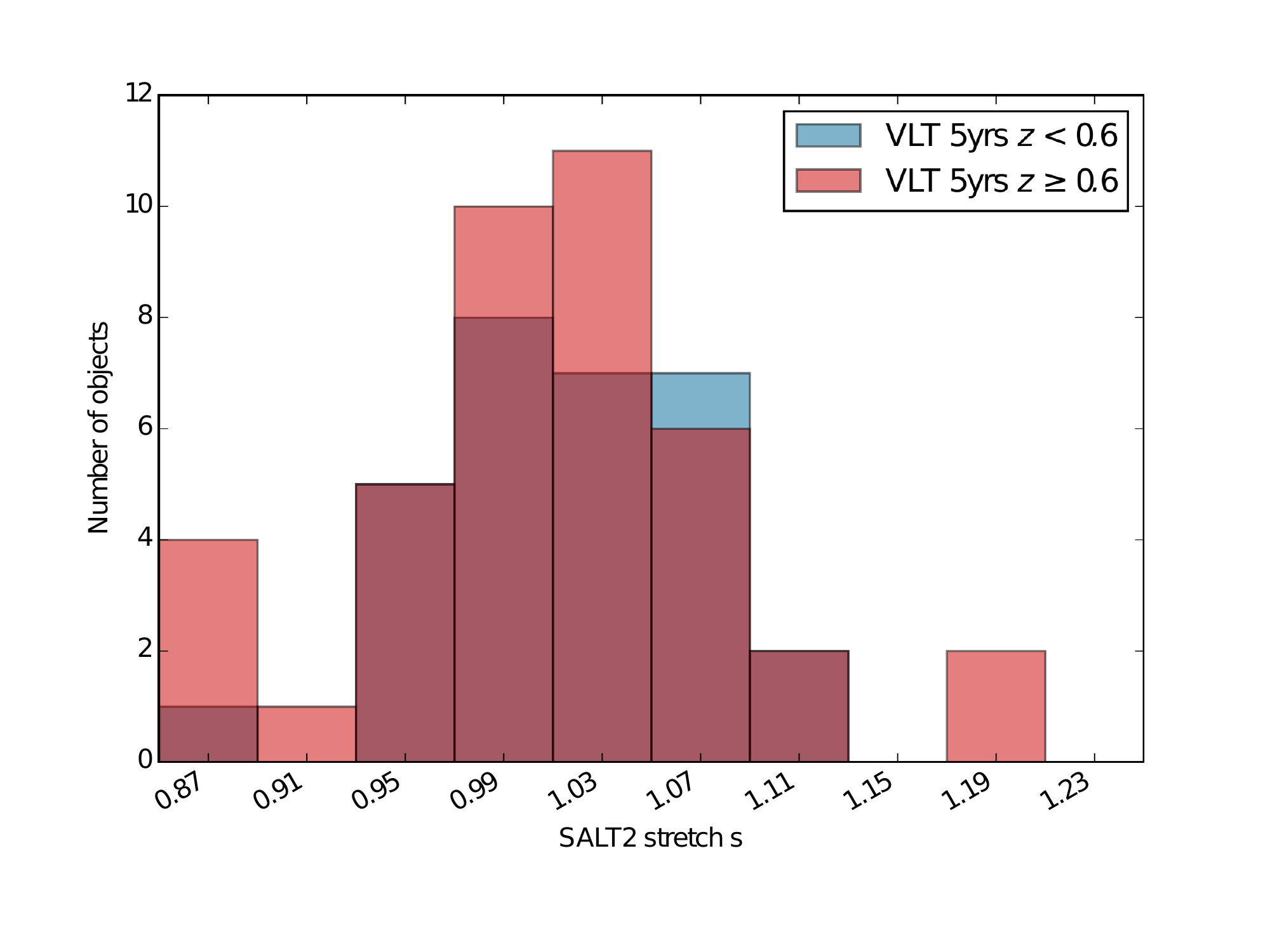}}
		\caption{SALT2 stretch distribution of low ($z<0.6$,
                  in blue) and high ($z\geq0.6$, in red) redshift
                  \Iae~subsamples used to build composite
                  spectra. The mean stretch is $\langle s
                  \rangle_{z<0.6} = 1.018 \pm 0.010$ for the 30
                  low-$z$ \Iae~and $\langle s \rangle_{z\geq0.6} =
                  1.010 \pm 0.011$ for the 41 high-$z$ \Iae}
		\label{fig:meanspec_stretchdistrib}
\end{figure}

\begin{figure}[!ht]\centering
\includegraphics[width=0.8\textwidth]{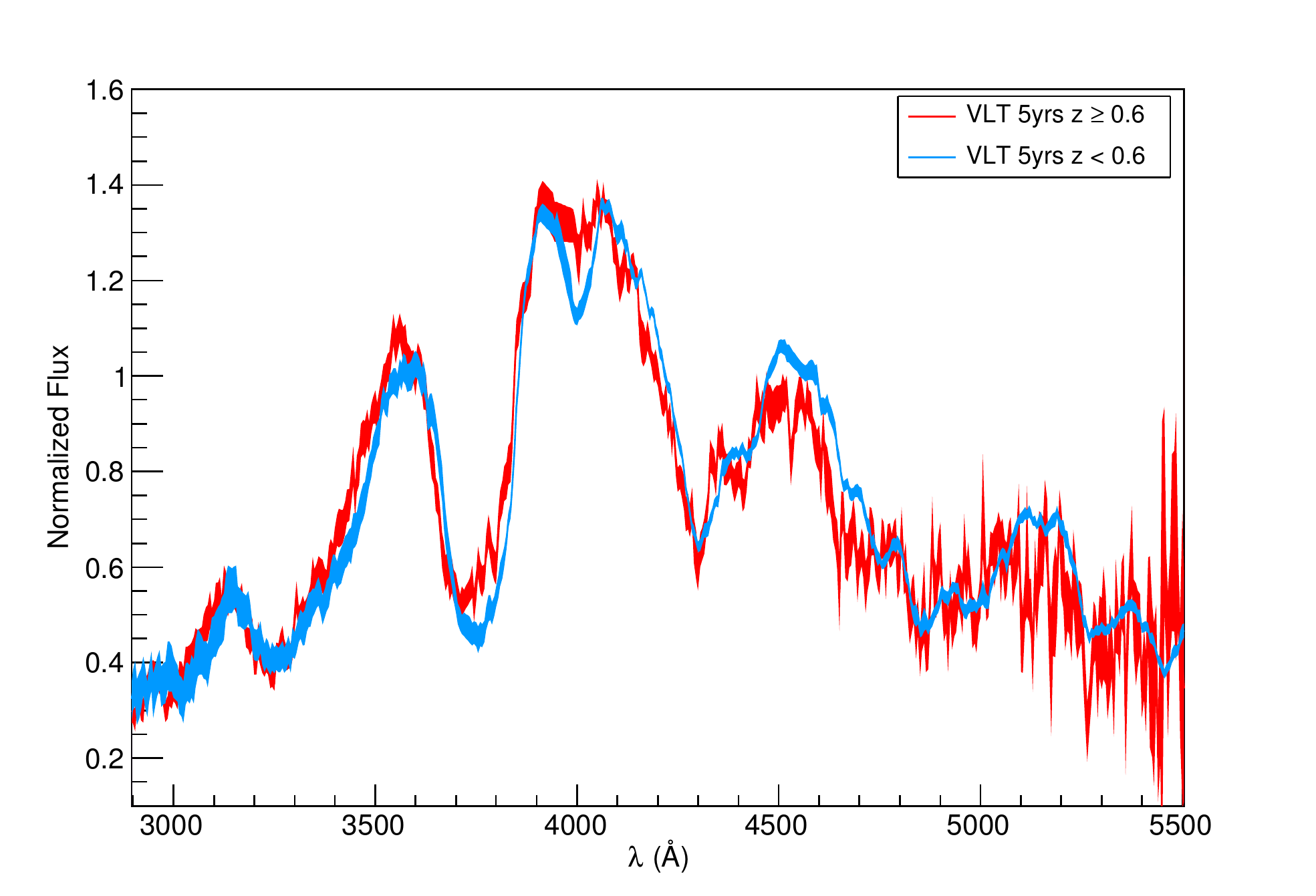}
\includegraphics[width=0.8\textwidth]{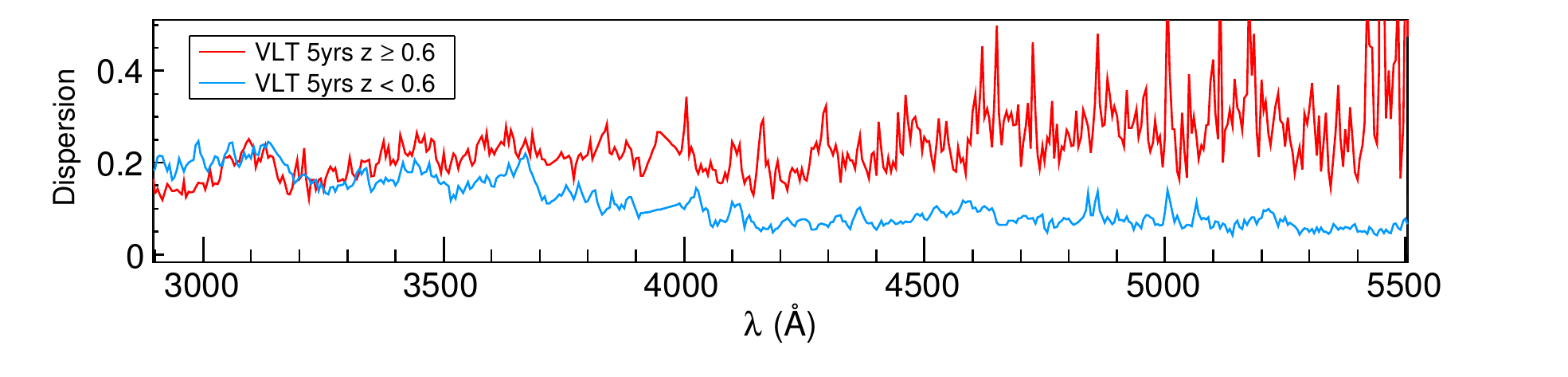}
\includegraphics[width=0.8\textwidth]{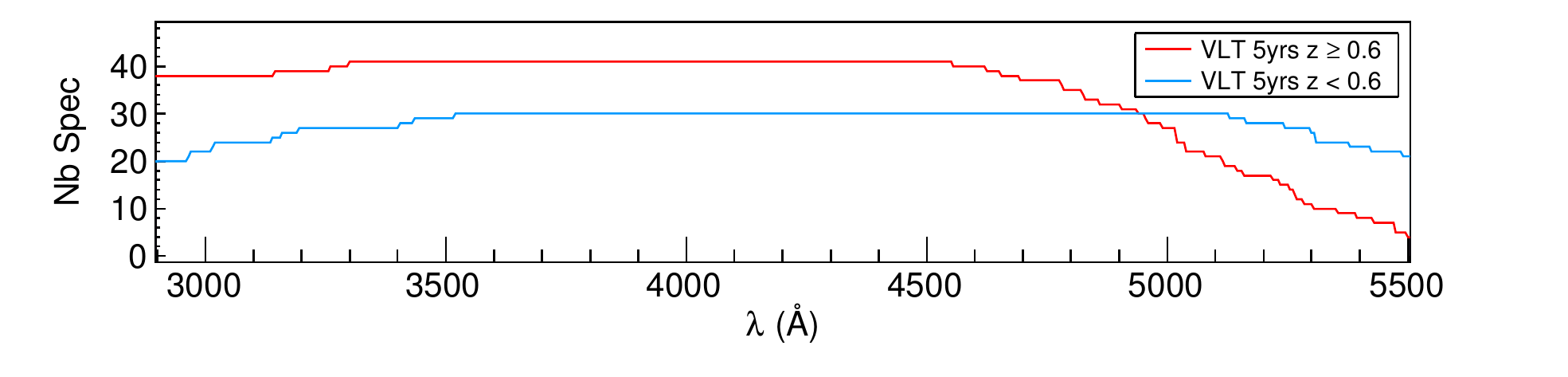}

\caption{Top panel : Low ($z<0.6$, in blue) and high ($z\geq0.6$, in
  red) redshift composite spectra around maximum light built from the
  VLT five year sample. A $\pm 1\sigma$ range is plotted. Individual spectra are
  color corrected (using SALT2 color law) and recalibrated. Residual
  host-galaxy lines have been removed. Middle panel : dispersion of the low
  (blue) and high (red) redshift composite spectra. Bottom panel :
  number of spectra entering the low (blue) and high (red) redshift
  composite spectra.}
\label{fig:meanspec_z}
\end{figure}

We then divide the \Ia~VLT sample into two redshift sets. We define a
low redshift sample with $z<0.6$ and a high redshift with $z\geq 0.6$\footnote{Splitting the sample at the average redshift insures a similar number of spectra in the two redshift bins (as the average redshift is close to the median redshift in our sample) and allows one for a direct comparison with previous works doing a similar split.}.
They contain 30 low redshift spectra and 41 high redshift ones,
respectively, with a similar phase distribution (see Table
\ref{table:evo_z}).  The redshift, phase, color and stretch
distributions of these two subamples are shown in
Fig.~\ref{fig:meanspec_zdistrib} to
Fig.~\ref{fig:meanspec_stretchdistrib}.  The mean spectra, built this
time after applying recalibration and color correction to individual
spectra, are displayed in Fig.~\ref{fig:meanspec_z}.  host-galaxy residual
lines are removed using the technique of Sect. \ref{subsec:color}.  In
Fig. \ref{fig:meanspec_z}, the two lower panels show the dispersion of
the mean spectra and the number of spectra used at each wavelength
bin, respectively.  They help to assess reality of observed spectral
differences.

We note spectral differences between low and high redshift spectra
around the \ion{Ca}{ii} H\&K and \ion{Si}{ii} $\lambda4130$ IME absorptions.  The low
redshift spectrum has deeper IME absorptions
($EW(\ion{Ca}{ii})_{z<0.6}=110\pm 3$ \AA~and
$EW(\ion{Si}{ii})_{z<0.6}=12\pm 3$ \AA) than the high redshift
spectrum ($EW(\ion{Ca}{ii})_{z\geq0.6}=102 \pm 3$ \AA~and
$EW(\ion{Si}{ii})_{z\geq0.6}=5\pm 3$ \AA).

To investigate whether these differences are significant, we select a
random number of spectra in each of the two subsamples (between 15 and
30 for the low-$z$ sample and between 20 and 41 for the high-$z$
sample).  We build the corresponding mean spectra and compute their
flux in spectral regions where the differences are most pronounced
(between 3500-3900 \AA~for the \ion{Ca}{ii} H\&K feature and between
3900-4100 \AA~for the \ion{Si}{ii} $\lambda4130$ feature).  We repeat
this process 5000 times.  Using this bootstrap technique, we find that
the flux is higher for the high-$z$ mean spectrum in 95.9\% of cases
($2\sigma$) in the \ion{Ca}{ii} H\&K region and in 98.9\% of cases
($2.5\sigma$) in the \ion{Si}{ii} $\lambda4130$ region.  The observed
increase of flux with redshift in the range [3300 - 3600] \AA, due to 
lessened line-blanketing from iron-group
elements (IGEs) with decreasing metallicity at high $z$, has been noted in all previous similar
studies using independent samples (e.g.,
\citealt{Sullivan09,Balland09,Maguire12,Foley12})

We note an increase of the dispersion for $\lambda < 3800$ \AA~
compared to optical wavelengths for low redshift spectra (second plot
of Fig.~\ref{fig:meanspec_z}).  This tendency is also found in
\citet{Ellis08} and \citet{Maguire12}.  This \Ia~variability might be
traced back to the host-galaxy properties.  Indeed, this spectral zone is
very sensitive to the chemical composition of the ejecta.  Thanks to
the \Ia~spectral synthesis models of \citet{Walker12},
\citet{Maguire12} show that the UV dispersion is consistent with
metallicity variation in the SN population.  For the high redshift
spectrum, the dispersion is generally larger than that of the low
redshift one. No increase of the UV dispersion with respect to
the one at optical wavelengths is seen in the high redshift spectrum.

\subsection{Split in stretch}
\label{subsec:stretch}

From table \ref{table:evo_z}, it can be seen that there is no
significant difference in stretch between the high and low redshift
samples, given the error on the mean, whereas previous studies (e.g.,
\citealt{Howell07,Sullivan09}) did find a change with redshift that
was consistent with a demographic evolution of \Ia~populations.  This
is expected as the redshift range considered in the present work is
much narrower, and the average star formation rate, on which \Ia~properties depend \citep{Sullivan06b, Howell07}, does not change
much over this range. However, it is interesting to investigate the
existence of spectral differences in our sample that cannot be
explained by mere selection effects. For this purpose, we split the
VLT sample according to the stretch values of the \Iae.  We use
$s=1.013$, the mean stretch value of the sample, as a cut.  We obtain
34 low and 37 high stretch spectra whose properties are summarized in
Table~\ref{table:stretch}.  The two subsamples have similar properties
on average, except for their mean stretch by construction.

\begin{figure}[!ht]\centering
\includegraphics[width=0.7\textwidth]{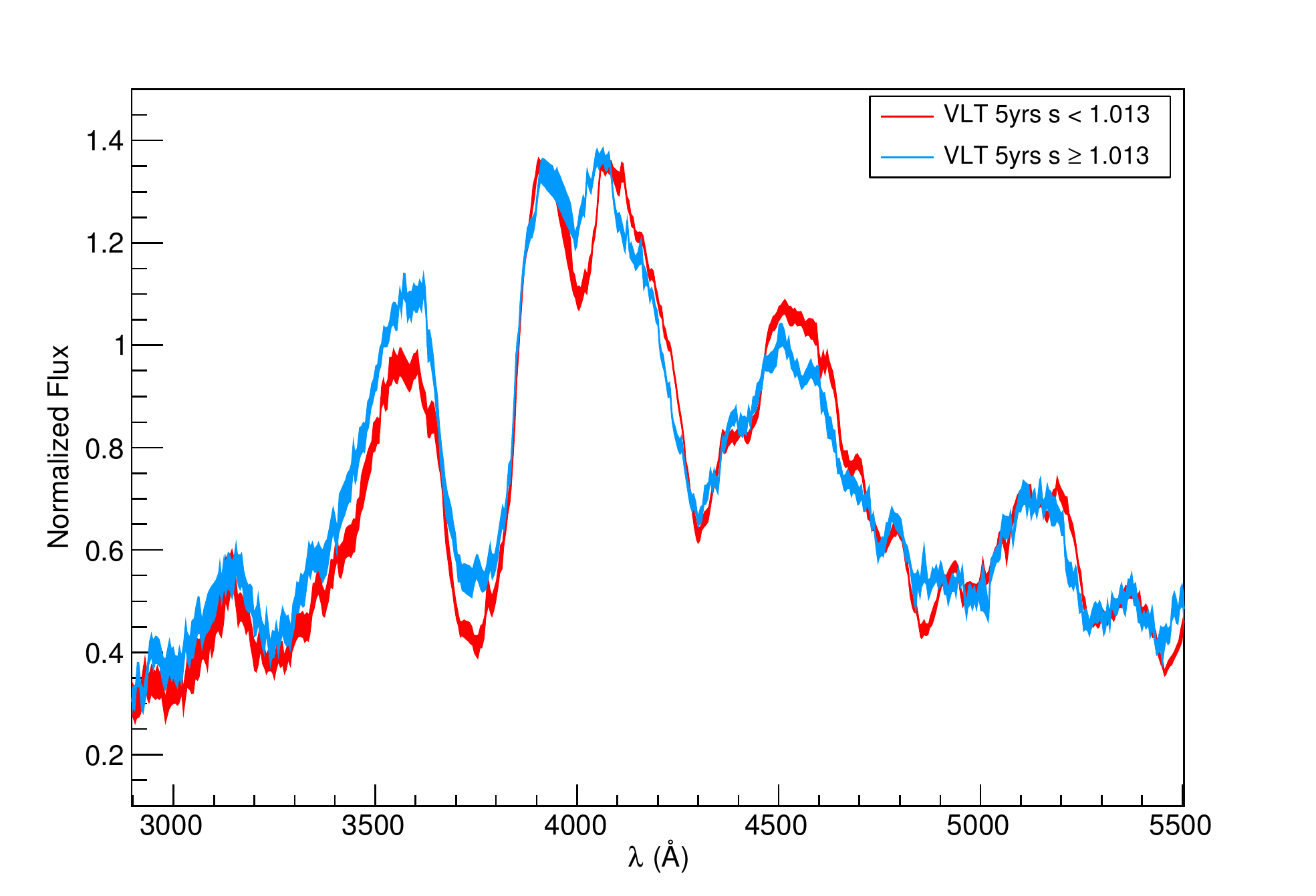}
\caption{Comparison of composite spectra from the full VLT five year
  sample split in stretch. The red spectrum is for $s<1.013$ and the
  blue spectrum for $s\geq1.013$. Individual spectra have been color
  corrected using the SALT2 color law and recalibrated. Residual host-galaxy
  lines have been removed.}
\label{fig:stretch}
\end{figure}

All individual spectra are color corrected (using the SALT2 color
law), recalibrated and lines from the host galaxy removed.  We build
composite spectra that are shown in Fig.~\ref{fig:stretch}.  High
stretch spectra are brighter in the UV below 3800 \AA~and have weaker absorption
lines. In particular, the \ion{Ca}{ii} H\&K and \ion{Si}{ii}
$\lambda4130$ absorption lines for the low stretch spectrum are deeper
($<EW>(\ion{Ca}{ii})_{s<1.013}=119\pm 3 $ \AA~and
$<EW>(\ion{Si}{ii})_{s<1.013}= 16 \pm 3 $ \AA) than for the high
stretch spectrum ($<EW>(\ion{Ca}{ii})_{s\geq1.013}= 108\pm 3 $ \AA~and
$<EW>(\ion{Si}{ii})_{s\geq1.013}= 6 \pm 3$ \AA).

\begin{figure}\centering
\includegraphics[width=0.7\textwidth]{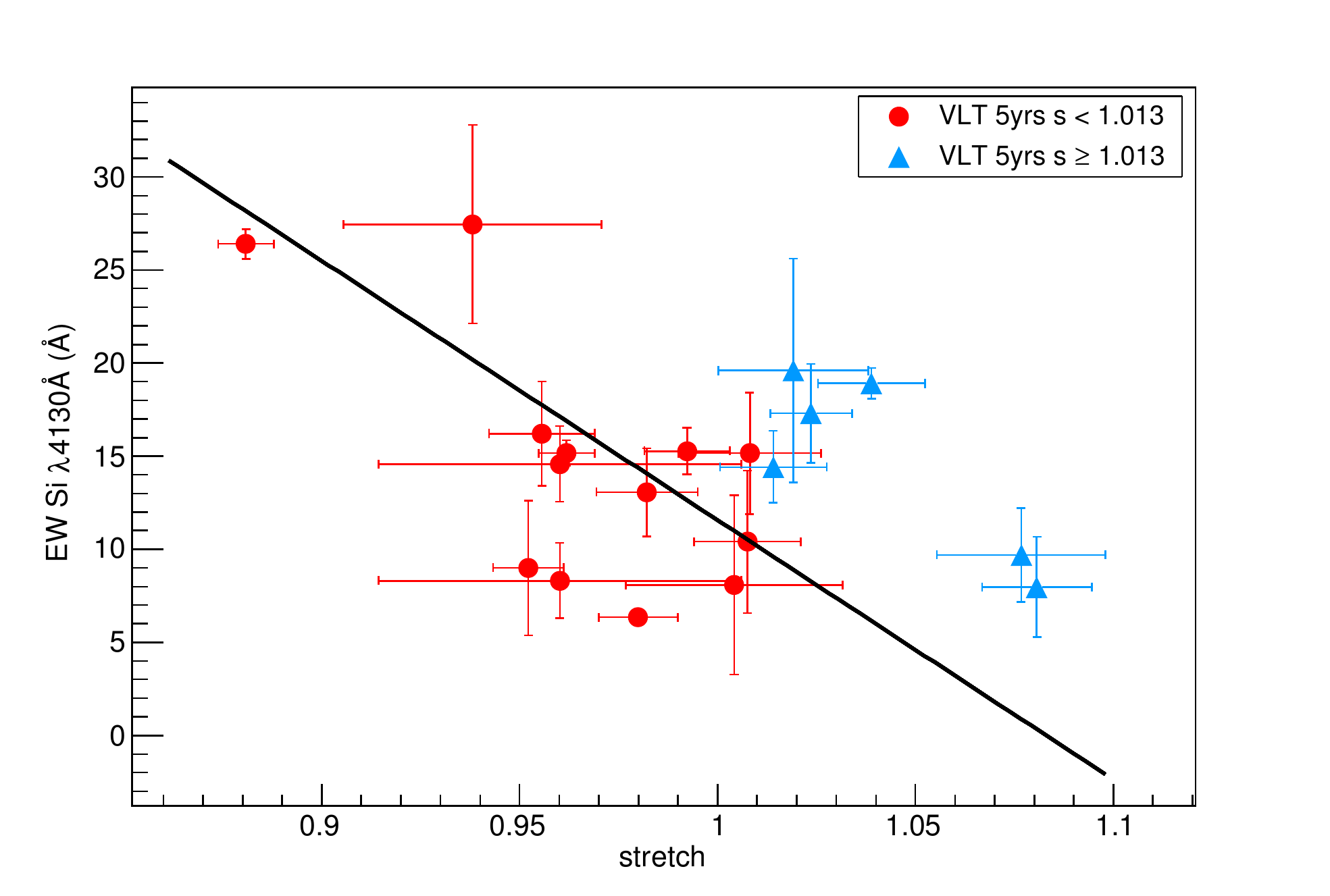} \caption{Correlation
between EW \ion{Si}{ii} $\lambda 4130$ and stretch measured for high
S/N \Iae~of the full VLT five year sample. The best-fit relation is
$EW(\ion{Si}{ii}) = (-139.3\pm 10.9)\times s + (150,9\pm 10.8)$. The low
stretch ($s<1.013$) and high stretch ($s\geq1.013$) \Iae~are shown as
red filled circles and blue triangles, respectively.}  \label{fig:stretch-ewsi} \end{figure}

We compute the equivalent width of the \ion{Ca}{ii} H\&K and
\ion{Si}{ii} $\lambda4130$ features, selecting now a subset of spectra
with high $S/N$ for which these lines are well defined.  We find a
small anti-correlation between the \ion{Si}{ii} $\lambda4130$
equivalent width and the stretch parameter (Fig. \ref{fig:stretch-ewsi}):
\begin{equation}
EW(\ion{Si}{ii}) = (-139.3\pm 10.9)\times s + (150.9\pm 10.8)\ \AA
\end{equation}

In a similar analysis, \citet{Maguire12} and \citet{Foley12} note
ejecta velocity differences as a function of stretch, high stretch
\Iae~spectra showing higher ejecta velocities on average. In
particular, a blueshift is seen in \ion{Ca}{ii}~H\&K,
\ion{Si}{ii} $\lambda4130$ \citep{Maguire12} or in \ion{Fe}{ii}
$\lambda$3250 absorption features \citep{Foley12}. This correlation
between stretch and \ion{Ca}{ii}~H\&K expansion velocity is not observed in the present
analysis\footnote{We note however a $\sim 25$ \AA~blueshift in the
  \ion{Si}{ii} absorption mimimum in the high stretch spectrum with
  respect to its low stretch counterpart at a level comparable to the one seen by \citet{Maguire12}}.

\subsection{Split in host-galaxy mass}
\label{subsec:mass}

In this section, we investigate the impact of host-galaxy mass (considered as
a crude proxy for metallicity or age) on our \Iae~spectra.  For each
supernova in our sample, we derive its host-galaxy properties from CFHT deep
reference images.  Galaxy colors are fitted with a spectro-photometric
code using PEGASE2 \citep{Fioc97,Fioc99} templates and trained on
galaxies of the DEEP-2 survey \citep{Davis03,Davis07}.  For each
supernova, the host-galaxy is identified on SNLS deep reference images and is
fitted to derive its type and age, yielding an estimate of the stellar
mass M$_{stellar}$ and specific star formation rate (sSFR).  Details
on this method can be found in \citet{Kronborg10}, \citet{Hardin17}
and \citet{Roman17}.  Among the 71 spectra used in Sect. 
\ref{sec:evo}, 62 have a reliable host-galaxy stellar mass. For the remaining 9 \Iae,
either no host-galaxy could be associated with the SN on reference images or
the spectro-photometric fit failed.

We compute the mean host-galaxy stellar mass for the two redshift subsamples
of Sect. \ref{subsec:evo} and for the two stretch subsamples of Sect.
\ref{subsec:stretch}.  Differences are marginal (at the $1\sigma$
level) for the redshift subsamples
($\log{(M_{stellar})}_{z<0.6}=9.96\pm0.14$~M$_{\odot}$ and
$\log{(M_{stellar})}_{z\geq0.6}=10.12\pm0.11$~M$_{\odot}$) as well as
for the stretch subsamples
($\log{(M_{stellar})}_{s<1.013}=10.14\pm0.12$~M$_{\odot}$ and
$\log{(M_{stellar})}_{s\geq1.013}=9.97\pm0.12$~M$_{\odot}$).  These
mass differences are not significant and are not likely to be the
cause of the differences observed between low and high redshift or
stretch \Ia~composite spectra.

\begin{figure}[!ht]\centering
\includegraphics[width=0.8\textwidth]{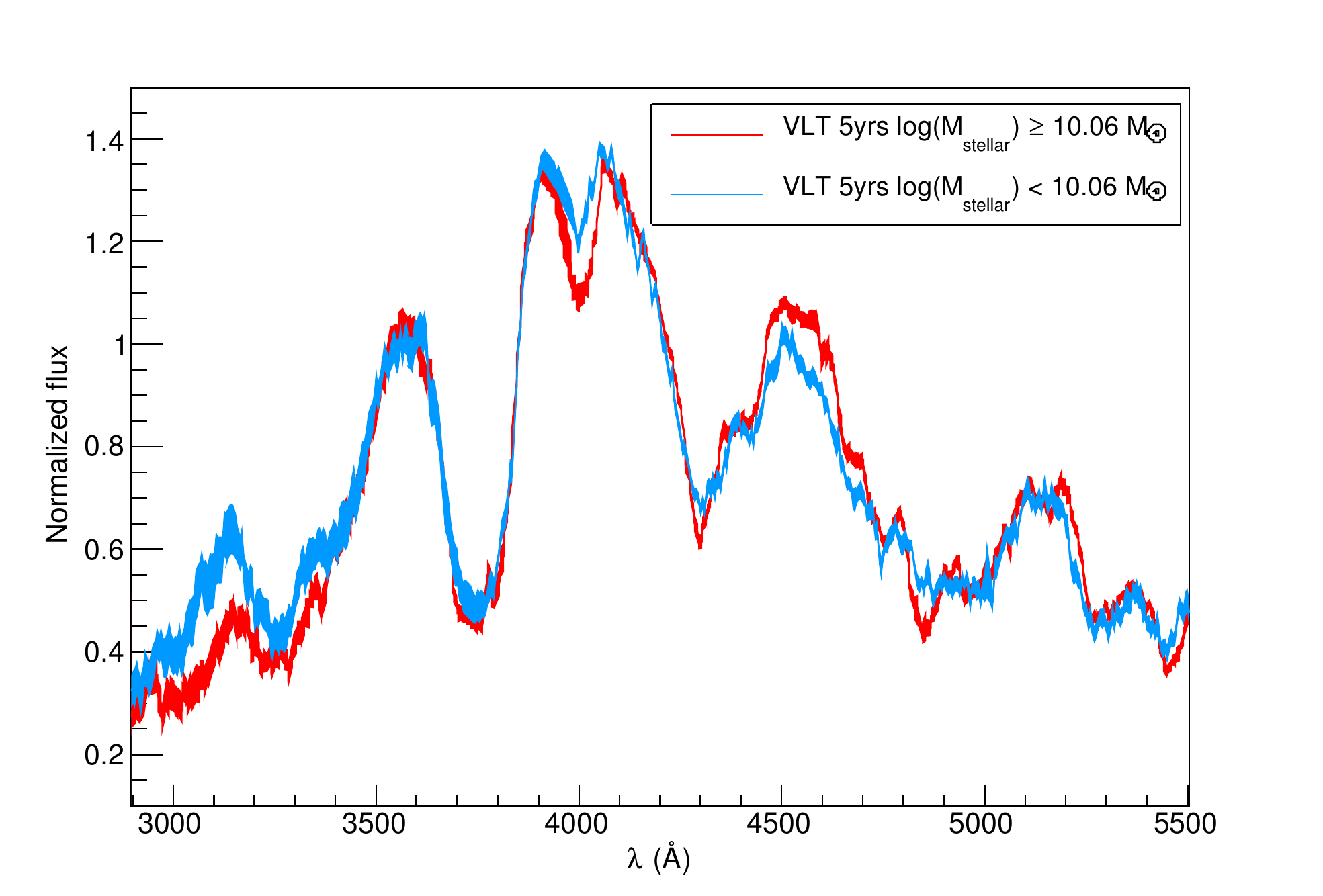}
\caption{Comparison of composite spectra at maximum light from the
  full VLT five year sample host-galaxy split in stellar mass. A $\pm 1\sigma$
  range is plotted. Individual spectra have been color corrected using
  the SALT2 color law and recalibrated. Residual host-galaxy lines have been
  removed.}
\label{fig:mass}
\end{figure}

Dividing now the \Ia~sample into two host-galaxy mass bins using the average
galaxy mass of the full sample as a cut
($\log{(M_{stellar})}=10.06$~M$_{\odot}$), we end up with 27 spectra
in the low mass bin and 35 in the high mass bin.  These two subsamples
differ in host-galaxy stellar mass (by construction) and stretch, all other
parameters being equal on average (Table~\ref{table:mass}).  As
before, color-correction (SALT2), recalibration and removal of host-galaxy
galaxy emission lines have been applied.  The low and high mass
composite spectra are shown in Fig.~\ref{fig:mass}.  Spectral
differences between low and high stellar mass spectra are clearly
visible in particular in the bluest part of the spectra. The low
stellar mass spectrum has an excess of flux for $\lambda<3400$ \AA~and around the \ion{Si}{ii} $\lambda4130$
absorption line.  Contrary to what was observed previously when
spliting the sample in stretch or redshift, the two mean spectra match
around the \ion{Ca}{ii} H\&K feature.

The spectral differences between low and high stellar mass average
spectra might be at least partially explained by the mean stretch
difference of the two subsamples ($0.031\pm0.016$, see
Table~\ref{table:mass}). Indeed, a stretch difference impacts the bluest parts of the mean spectra as seen in Sect.~\ref{subsec:stretch}.
Hence, the difference observed in the depth \ion{Si}{ii} $\lambda4130$
absorption, which is shallower in the low stellar mass spectrum
($EW(\ion{Si}{ii})_{\log{(M_{stellar})}<10.06M_{\odot}}= 7\pm 1 $~\AA)
than in its high stellar mass counterpart
($EW(\ion{Si}{ii})_{\log{(M_{stellar})}\geq10.06M_{\odot}}= 14\pm
1$~\AA), might be due to stretch differences.

\begin{figure}[!ht]\centering
\includegraphics[width=0.8\textwidth]{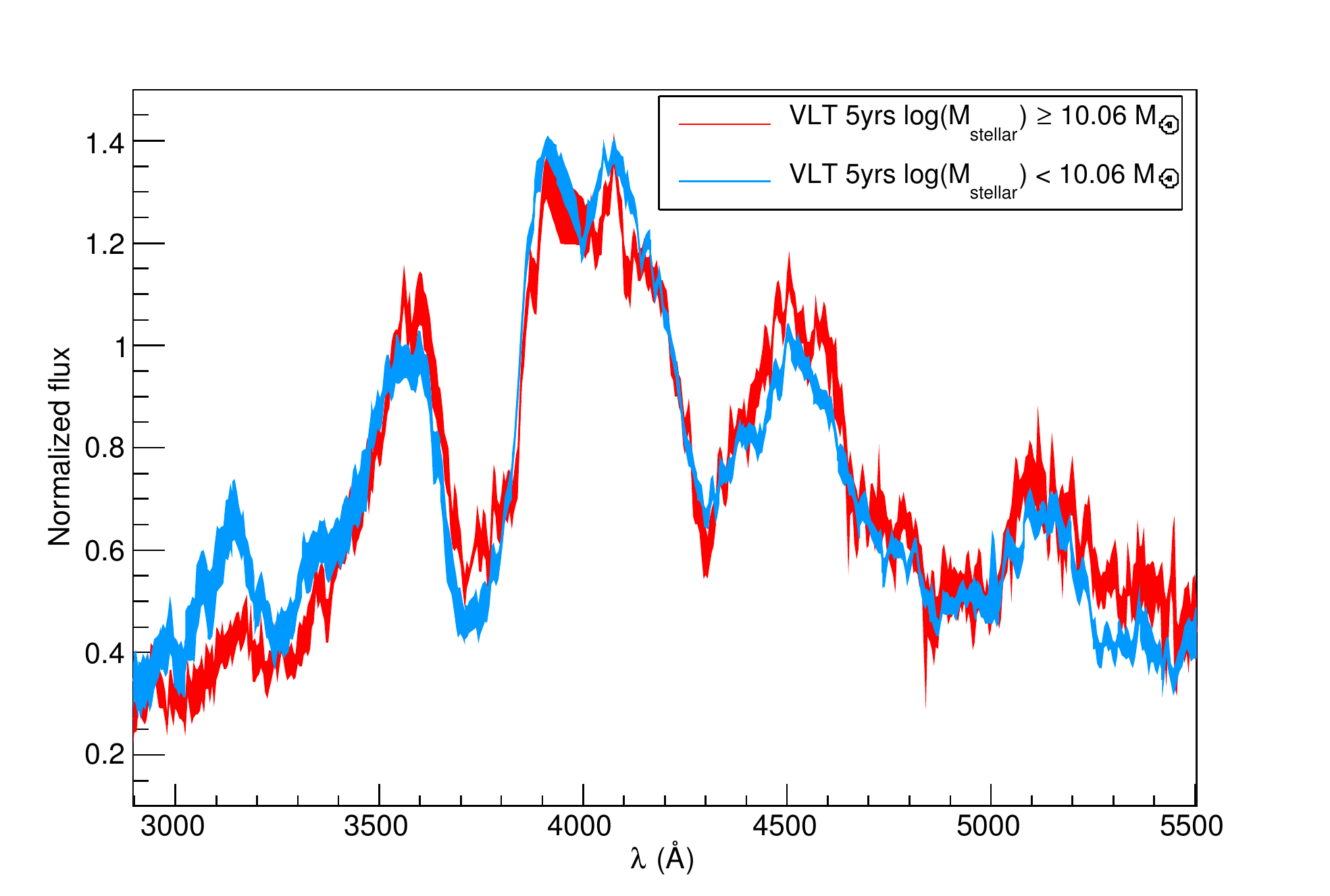}
\caption{Same as Fig. \ref{fig:mass}, with composite spectra built
  from two subsamples with the constraint that the stretch and other
  spectro-photometric parameter distributions of the two subsamples
  match.}
\label{fig:mass-same}
\end{figure}

Selecting two new subsamples split in host-galaxy mass with the constraint
that the stretch (and other spectro-photometric parameters) of the two
subsamples match on average (Table~\ref{table:mass-Same}) allows one
to test the effect of the mass split alone.  The resulting composite
specta are shown overlapping in Fig.~\ref{fig:mass-same}.  The
differences around the \ion{Si}{ii} $\lambda 4130$ feature are
significantly reduced and the equivalent widths are now similar given
the error ($EW(\ion{Si}{ii})_{\log{(M_{stellar})}<10.06M_{\odot}}=
8\pm 1 $ \AA~and
$EW(\ion{Si}{ii})_{\log{(M_{stellar})}\geq10.06M_{\odot}}= 9 \pm
2$~\AA), illustrating the impact of the stretch parameter on spectral
features.  However, the differences at $\lambda < 3400$ \AA~ are
still significantly present and this can be this time traced back to
the difference in host-galaxy mass. A similar result has been obtained by
\citet{Milne15} using pairs of SNLS (and other survey) spectra.  If
confirmed, this UV flux difference might be used as a third
\Iae~standardization parameter beyond stretch and color (see Sect.
\ref{subsec:prophost} for discussion).

\section{Discussion}

\subsection{Origin of the spectral differences with redshift}

From Table~\ref{table:evo_z}, we note that the high redshift composite
spectrum of Sect. \ref{subsec:evo} is made of \Iae~that are on average
bluer than those entering the low redshift spectrum, with a
2.7$\sigma$ color difference ($\Delta c=-0.056\pm0.021$).  This
corresponds to a 2.9$\sigma$ (distance corrected) magnitude difference
of $\Delta(m_{B}^{*\;c})=0.169 \pm 0.059$, the higher redshift
\Iae~being brighter than their low redshift counterparts.  We note
that this color difference is fully consistent with the color
difference of $\Delta c=-0.05$ expected in the SNLS samples due to the
spectroscopic selection of bluer and brighter \Iae~at higher redshift
\citep{Perrett10}. As stated in Sect. \ref{subsec:stretch}, average
stretch values are similar in the two samples given the error: $\Delta
s=-0.008 \pm 0.015$.

Using $\{\alpha\;,\;\beta\}=\{1.295\pm0.112\;,\;3.181\pm0.131\}$
\citep{Guy10}, where $\alpha$ and $\beta$ are the slopes of the light curve shape - luminosity and color - luminosity relationships used to standardize \Iae~in cosmological analyses \citep{Astier06}, and given the above differences in stretch and color, we
expect an intrinsic magnitude difference between the two
\Ia~subsamples of $|\alpha \Delta s - \beta \Delta c|= 0.168 \pm
0.070$. This number is consistent with the magnitude difference seen
in the two samples (see Table~\ref{table:evo_z}).

\begin{figure}[!ht]\centering
\includegraphics[width=0.8\textwidth]{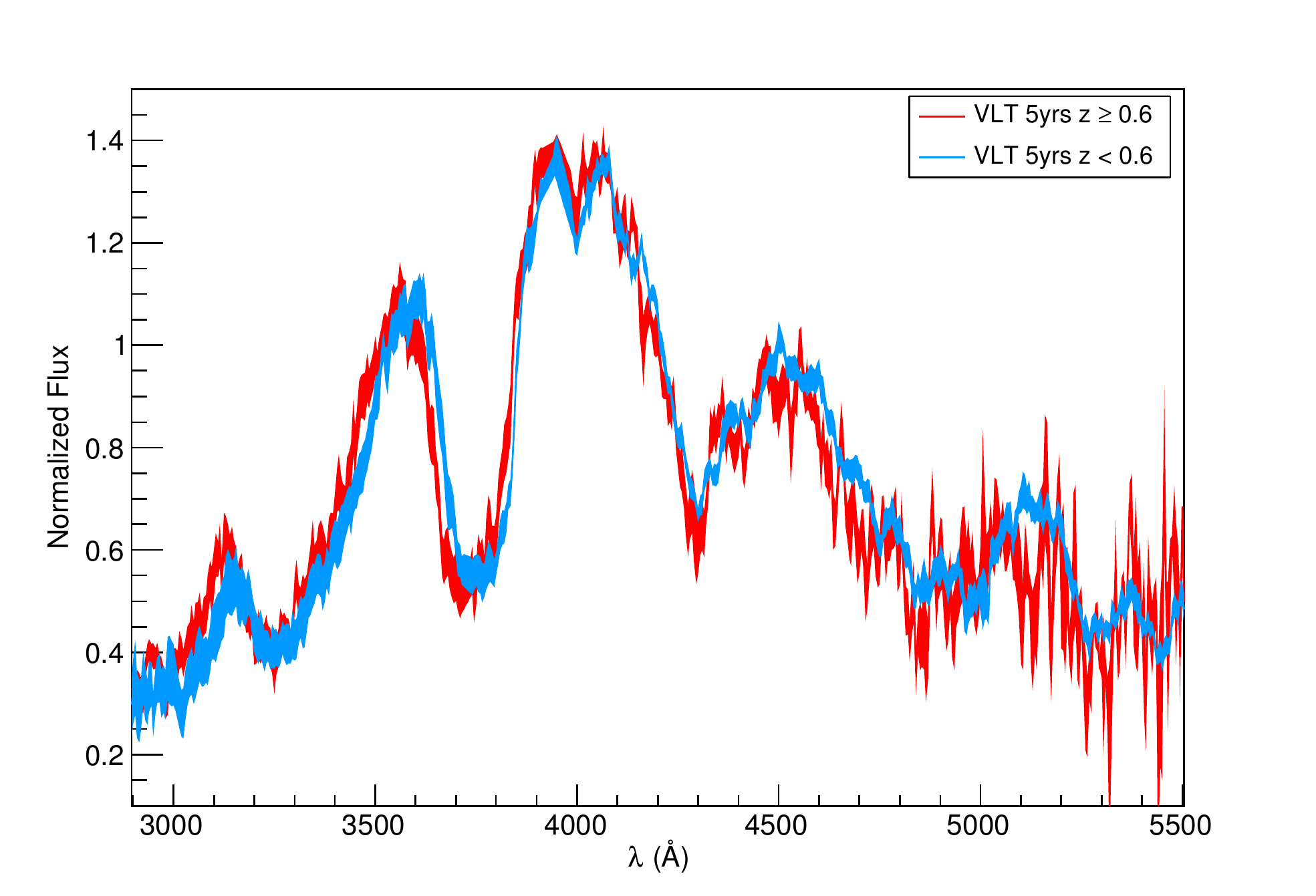}
\includegraphics[width=0.8\textwidth]{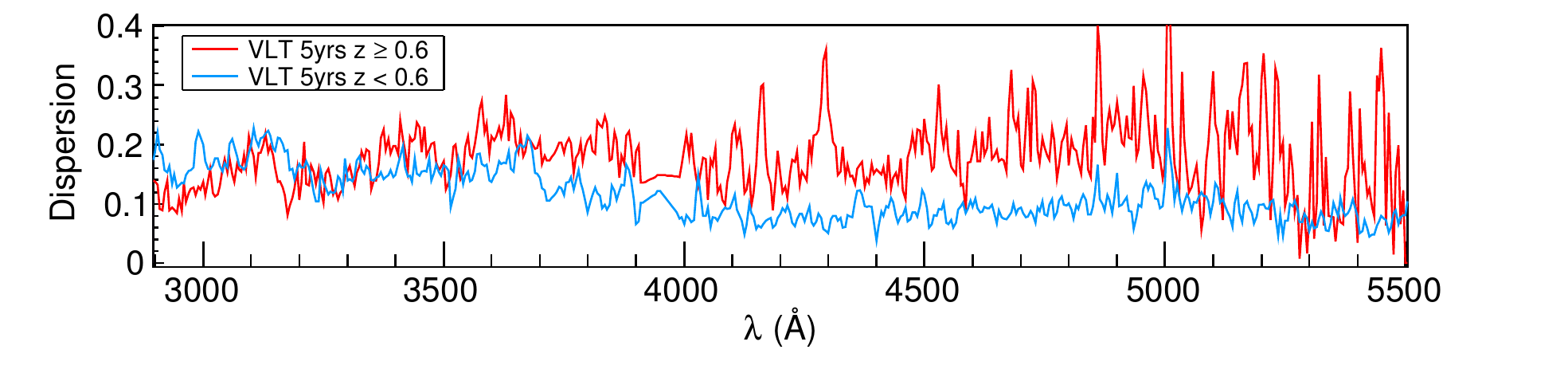}
\includegraphics[width=0.8\textwidth]{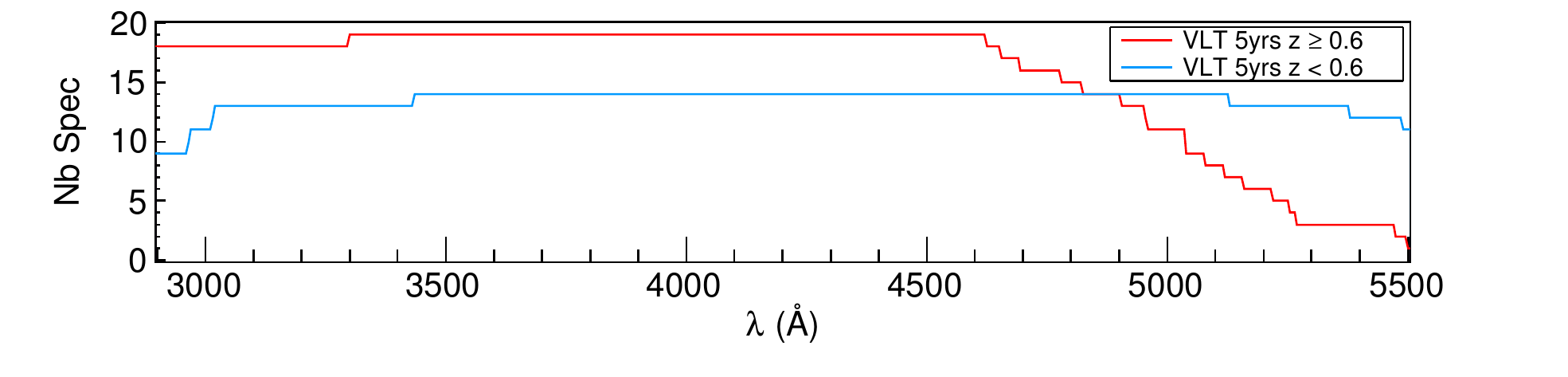}

\caption{Same as Fig. \ref{fig:meanspec_z} with composite spectra
  built from two subsamples split with matching spectro-photometric
  parameter distributions. Top panel : low ($z<0.6$, in blue) and high
  ($z\geq0.6$, in red) redshift composite spectra. Middle panel :
  dispersion of the low (blue) and high (red) redshift composite
  spectra. Bottom panel : number of spectra entering the low (blue)
  and high (red) redshift composite spectra.}
\label{fig:evo_z-same}
\vspace{1cm}
\end{figure}

We now build two new subsamples at low and high redshift with the
constraint that their color and stretch distributions are similar
(i.e., have consistent mean and variance values). For this purpose, we
identify pairs of spectra, one in each redshift bin, with matching
color and stretch.  We obtain 14 low and 19 high redshift spectra
whose properties are shown in Table~\ref{table:evo_z-same}.  The
corresponding mean spectra are presented in Fig. \ref{fig:evo_z-same}.
The spectral differences observed in Fig. \ref{fig:meanspec_z} are now
significantly reduced and the equivalent widths of the IMEs are
consistent within error ($EW(\ion{Ca}{ii})_{z<0.6}=100\pm 4$ \AA~and
$EW(\ion{Si}{ii})_{z<0.6}=6\pm 3$ \AA~ compared to
$EW(\ion{Ca}{ii})_{z\geq0.6}=105\pm 5$ \AA~and
$EW(\ion{Si}{ii})_{z\geq0.6}=4\pm 2$ \AA).

Hence the differences previously observed tend to vanish when two
populations with matching photometric properties are considered. This
shows that the spectral differences observed can be attributed to a
difference in the mixture of populations present in the sample at low
and high redshifts, rather than to evolution with redshift. This
'demographic' bias can be entirely attributed to the selection of
bluer and brighter \Iae~ at higher redshift in the SNLS spectroscopic
sample.

Residual differences between the two composite spectra are
nevertheless observed in Fig. \ref{fig:evo_z-same}.  Small velocity
shifts are seen in the \ion{Ca}{ii} absorption, as well as for the two
UV peaks blueward. A flux excess is also noticeable between 4500 and
5000 \AA~in the low redshift spectrum. Whether these differences are
real or are due to residual differences in the photometric parameters
of the two samples is unclear.

After selecting two subsamples with matching phase, color and stretch
distributions, \citet{Maguire12} note modest but significant
differences between the low ($z\sim0$) and high ($z\sim0.6$) redshift
average spectra.  Using a bootstrap technique, they show that the low
redshift spectrum has a depressed flux between 2900 and 3300
\AA~with respect to the high redshift one (a difference at the
$3.1\sigma$ level).  Using the \Ia~spectral synthesis models of
\cite{Walker12}, they show that this difference could result from a
metallicity evolution with redshift, the metallicity decreasing with
increasing redshift.  We look for comparable trends in our sample.
For this purpose, we use again the bootstrap technique described in
Sect.~\ref{subsec:evo} in the same wavelength regions, this time with the
two subsamples with similar photometric properties.  We find that the
high-redshift composite spectrum has a lower flux in the UV than the
low redshift counterpart in 10.1\% of cases (a significance of
$1.6\sigma$). We find that the high redshift composite spectrum has a
lower \ion{Ca}{ii} H\&K velocity in 13.1\% of cases
($1.5\sigma$). Thus, the spectra of our sample corresponding to the
same underlying supernova populations (having similar photometric
parameter distributions) at low and high redshift show less
significant differences than in \citet{Maguire12}, including in the UV
region. This is not unexpected, as the difference in lookback times involved between
our low and high redshift samples is much lower than the one in the
\citet{Maguire12} samples.

In the low and intermediate redshift samples of \cite{Maguire12}, the
difference between the highest redshift of the low-$z$ sample and the
lowest redshift of the high-$z$ sample is $\Delta z\sim 0.4$. This
'redshift gap' is not present in our sample in which the redshift
distribution is more uniform. In an attempt to understand the effect
of such a redshift gap and check the consistency of the trends seen
when splitting our sample by redshift, we re-build composite spectra,
excluding successively those spectra in the range [0.55,0.65],
[0.5,0.7], [0.45,0.75] and [0.4,0.8] (i.e., we impose gaps of $\Delta
z$ = 0.1, 0.2, 0.3 and 0.4 centered on $z=0.6$). The differences in
spectro-photometric properties between the low and high redshift
subsamples are shown in Table~\ref{table:gap} for each value of the
redshift gap considered. High redshift \Iae~are bluer (as well as
brighter) with increasing gap value compared to the low redshift
\Iae. Again, the observed differences are consistent with selecting
brighter and bluer \Iae~at higher redshift.  For each gap value, we
find an excess of UV flux in the high redshift spectra.  Every
comparison between low and high redshift spectra thus shows the same
trends, independently of the subsamples used.

\subsection{Stretch evolution}
\label{subsec:Dstretch}

When our sample is split acccording to stretch value, all other
parameters being equal on average, UV flux differences are observed
between low stretch and high stretch composite spectra. Namely, we
find a small anticorrelation between the mean stretch of the sample
and the depth of the \ion{Si}{ii} $\lambda 4130$ absorption, the
latter being shallower when the stretch is higher (see Sect.
\ref{subsec:stretch}). This correlation has also been noticed in
\citet{Arsenijevic08} and \citet{Walker11}. It originates as a
consequence of the width-luminosity relation of \Iae~(the so-called  brigther-slower relation; \citealt{Phillips93}). 
More luminous \Iae~have higher ejecta temperatures, and
\ion{Si}{ii} is partially ionized to \ion{Si}{iii} \citep{Nugent95}.

The kinetic energy of the explosion is higher for higher $^{56}$Ni
mass and a correlation between the expansion scale and stretch is
expected (see, e.g., \citealt{Howell06}). As explained earlier,
\citet{Maguire12} do observe such a correlation: the \ion{Ca}{ii} H\&K
velocity increases with stretch (see their Figs. 7 and 10) producing a
spectral difference at the $3.4\sigma$ level in the calcium
region. However, neither the present work nor \cite{Foley12} find
evidence for this effect. As noted by \citet{Maguire12}, the absence
of spectra with $s>1.05$ in the \cite{Foley12} sample might be
responsible for this, the higher stretch spectra contributing the most
to a shift in the ejecta velocity. When the 17 HST spectra with
$s>1.05$ are excluded from the \citet{Maguire12} sample, the
difference between the composite spectra at low and high stretch in
the \ion{Ca}{ii} region is considerably reduced and becomes
statistically insignificant (less than $1\sigma$ significance). The
observation of a stretch-velocity correlation might thus be due to the
large mean stretch difference in the original \citet{Maguire12}
samples ($\Delta<s> \sim 0.12$). This argument might explain as well
why we do not see the effect in our sample: the fraction of high
stretch spectra in our sample (25\%) is indeed much lower than in
\citet{Maguire12}'s (41\%).

\begin{figure}[!ht]\centering
\includegraphics[width=0.8\textwidth]{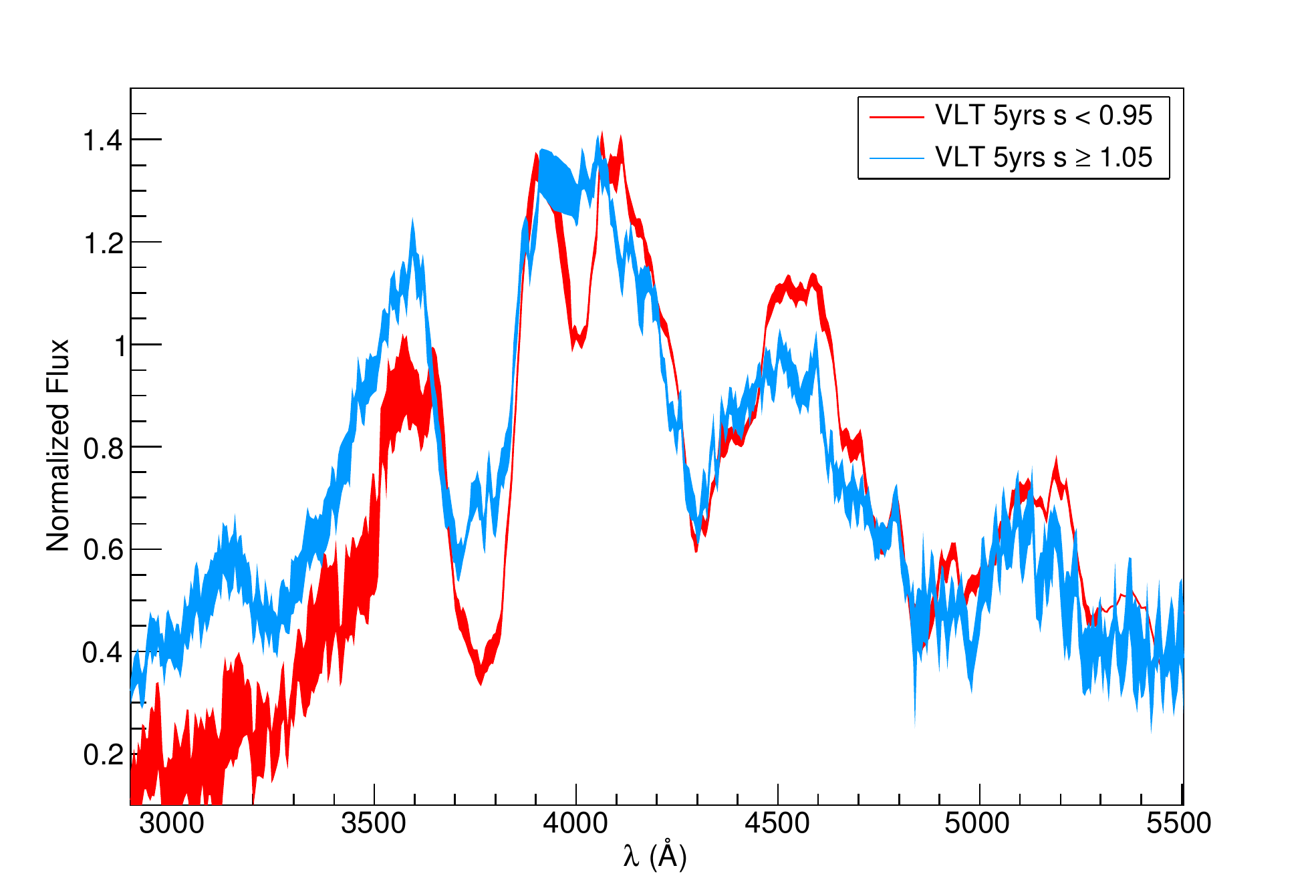}
\caption{Comparison of composite spectra from the full VLT five year
  sample split by stretch while imposing a stretch gap (no spectra)
  between 0.95 and 1.05 in order to exacerbate the effects due to the
  stretch. Individual spectra have been color corrected using the
  SALT2 color law and recalibrated. Residual host-galaxy lines have been
  removed.}
\label{fig:meanspec_stretch_gap}
\end{figure}

To test this hypothesis, we create two new subsamples by excluding all
spectra with stretches in the range $0.95<s<1.05$ yielding a mean
stretch difference comparable to the one of \citet{Maguire12}. Figure
\ref{fig:meanspec_stretch_gap} shows an overlap of the two composite
spectra computed from our new samples. The difference in the IME
absorption regions are stronger than before, illustrating the impact
of the mean stretch on the composite spectra; however, no significant 
differences beyond the ones seen in Fig. \ref{fig:stretch} (see Sect. \ref{subsec:stretch})
 are observed in the velocities of IME ejecta. This leaves
us either with the possibility that the mean stretch difference
between our samples is not large enough or that the effect is blurred
by the IME peculiar velocities in spectra for which the redshift has
been estimated from SN features. Indeed, in the construction of the
mean spectra, we use both \Iae~whose redshift has been estimated from
host-galaxy lines and from our spectral template fitting. If there are subtle
velocity effects in the spectra, they may be washed out by using the
template fitting. As about 25\% of redshifts have been obtained from
this technique in the present sample, as opposed to 6\% in
\citet{Maguire12}, this might be a non negligible effect and a
limitation to the present comparison.

A possible explanation of the spectral differences observed as a
function of stretch could be the existence of two (or more) distinct
\Ia~ populations, with distinct stretch and spectral properties. Based
on the A+B model\footnote{ This model postulates the existence of two groups of \Iae: a 'prompt' population of intrinsically more luminous \Iae~with broad light curves and a 'delayed' component of intrinsically fainter \Iae~with narrower light curves.} \citep{Scannapieco05}, \citet{Howell07} identify in a
sample of \Iae~obtained from various sources in the range $0.1<z<0.75$
a delayed and a prompt component, with a mean stretch of $s\sim 0.95$
and $s\sim 1.08$, respectively. The mean stretch of our two subsamples
are indeed very similar to those values ($\sim 0.96$ and $s\sim 1.06$,
see Table \ref{table:stretch}) and it is tempting to invoke the
existence of these two populations in our samples.

However, if such two populations coexist at different redshifts in
proportions predicted by the A+B model (in agreement with the ratios
of prompt to delayed \Iae~observed in SNfactory \citep{Aldering02} data
by \citealt{Childress13b}), one would expect a stretch increase of the
whole \Ia~ population of about 8\% from $z=0$ to $1.1$
\citep{Howell07}. Assuming the increase is linear in cosmic time and
provided that our sample selection reflects the global properties of
the underlying \Ia~population, one expects a $\sim 2$\% increase in
the stretch of the spectra used to create composites ($z=0.47$ to
$0.73$).  We find instead a moderate decrease of the mean stretch of
$\lesssim 1$\% from $z=0.47$ to $0.73$.  This tends to show that the
differences observed in our sample trace back to the selection effects
of the survey, rather than to a demographic shift in the
\Ia~populations with redshift.

\subsection{Spectral differences with host-galaxy properties}
\label{subsec:prophost}

\Iae~photometric properties correlate with their environment (e.g.,
\citealt{Sullivan06b, Rigault13,Childress13b}). Brighter supernovae
with higher stretch explode preferentially in late-type star-forming
spiral galaxies (e.g.,
\citealt{Hamuy95,Hamuy00,Sullivan06b,Howell07}).  Moreover, once
corrected for stretch and color, \Iae~are 10\% brighter on average in
massive later type galaxies, which also tend to have higher metalicity
(e.g. \citealt{Sullivan10,Lampeitl10,Kelly10}). From Table
\ref{table:mass-Same}, we see that \Iae~are marginally brighter in
high mass host-galaxy galaxies by 0.1 mag, in agreement with this trend.

In terms of spectral shape, we find that the high stretch and low host-galaxy
stellar mass \Iae~have weaker \ion{Si}{ii} $\lambda4130$ absorptions
(Sect. \ref{subsec:stretch}), in agreement, for example, with \citet{Bronder08}
who showed that \Iae~in spiral galaxies have weaker IME absorptions
than those in elliptical galaxies.  We also find that \Iae~with high
stellar mass host-galaxies (all other parameters matching on average) have a
significantly lower flux in the 3000-3500 \AA~UV region of their
spectra. Interestingly, this is the region where \citet{Ellis08} and
\citet{Maguire12} observe an increased dispersion in their spectral
samples, whether split in stretch or redshift. Part of this effect
could be explained by the diversity of host-galaxy properties of the
\Iae~entering the composite spectra in these studies. However, while
some theoretical studies have shown that UV spectra are indeed
affected by host-galaxy metallicity in a way that causes an increase in
\Ia~variability in this region of the spectrum (e.g.,
\citealt{Hoeflich98, Lentz00,Sauer08}), it cannot be to such an extent as to
account for the full increase in dispersion observed, as noticed by
\citet{Ellis08} and \citet{Maguire12}. Nevertheless, the analysis of
the present VLT spectral set supports the importance of host-galaxy
parameters in understanding \Iae~properties.

Recently, \citet{Roman17} have estimated the host-galaxy restframe $U-V$
color at the location of the supernova explosion, using a sample of
882 \Iae~host-galaxies from the SNLS, the SDSS and local surveys. They show
that there is a significant difference between this local color and
the global restframe $U-V$ color, the latter being bluer at all
redshifts. Moreover, performing a cosmological fit to the \Iae~JLA
data \citep{Betoule14}, they find the Hubble residuals to be more
correlated with local color than with the host-galaxy stellar mass or global
color and conclude that local color conveys more physical information
on the lightcurve properties than the host-galaxy stellar mass. Using the
local color as a third lightcurve standardization variable reduces the
total dispersion in the Hubble diagram by $\sim 7$ \% relative to
using only stretch and color for standardization.

\begin{figure}[!ht]\centering
\includegraphics[width=0.8\textwidth]{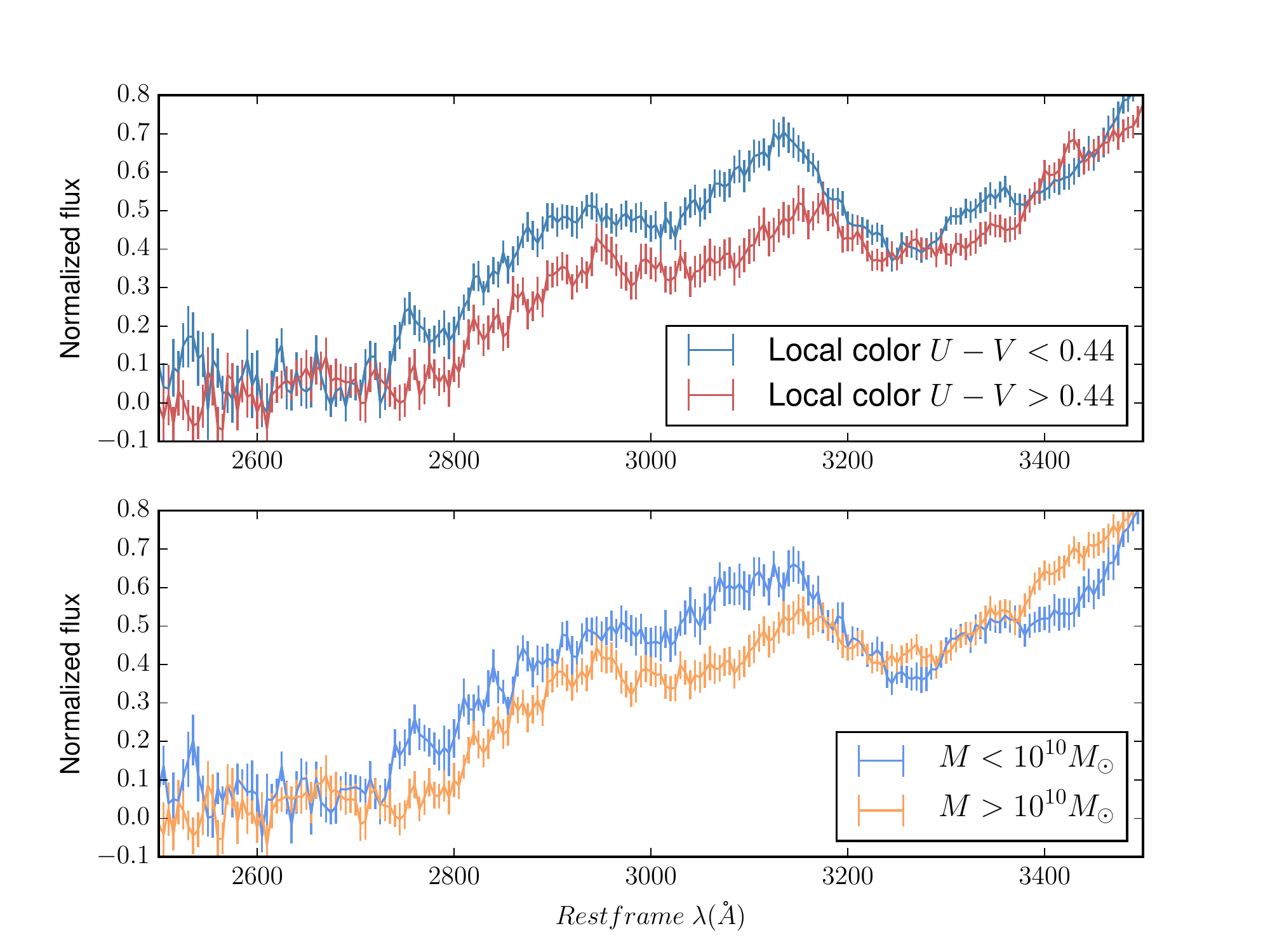}
\caption{Comparision of composite spectra obtained when spliting the
  sample according to local color (top panel) or mass (bottom
  panel). The cut are made at the sample average value.}
\label{fig:comp_loc_col_with_mass}
\end{figure}

The \citet{Roman17} analysis involves 397 SNLS \Iae~host galaxies. For
a fraction of these, the supernova spectrum has been observed at the
VLT and belongs to the sample presented in this paper. We use these
spectra along with the measurement of the corresponding host-galaxy restframe
$U-V$ local color provided by \citet{Roman17} to compute two composite
spectra in two bins of local color. As before, the split value is the
sample average of the relevant variable (here $<U-V>_{local}=0.439$)
and only spectra with phases in the range $-4<\phi<+4$ days are
considered. Spectra are color corrected and recalibrated
individually. In Fig. \ref{fig:comp_loc_col_with_mass}, we show the
two spectra in the 2500-3500 \AA~region (top panel). For comparison,
we also show the spectra obtained in the same spectral region when
splitting our sample according to the host-galaxy mass (bottom panel, see
\ref{subsec:mass}). The difference obtained with a cut in local $U-V$
is slightly more pronounced than the one with a cut in host-galaxy
mass. Chosing the sample median $U-V$ rather than the average value
has basically no impact on the difference seen. Moreover, excluding
from the composites those spectra for which the redshift has been
obtained from the SN features rather than from host-galaxy lines does not
modify the result. This analysis supports the existence of a variation
in the UV spectral properties of the \Iae~populations linked to their
environment. As shown in \citet{Roman17}, this link can be exploited
to further standardize \Iae~properties and reduce Hubble residuals.

\section{Conclusion}
\label{sec:conclu}

The SNLS experiment benefited from exceptional spectroscopic surveys,
with a total of $\sim 1500$ hours of observation on 8-10m class
telescopes, including the VLT with two ESO large programs. Special
attention was paid to assessing the type and determining the redshift
of the \Ia~candidates using a combination of SALT2 and visual
inspection.

In the present paper, we publish for the first time the spectra of the
\Iae~measured at the VLT during the last two years of SNLS\footnote{We
  add eight \Iae~observed in MOS mode during the first three years of
  operation of the survey plus one \Ia~not identified in the three year
  analysis.}.  51 \Iae~ were identified as \Ia~(certain \Iae), 16 as
\Ias~ (probable \Iae) and one object was found to be peculiar (SN~1991T
like).  These \Iae~and \Iaes~subsamples have on average very similar
photometric and spectroscopic properties (color, stretch and $B$-band
absolute magnitude at maximum light), which suggests that the
SNe~Ia$\star$ sample is not contaminated by non-Ia-objects.  The host-galaxy
subtraction is well under control, spectra with low and high host-galaxy
contamination being remarkably similar.

This new VLT sample completes the VLT three year spectral data set which
contains spectra of identified \Iae~measured in LSS mode at the VLT
during the first three years of operation of the survey
\citep{Balland09}.  We find that the two VLT samples have similar
spectro-photometric properties and spectra on average.  Once combined
to build the final VLT spectroscopic sample of SNLS, there are 192
\Iae~for a total of 209 spectra.  When Gemini and Keck spectra are
added to the VLT spectral set, the final SNLS spectroscopic sample
contains 427 identified SNe~Ia and SNe~Ia$\star$.  This is the largest
intermediate to high redshift SN~Ia sample to date and the final SNLS
cosmology analysis will rely on it \citep{Betoule17}.

With this exceptional sample in hand, we have studied \Iae~spectral
and photometric properties at intermediate to high redshift. We
have reassesed the key question of a possible evolution of SNe~Ia
properties with redshift in the light of this new sample.  Using color
corrected VLT spectra around maximum light ($-4 < \Phi < 4$ days), we
find that:
\begin{enumerate}
\item the spectral comparison between low and high redshift SNe~Ia
  shows an increase of their UV flux with redshift.  Brighter and
  bluer \Iae~are observed at higher redshift, in full quantitative
  agreement with what one expects from the spectroscopic selection
  process of SNLS \citep{Perrett10}. No definite sign of intrinsic
  evolution of the \Iae~properties with redshift is seen in the
  present sample.
\item \Iae~in more massive galaxies have a higher stretch on
  average. Once this stretch difference is accounted for, a residual
  flux excess is found in the [3000-3400] \AA~region for \Iae~with
  low mass host-galaxies. The same trend is observed for \Iae~with low local
  $U-V$ color. If the \Iae~UV flux is indeed indexed on the host-galaxy
  stellar mass or local color, this could open the way toward using
  the latter as a third parameter beyond stretch and color in the
  standardization process of \Iae~\citep{Roman17}.
\end{enumerate}
The full VLT spectral set of the SNLS experiment is remarkable for its
size and homogeneity in the intermediate to high redshift window.  It
will be combined with the other SNLS spectroscopic samples from Gemini
\citep{Howell05, Bronder08, Walker11} and Keck \citep{Ellis08} and
used in the final SNLS cosmology analysis.

\begin{acknowledgements}
We gratefully acknowledge the assistance of the VLT Queue Scheduling
Observing Team. Part of this research was conducted by the Australian
Research Council Centre of Excellence for All-sky Astrophysics
(CAASTRO), through project number CE110001020.  French authors
acknowledge support from CNRS-IN2P3, CNRS-INSU and PNCG. MS
acknowledges support from EU/FP7-ERC grant No. [615929]. AMM and VA
acknowledge support from Funda\c{c}\~ao para a Ci\^encia e Tecnologia,
Portugal.
\end{acknowledgements}

\bibliographystyle{aa}
\bibliography{bibi_these}

\onecolumn

\input{Table_ObsCond.tex}

\clearpage

\input{Table_Result.tex}

\clearpage

\input{Table_MOS_short.tex}

\clearpage

\input{Tables_mean.tex}

\clearpage

\twocolumn

\clearpage
\onecolumn

\clearpage

\twocolumn

\clearpage
\onecolumn

\clearpage

\clearpage

\clearpage


\onecolumn
\input{BookSpec_SNLSVLT5yrs.tex}

\end{document}

%% file: Table_ObsCond.tex
\begin{longtable}{cccccccc}
\caption{\label{table:obs} Observing log of the VLT \Iae~of the last two years of SNLS. The $i_M$ magnitude is the magnitude at observation time.}\\
\hline\hline
SN name & RA (J2000) & Dec (J2000) & Spectrum date (UTC) & Exp. time (s) & Seeing ($\arcsec$) & Air Mass & $i_M$ \\
\hline
\endfirsthead
\caption{continued.}\\
\hline
   \hline
SN name & RA (J2000) & Dec (J2000) & Spectrum date (UTC) & Exp. time (s) & Seeing ($\arcsec$) & Air Mass & $i_M$ \\
\hline
\endhead
\hline
\endfoot  
05D1dx \tablefootmark{a} & 02:27:47.00 & -04:01:57.1 & 2005-10-10 & 4 x 900 & 1.15 & 1.20 & 23.27 \\ 
05D1dx \tablefootmark{a} & 02:27:47.00 & -04:01:57.1 & 2005-11-12 & 4 x 900 & 0.89 & 1.08 & 23.32 \\ 
05D1hm \tablefootmark{a} & 02:27:46.19 & -04:43:02.0 & 2005-11-29 & 4 x 900 & 0.86 & 1.19 & 23.51 \\ 
05D1if \tablefootmark{a} & 02:24:29.72 & -04:34:13.0 & 2005-12-01 & 4 x 900 & 1.12 & 1.11 & 23.50 \\ 
05D2le \tablefootmark{a} & 10:01:54.86 & +02:05:34.8 & 2005-12-01 & 4 x 900 & 0.64 & 1.35 & 23.56 \\ 
06D1bg & 02:25:20.73 & -04:06:58.2 & 2006-08-23 & 4 x 900 & 0.75 & 1.38 & 23.60 \\ 
06D1bo & 02:26:15.53 & -04:20:58.6 & 2006-08-23 & 4 x 900 & 0.98 & 1.13 & 23.52 \\
06D1cm \tablefootmark{b} & 02:25:01.91 & -04:28:43.2 & 2006-09-04 & 8 x 900 & 0.63 & 1.18 & 23.53 \\
06D1cx & 02:26:02.59 & -04:14:43.7 & 2006-09-01 & 4 x 900 & 1.48 & 1.08 & 23.67 \\ 
06D1dc & 02:25:09.01 & -04:33:13.2 & 2006-09-22 & 4 x 900 & 0.98 & 1.14 & 23.38 \\ 
06D1dl & 02:27:38.39 & -04:32:36.6 & 2006-09-22 & 4 x 900 & 1.08 & 1.08 & 22.97 \\ 
06D1du \tablefootmark{b} & 02:25:40.88 & -04:12:23.3 & 2006-09-20 & 3 x 750 & 0.76 & 1.07 & 20.80 \\
06D1eb & 02:25:03.05 & -04:06:25.2 & 2006-09-26 & 6 x 900 & 0.74 & 1.08 & 23.45 \tablefootmark{e}\\ 
06D1eb & 02:25:03.05 & -04:06:25.2 & 2006-10-01 & 3 x 750 & 1.06 & 1.07 & 23.38 \tablefootmark{e}\\ 
06D1ez \tablefootmark{b} & 02:25:41.00 & -04:14:49.3 & 2006-10-21 & 3 x 750 & 0.89 & 1.07 & 23.44 \\
06D1fd \tablefootmark{b} & 02:25:54.43 & -04:36:28.9 & 2006-10-27 & 3 x 750 & 0.65 & 1.27 & 22.37 \\ 
06D1fx & 02:27:41.89 & -04:46:43.3 & 2006-11-14 & 4 x 900 & 0.99 & 1.07 & 22.75 \\ 
06D1gl \tablefootmark{c} & 02:26:53.28 & -04:46:29.8 & 2006-11-18 & 9 x 900 & 1.06 & 1.08 & 24.06 \\ 
06D1hi & 02:24:30.50 & -04:11:39.4 & 2006-11-25 & 4 x 900 & 0.93 & 1.08 & 23.24 \\ 
06D1ix \tablefootmark{b} & 02:26:33.63 & -03:59:03.8 & 2006-12-18 & 4 x 900 & 0.68 & 1.44 & 22.85 \\ 
06D1jf & 02:24:17.47 & -04:25:52.5 & 2006-12-18 & 4 x 900 & 1.39 & 1.11 & 23.41 \\ 
06D1jz & 02:27:11.07 & -04:26:25.8 & 2006-12-23 & 4 x 900 & 0.79 & 1.25 & 21.73 \\ 
06D1kf & 02:26:49.27 & -04:10:10.1 & 2006-12-24 & 4 x 900 & 1.10 & 1.07 & 23.37 \\ 
06D1kg & 02:24:32.57 & -04:15:02.0 & 2007-01-17 & 3 x 750 & 1.25 & 1.29 & 21.91 \\ 
06D1kh \tablefootmark{c} & 02:24:50.09 & -04:42:30.6 & 2007-01-23 & 6 x 900 & 1.66 & 1.32 & 23.74 \\ 
06D2ag \tablefootmark{a} & 10:01:43.36 & +01:51:37.3 & 2006-01-26 & 3 x 750 & 0.70 & 1.12 & 21.97 \\ 
06D2bo \tablefootmark{b,d} & 10:00:52.54 & +02:03:22.9 & 2006-02-09 & 6 x 900 & 0.64 & 1.13 & 24.15 \\ 
06D2hm & 09:58:44.57 & +02:19:58.7 & 2006-12-18 & 4 x 900 & 0.64 & 1.16 & 23.16 \\ 
06D2hu & 09:59:56.99 & +02:08:03.3 & 2006-12-20 & 4 x 900 & 0.84 & 1.30 & 22.16 \\ 
06D2jw \tablefootmark{c} & 09:59:37.59 & +02:34:18.4 & 2006-12-27 & 6 x 900 & 0.82 & 1.26 & 24.13 \\ 
06D4ba \tablefootmark{a} & 22:15:35.72 & -18:13:44.7 & 2006-07-04 & 4 x 900 & 0.77 & 1.16 & 23.58 \\ 
06D4bo \tablefootmark{a} & 22:15:28.12 & -17:24:33.2 & 2006-07-04 & 4 x 900 & 0.71 & 1.03 & 22.73 \\ 
06D4bw \tablefootmark{a} & 22:15:03.70 & -17:53:00.2 & 2006-07-03 & 6 x 900 & 0.74 & 1.05 & 23.41 \\ 
06D4gs \tablefootmark{b} & 22:15:14.80 & -17:14:52.5 & 2006-09-20 & 3 x 750 & 0.84 & 1.28 & 21.59 \\
06D4jh & 22:15:31.24 & -18:04:22.1 & 2006-11-14 & 4 x 900 & 0.59 & 1.09 & 23.02 \\
06D4jt & 22:14:45.52 & -18:00:56.8 & 2006-11-19 & 4 x 900 & 1.13 & 1.39 & 22.96 \\ 
07D1ab & 02:26:44.88 & -04:01:00.7 & 2007-01-23 & 4 x 900 &  - \tablefootmark{f} & 1.40 & 22.19 \\ 
07D1ad & 02:27:44.72 & -04:57:42.3 & 2007-01-24 & 4 x 900 & 1.60 & 1.60 & 21.80 \\ 
07D1ah & 02:27:33.33 & -04:06:54.3 & 2007-08-27 & 4 x 900 & 1.10 & 1.17 & 24.12 \\ 
07D1bl & 02:27:30.62 & -04:40:21.2 & 2007-09-04 & 4 x 900 & 1.12 & 1.08 & 25.55 \\ 
07D1bs & 02:26:04.32 & -04:54:27.6 & 2007-09-08 & 4 x 900 & 1.06 & 1.37 & 23.15 \tablefootmark{e}\\ 
07D1bu & 02:27:00.37 & -04:32:32.9 & 2007-09-08 & 4 x 900 & 1.59 & 1.14 & 22.93 \tablefootmark{e}\\ 
07D1by & 02:24:05.44 & -04:32:00.6 & 2007-09-12 & 4 x 900 & 0.91 & 1.07 & 23.44 \tablefootmark{e}\\ 
07D1ca & 02:24:47.20 & -04:50:56.0 & 2007-09-16 & 6 x 900 & 1.19 & 1.10 & 24.00 \tablefootmark{e}\\ 
07D1cc & 02:25:16.94 & -04:06:49.7 & 2007-09-16 & 6 x 900 & 0.81 & 1.09 & 24.14 \tablefootmark{e}\\ 
07D1cd & 02:25:33.96 & -04:45:06.5 & 2007-09-21 & 6 x 900 & 0.99 & 1.11 & 24.22 \tablefootmark{e}\\ 
07D1cf & 02:26:34.20 & -04:58:52.3 & 2007-09-20 & 4 x 900 & 2.24 & 1.08 & 25.30 \\ 
07D2aa \tablefootmark{b,c} & 10:02:05.50 & +02:25:43.4 & 2007-01-27 & 6 x 900 & 0.52 & 1.13 & 24.11 \\ 
07D2ae & 10:01:50.58 & +01:52:33.6 & 2007-01-25 & 4 x 900 & 0.60 & 1.13 & 22.84 \\ 
07D2ag & 10:00:09.01 & +02:09:59.2 & 2007-01-25 & 4 x 900 & 0.73 & 1.16 & 21.48 \\ 
07D2ah & 09:59:58.16 & +01:53:21.9 & 2007-01-26 & 4 x 900 & 0.64 & 1.22 & 23.49 \\ 
07D2aw & 10:02:21.66 & +02:27:06.6 & 2007-02-24 & 4 x 900 & 0.82 & 1.23 & 23.94 \\ 
07D2bd & 09:58:38.09 & +02:07:35.4 & 2007-02-19 & 4 x 900 & 0.88 & 1.12 & 23.15 \\ 
07D2be & 10:02:08.40 & +02:40:00.6 & 2007-02-19 & 4 x 900 & 0.90 & 1.20 & 24.03 \\ 
07D2bi & 09:58:46.65 & +02:40:29.9 & 2007-02-23 & 4 x 900 & 1.00 & 1.13 & 23.58 \\ 
07D2bq & 10:01:51.86 & +02:00:48.4 & 2007-02-27 & 4 x 900 & 1.33 & 1.13 & 23.10 \\ 
07D2cb & 10:01:27.40 & +01:55:47.7 & 2007-03-17 & 4 x 900 & 0.96 & 1.12 & 23.24 \\ 
07D2cq & 10:00:47.03 & +01:52:04.1 & 2007-03-20 & 4 x 900 & 0.94 & 1.26 & 23.34 \\ 
07D2ct & 10:00:17.00 & +02:17:13.8 & 2007-03-21 & 6 x 900 & 0.82 & 1.35 & 23.88 \\ 
07D2du & 10:00:30.20 & +01:51:30.4 & 2007-04-20 & 4 x 900 & 1.24 & 1.34 & 22.88 \\ 
07D2fy & 09:59:32.36 & +02:18:00.9 & 2007-05-15 & 4 x 900 & 0.58 & 1.16 & 23.51 \\ 
07D2fz & 10:00:13.41 & +02:24:16.9 & 2007-05-15 & 4 x 900 & 0.50 & 1.15 & 23.21 \\ 
07D4aa & 22:16:59.49 & -17:52:03.3 & 2007-06-19 & 4 x 900 & 0.59 & 1.02 & 21.18 \\ 
07D4cy & 22:15:02.47 & -17:37:43.2 & 2007-08-22 & 4 x 900 & 1.20 & 1.20 & 23.35 \tablefootmark{e}\\ 
07D4dp & 22:14:33.80 & -17:25:58.9 & 2007-09-10 & 4 x 900 & 0.82 & 1.23 & 23.28 \tablefootmark{e}\\ 
07D4dq & 22:14:02.17 & -17:48:43.4 & 2007-09-10 & 4 x 900 & 0.88 & 1.06 & 22.78 \tablefootmark{e}\\ 
07D4dr & 22:14:43.06 & -17:18:34.1 & 2007-09-10 & 4 x 900 & 1.12 & 1.01 & 23.10 \tablefootmark{e}\\ 
07D4ec & 22:16:09.48 & -18:02:18.8 & 2007-09-19 & 4 x 900 & 1.33 & 1.21 & 25.08 \\ 
07D4ed & 22:15:18.55 & -18:09:52.5 & 2007-09-28 & 4 x 900 & 0.43 & 1.14 & 24.12 \\ 
07D4ei & 22:16:29.93 & -17:32:05.1 & 2007-09-22 & 4 x 900 & 0.76 & 1.18 & 23.28 \tablefootmark{e}\\ 
\hline
\end{longtable}
\tablefoot{
\tablefoottext{a}{Spectrum measured in MOS mode during the first three years of SNLS. The SN is included in the SNLS three-year sample of \cite{Guy10} but the spectrum is not in the VLT three-year spectral set of \citet{Balland09}.}\\
\tablefoottext{b}{Spectrum taken in LSS mode.}\\
\tablefoottext{c}{Observed with Grism 300I and order sorting filter OG590.}\\
\tablefoottext{d}{Identified as a \Ia~by \cite{Bazin11} after a new extraction.}\\
\tablefoottext{e}{Value obtained from online logs.}\\
\tablefoottext{f}{Seeing not available for this object}
}

%% file: Table_Result.tex
   \begin{longtable}{cccccccc}
   \caption{\label{table:result} Spectral properties of the VLT \Iae~of the last two years of SNLS}\\
   \hline
   \hline
   SN name & Type & z\tablefootmark{a} & z source & $\Phi$ & Host model & Host fraction & $\langle S/N \rangle$\tablefootmark{b}\\
\hline
\endfirsthead
\caption{continued.}\\
\hline
   \hline
   SN name & Type & z\tablefootmark{a} & z source & $\Phi$ & Host model & Host fraction & $\langle S/N \rangle$\tablefootmark{b} \\
\hline
\endhead
\hline
\endfoot  
               05D1dx & SNIa & $0.58\pm0.01$ & S & -8.5 & S0(1) & 0.21 & 3.2 \\ 
               05D1dx & SNIa & $0.58\pm0.01$ & S & 12.4 & S0(12) & 0.63 & 0.9 \\ 
               05D1hm & SNIa & $0.587\pm0.001$ & H & 4.5 & E(1) & 0.83 & 1.5 \\ 
               05D1if & SNIa & $0.763\pm0.001$ & H & -5.9 & S0-Sa & 0.54 & 1.0 \\ 
               05D2le & SNIa & $0.700\pm0.001$ & H & 5.9 & NoGalaxy & 0 & 1.2 \\
               06D1bg & SNIa$\star$ & $0.76\pm0.01$ & S & 8.0 & S0(1) & 0.39 & 1.6 \\ 
               06D1bo & SNIa & $0.62\pm0.01$ & S & -3.0 & Sd(1) & 0.2 & 2.4 \\ 
               06D1cm & SNIa & $0.619\pm0.001$ & H & 8.3 & NoGalaxy & 0 & 1.5 \\
               06D1cx & SNIa & $0.860\pm0.001$ & H & -4.2 & NoGalaxy & 0 & 1.5 \\ 
               06D1dc & SNIa$\star$ & $0.767\pm0.001$ & H & 3.8 & E-S0 & 0.77 & 2.8 \\
               06D1dl & SNIa & $0.514\pm0.001$ & H & -5.2 & E(1) & 0.68 & 3.5 \\ 
               06D1du & SNIa & $0.24\pm0.01$ & S & -0.2 & E(1) & 0.04 & 23.3 \\
               06D1eb & SNIa & $0.704\pm0.001$ & H & -5.2 & Sd(1) & 0.42 & 4.8 \\ 
               06D1eb & SNIa & $0.704\pm0.001$ & H & -2.3 & Sd(7) & 0.49 & 2.6 \\
               06D1ez & SNIa & $0.692\pm0.001$ & H & 7.3 & S0(1) & 0.31 & 0.8\\
               06D1fd & SNIa & $0.350\pm0.001$ & H & 4.9 & Sd(13) & 0.33 & 6.9 \\
               06D1fx & SNIa & $0.524\pm0.001$ & H & 6.8 & Sa-Sb & 0.7 & 5.3 \\ 
               06D1gl & SNIa & $0.98\pm0.01$ & S & 4.3 & S0-Sa & 0.31 & 2.3 \\ 
               06D1hi & SNIa$\star$ & $0.803\pm0.001$ & H & -3.3 & E(4) & 0.75 & 3.8 \\ 
               06D1ix & SNIa & $0.65\pm0.01$ & S & 3.8 & Sd(1) & 0.09 & 2.9 \\ 
               06D1jf & SNIa & $0.641\pm0.001$ & H & 1.5 & Sc(4) & 0.6 & 1.9 \\ 
               06D1jz & SNIa & $0.346\pm0.001$ & H & 3.3 & S0(7) & 0.73 & 23.6 \\ 
               06D1kf & SNIa & $0.561\pm0.001$ & H & -6.5 & Sd(1) & 0.26 & 2.3 \\ 
               06D1kg & SNIa & $0.32\pm0.01$ & S & 6.1 & S0(2) & 0.5 & 3.7 \\ 
               06D1kh & SNIa$\star$ & $0.882\pm0.001$ & H & 7.3 & E(1) & 0.37 & 1.1 \\ 
               06D2ag & SNIa & $0.310\pm0.001$ & H & 4.0 & NoGalaxy & 0 & 13.8 \\ 
               06D2bo & SNIa$\star$ & $0.82\pm0.01$ & S & 2.6 & Sa(1) & 0.54 & 0.6 \\ 
               06D2hm & SNIa & $0.56\pm0.01$ & S & 7.9 & Sa(3) & 0.02 & 4.9 \\ 
               06D2hu & SNIa & $0.342\pm0.001$ & H & 7.2 & E-S0 & 0.71 & 16.1 \\ 
               06D2jw & SNIa$\star$ & $0.90\pm0.01$ & S & -0.1 & E(1) & 0.47 & 1.7 \\ 
               06D4ba & SNIa & $0.70\pm0.01$ & S & 9.2 & Sd(2) & 0.17 & 1.6 \\ 
               06D4bo & SNIa & $0.552\pm0.001$ & H & 1.0 & S0-Sb & 0.52 & 5.6 \\ 
               06D4bw & SNIa & $0.732\pm0.001$ & H & 5.8 & Sa(1) & 0.48 & 2.0 \\ 
               06D4gs & SNIa & $0.31\pm0.01$ & S & -4.2 & E(1) & 0.24 & 6.1 \\
               06D4jh & SNIa & $0.566\pm0.001$ & H & 3.7 & Sd(2) & 0.49 & 3.3 \\ 
               06D4jt & SNIa$\star$ & $0.76\pm0.01$ & S & 2.9 & Sd(1) & 0.44 & 1.4 \\ 
               07D1ab & SNIa & $0.328\pm0.001$ & H & -0.2 & E(1) & 0.51 & 8.0 \\ 
               07D1ad & SNIa & $0.297\pm0.001$ & H & 6.9 & S0(12) & 0.69 & 8.8 \\ 
               07D1ah & SNIa-pec & $0.342\pm0.001$ & H & -0.6 & E(1) & 0.26 & 7.6 \\ 
               07D1bl & SNIa & $0.636\pm0.001$ & H & 2.0 & E(2) & 0.50 & 4.3 \\ 
               07D1bs & SNIa$\star$ & $0.617\pm0.001$ & H & 0.7 & Sa-Sb & 0.8 & 1.6 \\ 
               07D1bu & SNIa & $0.626\pm0.001$ & H & -2.8 & Sd(5) & 0.47 & 4.7 \\ 
               07D1by & SNIa & $0.73\pm0.01$ & S & -0.5 & Sd(1) & 0.05 & 2.6 \\ 
               07D1ca & SNIa$\star$ & $0.835\pm0.001$ & H & 1.4 & Sa(1) & 0.42 & 2.0 \\ 
               07D1cc & SNIa & $0.853\pm0.001$ & H & 1.2 & Sa-Sb & 0.49 & 1.9 \\ 
               07D1cd & SNIa$\star$ & $0.873\pm0.001$ & H & 4.1 & S0-Sa & 0.88 & 0.6 \\ 
               07D1cf & SNIa & $0.500\pm0.001$ & H & -8.4 & E(1) & 0.33 & 3.1 \\ 
               07D2aa & SNIa & $0.899\pm0.001$ & H & -1.9 & S0(12) & 0.69 & 3.3 \\ 
               07D2ae & SNIa & $0.501\pm0.001$ & H & 1.7 & S0(1) & 0.43 & 7.9 \\ 
               07D2ag & SNIa & $0.25\pm0.01$ & S & -2.6 & S0(5) & 0.19 & 19.6 \\ 
               07D2ah & SNIa & $0.780\pm0.001$ & H & -0.6 & S0(1) & 0.32 & 3.9 \\ 
               07D2aw & SNIa$\star$ & $0.610\pm0.001$ & H & 10.0 & E(1) & 0.65 & 1.1 \\ 
               07D2bd & SNIa & $0.572\pm0.001$ & H & 2.1 & Sa-Sb & 0.66 & 3.5 \\ 
               07D2be & SNIa$\star$ & $0.793\pm0.001$ & H & 7.0 & Sc(1) & 0.54 & 1.3 \\ 
               07D2bi & SNIa & $0.551\pm0.001$ & H & 0.9 & S0(1) & 0.64 & 1.6 \\ 
               07D2bq & SNIa & $0.535\pm0.001$ & H & -3.5 & E(1) & 0.6 & 2.0 \\ 
               07D2cb & SNIa & $0.694\pm0.001$ & H & 1.8 & Sd(1) & 0.38 & 2.7 \\ 
               07D2cq & SNIa$\star$ & $0.746\pm0.001$ & H & 1.1 & E(2) & 0.62 & 2.6 \\ 
               07D2ct & SNIa$\star$ & $0.94\pm0.01$ & S & 1.9 & Sa-Sb & 0.63 & 0.6 \\ 
               07D2du & SNIa & $0.538\pm0.001$ & H & -1.4 & E(1) & 0.39 & 3.3 \\ 
               07D2fy & SNIa & $0.72\pm0.01$ & S & 0.3 & NoGalaxy & 0 & 2.4 \\ 
               07D2fz & SNIa & $0.743\pm0.001$ & H & -1.4 & E-S0(1.0) & 0.18 & 4.7 \\ 
               07D4aa & SNIa & $0.207\pm0.001$ & H & 13.9 & Sb-Sc & 0.28 & 32.9 \\ 
               07D4cy & SNIa$\star$ & $0.456\pm0.001$ & H & -0.1 & Sd(9) & 0.9 & 0.6 \\ 
               07D4dp & SNIa$\star$ & $0.743\pm0.001$ & H & -1.8 & Sd(11) & 0.7 & 2.7 \\ 
               07D4dq & SNIa & $0.554\pm0.001$ & H & 1.9 & E(3) & 0.78 & 5.7 \\ 
               07D4dr & SNIa & $0.772\pm0.001$ & H & 2.2 & E(4) & 0.69 & 2.7 \\ 
               07D4ec & SNIa & $0.653\pm0.001$ & H & -4.0 & Sa-Sb & 0.76 & 2.0 \\ 
               07D4ed & SNIa & $0.52\pm0.01$ & S & -1.5 & NoGalaxy & 0 & 3.0 \\ 
               07D4ei & SNIa & $0.37\pm0.01$ & S & -6.7 & S0(1) & 0.39 & 1.5 \\ 
\hline
\end{longtable}  
\tablefoot{
\tablefoottext{a}{Not corrected to the heliocentric reference frame.}\\
\tablefoottext{b}{Computed in 5\AA~bins.}\\
}

%% file: Table_MOS_short.tex
   \begin{longtable}{ccccc}
   \caption{\label{table:MOS} Catalog of redshifts and identifications
     of the non-SN~Ia objects measured in MOS mode at the VLT during
     the SNLS survey. We use the ID column to distinguish between host
     galaxies that were targeted after the transient had faded from
     view, live transients (using the labels SNIbc, SNII, SNII? or ?),
     and random field galaxies. An asterisk next to a redshift value denotes a redshift obtained from a single identified line.}\\ \hline \hline Name & RA (J2000) &
   Dec (J2000) & z & ID \\ \hline \endfirsthead
\caption{continued.}\\
\hline
   \hline
   Name & RA (J2000) & Dec (J2000) & z & ID \\
\hline
\endhead
\hline
\endfoot  
              & 02:25:51.01 & -04:38:38.6 & 0.996* & 	Field Galaxy                   \\ 
       05D1hb & 02:24:28.93 & -04:45:23.7 & 0.764	 & ?                              \\ 
              & 22:16:44.39 & -17:20:18.8 & 0.784	 & Field Galaxy                   \\ 
       05D4jy & 22:16:23.06 & -17:47:01.6 & 0.869	 & ?                              \\ 
       06D1aq & 02:26:57.80 & -04:03:43.6 & 0.332	 & ?                              \\ 
              & 02:26:15.82 & -04:20:49.9 & 0.178	 & Field Galaxy                   \\ 
              & 02:25:09.01 & -04:33:13.2 & 0.269* & 	Field Galaxy                   \\ 
              & 02:25:09.01 & -04:33:13.2 & 0.767	 & Field Galaxy                   \\ 
       06D1dv & 02:26:36.85 & -04:44:37.5 & 0.062	 & SNIbc                          \\ 
              & 02:26:36.70 & -04:44:33.7 & 0.700	 & Field Galaxy                   \\ 
       06D1hc & 02:24:48.25 & -04:56:03.6 & 0.555* & 	SNII?                          \\ 
       06D1jd & 02:27:36.19 & -04:31:56.6 & 0.324	 & SNII                           \\ 
       06D1jx & 02:24:34.75 & -04:57:51.7 & 0.14 	 & SNII                           \\ 
              & 10:01:43.28 & +01:51:35.1 & 0.309	 & Field Galaxy                   \\ 
              & 10:01:43.67 & +01:51:37.3 & 0.794* & 	Field Galaxy                   \\ 
              & 10:01:43.74 & +01:51:38.7 & 0.682* & 	Field Galaxy                   \\ 
       06D2bb & 09:59:35.02 & +02:17:10.1 & 0.313	 & SNII?                          \\ 
       06D2bt & 09:59:01.76 & +02:36:59.1 & 0.079	 & SNII                           \\ 
              & 09:59:57.22 & +02:07:57.2 & 0.658* & 	Field Galaxy                   \\ 
       06D2iy & 10:01:35.56 & +02:26:46.8 & 0.392	 & SNII?                          \\ 
       06D4eu & 22:15:54.29 & -18:10:45.6 & 1.588	 & SLSN                           \\ 
       07D1bw & 02:27:57.03 & -04:37:27.6 & 0.286	 & SNII?                          \\ 
       07D1cd & 02:25:33.96 & -04:45:06.5 & 0.873	 & ?                              \\ 
       07D1ci & 02:25:36.71 & -04:43:26.0 & 0.319	 & SNII?                          \\ 
              & 02:25:36.54 & -04:43:26.9 & 0.319	 & Field Galaxy                   \\ 
       07D2ab & 10:02:25.74 & +02:19:39.8 & 0.312	 & SNII                           \\ 
       07D2an & 10:00:08.93 & +02:36:14.1 & 0.135	 & SNII                           \\ 
       07D2at & 10:02:15.70 & +02:05:26.1 & 0.216	 & SNII?                          \\ 
       07D2bv & 10:00:06.63 & +02:38:35.8 & 1.50 	 & SLSN                           \\ 
       07D2ca & 09:58:49.62 & +02:31:17.3 & 0.507	 & SNII?                          \\ 
       07D2ge & 10:01:28.26 & +02:42:31.9 & 0.084	 & SNII                           \\ 
       07D4af & 22:16:23.34 & -18:12:35.8 & 0.135	 & SNII                           \\ 
       07D4ck & 22:16:02.21 & -17:39:42.4 & 0.581	 & SNII?                          \\ 
              & 22:14:33.72 & -17:25:57.7 & 0.743	 & Field Galaxy                   \\ 
       07D4ds & 22:15:54.73 & -17:44:54.8 & 0.338	 & ?                              \\ 
       07D4dt & 22:13:50.13 & -17:36:51.8 & 0.677	 & ?                              \\ 
       07D4ee & 22:15:29.69 & -18:04:40.1 & 0.470	 & ?                              \\ 
       03D1ad & 02:27:32.66 & -04:29:23.7 & 0.524	 & Host galaxy                    \\ 
              & 02:27:32.66 & -04:29:25.2 & 0.525	 & Field Galaxy                   \\ 
       03D1af & 02:24:12.67 & -04:26:14.2 & 0.603	 & Host galaxy                    \\ 
       03D1am & 02:24:13.84 & -04:26:02.0 & 0.556	 & Host galaxy                    \\ 
              & 02:24:13.97 & -04:25:53.1 & 0.958* & 	Field Galaxy                   \\ 
       03D1ap & 02:26:35.46 & -04:46:03.8 & 0.513	 & Host galaxy                    \\ 
       03D1aq & 02:25:03.08 & -04:05:01.8 & 0.706	 & Host galaxy                    \\ 
       03D1aw & 02:24:14.72 & -04:31:01.4 & 0.582	 & Host galaxy                    \\ 
       03D1bc & 02:27:38.48 & -04:41:48.5 & 0.383	 & Host galaxy                    \\ 
              & 02:27:37.97 & -04:41:59.0 & 0.384	 & Field Galaxy                   \\ 
       03D1bg & 02:27:05.91 & -04:47:34.4 & 0.512	 & Host galaxy                    \\ 
       03D1by & 02:27:54.03 & -04:03:04.2 & 0.378	 & Host galaxy                    \\ 
       03D1ch & 02:24:29.12 & -04:09:54.9 & 0.265	 & Host galaxy                    \\ 
       03D1da & 02:25:03.19 & -04:05:39.3 & 0.785	 & Host galaxy, AGN               \\ 
              & 02:25:04.47 & -04:05:34.4 & 0.172	 & Field Galaxy                   \\ 
       03D1dg & 02:25:19.84 & -04:30:46.0 & 0.496	 & Host galaxy                    \\ 
       03D1ea & 02:27:50.37 & -04:05:01.9 & 0.312	 & Host galaxy                    \\ 
       03D1et & 02:24:26.93 & -04:47:54.4 & 0.855	 & Host galaxy                    \\ 
       03D1ft & 02:27:07.44 & -04:04:38.7 & 0.491	 & Host galaxy                    \\ 
       03D1fy & 02:27:16.05 & -04:24:33.4 & 0.177	 & Host galaxy                    \\ 
       03D1gi & 02:25:18.13 & -04:31:55.5 & 0.525* & 	Host galaxy                    \\ 
       03D1gl & 02:27:27.82 & -04:08:07.0 & 0.634	 & Host galaxy, AGN               \\ 
       03D4au & 22:16:09.92 & -18:04:39.0 & 0.468	 & Host galaxy                    \\ 
       03D4az & 22:15:47.78 & -18:07:51.2 & 0.409* & 	Host galaxy                    \\ 
       03D4cb & 22:15:41.48 & -18:12:44.8 & 0.517	 & Host galaxy                    \\ 
       03D4cl & 22:15:38.20 & -18:06:26.8 & 0.90*	 & Host galaxy, AGN               \\ 
       03D4dl & 22:13:35.29 & -17:18:03.2 & 0.305	 & Host galaxy                    \\ 
       03D4ec & 22:14:43.72 & -17:21:40.7 & 1.016* & 	Host galaxy                    \\ 
       03D4ed & 22:16:19.76 & -17:31:27.5 & 0.860* & 	Host galaxy                    \\ 
       03D4ev & 22:16:51.40 & -17:20:03.1 & 0.538	 & Host galaxy                    \\ 
       03D4fb & 22:14:27.22 & -17:22:40.2 & 0.291	 & Host galaxy                    \\ 
       03D4gj & 22:16:01.40 & -18:05:20.7 & 0.318	 & Host galaxy                    \\ 
       04D1aa & 02:26:06.22 & -04:22:33.8 & 0.526	 & Host galaxy                    \\ 
       04D1ab & 02:25:37.75 & -04:42:40.2 & 0.241	 & Host galaxy                    \\ 
              & 02:25:37.85 & -04:42:36.0 & 0.265	 & Field Galaxy                   \\ 
\end{longtable}
\noindent
See online material for full table.

%% file: Tables_mean.tex

\begin{table}[ht!]
\caption{Mean properties of the VLT \Ia~and \Ia$\star$~subsamples of the last two years of SNLS. $<S/N>$ are computed in 5 \AA~bins. Errors are 1$\sigma$ on the mean (dispersion is given in parentheses).}
\label{table:VLT5_propri_IaIa?}
\centering
\begin{tabular}{c c c c}
\hline
\hline
& 51 \Iae & 16 \Iae$\star$ & 67 \Iae~+ \Iae$\star$ \\
& 53 spectra & 16 spectra & 69 spectra\\
\hline
$\langle z \rangle\;(\sigma_{z})$ & $0.57 \pm 0.03$ (0.18) &  $0.77 \pm 0.03$ (0.12) & $0.62 \pm 0.02$ (0.19) \\
$\langle \Phi \rangle\;(\sigma_{\Phi})$ & $1.0 \pm 0.7$ (5.1) & $2.8 \pm 0.9$ (3.7) & $1.4 \pm 0.6$ (4.9) \\
$\langle f_{gal}\rangle\;(\sigma_{f_{gal}})$ & $0.39 \pm 0.03$ (0.25) & $0.61 \pm 0.04$ (0.17) & $0.44 \pm 0.03$ (0.25) \\
$\langle S/N \rangle\;(\sigma_{S/N})$ & $5.5 \pm 0.9$ (6.5) &  $1.6 \pm 0.2$ (0.9) & $4.6 \pm 0.7$ (5.9)\\
$\langle \gamma_{1} \rangle\;(\sigma_{\gamma_{1}})$ & $0.42\pm0.12$ (0.86) & $0.66\pm0.39$ (1.57) & $0.48\pm0.13$ (1.06)\\
$\langle m_{B}^{*}\rangle\;(\sigma_{m_{B}^{*}})$ & $23.285 \pm 0.127$ (0.895) & $24.217 \pm 0.086$ (0.342) & $23.511 \pm 0.110$ (0.890) \\
$\langle m_{B}^{*\;c} \rangle\;(\sigma_{m_{B}^{*\;c}})$ & $ 23.965 \pm 0.045$ (0.319) & $ 23.981 \pm 0.089$ (0.357) & $23.969 \pm 0.040$ (0.326) \\
$\langle c \rangle\;(\sigma_{c}) $ & $-0.016 \pm 0.014$ (0.101) & $-0.045 \pm 0.033$ (0.131) &  $-0.023 \pm 0.013$ (0.109) \\
$\langle x_{1} \rangle\;(\sigma_{x_{1}})$ & $0.215 \pm 0.109$ (0.772) &  $0.065 \pm 0.226$ (0.903) & $0.180 \pm 0.099$ (0.801) \\
$\langle s \rangle\;(\sigma_{s})$ &  $1.001 \pm0.010$ (0.070) & $0.988 \pm 0.020$ (0.081) &  $0.998 \pm 0.009$ (0.072)\\
\hline
\end{tabular}
\end{table}

%

\begin{table}[h!]
\caption{Mean spectro-photometric properties of the SNLS - VLT
  \Ia~+\Ias~subsamples.  $<S/N>$ are computed in 5 \AA~bins. Errors are 1$\sigma$ on the mean (dispersion
  is given in parentheses). The spectra of \Iae~identified as
  \Ia$_{pec}$ (two SN in the first three years sample and one in the final
  two years sample) have been discarded.}
\label{table:VLT5_propri_3+5}
\centering
\begin{tabular}{c c c c}
\hline
\hline
 & VLT (first 3 years) & VLT (final 2 years) & VLT (all 5 years) \\
& 122 \Iae~+ \Iaes & 67 \Iae~+ \Iaes & 189 \Iae~+ \Iaes \\
& 137 spectra & 69 spectra & 206 spectra\\
\hline
$\langle z \rangle\;(\sigma_{z})$ & $0.64 \pm 0.02$ (0.21) & $0.62 \pm 0.02$ (0.19) & $0.63 \pm 0.01$ (0.20) \\
$\langle \Phi \rangle\;(\sigma_{\Phi})$ & $2.9 \pm 0.5$ (5.6) & $1.4 \pm 0.6$ (4.9) & $2.4 \pm 0.4$ (5.4) \\
$\langle f_{gal}\rangle\;(\sigma_{f_{gal}})$ & $0.24 \pm 0.02$ (0.28) & $0.44 \pm 0.03$ (0.25) & $0.31 \pm 0.02$ (0.29) \\
$\langle S/N \rangle\;(\sigma_{S/N})$ & $4.4 \pm 0.4$ (4.6) & $4.6 \pm 0.7$ (5.9) & $4.5 \pm 0.4$ (5.1) \\
$\langle m_{B}^{*}\rangle\;(\sigma_{m_{B}^{*}})$ & $23.621 \pm 0.080$ (0.832) & $23.511 \pm 0.110$ (0.890) & $23.582 \pm 0.062$ (0.852) \\
$\langle m_{B}^{*\;c} \rangle\;(\sigma_{m_{B}^{*\;c}})$ & $ 24.005 \pm 0.037$ (0.407) & $23.969 \pm 0.040$ (0.326) & $23.992 \pm 0.028$ (0.380) \\
$\langle c \rangle\;(\sigma_{c}) $ & $-0.016 \pm 0.011$ (0.121) & $-0.023 \pm 0.013$ (0.109) & $-0.019 \pm 0.009$ (0.117) \\
$\langle x_{1} \rangle\;(\sigma_{x_{1}})$ & $0.390 \pm 0.088$ (0.968) & $0.180 \pm 0.099$ (0.801) & $0.315 \pm 0.067$ (0.916) \\
$\langle s \rangle\;(\sigma_{s})$ & $1.018 \pm 0.008$ (0.087) & $0.998 \pm 0.009$ (0.072) & $1.011 \pm 0.006$ (0.083) \\
\hline
\end{tabular}
\end{table}

\begin{table}[h!]
\caption{Mean spectro-photometric properties of the blue ($c<0$) and red ($c\geq0$) \Iae~ samples used to build the VLT five year composite spectra. Errors are 1$\sigma$ on the mean.}
\label{table:color}
\centering
\begin{tabular}{c c c}
\hline
\hline
& $c<0$ & $c\geq0$ \\
\hline
Nb spec & 40 & 31 \\
$\langle z \rangle$ & $0.66\pm0.02$ & $0.57\pm0.03$ \\
$\langle \Phi \rangle$ & $0.3\pm0.4$& $0.3\pm0.4$ \\
$\langle c \rangle$ & $-0.098\pm0.008$ & $0.056\pm0.007$ \\
$\langle s \rangle$ & $1.014\pm0.010$ &  $1.013\pm0.012$ \\
$\langle m_{B}^{*\;c} \rangle$ & $23.796\pm0.025$ & $24.091\pm0.046$ \\
\hline
\end{tabular}
\end{table}


\begin{table}[h!]
\caption{Mean spectro-photometric properties of the low ($z<0.6$) and high ($z\geq0.6$) redshift \Iae~ samples used to build the VLT five year composite spectra. Errors are 1$\sigma$ on the mean.}
\label{table:evo_z}
\centering
\vspace{0.5cm}
\begin{tabular}{c c c}
\hline
\hline
& $z < 0.6$ & $z\geq 0.6$ \\
\hline
Nb spec & 30 & 41 \\
$\langle z \rangle$ & $0.47\pm0.02$ & $0.73\pm0.01$\\
$\langle \Phi \rangle$ & $0.5\pm0.4$ & $0.1\pm0.4$\\
$\langle c \rangle$ & $0.001\pm0.015$ & $-0.055\pm0.014$\\
$\langle s \rangle$ & $1.018\pm0.010$ & $1.010\pm0.011$\\
$\langle m_{B}^{*\;c} \rangle$ & $24.022\pm0.048$ & $23.853\pm0.035$\\
\hline
\end{tabular}

\end{table}


\begin{table}[h!]
\caption{Mean spectro-photometric properties of the low ($s<1.013$) and high ($s\geq1.013$) stretch \Iae~samples used to build the VLT five year composite spectra. Errors are 1$\sigma$ on the mean.}
\label{table:stretch}
\centering
\begin{tabular}{c c c}
\hline
\hline
& $s<1.013$ & $s\geq1.013$ \\
\hline
Nb spec & 34 & 37 \\
$\langle z \rangle$ & $0.63\pm0.03$ & $0.61\pm0.03$ \\
$\langle \Phi \rangle$ & $0.0\pm0.4$& $0.6\pm0.3$ \\
$\langle c \rangle$ & $-0.033\pm0.014$ & $-0.029\pm0.016$ \\
$\langle s \rangle$ & $0.961\pm0.007$ &  $1.061\pm0.007$ \\
$\langle m_{B}^{*\;c} \rangle$ & $23.948\pm0.040$ & $23.903\pm0.044$ \\
\hline
\end{tabular}
\end{table}


\begin{table}[h!]
\caption{Mean spectro-photometric properties of the low ($\log{(M_{stellar})}<10.06$~M$_{\odot}$) and high ($\log{(M_{stellar})}\geq10.06$~M$_{\odot}$) stellar mass \Iae~samples used to build the VLT five year composite spectra. Errors are 1$\sigma$ on the mean.}
\label{table:mass}
\centering
\begin{tabular}{c c c}
\hline
\hline
& $\log{(M_{stellar})}<10.06$~M$_{\odot}$ & $\log{(M_{stellar})}\geq10.06$~M$_{\odot}$ \\
\hline
Nb spec & 27 & 35 \\
$\langle z \rangle$ & $0.58\pm0.03$ & $0.64\pm0.03$ \\
$\langle \Phi \rangle$ & $0.2\pm0.4$& $0.3\pm0.4$ \\
$\langle c \rangle$ & $-0.035\pm0.018$ & $-0.030\pm0.014$ \\
$\langle s \rangle$ & $1.027\pm0.011$ &  $0.996\pm0.012$ \\
$\langle m_{B}^{*\;c} \rangle$ & $23.940\pm0.057$ & $23.922\pm0.039$ \\
$\langle \log{M_{stellar}} \rangle$ & $9.31\pm0.10$~M$_{\odot}$ & $10.64\pm0.06$~M$_{\odot}$\\
\hline
\end{tabular}
\end{table}


\begin{table}[h!]
\caption{Mean spectro-photometric properties of the low ($\log{(M_{stellar})}<10.06$~M$_{\odot}$) and high ($\log{(M_{stellar})}\geq10.06$~M$_{\odot}$) stellar mass \Iae~samples with matching photometric parameter distributions. Errors are $1\sigma$ on the mean.}
\label{table:mass-Same}
\centering
\begin{tabular}{c c c}
\hline
\hline
& $\log{(M_{stellar})}<10.06$~M$_{\odot}$ & $\log{(M_{stellar})}\geq10.06$~M$_{\odot}$ \\
\hline
Nb spec & 18 & 18 \\
$\langle z \rangle$ & $0.59\pm0.03$ & $0.65\pm0.03$ \\
$\langle \Phi \rangle$ & $-0.1\pm0.6$& $0.6\pm0.6$ \\
$\langle c \rangle$ & $-0.038\pm0.023$ & $-0.036\pm0.023$ \\
$\langle s \rangle$ & $1.014\pm0.013$ &  $1.014\pm0.013$ \\
$\langle m_{B}^{*\;c} \rangle$ & $23.994\pm0.072$ & $23.904\pm0.054$ \\
$\langle \log{M_{stellar}} \rangle$ & $9.35\pm0.11$~M$_{\odot}$ & $10.65\pm0.08$~M$_{\odot}$\\
\hline
\end{tabular}
\end{table}


\begin{table}[h!]
\caption{Mean properties of the low ($z<0$) and high ($z\geq0.6$) redshift \Iae~samples with matching photometric parameter distribution. Errors are 1$\sigma$ on the mean.}
\label{table:evo_z-same}
\centering
\begin{tabular}{c c | c}
\hline
\hline
& $z<0.6$ & $z\geq0.6$ \\
\hline
Nb spec & 14 & 19\\
$\langle z \rangle$ & $0.49\pm0.03$ & $0.73\pm0.02$ \\
$\langle \Phi \rangle$ & $0.2\pm0.7$& $0.1\pm0.5$ \\
$\langle c \rangle$ & $-0.073\pm0.011$ & $-0.087\pm0.010 $ \\
$\langle s \rangle$ & $1.017\pm0.013$ &  $1.015\pm0.018$ \\
$\langle m_{B}^{*\;c} \rangle$ & $23.824\pm0.042$ & $23.792\pm0.030$ \\
\hline
\end{tabular}
\end{table}


\begin{table}[h!]
\caption{Differences in the mean spectro-photometric properties of the low and high redshift \Iae~ samples as a function of the redshift gap introduced between the two subsamples (see text for details). Errors are 1$\sigma$ on the mean.}
\label{table:gap}
\centering
\begin{tabular}{c | c c c c }
\hline
\hline
z-gap & 0.1 & 0.2 & 0.3 & 0.4\\
\hline
$\Delta z$ & $0.26\pm0.02$ & $0.40\pm0.04$ & $0.49\pm0.04$ & $0.60\pm0.3$ \\
$\Delta c$ & $-0.056\pm0.021$ & $-0.062\pm0.028$ & $-0.057\pm0.033$& $-0.089\pm0.044$\\
$\Delta s$ & $-0.008\pm0.015$ & $0.005\pm0.022$ & $0.029\pm0.026$ & $0.019\pm0.038$\\
$\Delta m_{B}^{*\;c}$ & $-0.169\pm0.059$ & $-0.190\pm0.096$ & $-0.218\pm0.093$ & $-0.363\pm0.124$\\
\hline
\end{tabular}
\end{table}

%% file: BookSpec_SNLSVLT5yrs.tex
    \begin{figure}
    \begin{center}
    \includegraphics[scale=0.45]{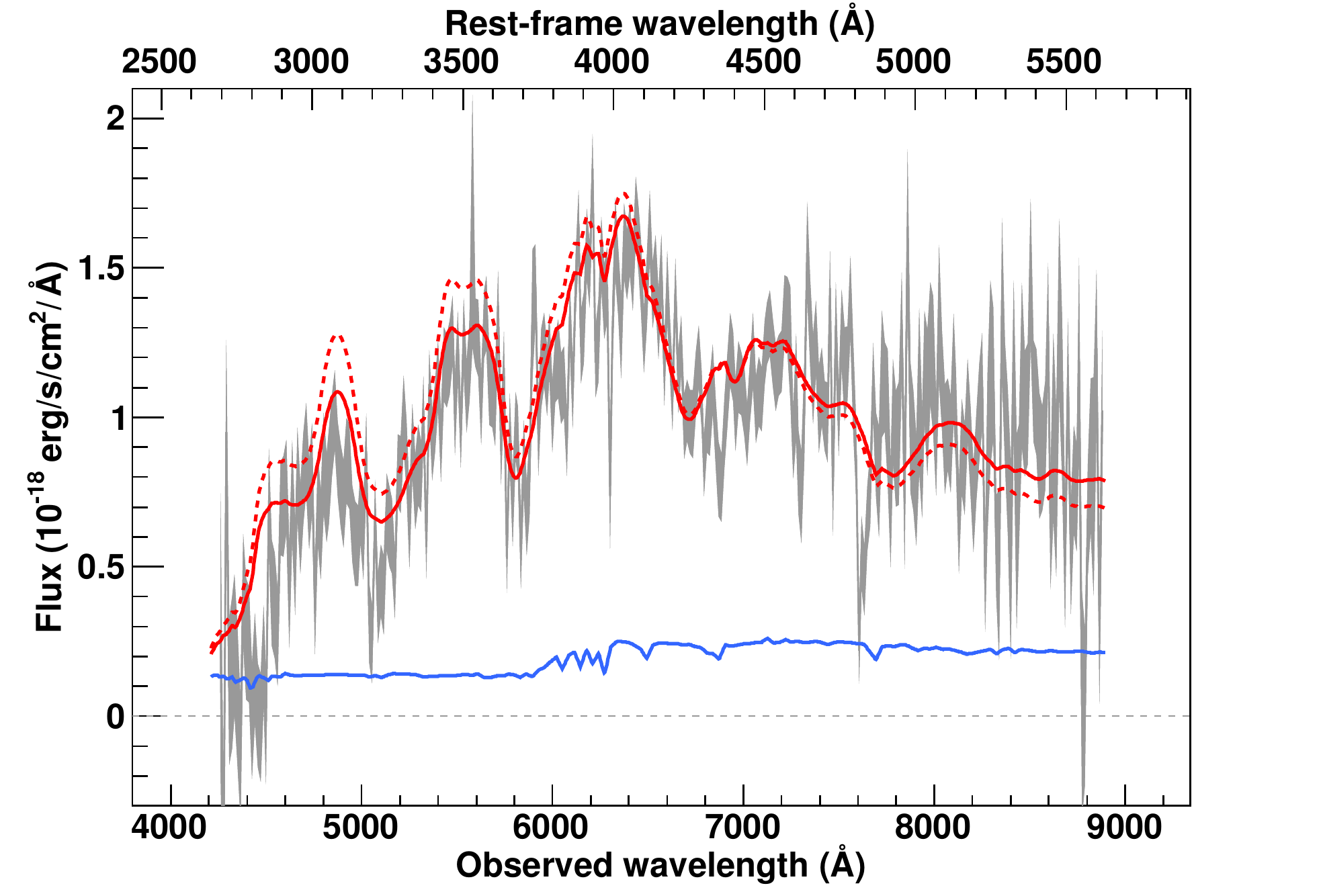}
    \includegraphics[scale=0.45]{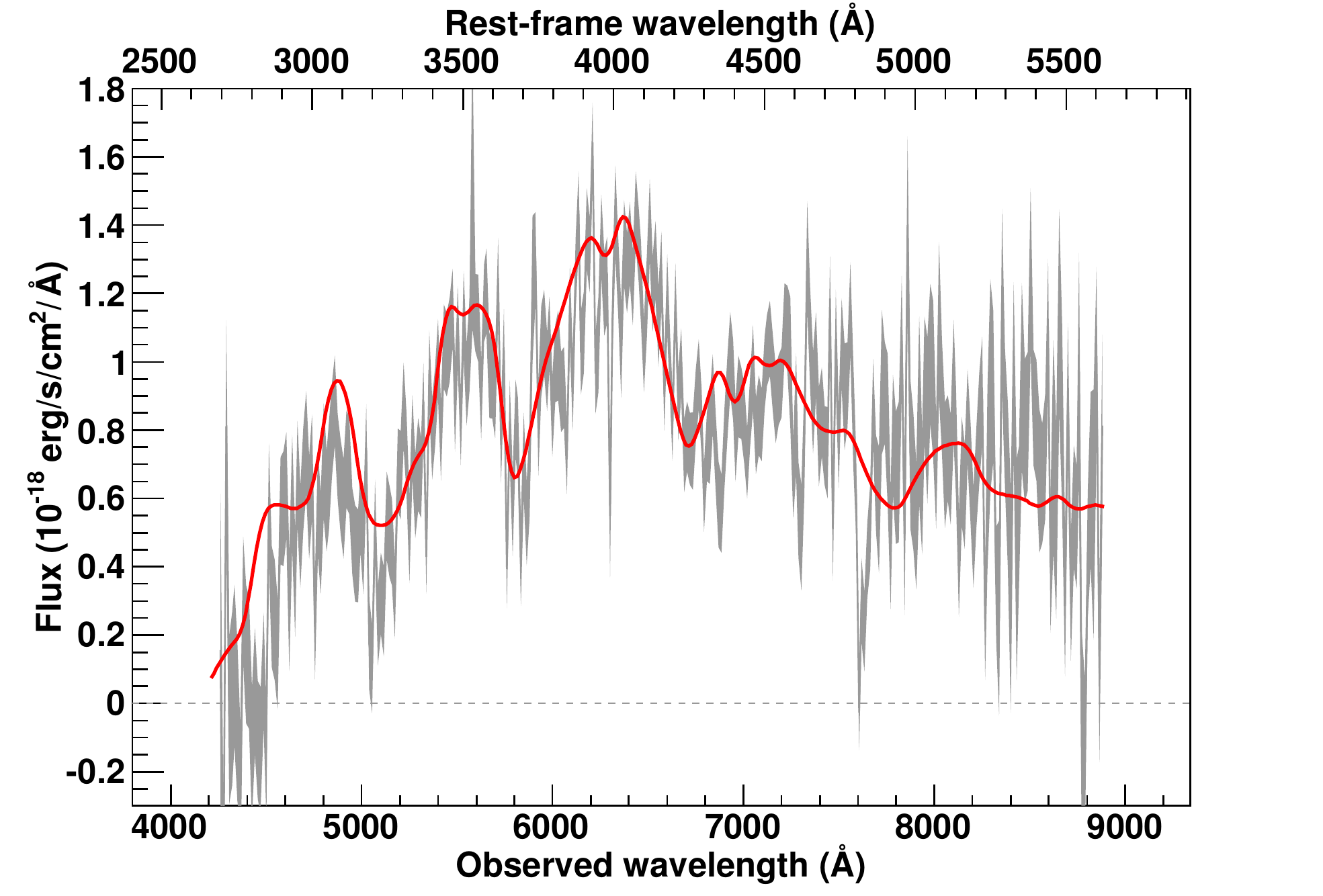}
    \end{center}
    \caption{The SNIa 05D1dx\_1013 spectrum measured at $z=0.58$ with a phase of -8.5 days. A S0(1) host model has been subtracted.}
    \label{fig:Spec05D1dx_1013}
    \end{figure}
    
    \begin{figure}
    \begin{center}
    \includegraphics[scale=0.45]{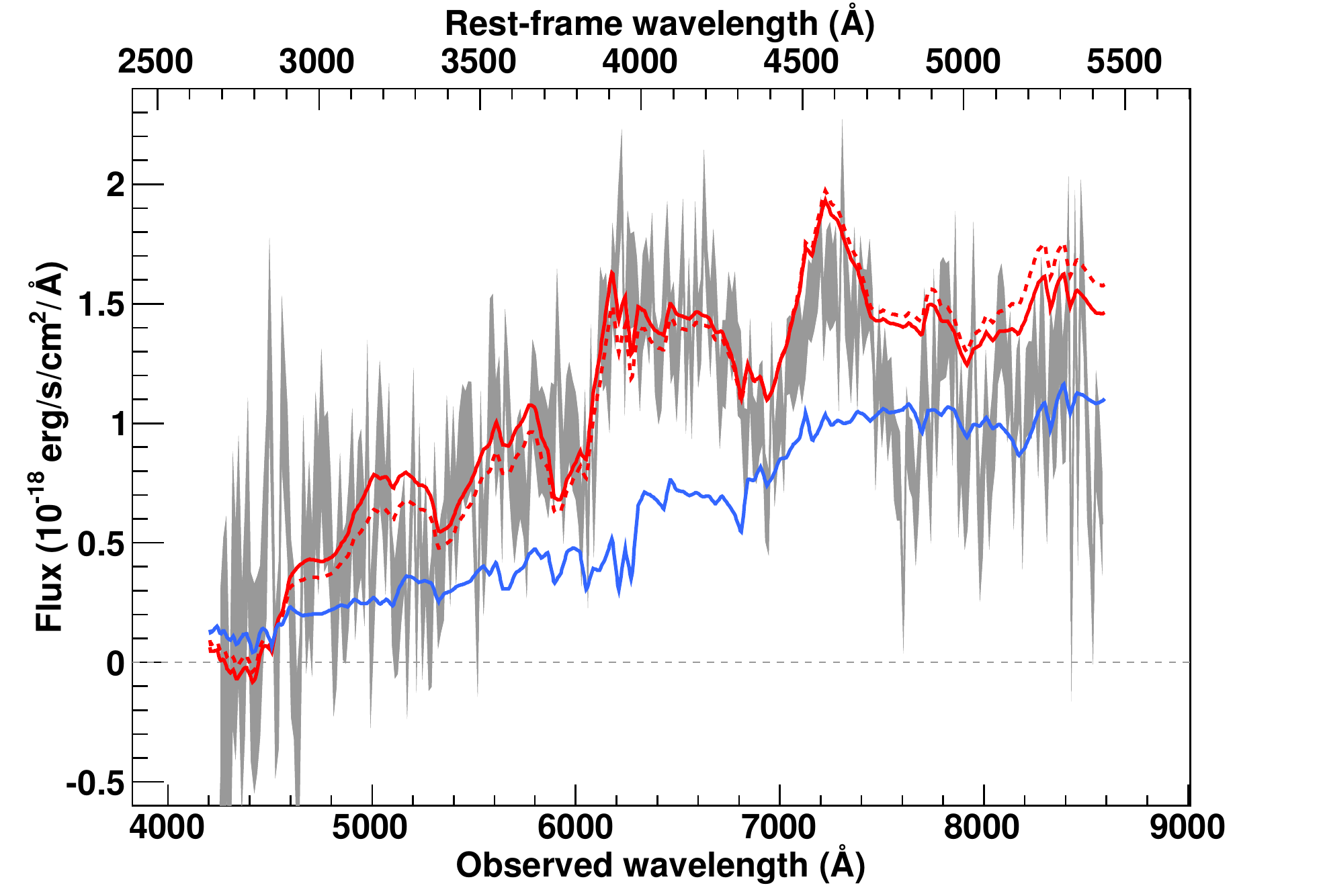}
    \includegraphics[scale=0.45]{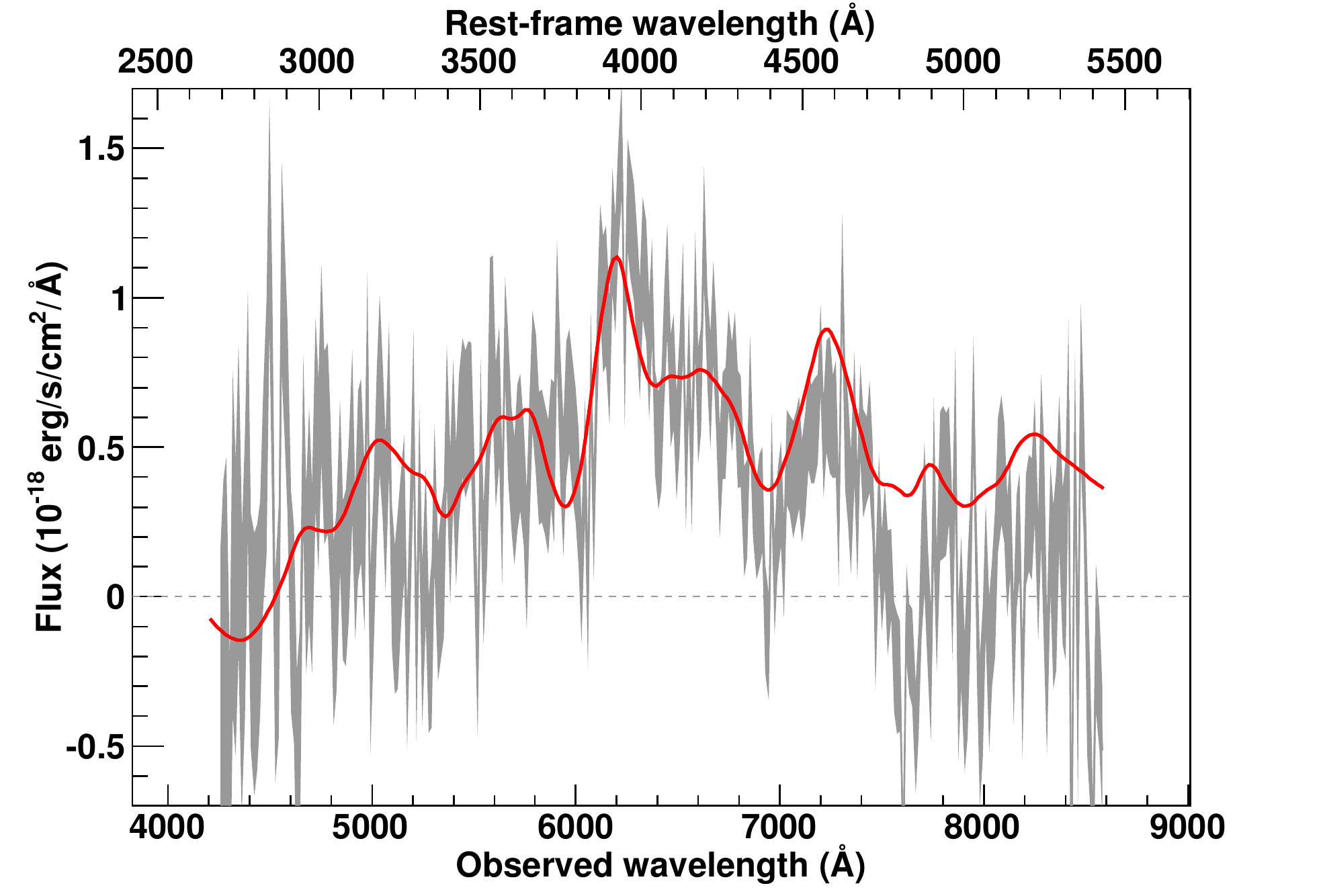}
    \end{center}
    \caption{The SNIa 05D1dx\_1046 spectrum measured at $z=0.58$ with a phase of 12.4 days. A S0(12) host model has been subtracted.}
    \label{fig:Spec05D1dx_1046}
    \end{figure}
    
    \begin{figure}
    \begin{center}
    \includegraphics[scale=0.45]{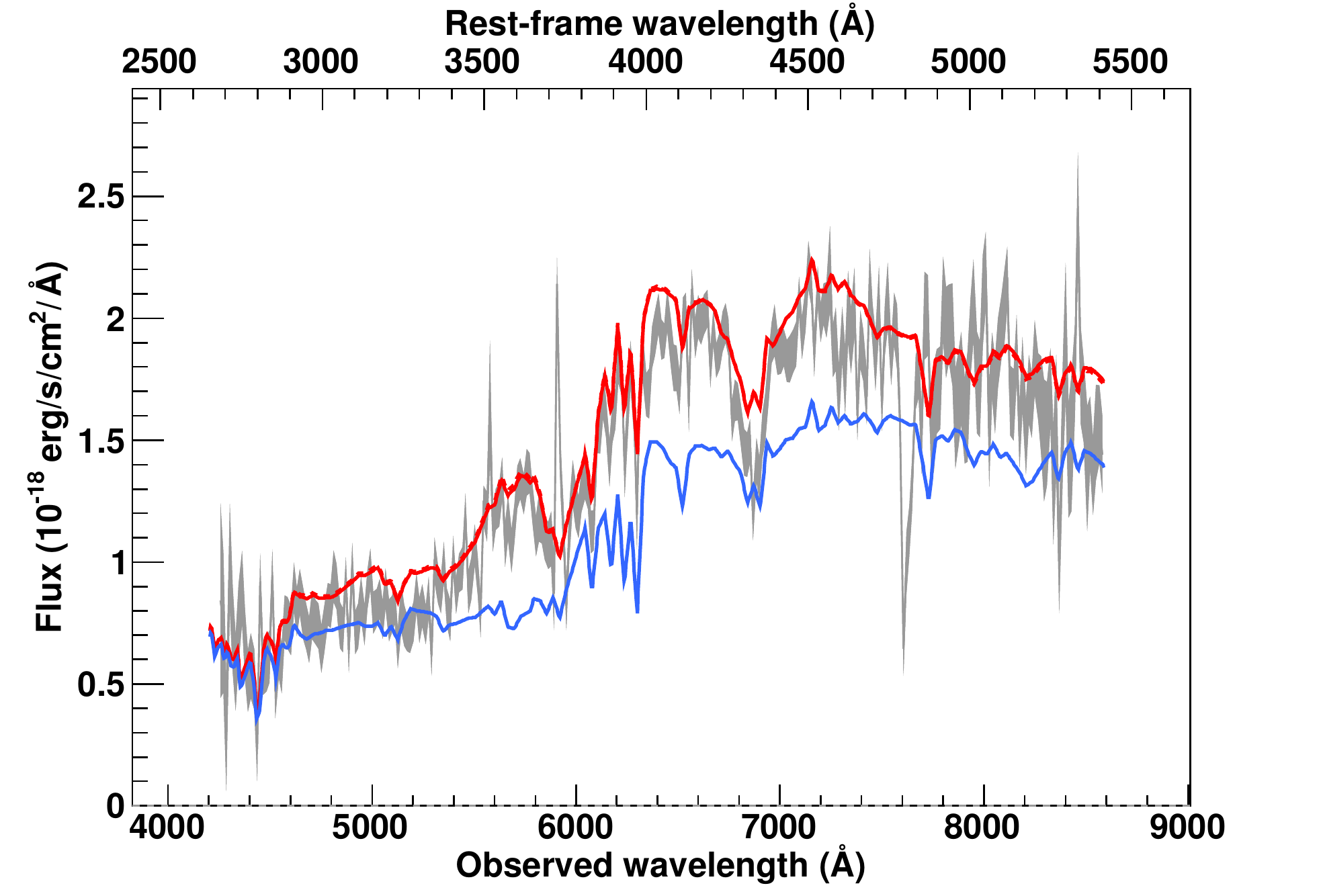}
    \includegraphics[scale=0.45]{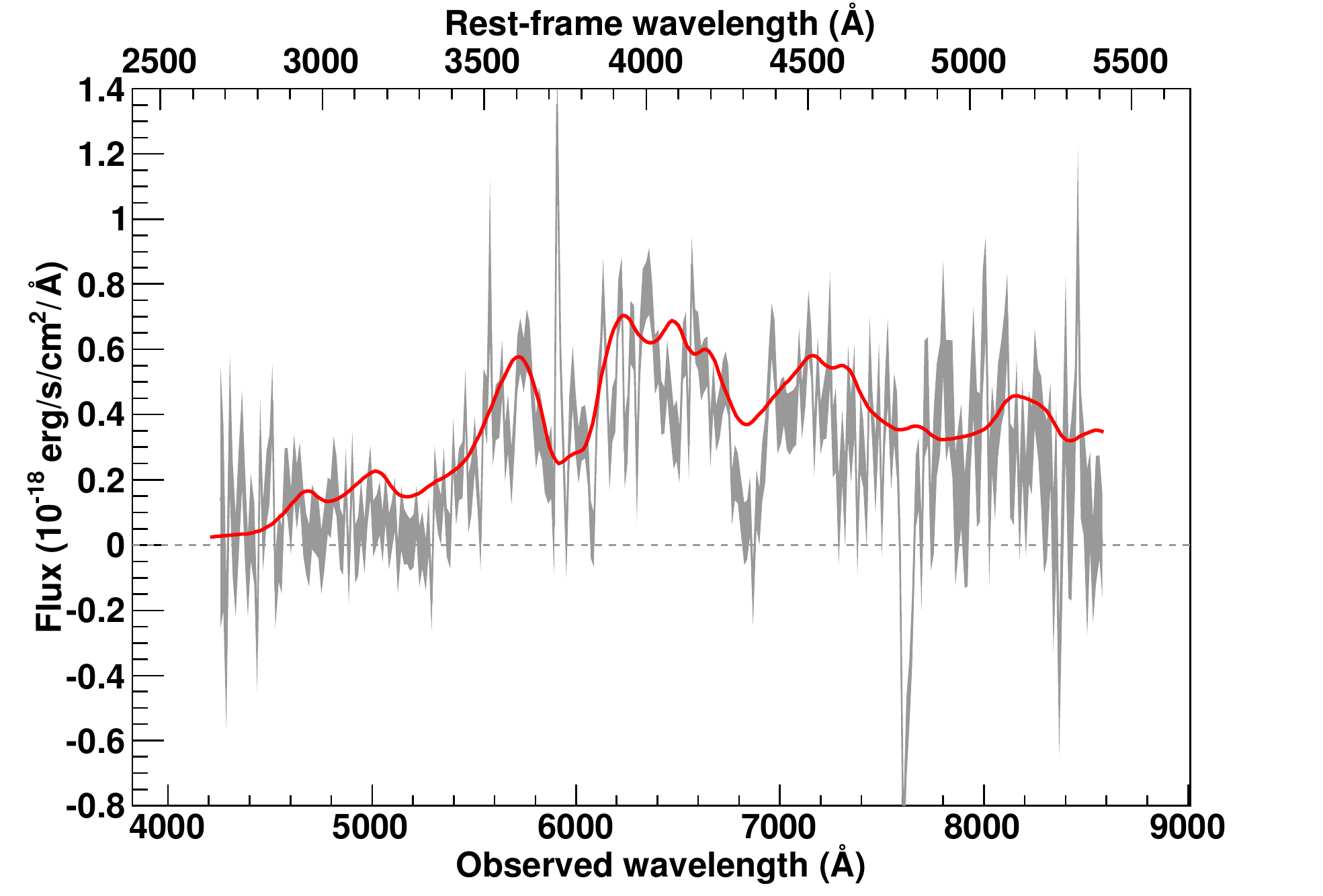}
    \end{center}
    \caption{The SNIa 05D1hm\_1063 spectrum measured at $z=0.587$ with a phase of 4.5 days. A E(1) host model has been subtracted.}
    \label{fig:Spec05D1hm_1063}
    \end{figure}
    
\clearpage    \begin{figure}
    \begin{center}
    \includegraphics[scale=0.45]{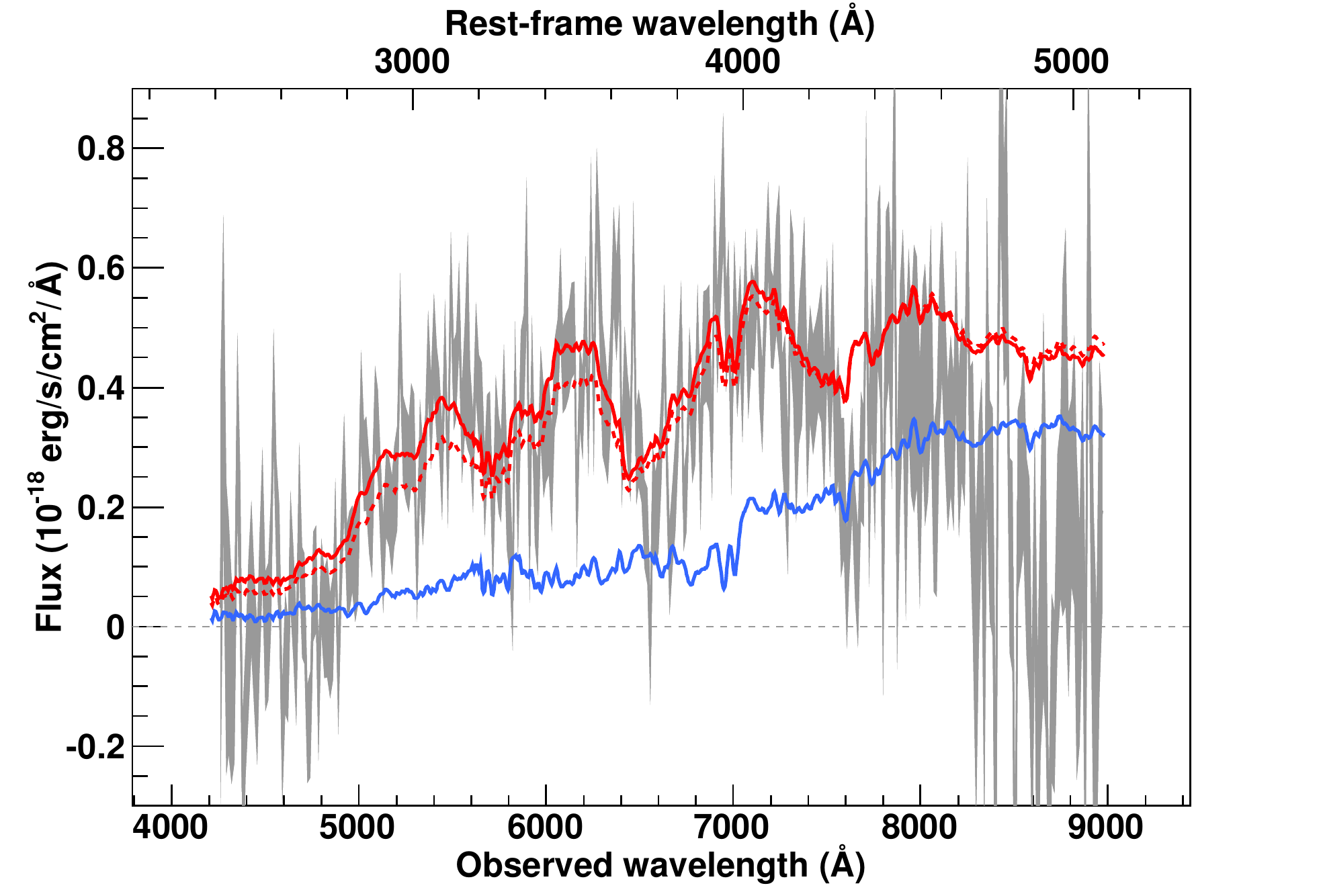}
    \includegraphics[scale=0.45]{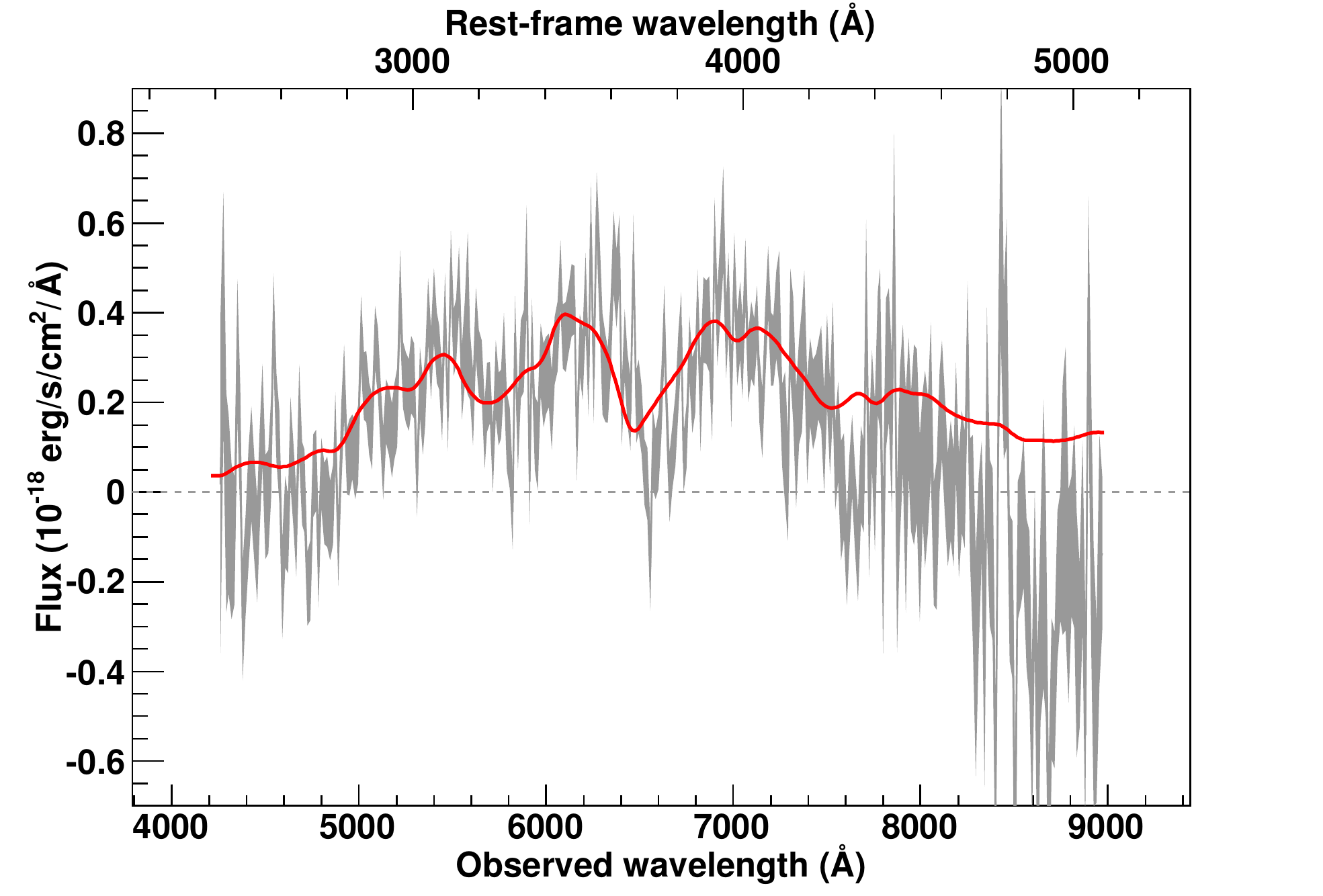}
    \end{center}
    \caption{The SNIa 05D1if\_1065 spectrum measured at $z=0.763$ with a phase of -5.9 days. A S0-Sa host model has been subtracted.}
    \label{fig:Spec05D1if_1065}
    \end{figure}
    
    \begin{figure}
    \begin{center}
    \includegraphics[scale=0.45]{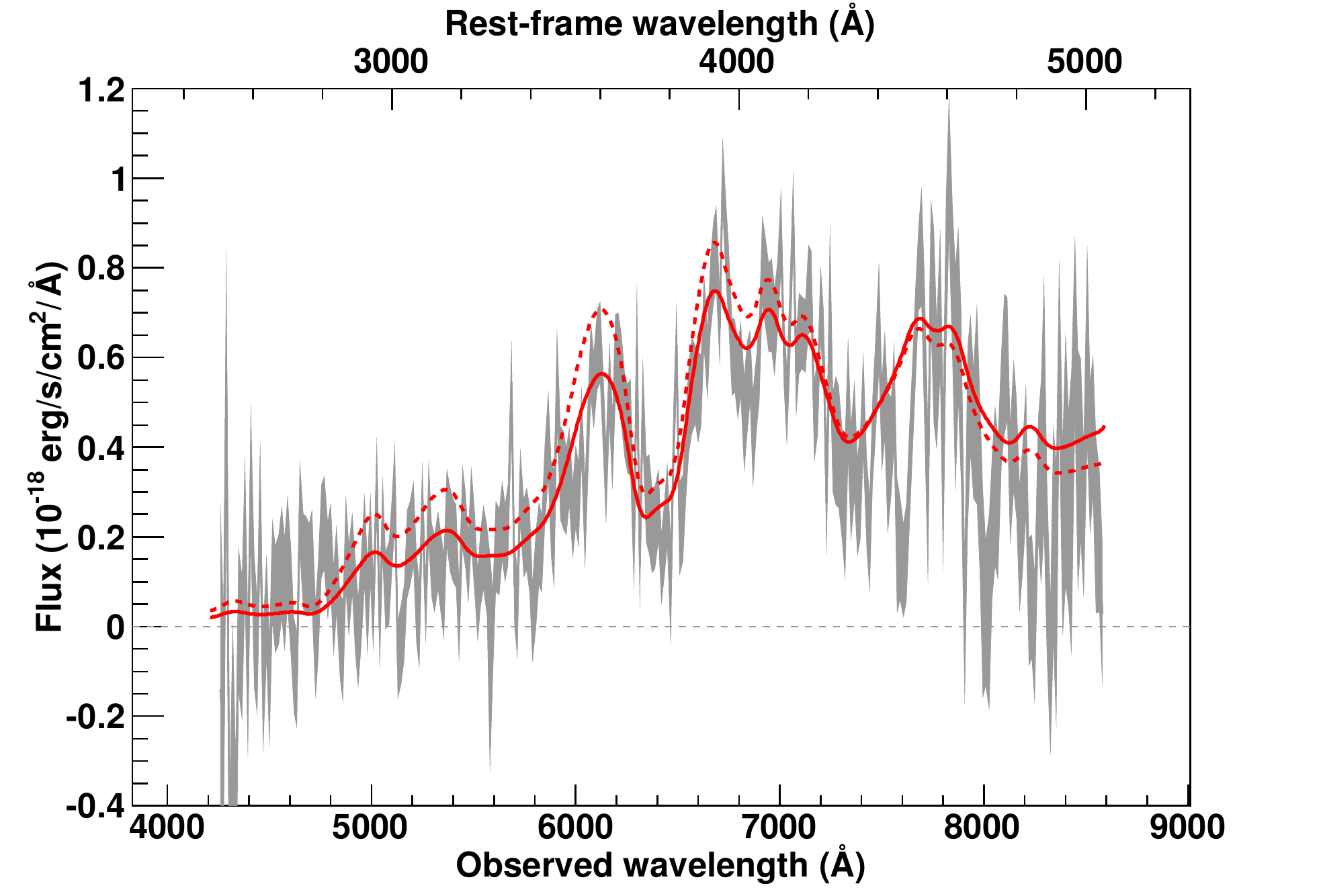}
    \end{center}
    \caption{The SNIa 05D2le\_1065 spectrum measured at $z=0.700$ with a phase of 5.9 days. Best fit is obtained without galactic component.}
    \label{fig:Spec05D2le_1065}
    \end{figure}
    
    \begin{figure}
    \begin{center}
    \includegraphics[scale=0.45]{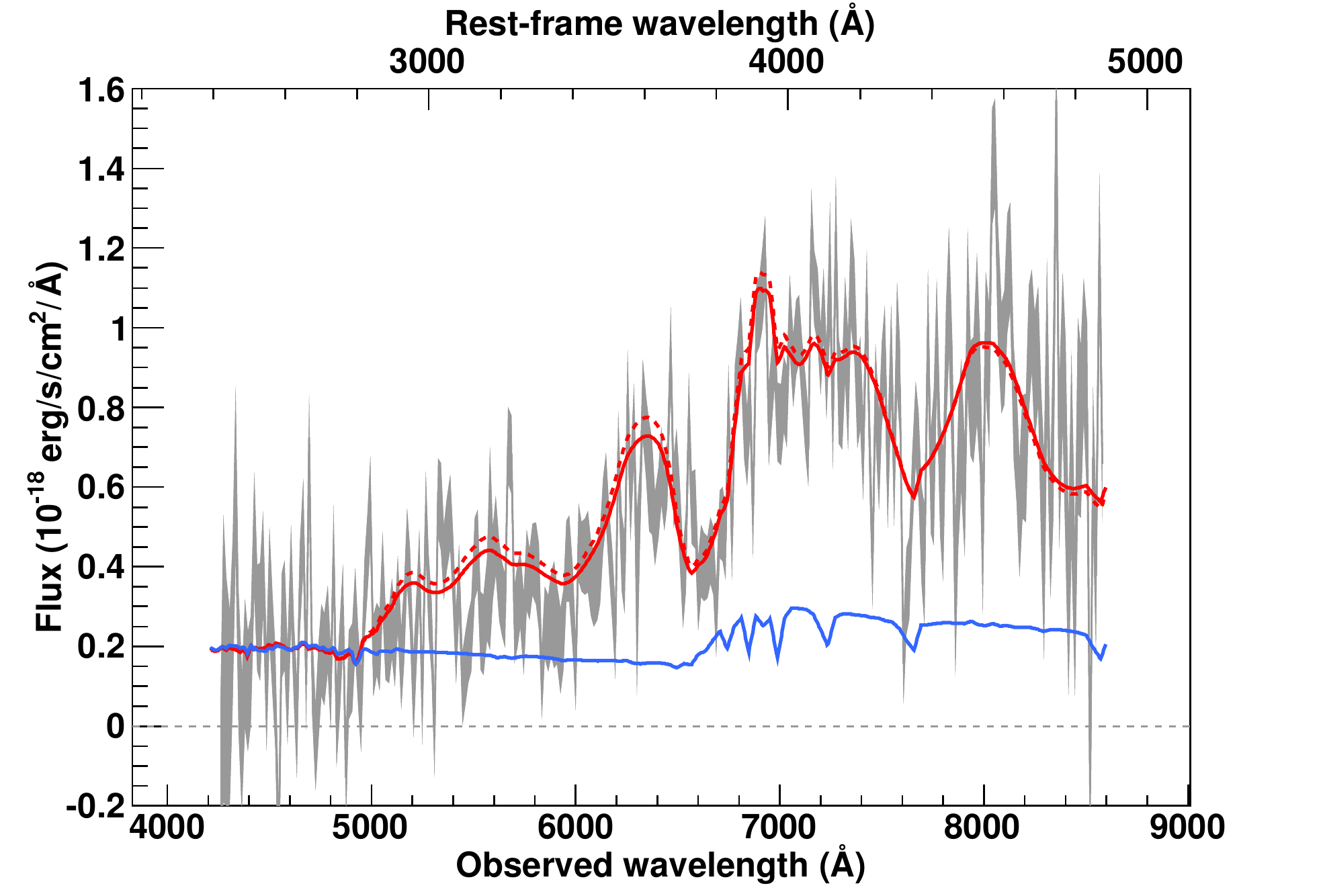}
    \includegraphics[scale=0.45]{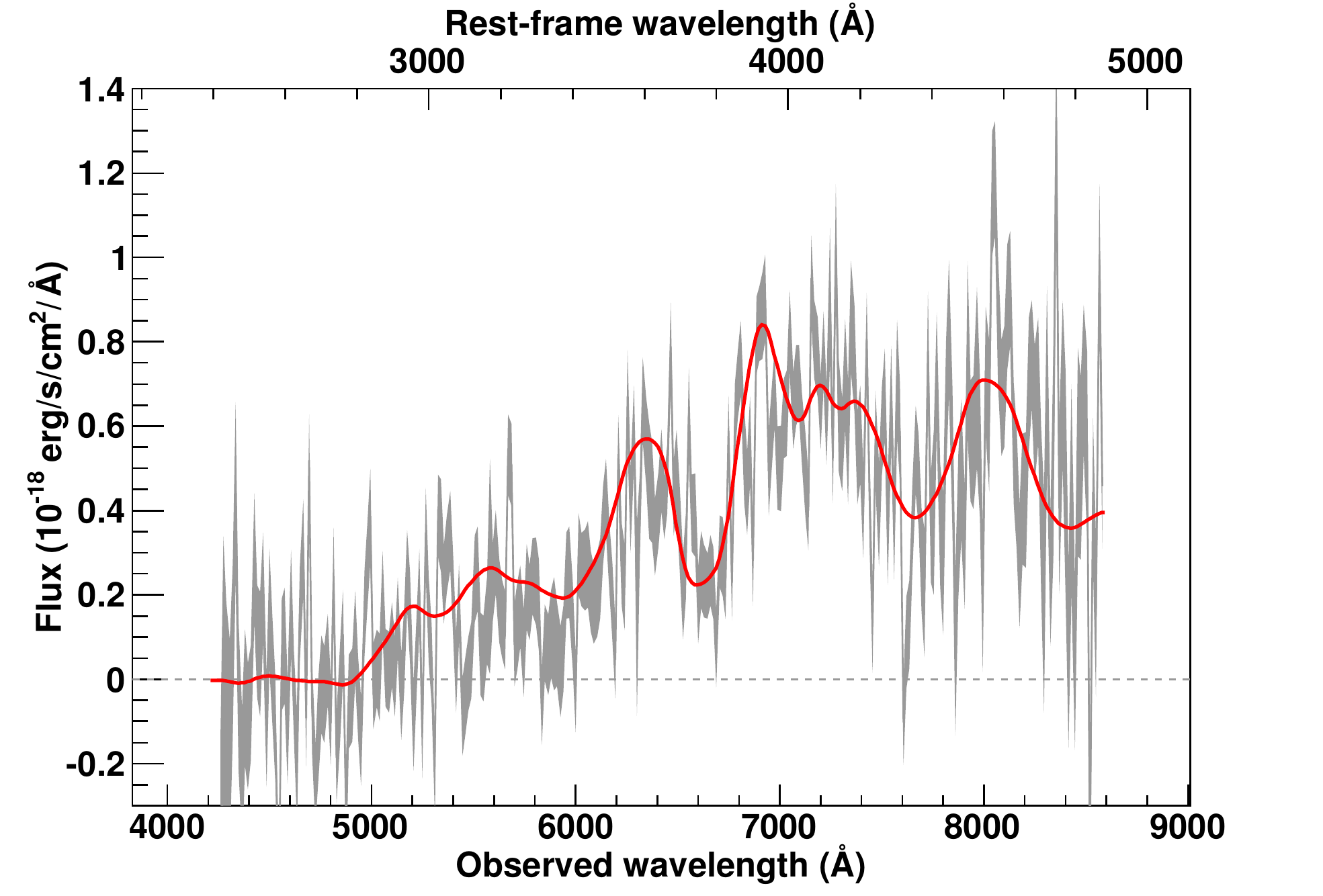}
    \end{center}
    \caption{The SNIa$\star$ 06D1bg\_1330 spectrum measured at $z=0.76$ with a phase of 8.0 days. A S0(1) host model has been subtracted.}
    \label{fig:Spec06D1bg_1330}
    \end{figure}
    
\clearpage    \begin{figure}
    \begin{center}
    \includegraphics[scale=0.45]{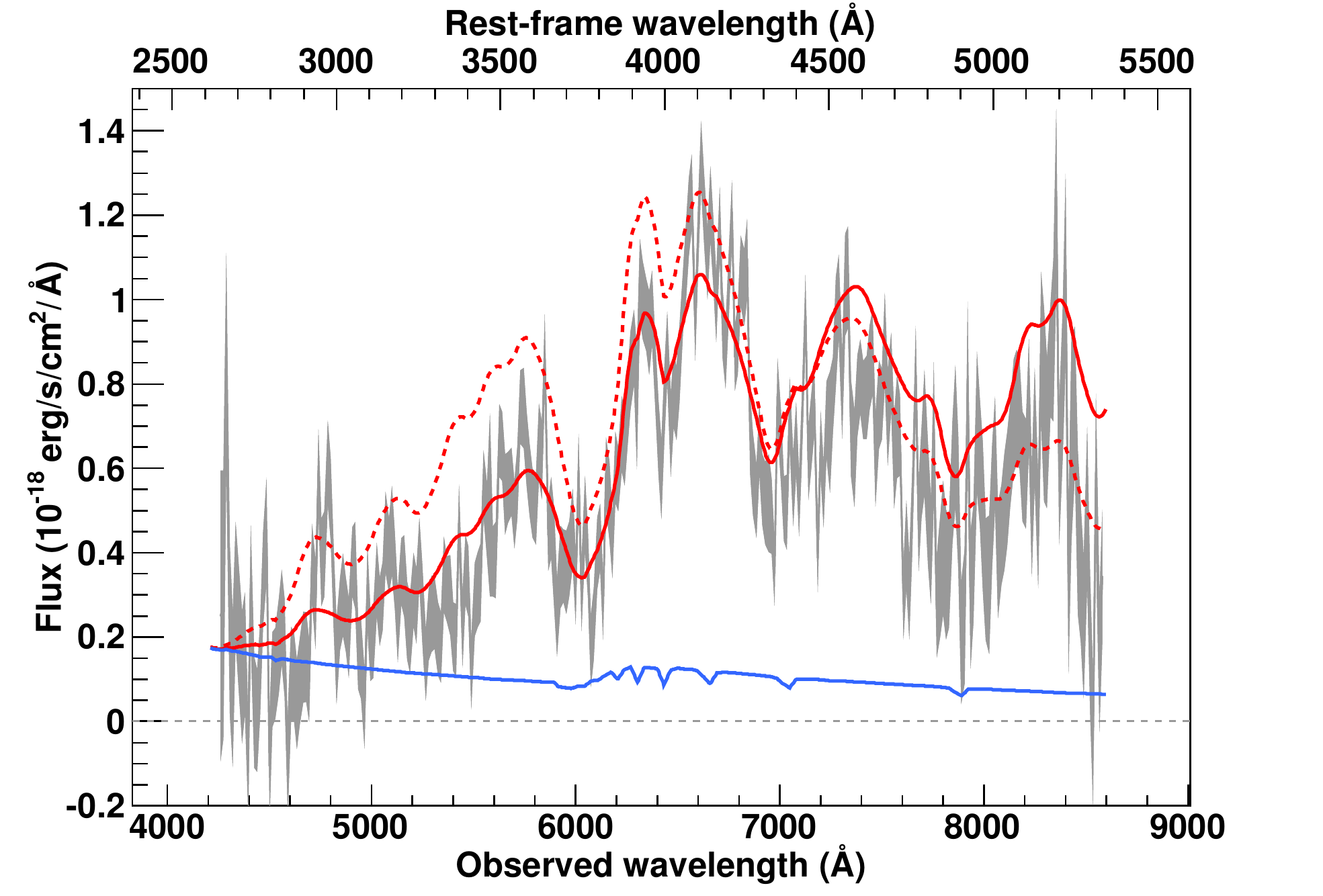}
    \includegraphics[scale=0.45]{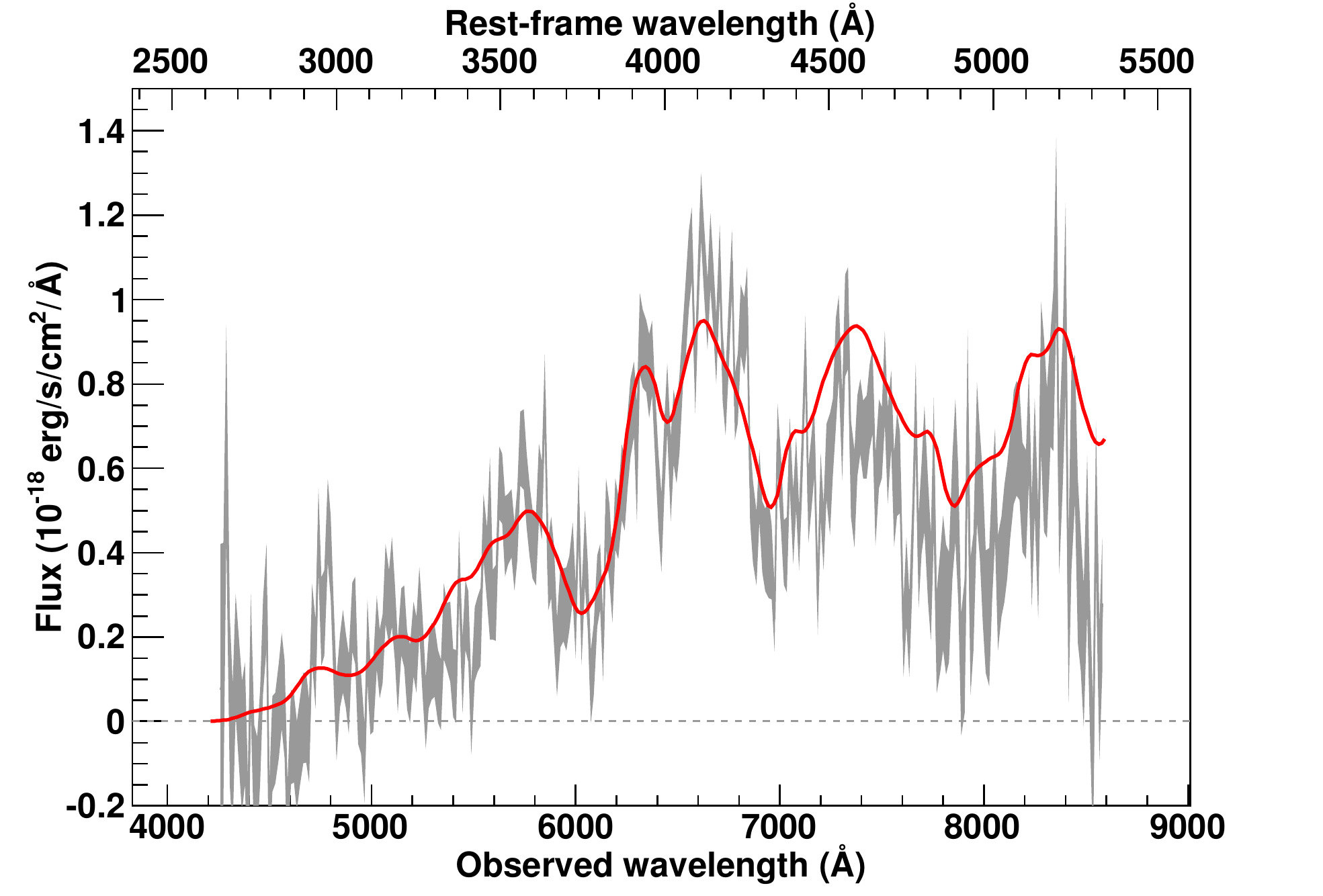}
    \end{center}
    \caption{The SNIa 06D1bo\_1330 spectrum measured at $z=0.62$ with a phase of -3.0 days. A Sd(1) host model has been subtracted.}
    \label{fig:Spec06D1bo_1330}
    \end{figure}
    
    \begin{figure}
    \begin{center}
    \includegraphics[scale=0.45]{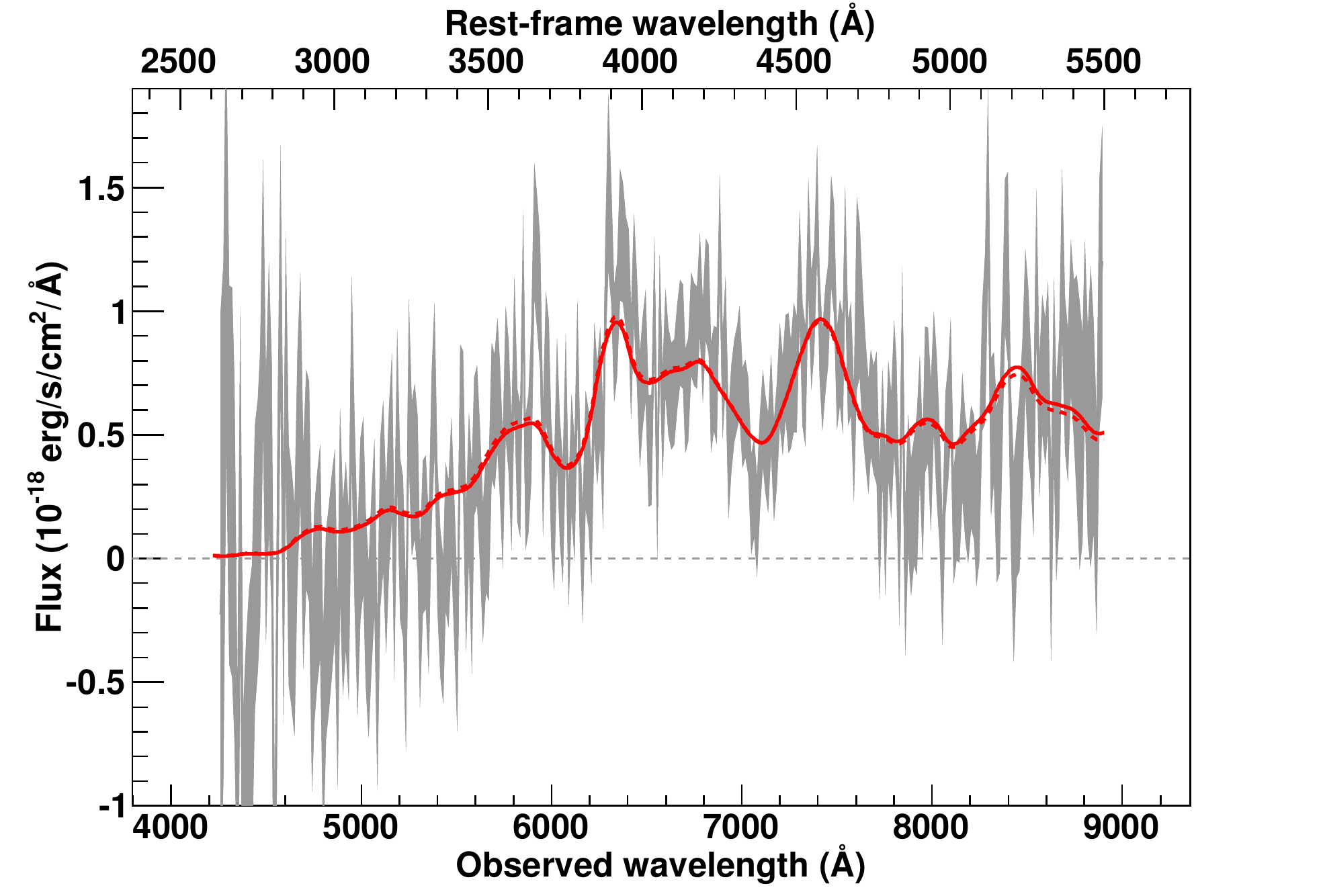}
    \end{center}
    \caption{The SNIa 06D1cm\_1342 spectrum measured at $z=0.619$ with a phase of 8.3 days. Best fit is obtained without galactic component.}
    \label{fig:Spec06D1cm_1342}
    \end{figure}
    
    \begin{figure}
    \begin{center}
    \includegraphics[scale=0.45]{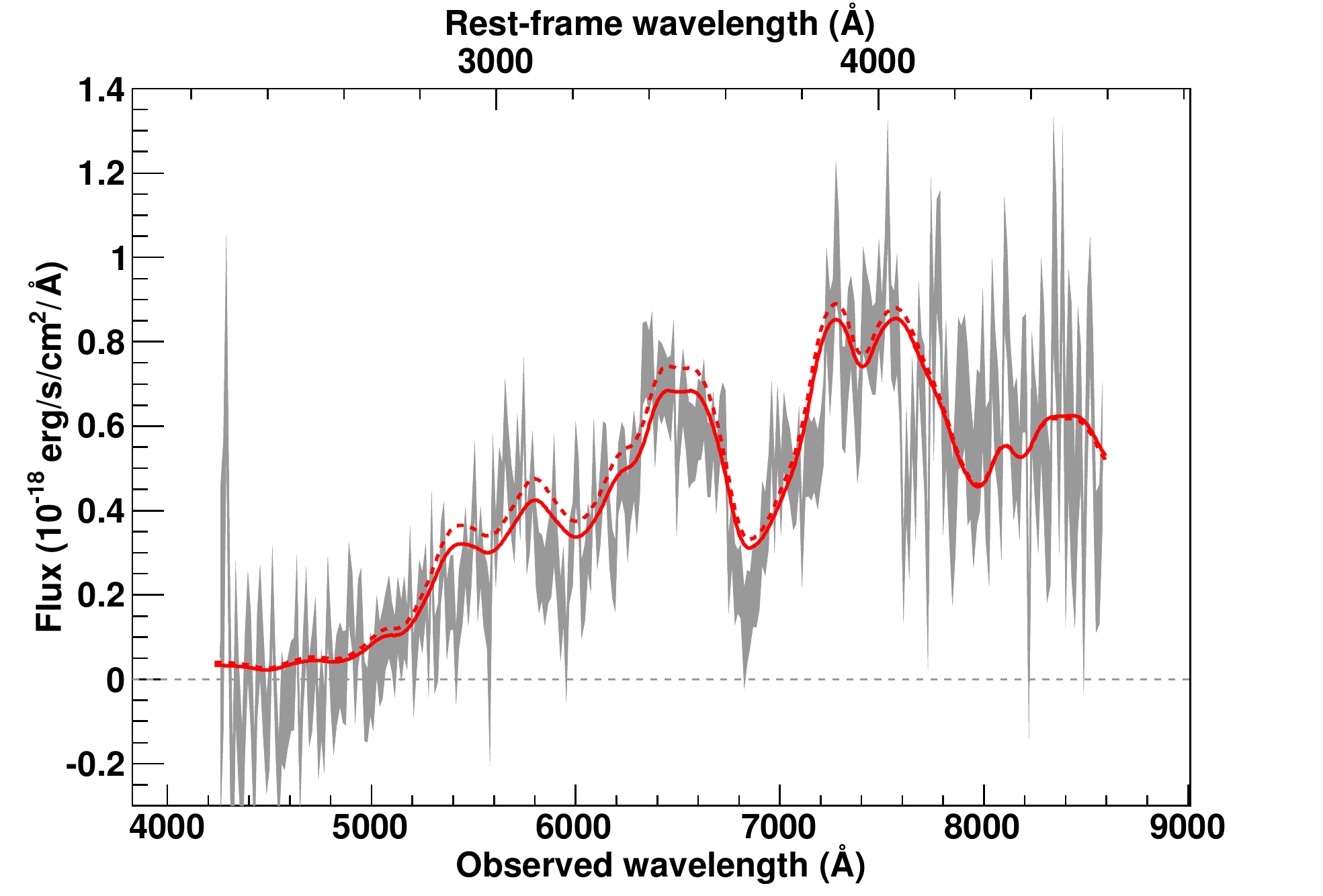}
    \end{center}
    \caption{The SNIa 06D1cx\_1339 spectrum measured at $z=0.860$ with a phase of -4.2 days. Best fit is obtained without galactic component.}
    \label{fig:Spec06D1cx_1339}
    \end{figure}
    
\clearpage    \begin{figure}
    \begin{center}
    \includegraphics[scale=0.45]{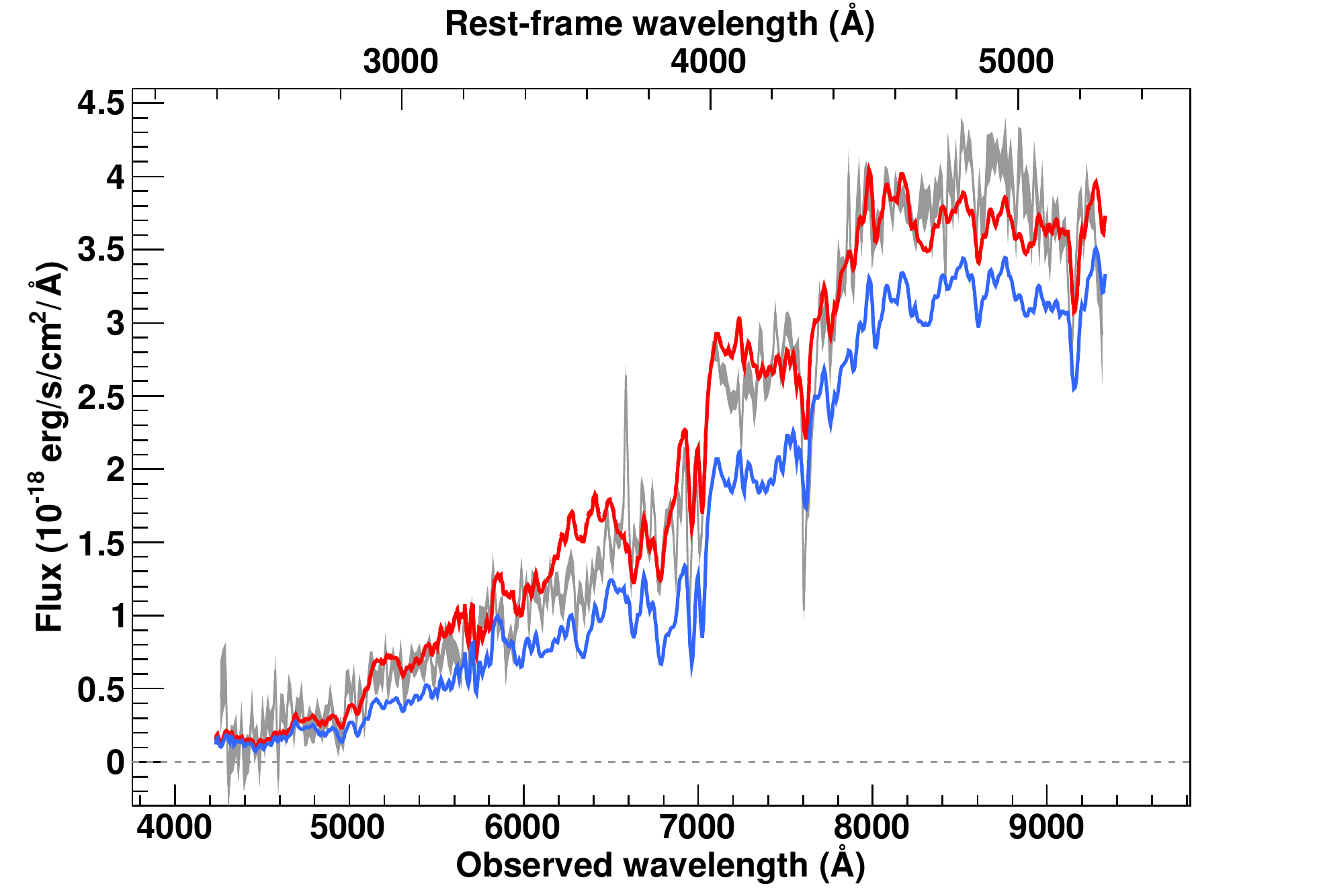}
    \includegraphics[scale=0.45]{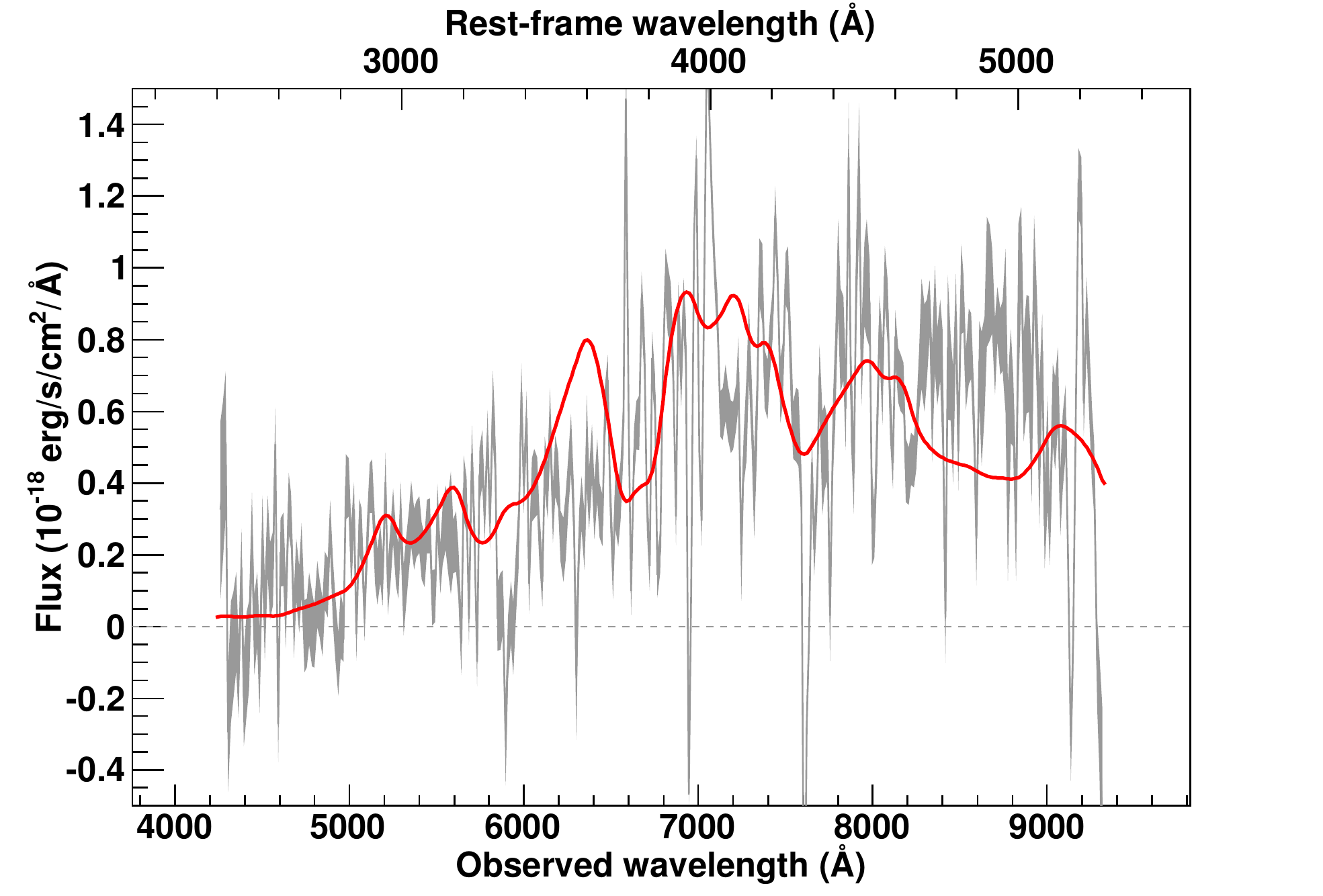}
    \end{center}
    \caption{The SNIa$\star$ 06D1dc\_1360 spectrum measured at $z=0.767$ with a phase of 3.8 days. A E-S0 host model has been subtracted.}
    \label{fig:Spec06D1dc_1360}
    \end{figure}
    
    \begin{figure}
    \begin{center}
    \includegraphics[scale=0.45]{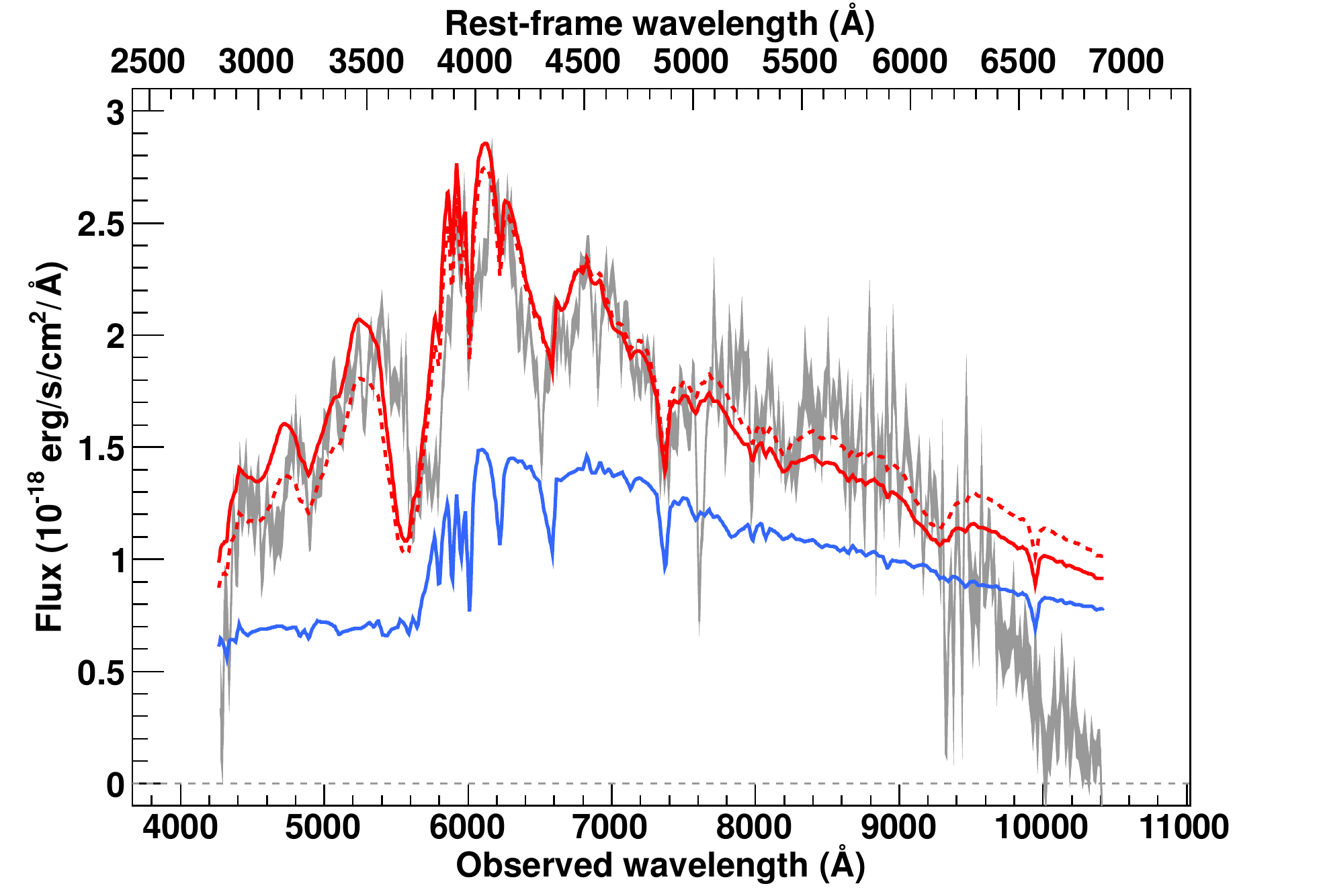}
    \includegraphics[scale=0.45]{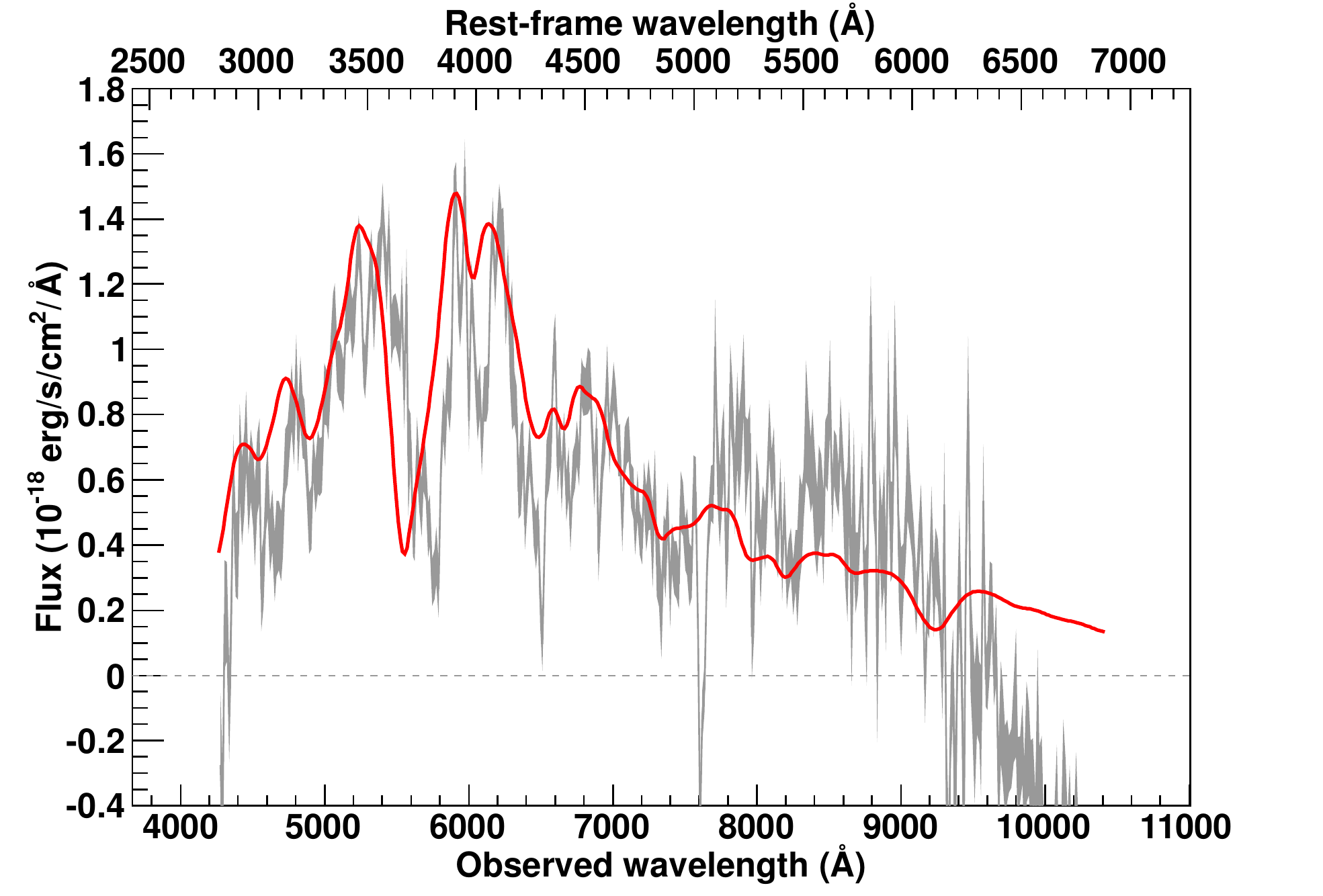}
    \end{center}
    \caption{The SNIa 06D1dl\_1360 spectrum measured at $z=0.514$ with a phase of -5.2 days. A E(1) host model has been subtracted.}
    \label{fig:Spec06D1dl_1360}
    \end{figure}
    
    \begin{figure}
    \begin{center}
    \includegraphics[scale=0.45]{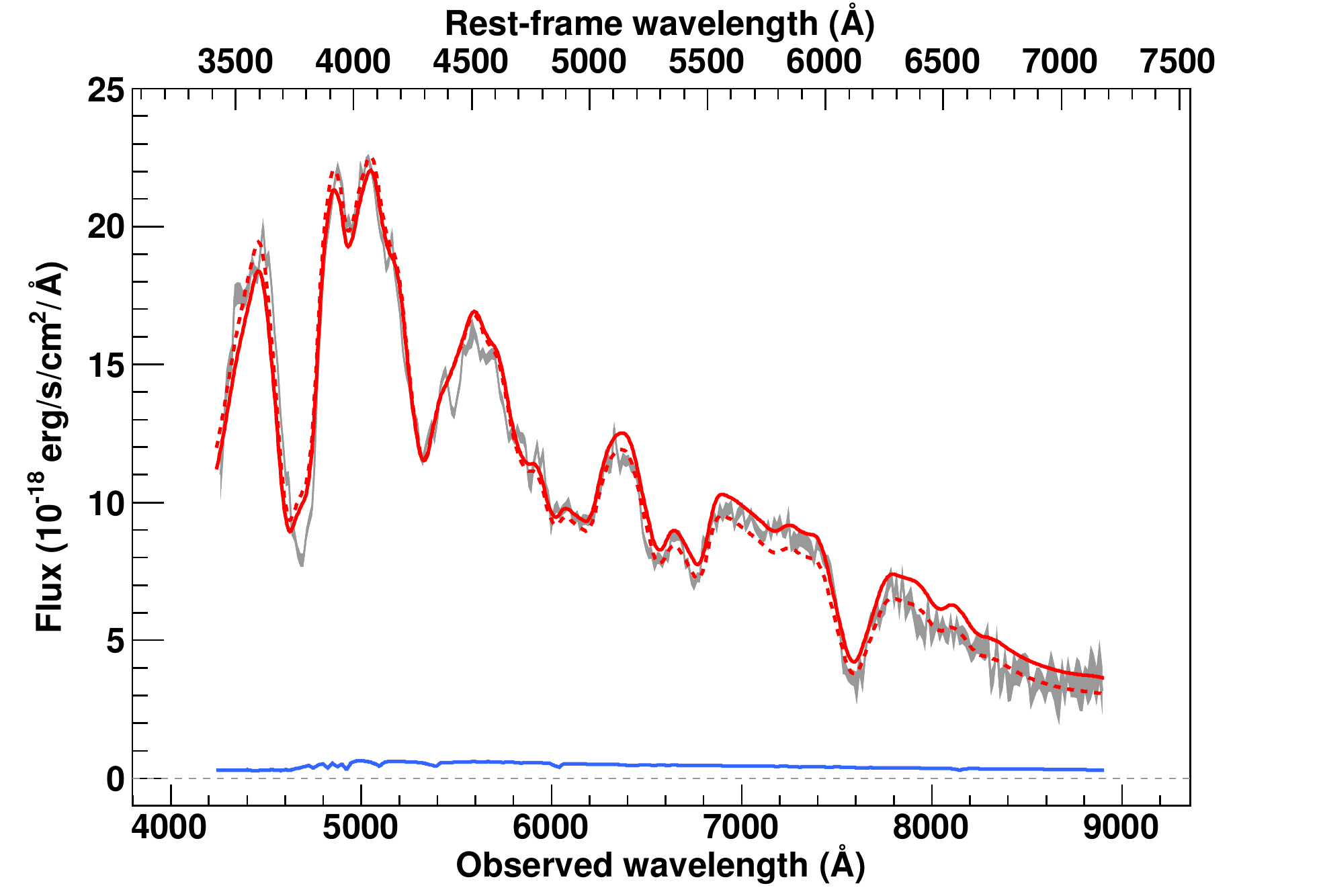}
    \includegraphics[scale=0.45]{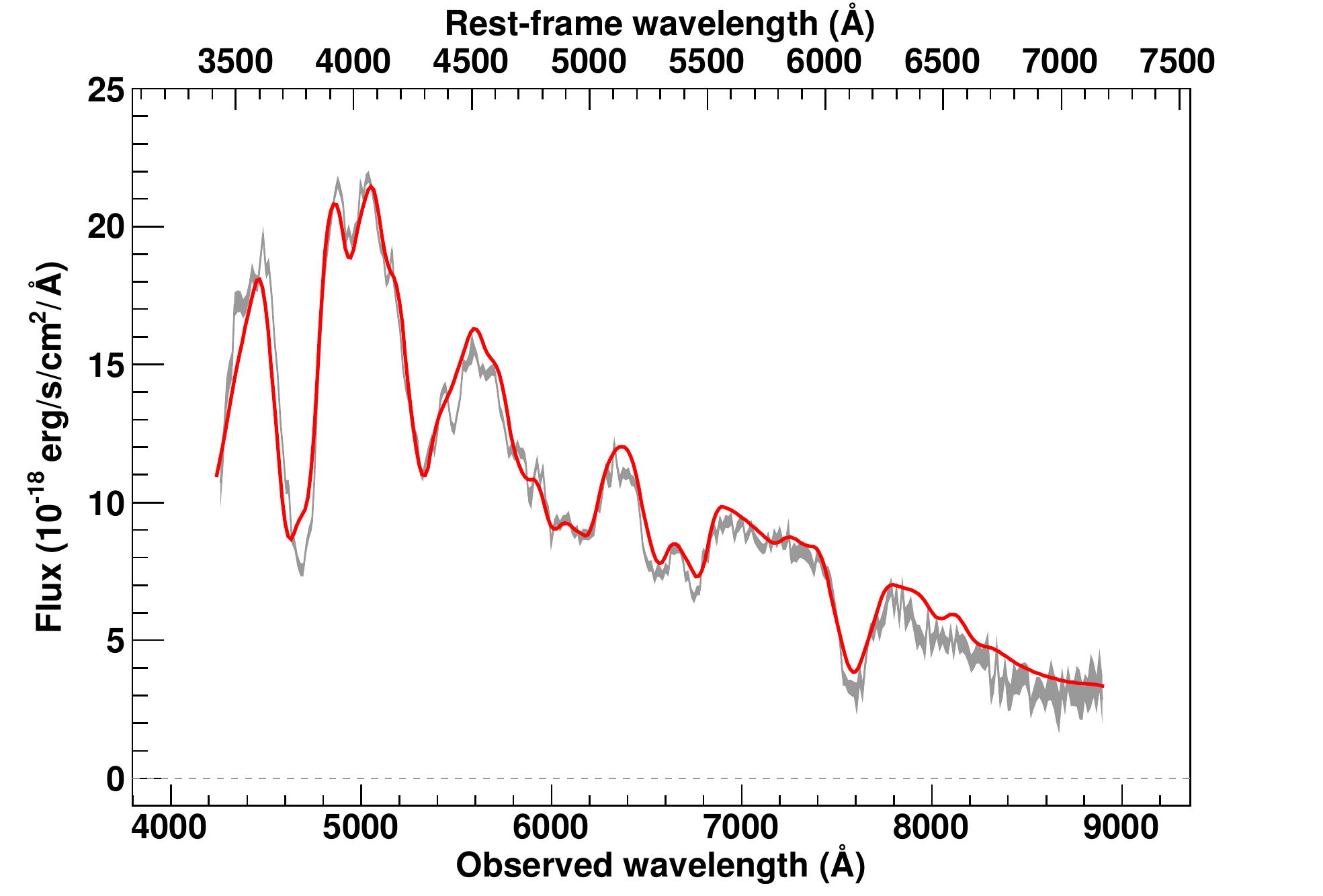}
    \end{center}
    \caption{The SNIa 06D1du\_1358 spectrum measured at $z=0.24$ with a phase of -0.2 days. A E(1) host model has been subtracted.}
    \label{fig:Spec06D1du_1358}
    \end{figure}
    
\clearpage    \begin{figure}
    \begin{center}
    \includegraphics[scale=0.45]{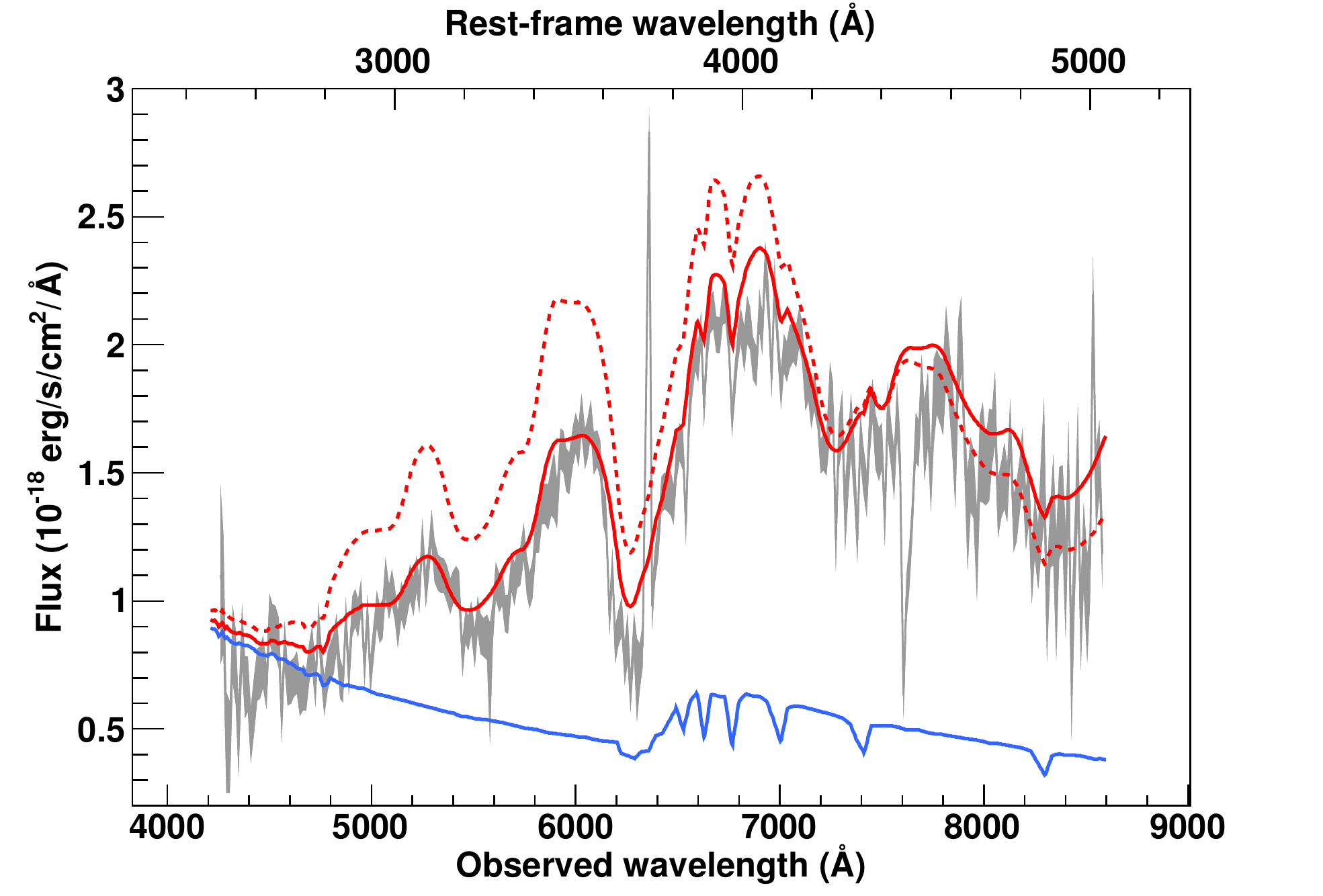}
    \includegraphics[scale=0.45]{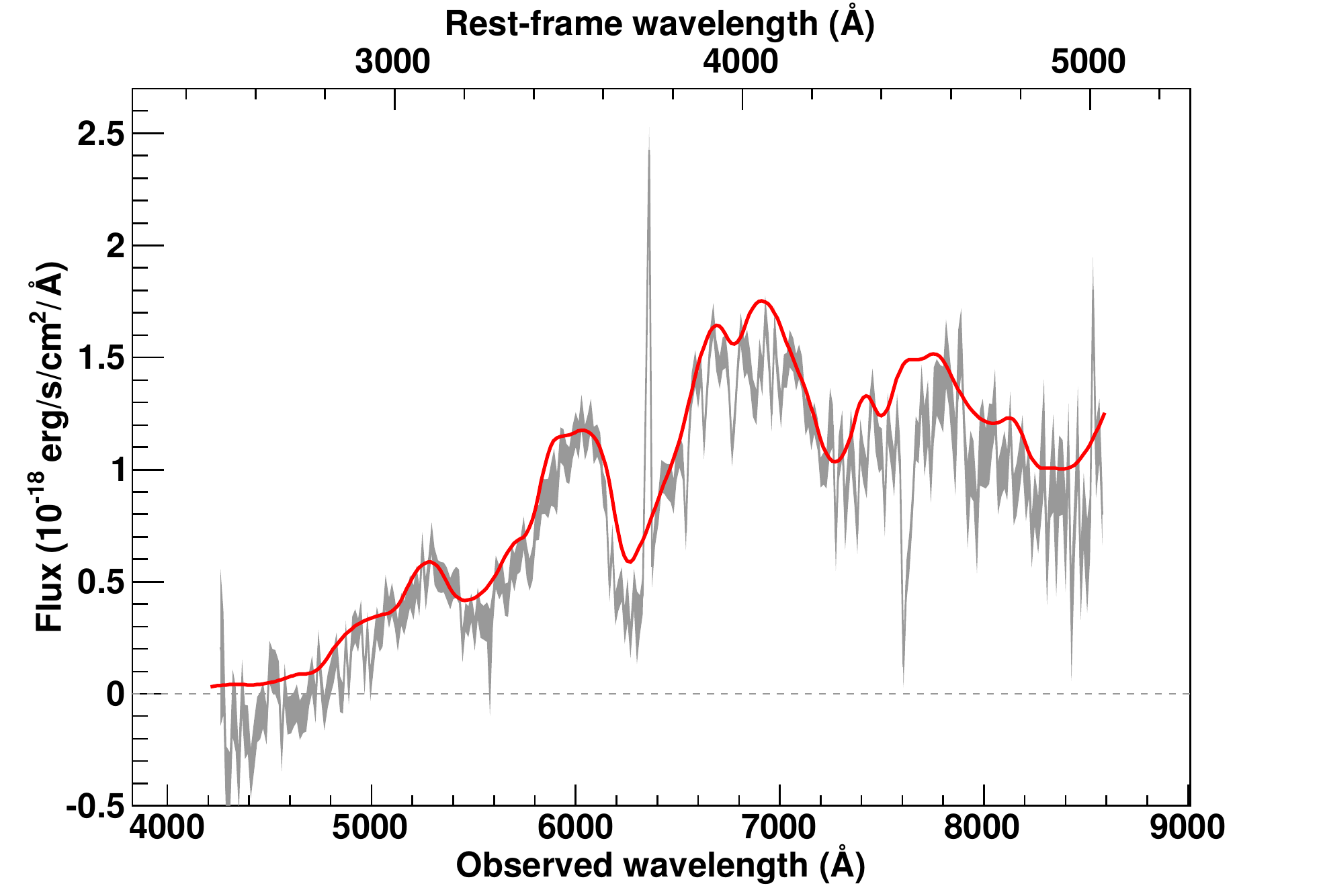}
    \end{center}
    \caption{The SNIa 06D1eb\_1364 spectrum measured at $z=0.704$ with a phase of -5.2 days. A Sd(1) host model has been subtracted.}
    \label{fig:Spec06D1eb_1364}
    \end{figure}
    
    \begin{figure}
    \begin{center}
    \includegraphics[scale=0.45]{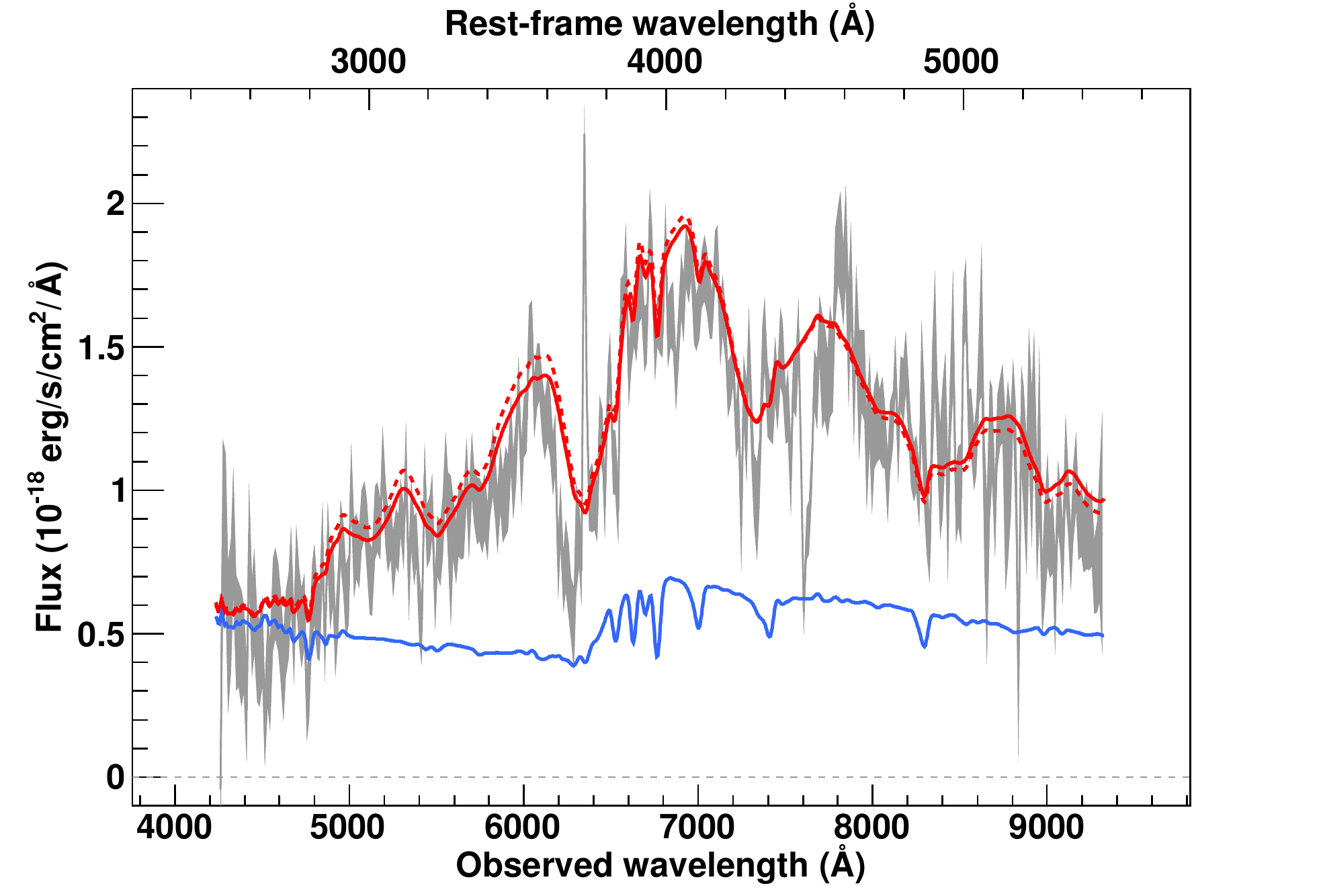}
    \includegraphics[scale=0.45]{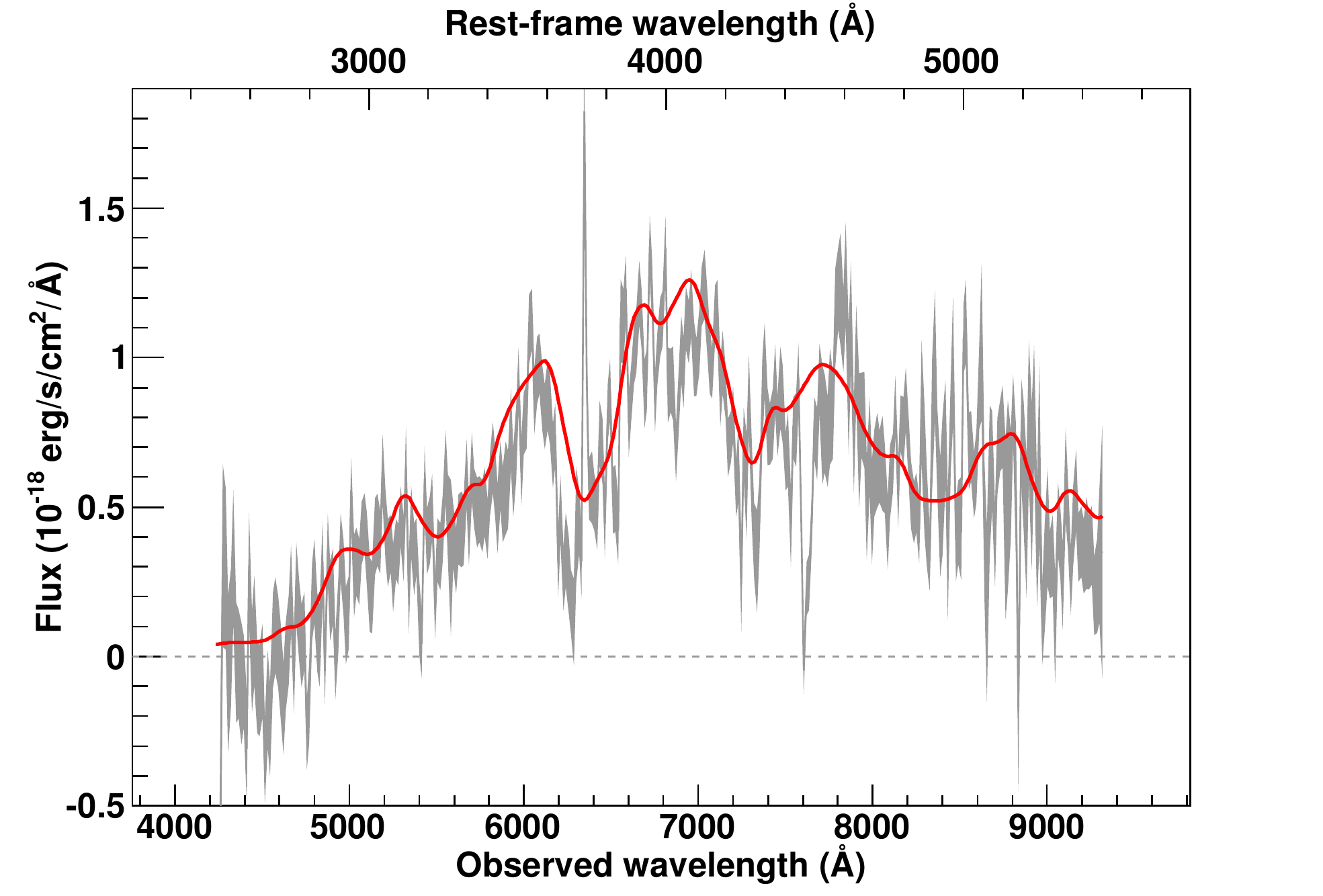}
    \end{center}
    \caption{The SNIa 06D1eb\_1369 spectrum measured at $z=0.704$ with a phase of -2.3 days. A Sd7 host model has been subtracted.}
    \label{fig:Spec06D1eb_1369}
    \end{figure}
    
    \begin{figure}
    \begin{center}
    \includegraphics[scale=0.45]{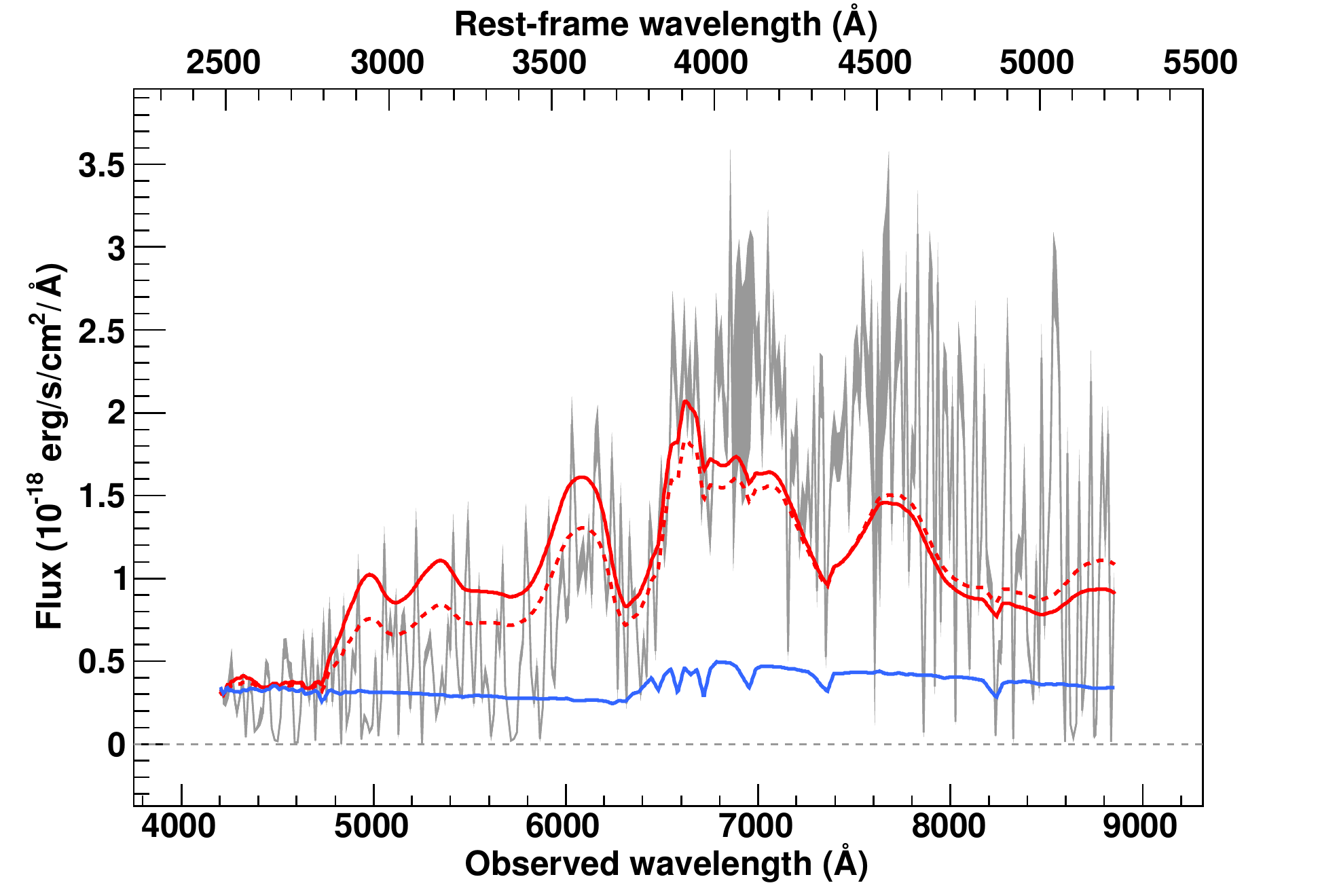}
    \includegraphics[scale=0.45]{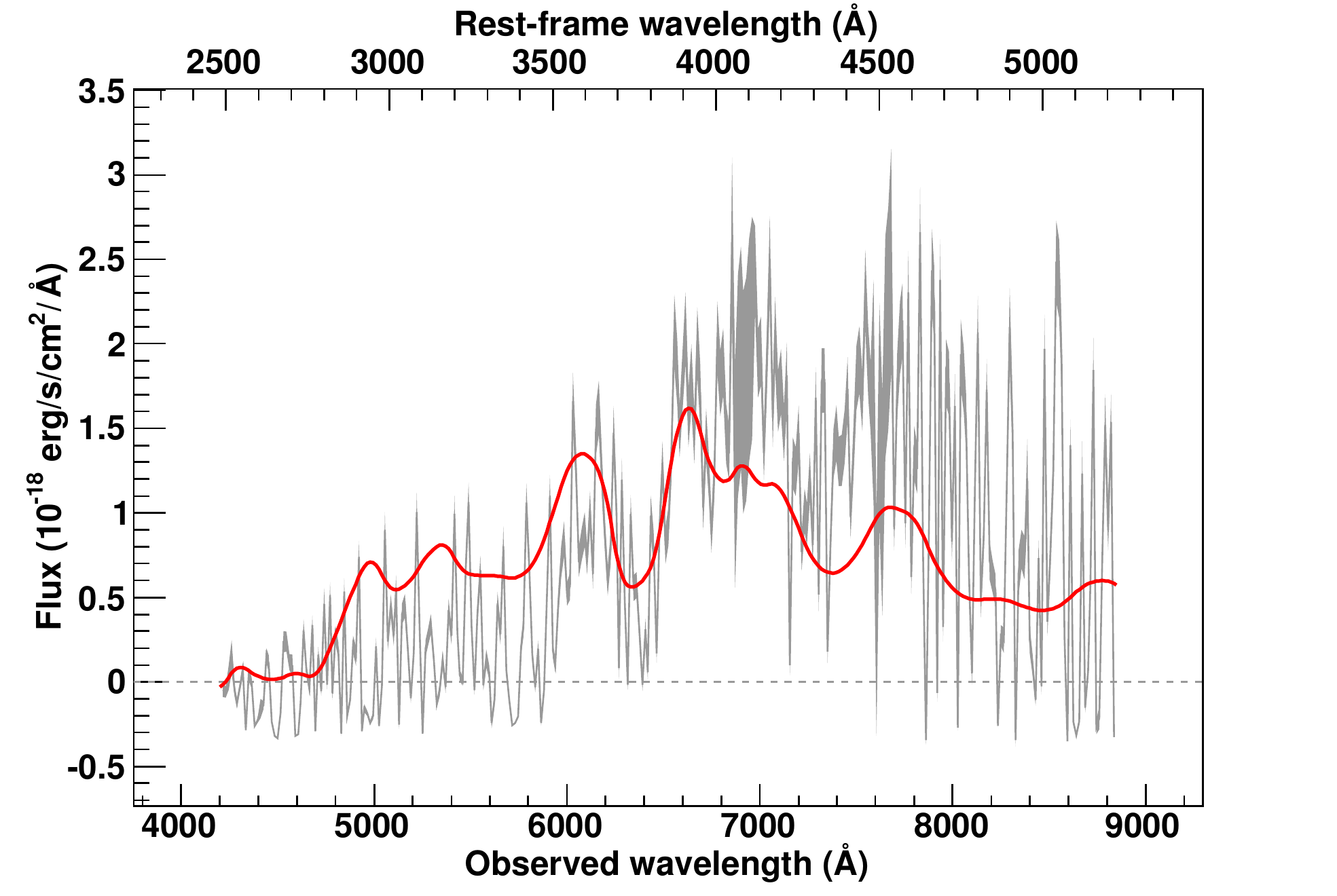}
    \end{center}
    \caption{The SNIa 06D1ez\_1389 spectrum measured at $z=0.692$ with a phase of 7.6 days. A S01 host model has been subtracted.}
    \label{fig:Spec06D1ez_1389}
    \end{figure}
    
    \clearpage
    \begin{figure}
    \begin{center}
    \includegraphics[scale=0.45]{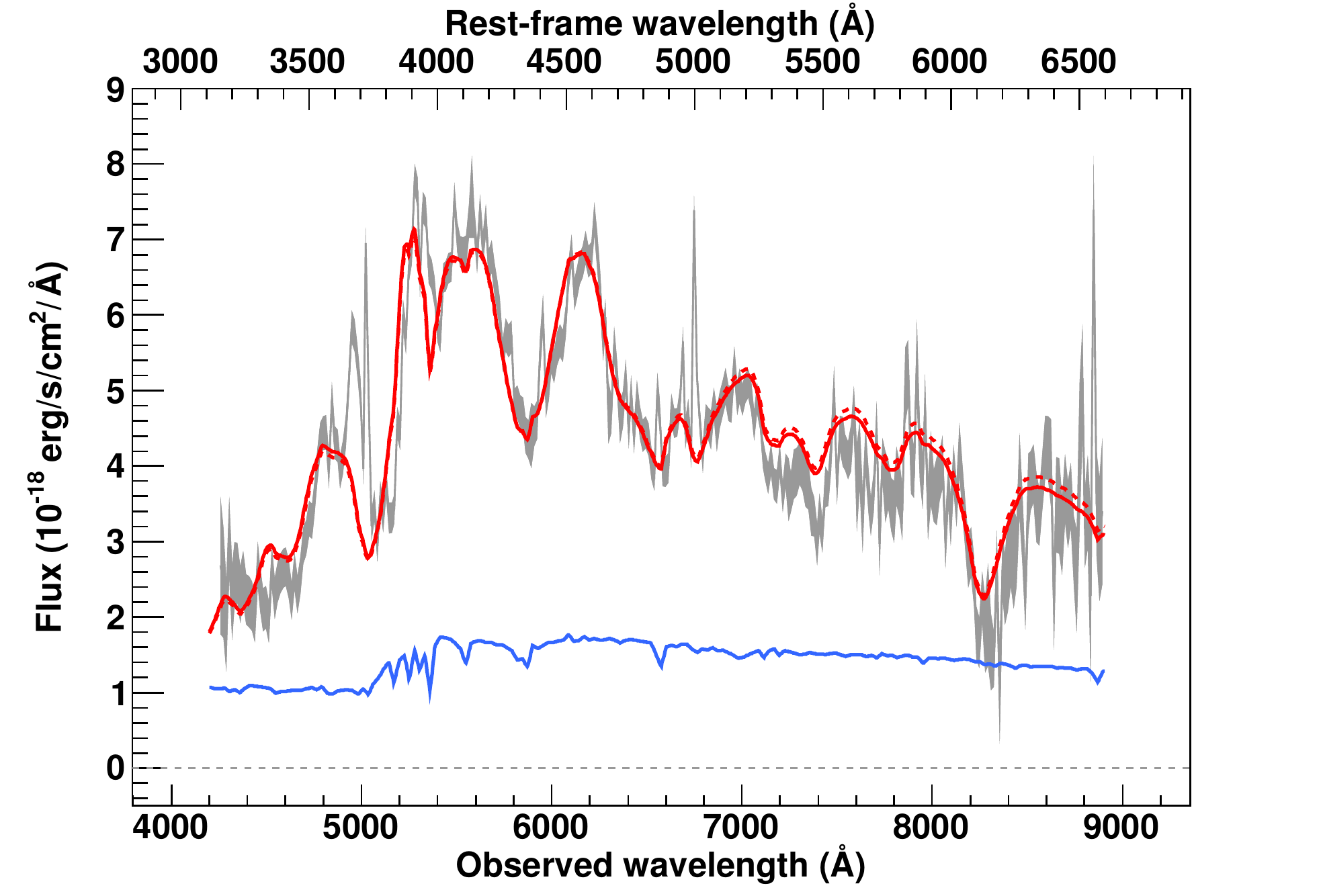}
    \includegraphics[scale=0.45]{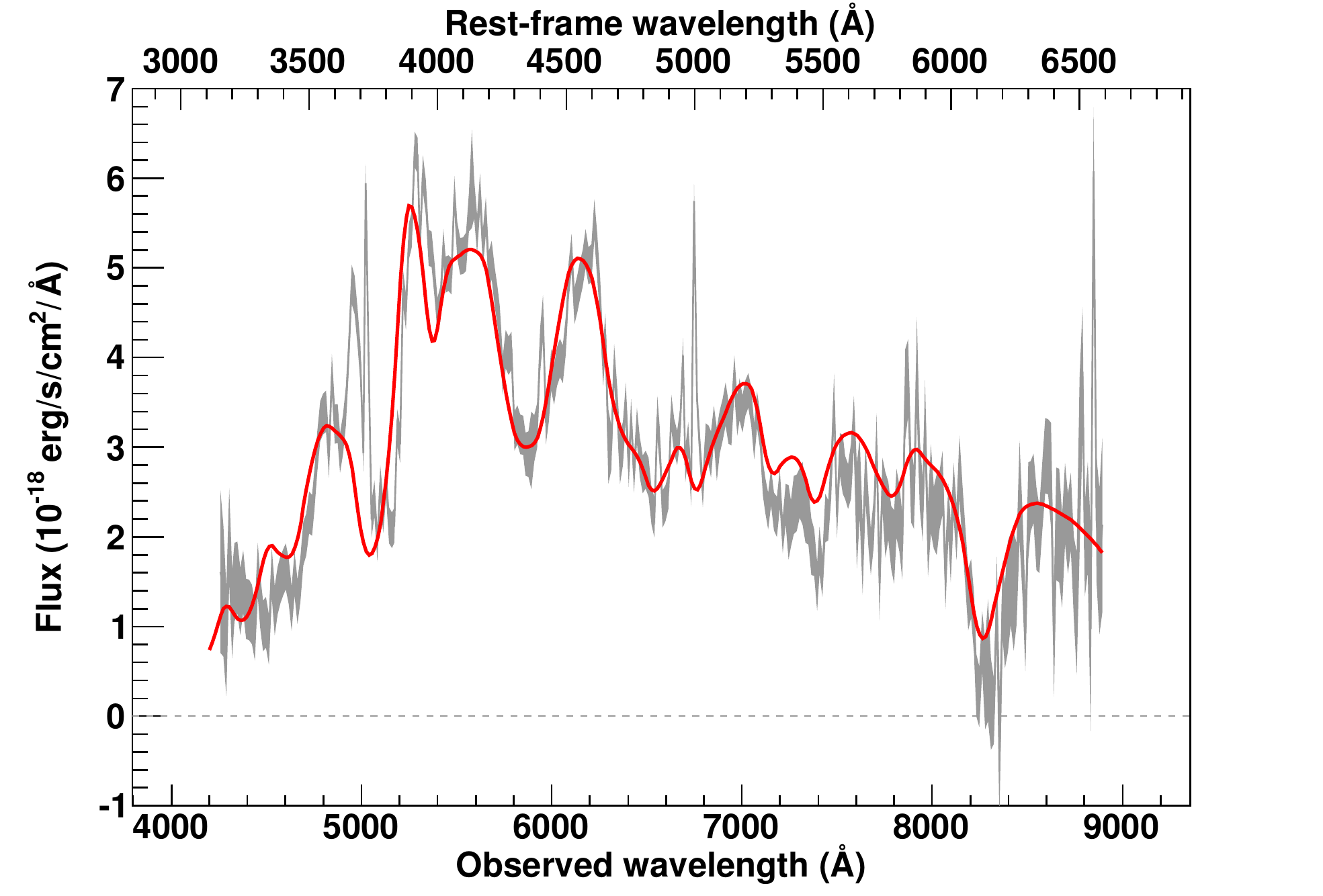}
    \end{center}
    \caption{The SNIa 06D1fd\_1395 spectrum measured at $z=0.350$ with a phase of 4.9 days. A Sd(13) host model has been subtracted.}
    \label{fig:Spec06D1fd_1395}
    \end{figure}
    
    \begin{figure}
    \begin{center}
    \includegraphics[scale=0.45]{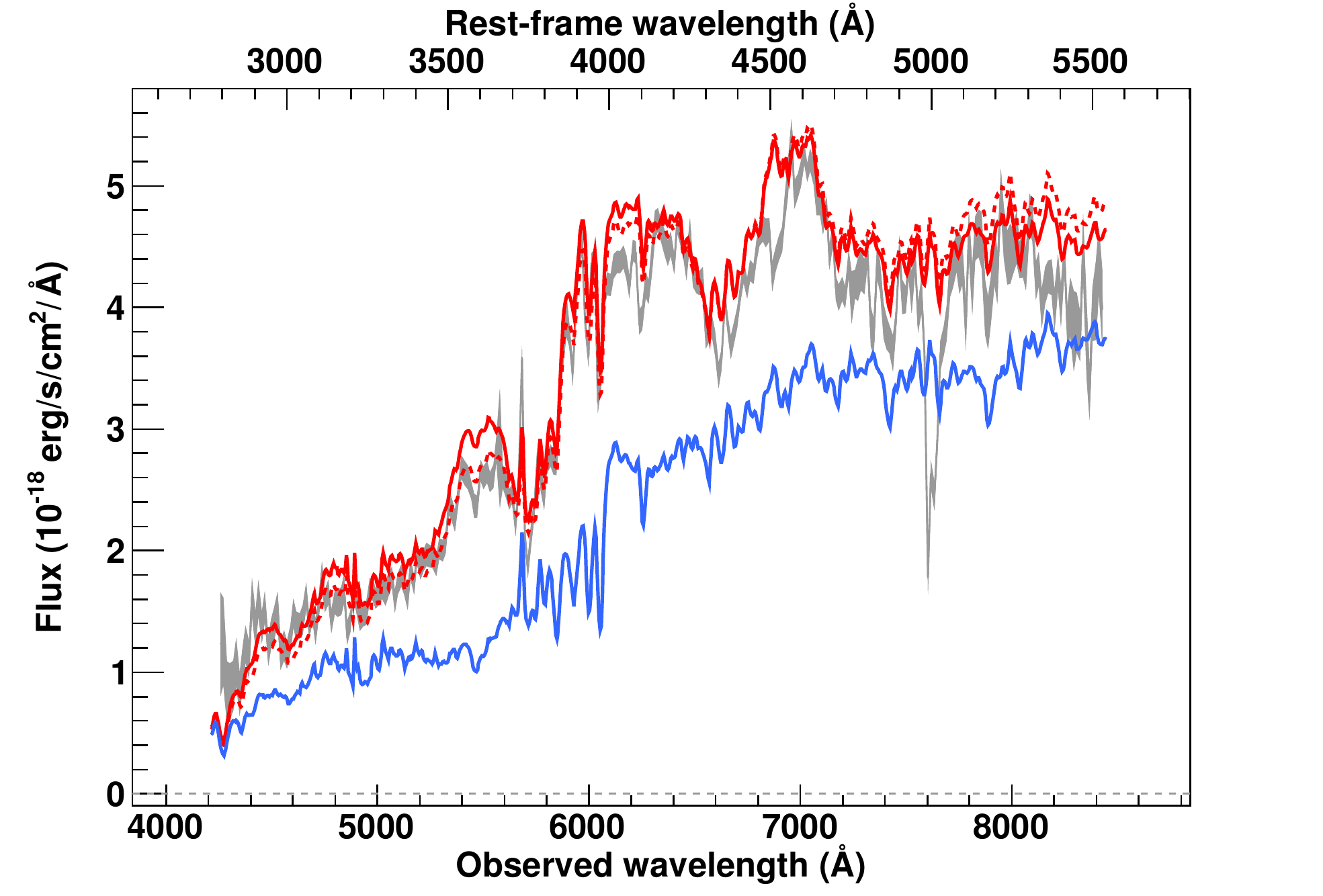}
    \includegraphics[scale=0.45]{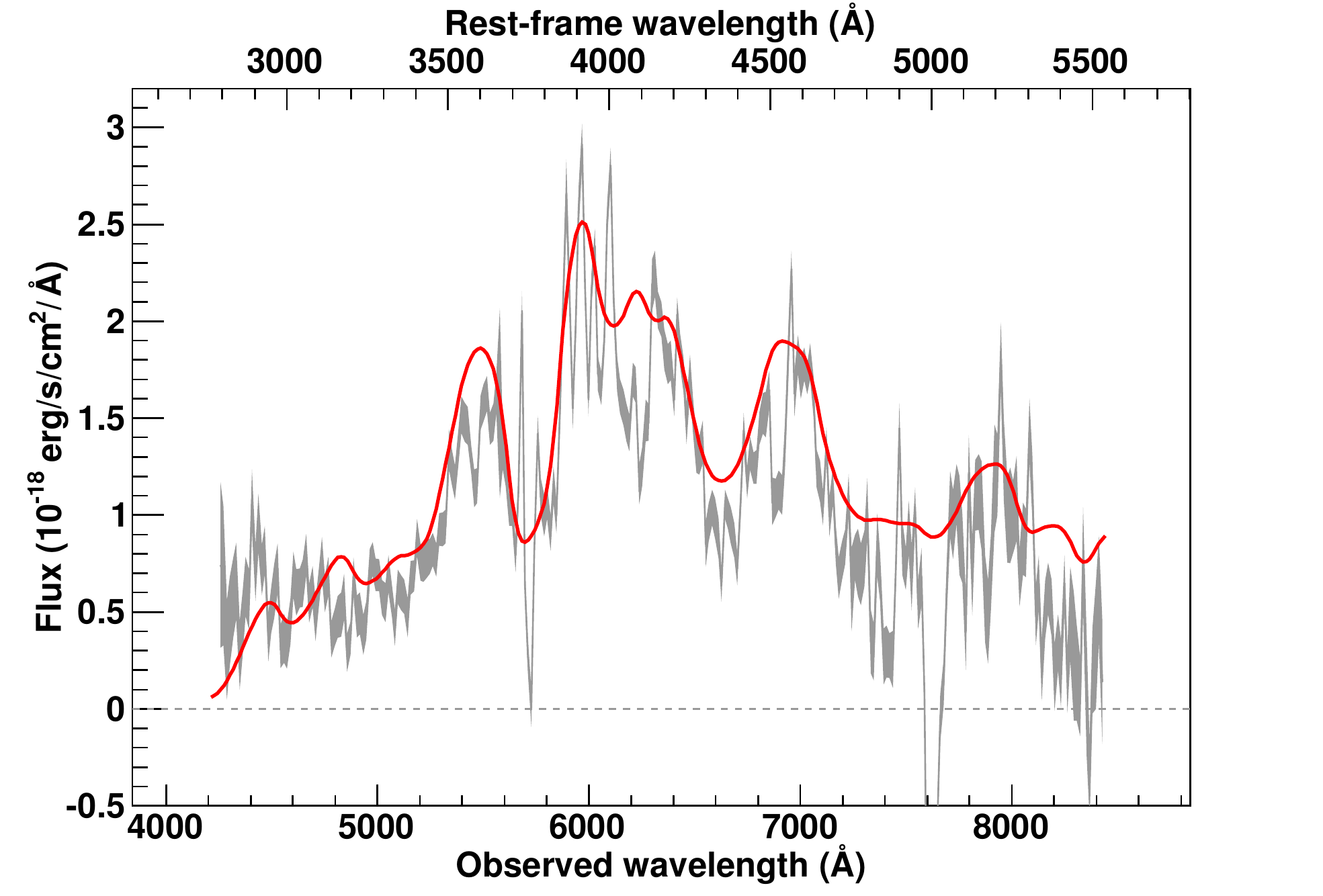}
    \end{center}
    \caption{The SNIa 06D1fx\_1413 spectrum measured at $z=0.524$ with a phase of 6.8 days. A Sa-Sb host model has been subtracted.}
    \label{fig:Spec06D1fx_1413}
    \end{figure}
    
    \begin{figure}
    \begin{center}
    \includegraphics[scale=0.45]{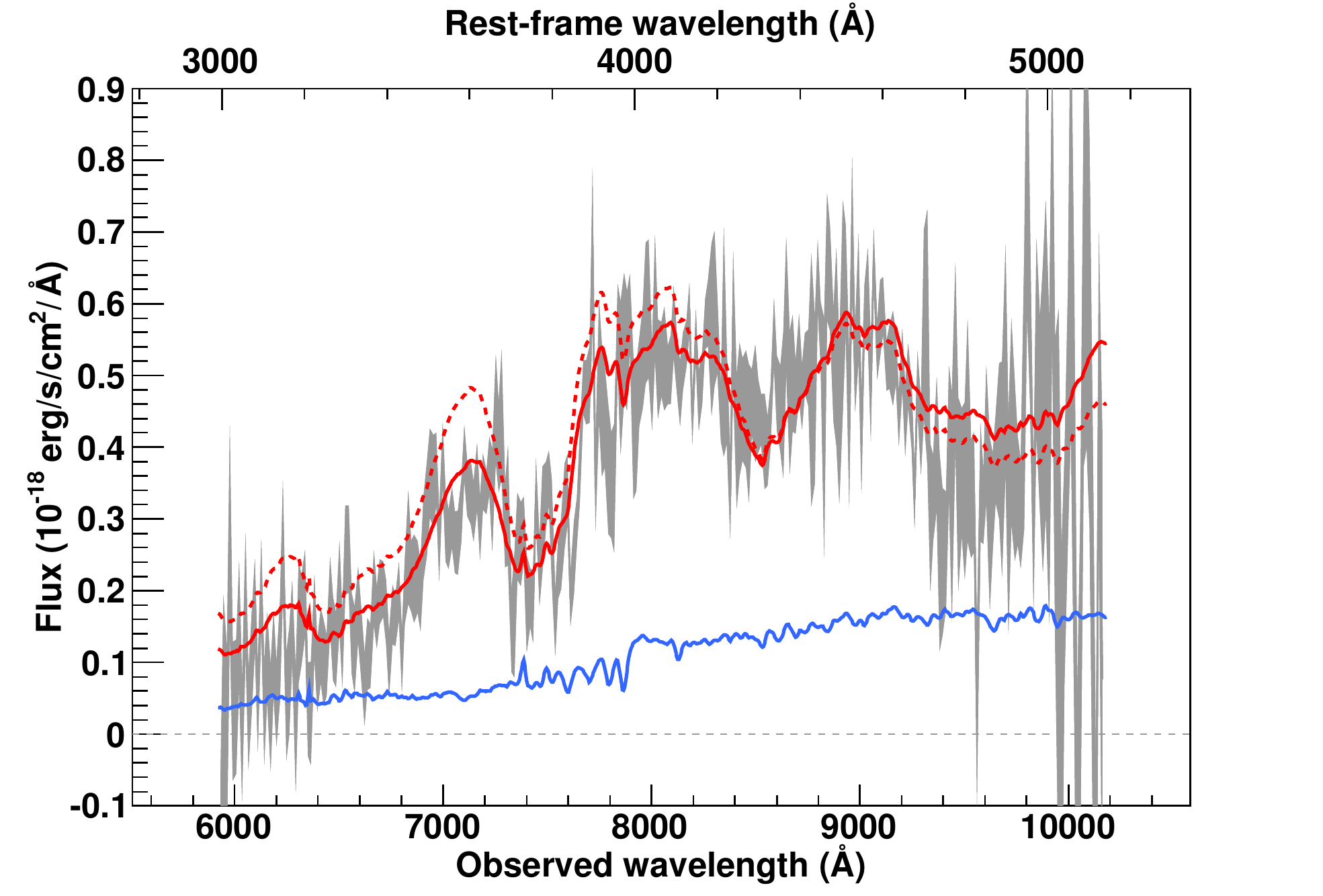}
    \includegraphics[scale=0.45]{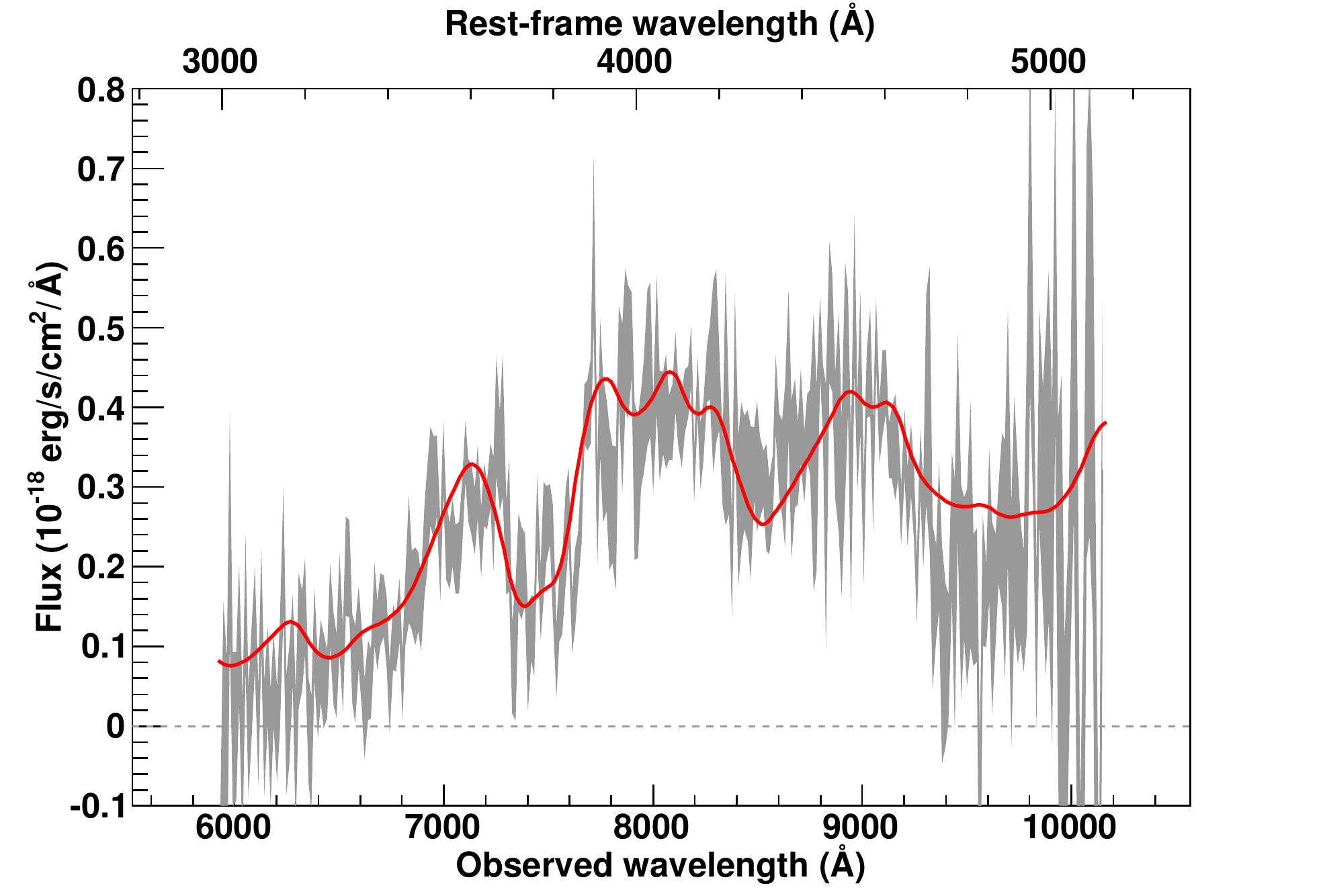}
    \end{center}
    \caption{The SNIa 06D1gl\_1417 spectrum measured at $z=0.98$ with a phase of 4.3 days. A S0-Sa host model has been subtracted.}
    \label{fig:Spec06D1gl_1417}
    \end{figure}
    
    \clearpage
    \begin{figure}
    \begin{center}
    \includegraphics[scale=0.45]{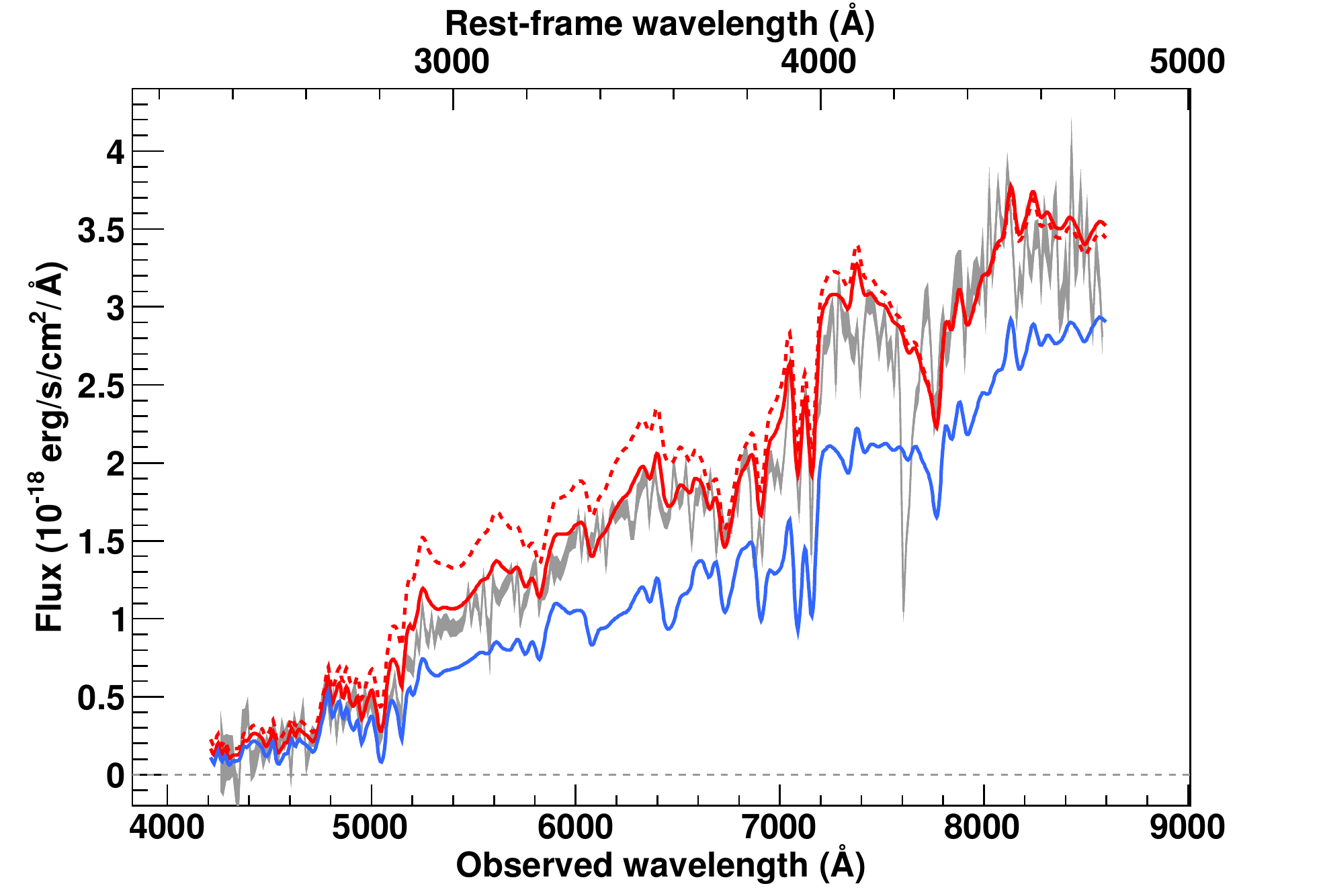}
    \includegraphics[scale=0.45]{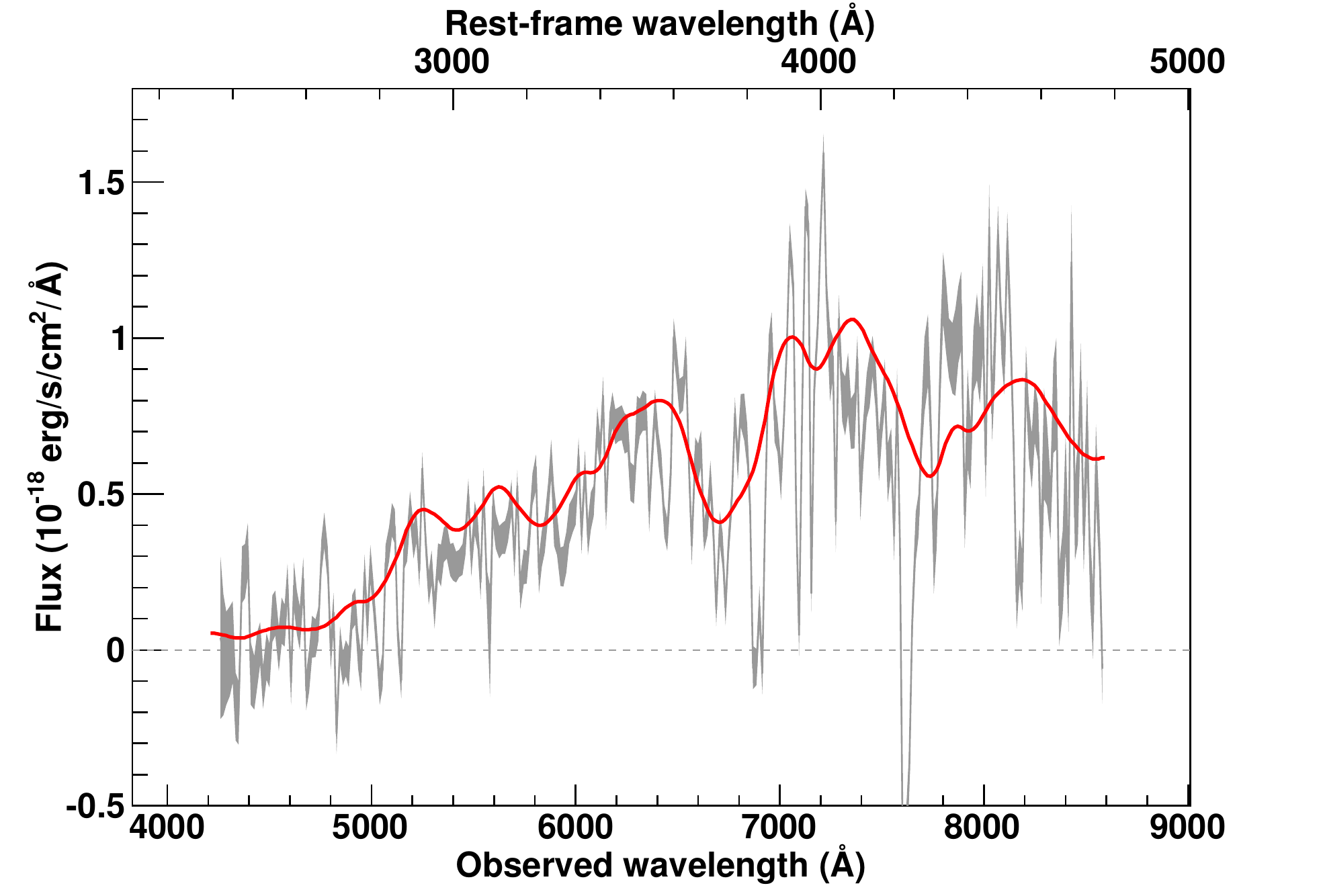}
    \end{center}
    \caption{The SNIa$\star$ 06D1hi\_1424 spectrum measured at $z=0.803$ with a phase of -3.3 days. A E(4) host model has been subtracted.}
    \label{fig:Spec06D1hi_1424}
    \end{figure}
    
    \begin{figure}
    \begin{center}
    \includegraphics[scale=0.45]{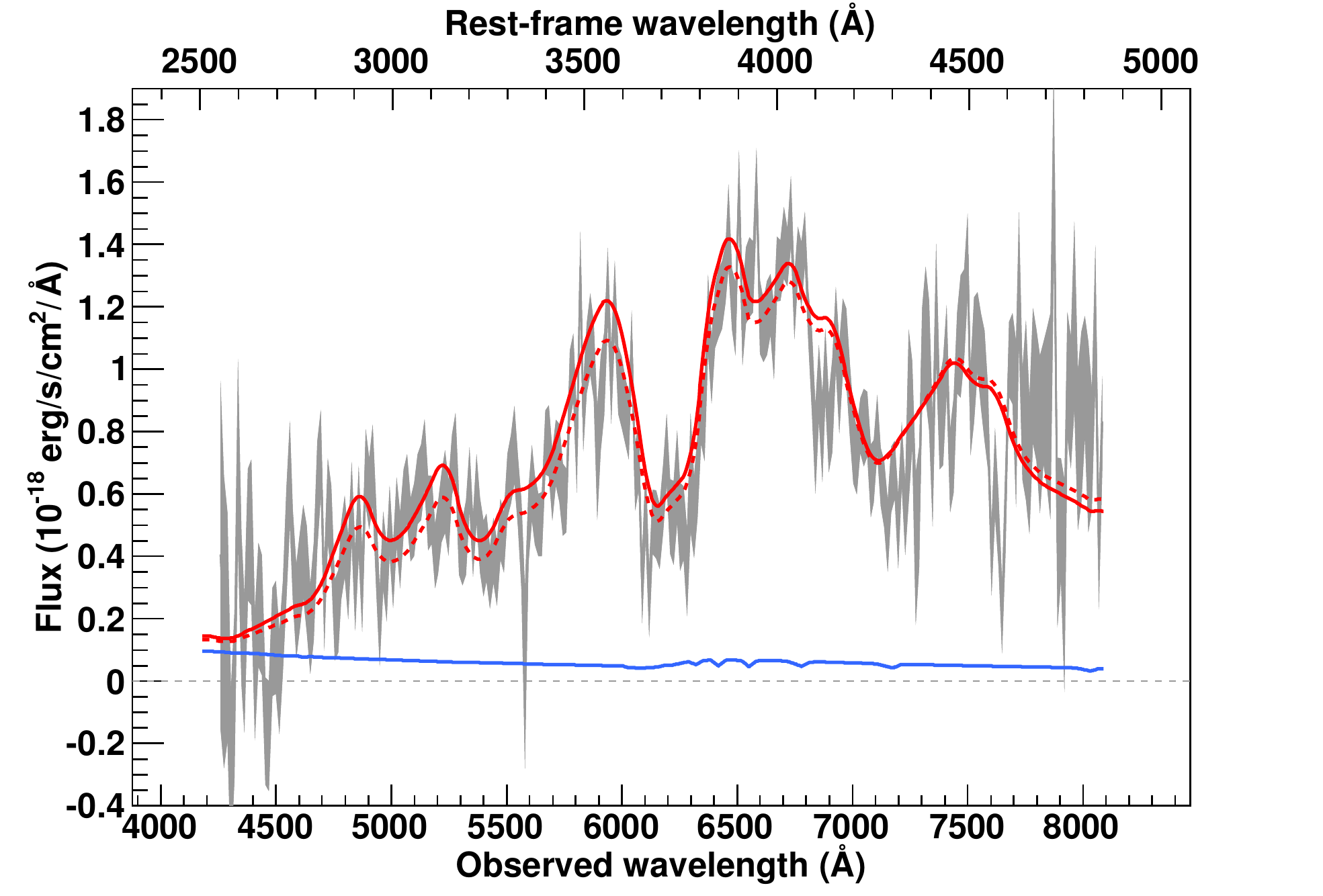}
    \includegraphics[scale=0.45]{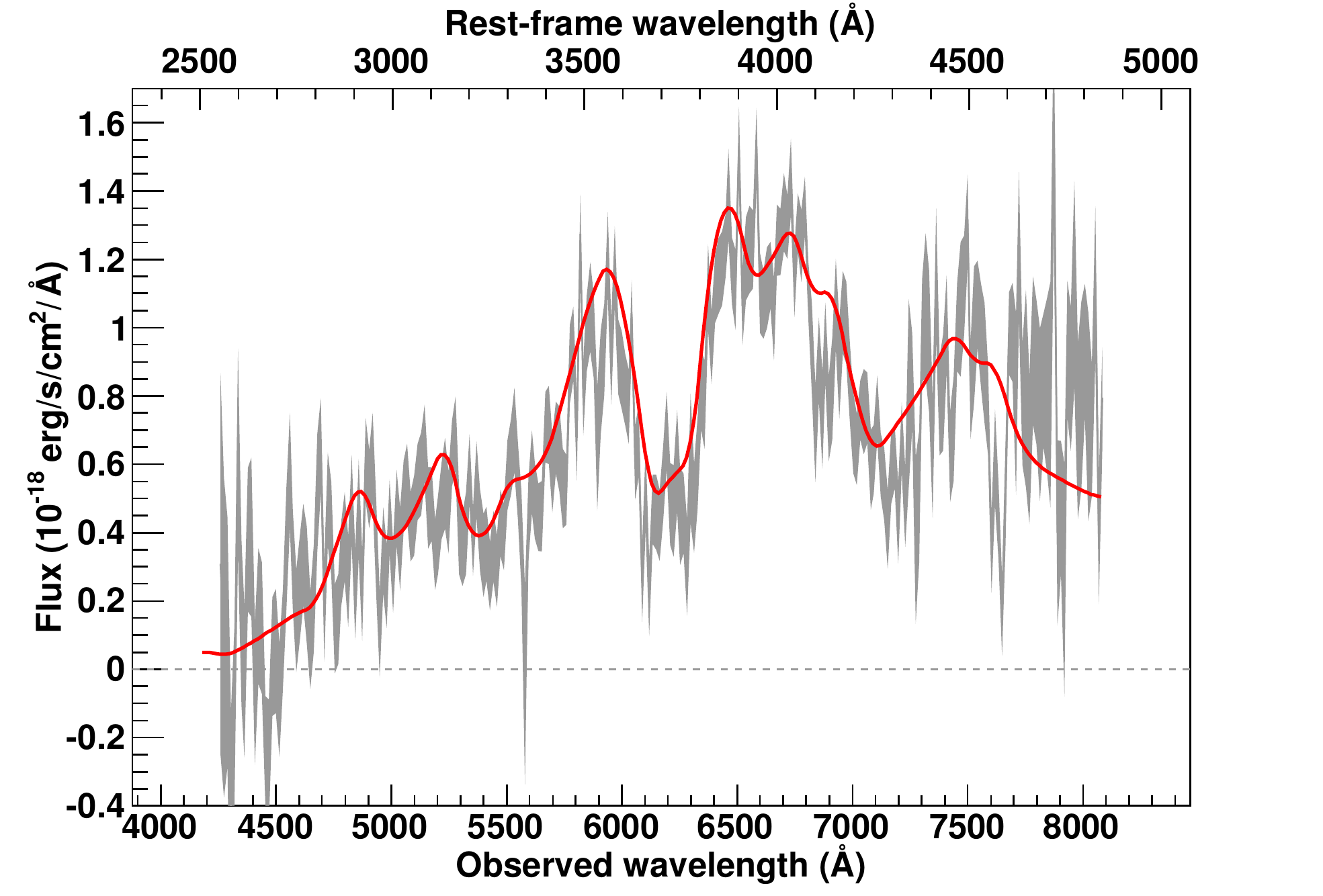}
    \end{center}
    \caption{The SNIa 06D1ix\_1447 spectrum measured at $z=0.65$ with a phase of 3.8 days. A Sd(1) host model has been subtracted.}
    \label{fig:Spec06D1ix_1447}
    \end{figure}
    
    \begin{figure}
    \begin{center}
    \includegraphics[scale=0.45]{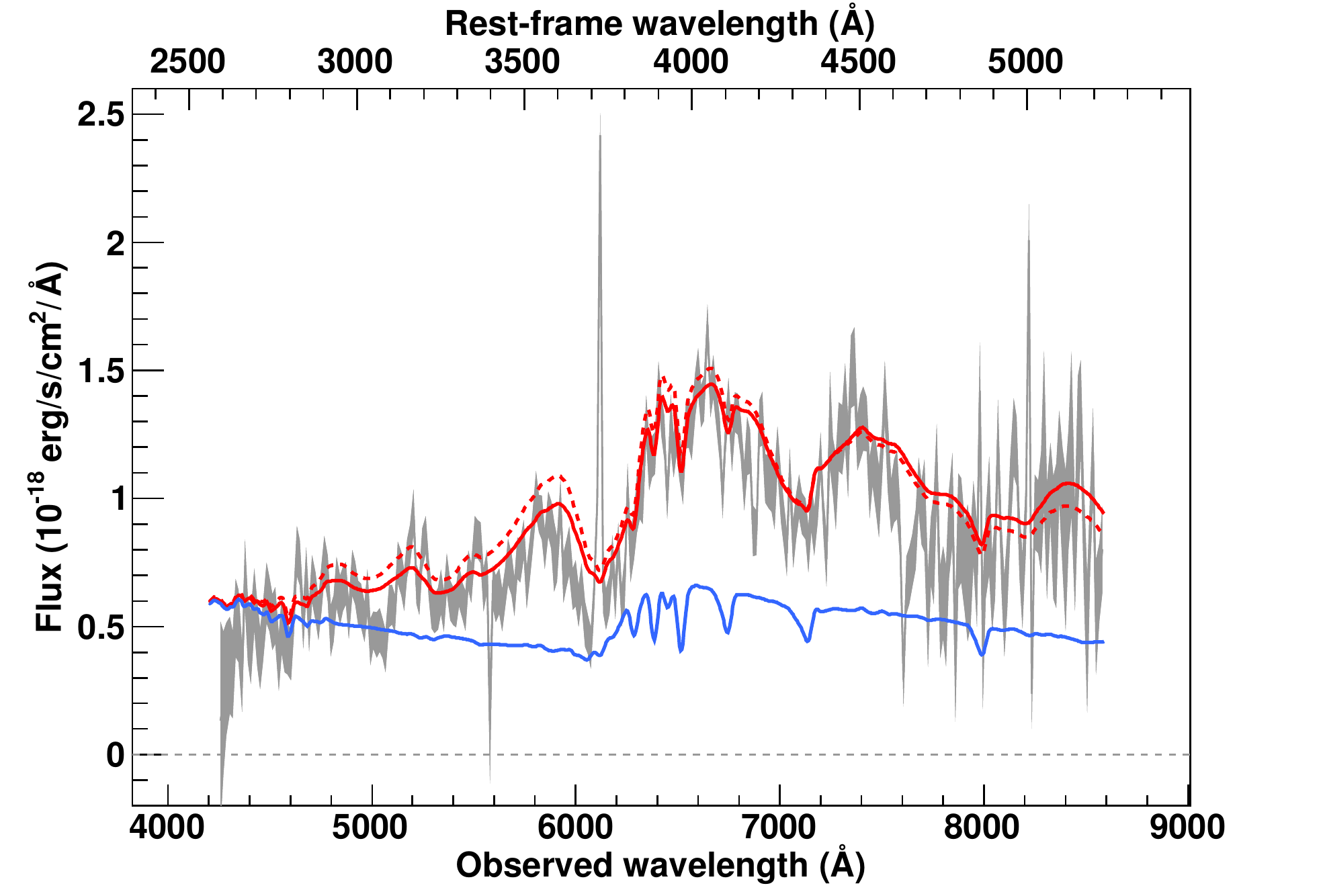}
    \includegraphics[scale=0.45]{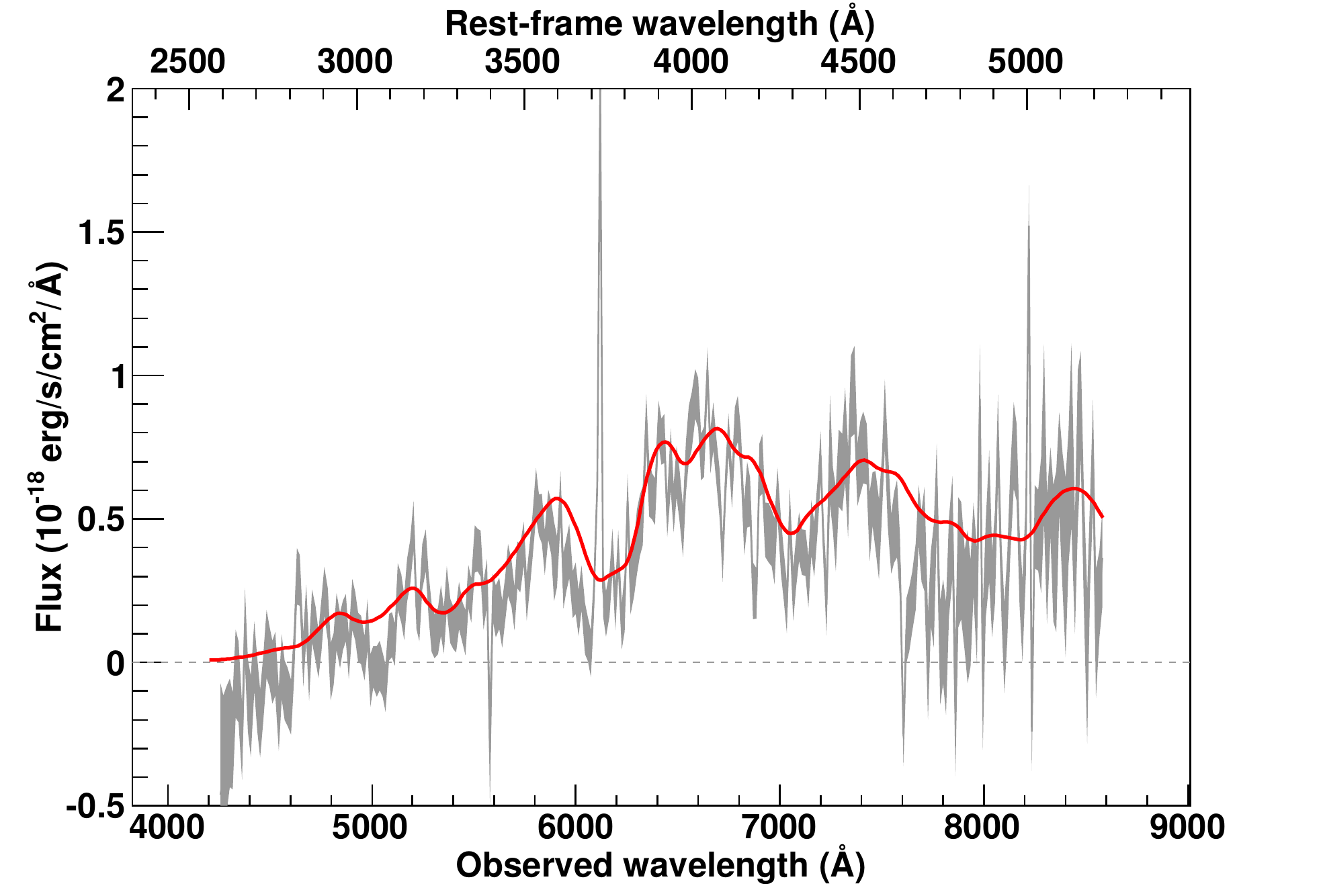}
    \end{center}
    \caption{The SNIa 06D1jf\_1447 spectrum measured at $z=0.641$ with a phase of 1.5 days. A Sc(4) host model has been subtracted.}
    \label{fig:Spec06D1jf_1447}
    \end{figure}
    
    \clearpage
    \begin{figure}
    \begin{center}
    \includegraphics[scale=0.45]{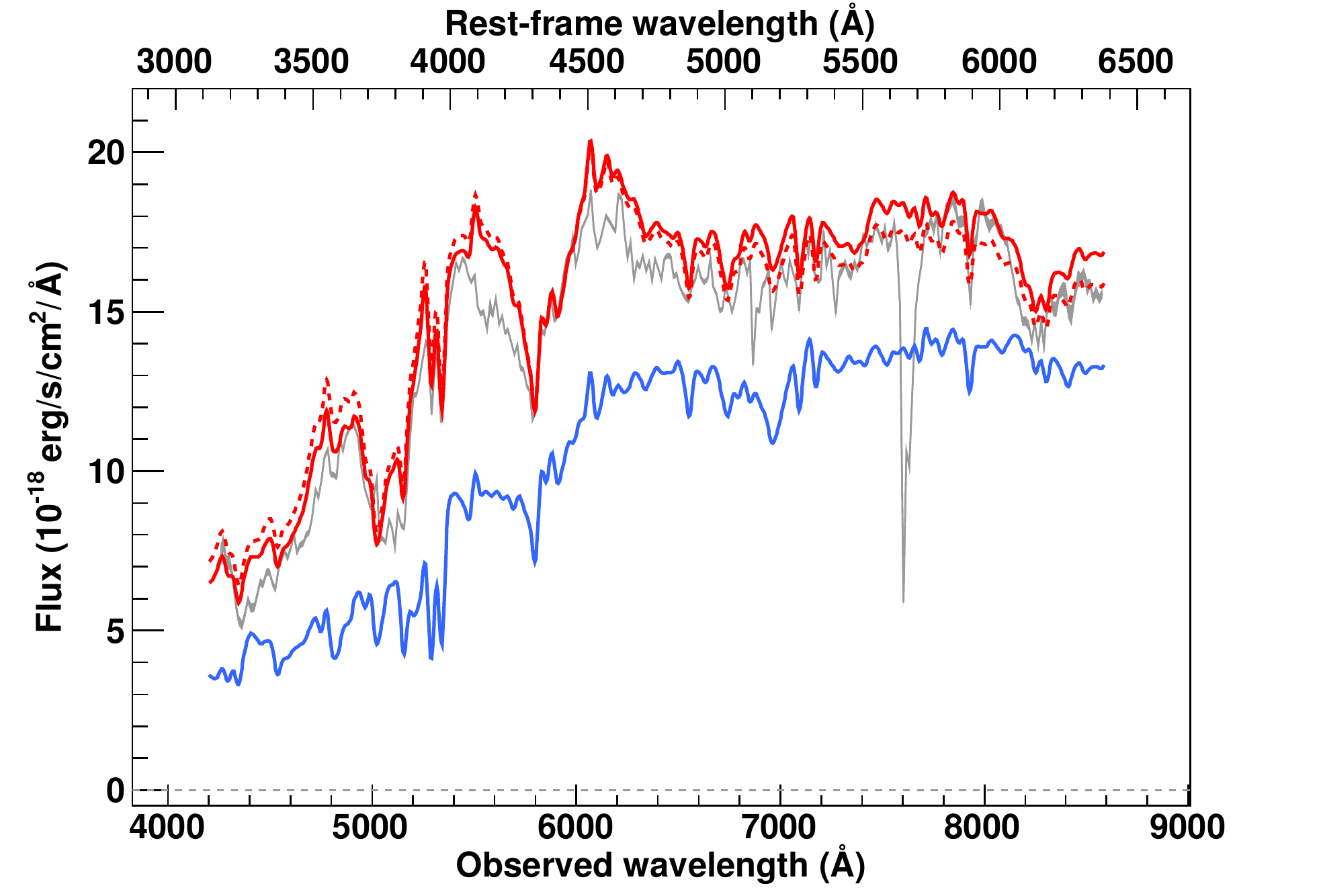}
    \includegraphics[scale=0.45]{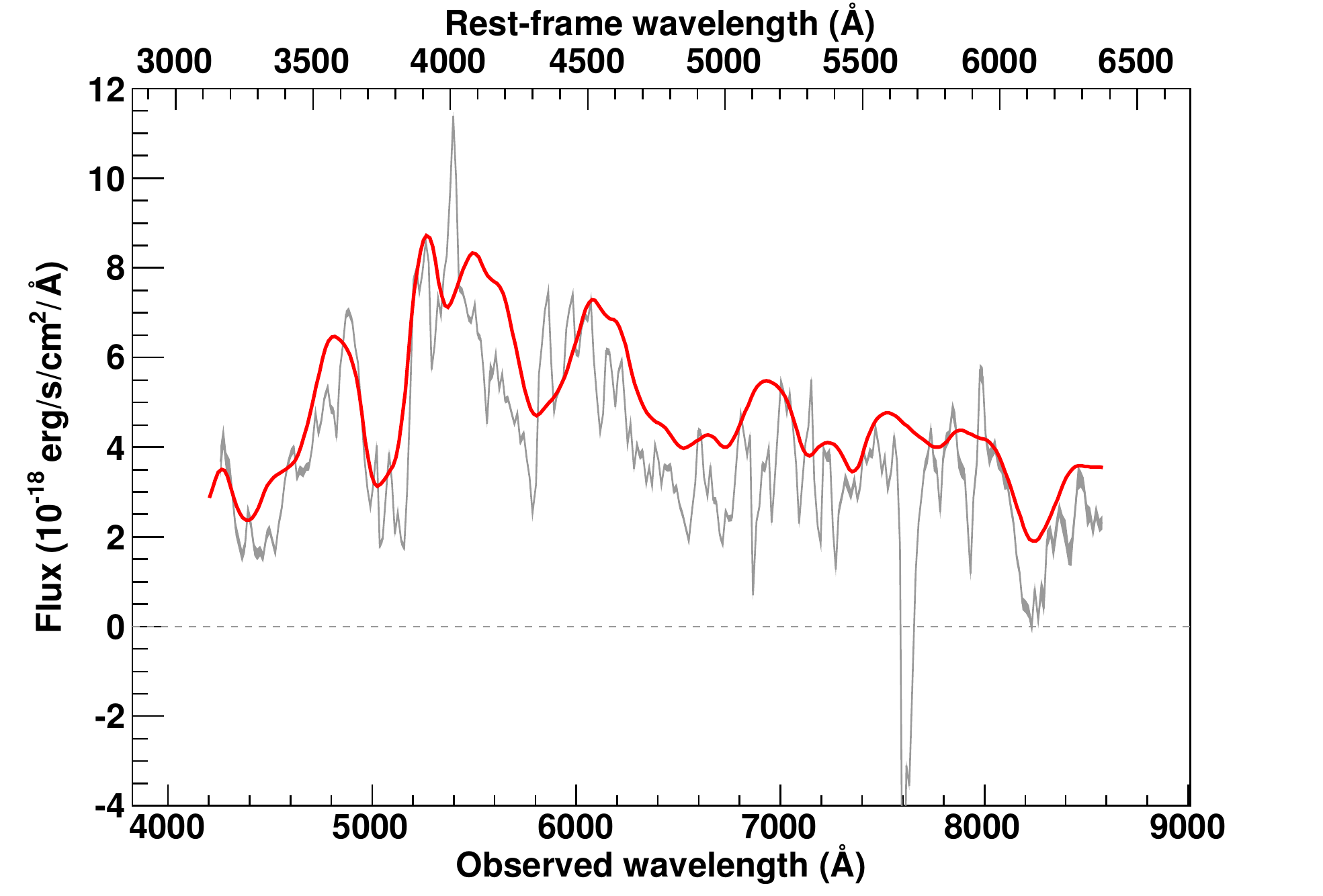}
    \end{center}
    \caption{The SNIa 06D1jz\_1452 spectrum measured at $z=0.346$ with a phase of 3.3 days. A S0(7) host model has been subtracted.}
    \label{fig:Spec06D1jz_1452}
    \end{figure}
    
    \begin{figure}
    \begin{center}
    \includegraphics[scale=0.45]{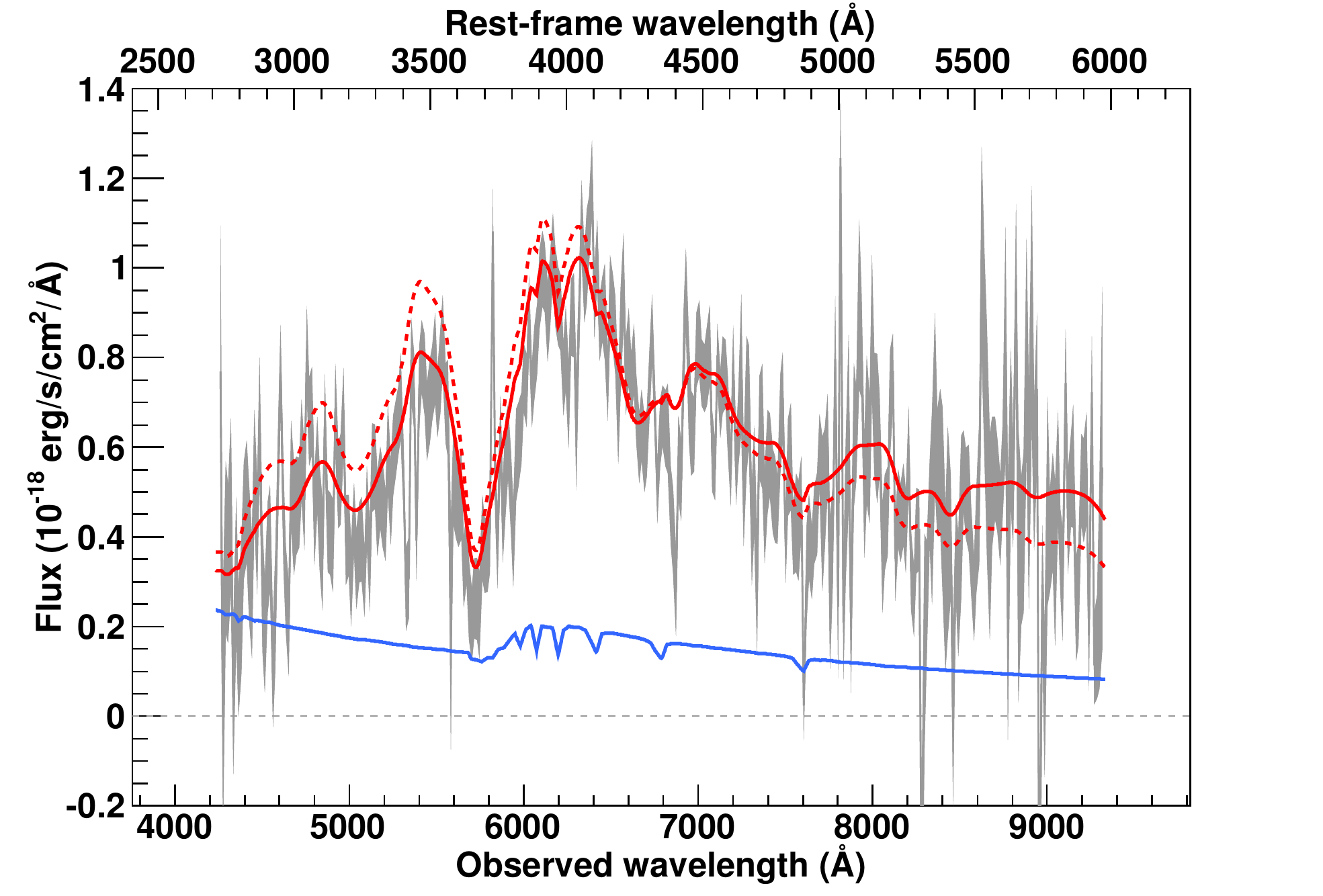}
    \includegraphics[scale=0.45]{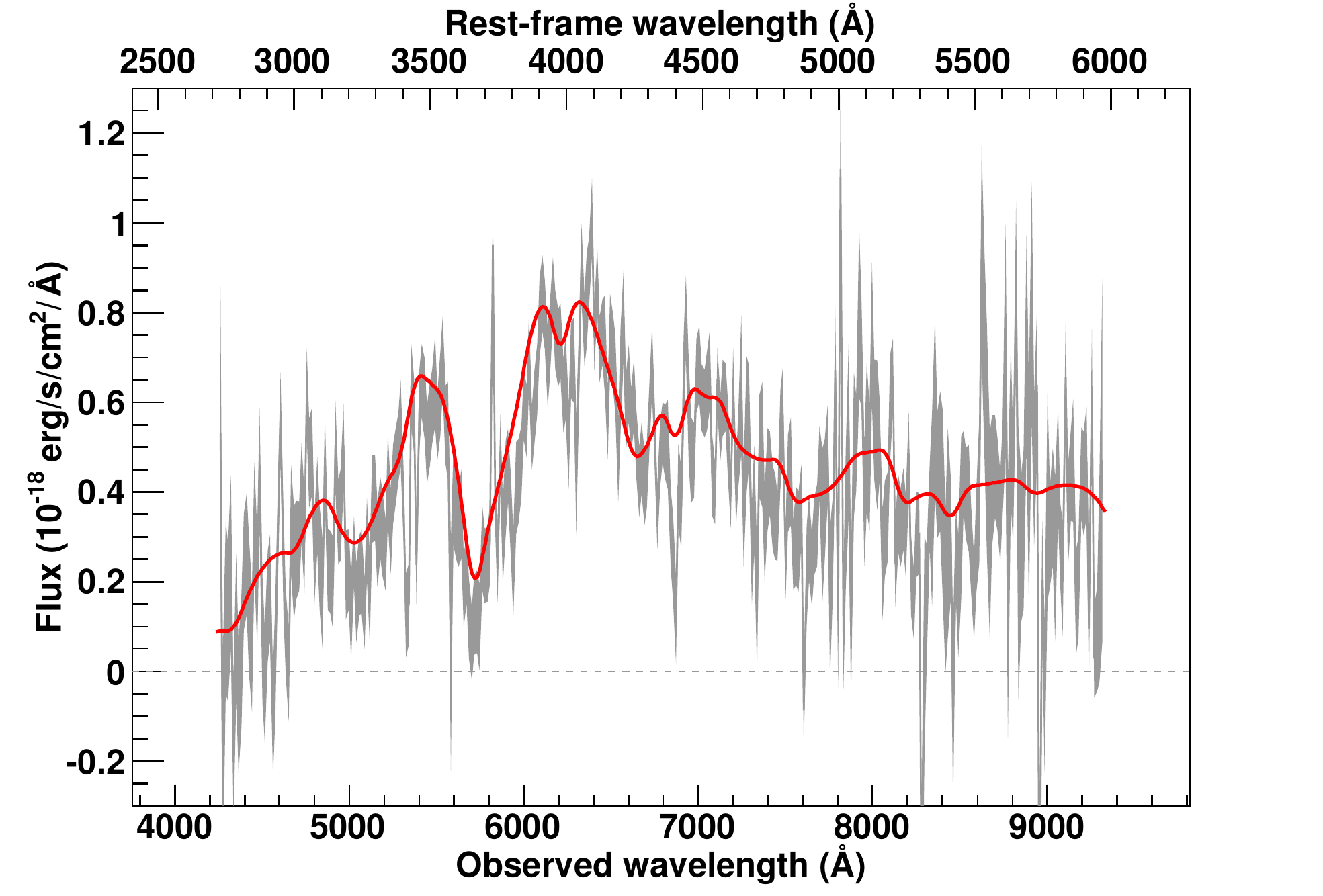}
    \end{center}
    \caption{The SNIa 06D1kf\_1453 spectrum measured at $z=0.561$ with a phase of -6.5 days. A Sd(1) host model has been subtracted.}
    \label{fig:Spec06D1kf_1453}
    \end{figure}
    
    \begin{figure}
    \begin{center}
    \includegraphics[scale=0.45]{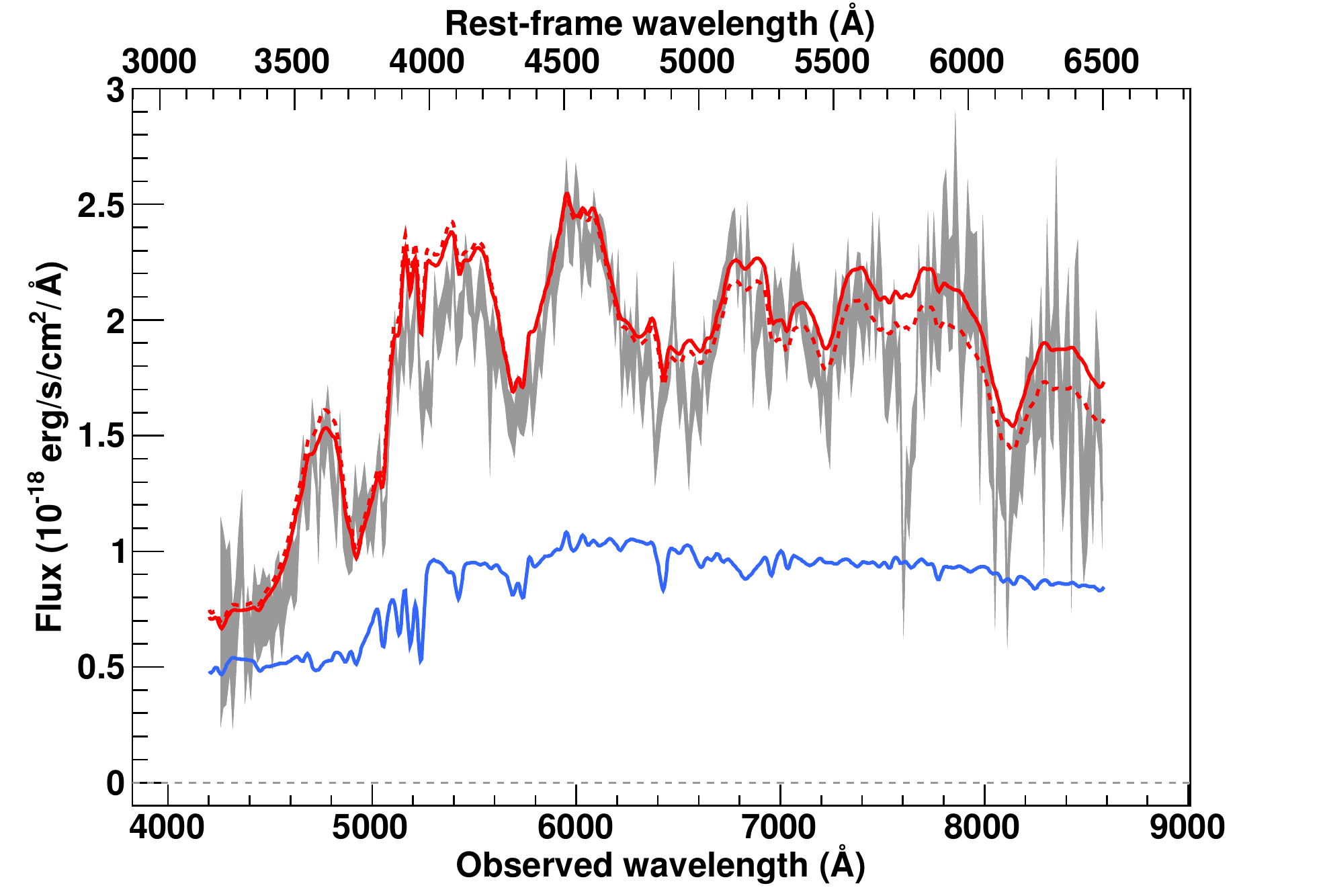}
    \includegraphics[scale=0.45]{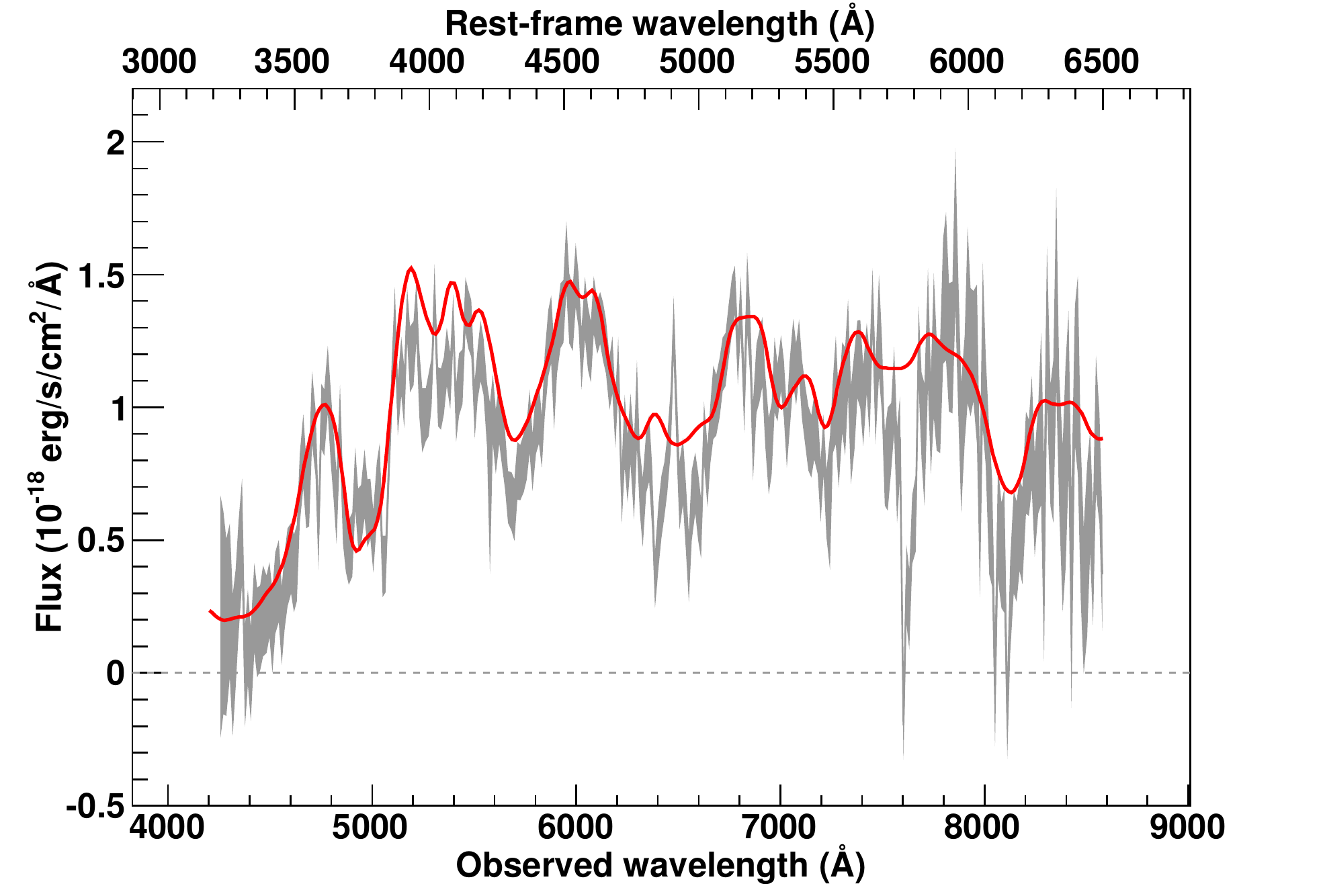}
    \end{center}
    \caption{The SNIa 06D1kg\_1477 spectrum measured at $z=0.32$ with a phase of 6.1 days. A S0(2) host model has been subtracted.}
    \label{fig:Spec06D1kg_1477}
    \end{figure}
    
    \clearpage
    \begin{figure}
    \begin{center}
    \includegraphics[scale=0.45]{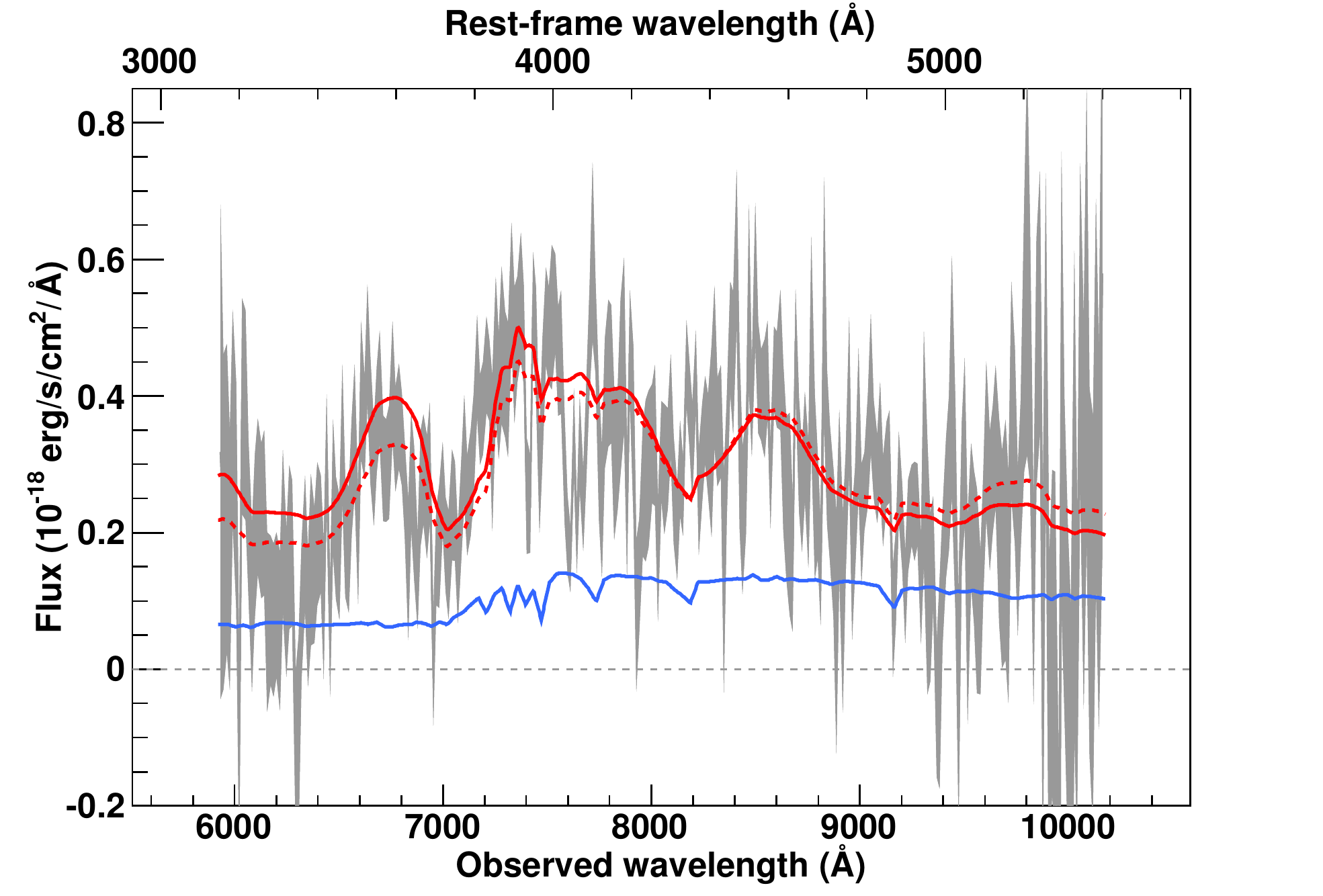}
    \includegraphics[scale=0.45]{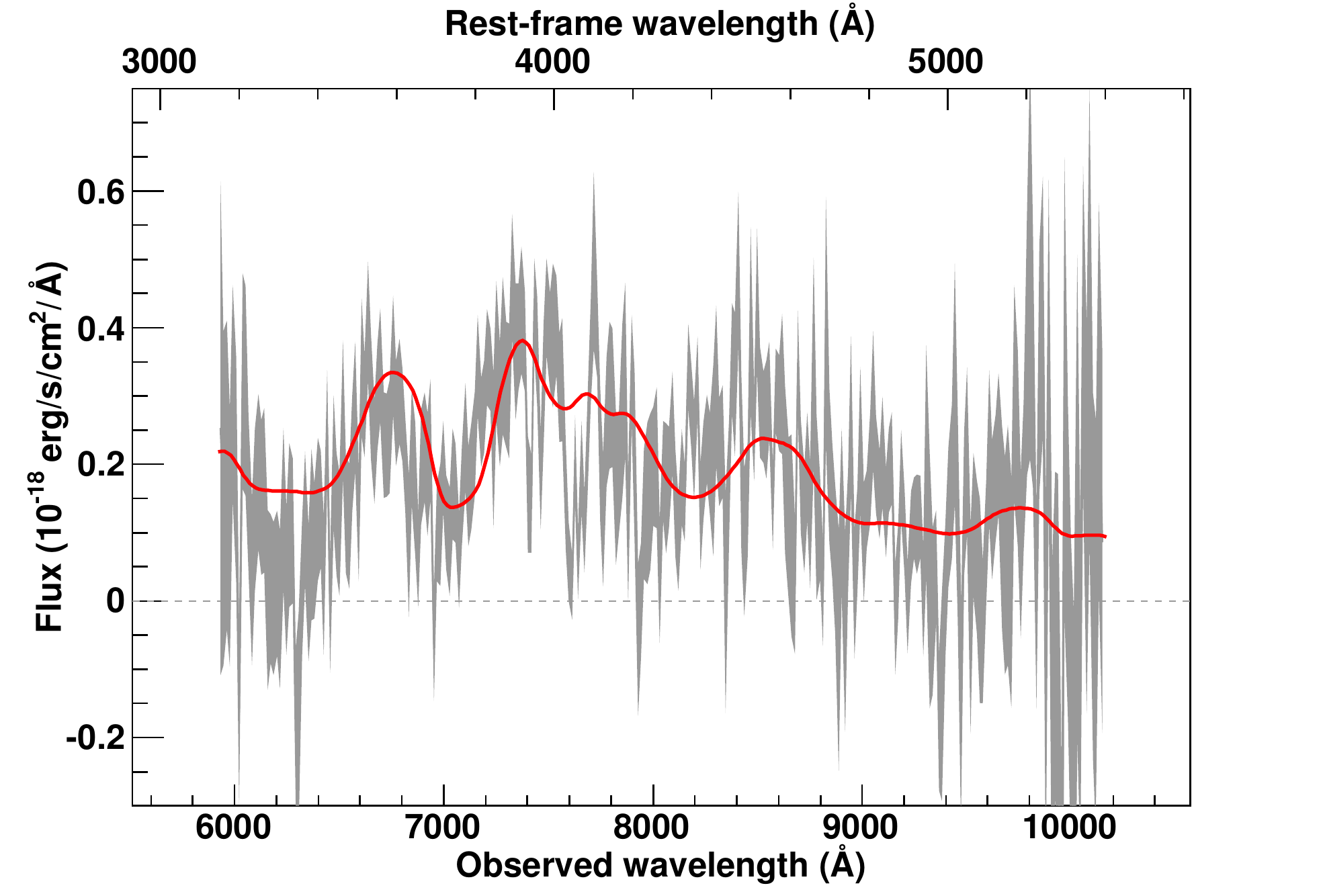}
    \end{center}
    \caption{The SNIa$\star$ 06D1kh\_1483 spectrum measured at $z=0.882$ with a phase of 7.3 days. A E(1) host model has been subtracted.}
    \label{fig:Spec06D1kh_1483}
    \end{figure}
    
    \begin{figure}
    \begin{center}
    \includegraphics[scale=0.45]{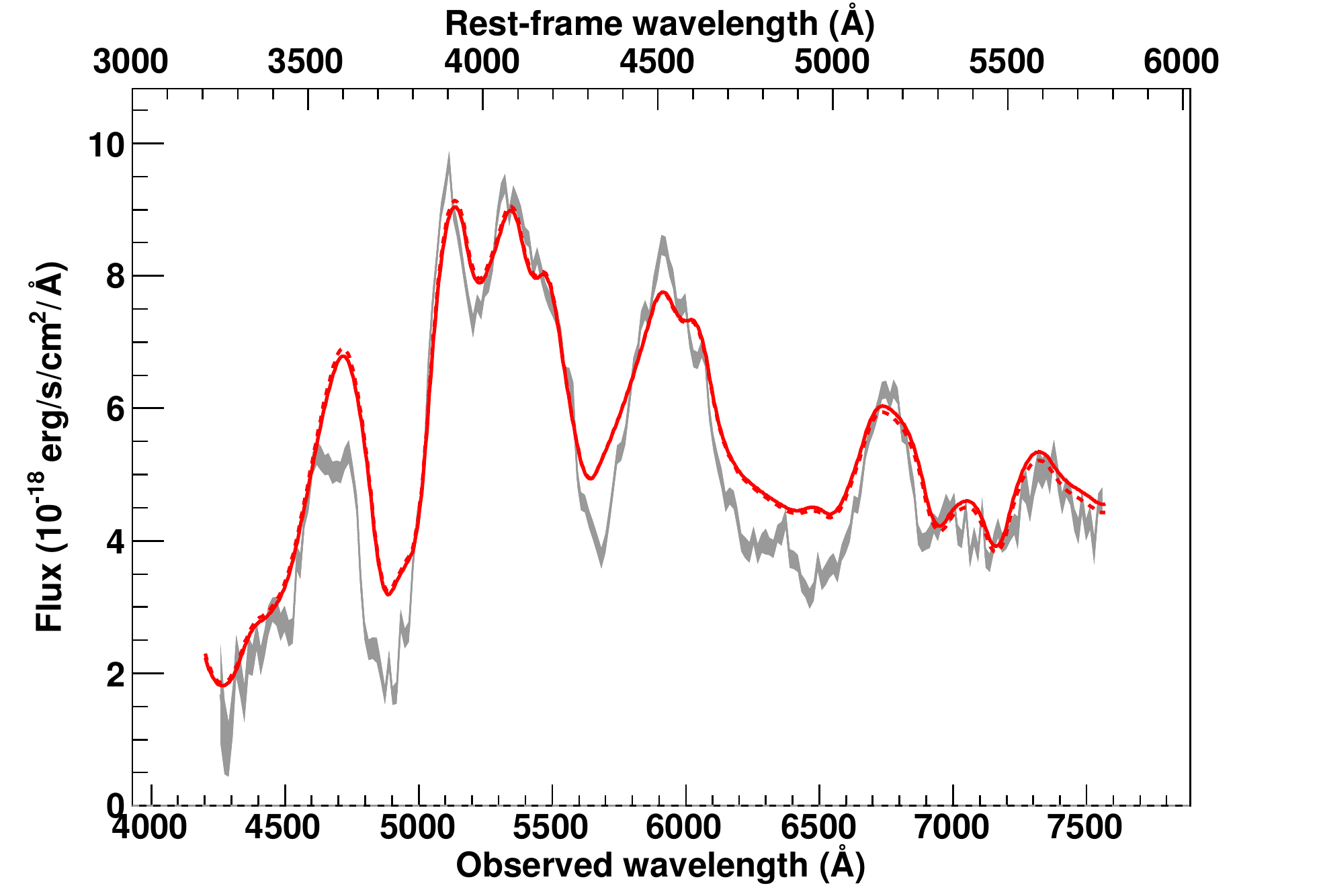}
    \end{center}
    \caption{The SNIa 06D2ag\_1121 spectrum measured at $z=0.310$ with a phase of 4.0 days. A Best fit is obtained without galactic component.}
    \label{fig:Spec06D2ag_1121}
    \end{figure}
    
    \begin{figure}
    \begin{center}
    \includegraphics[scale=0.45]{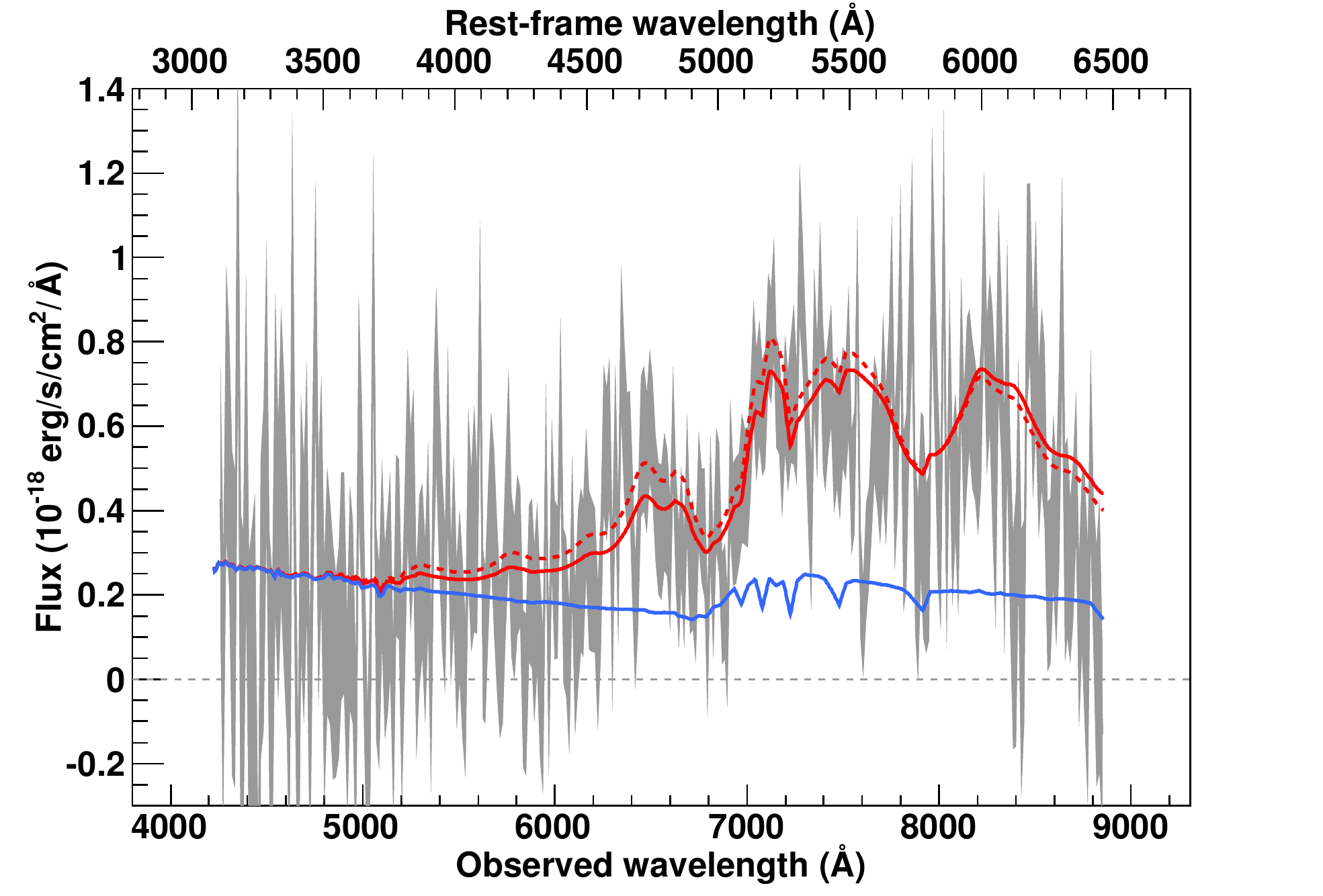}
    \includegraphics[scale=0.45]{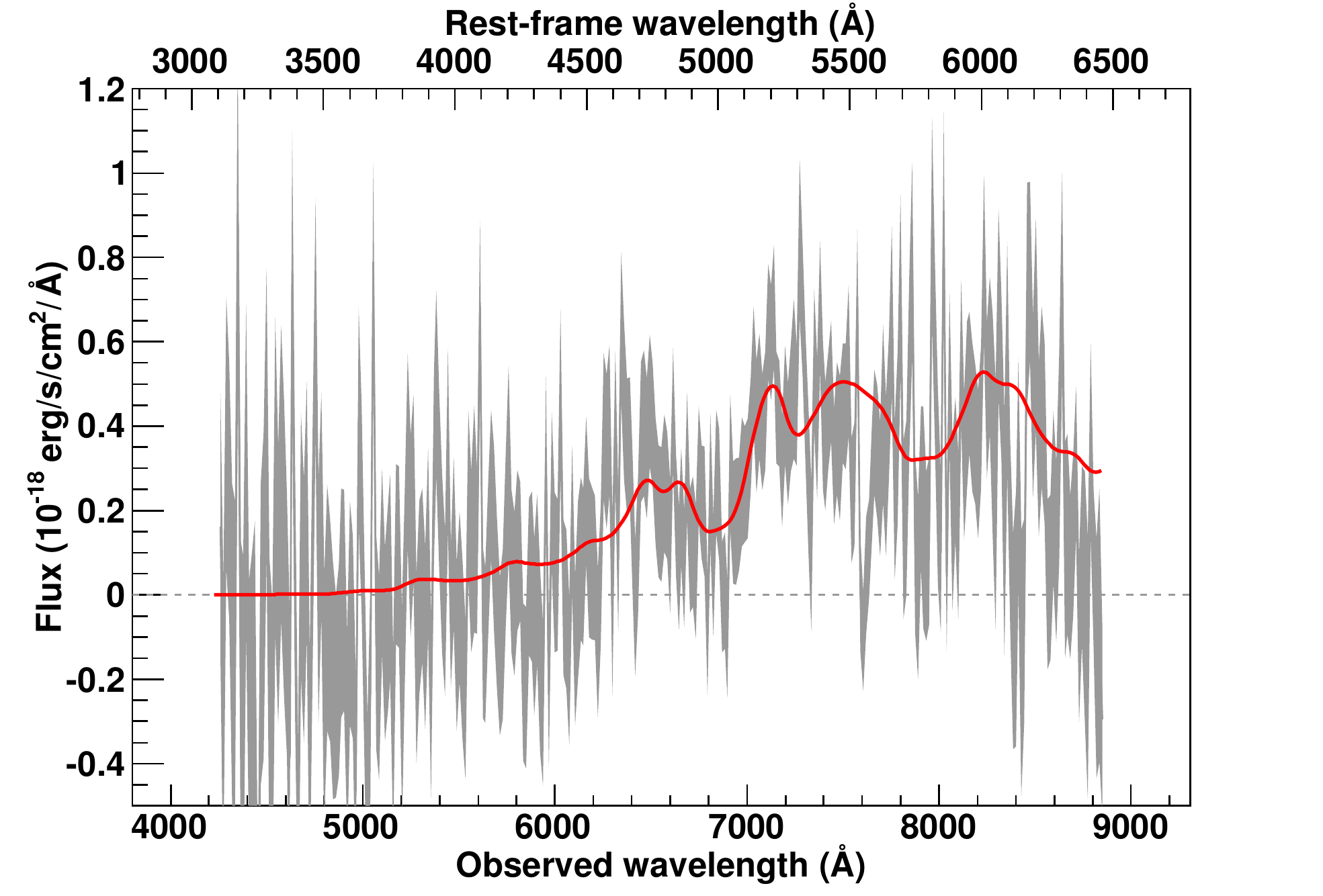}
    \end{center}
    \caption{The SNIa$\star$ 06D2bo\_1134 spectrum measured at $z=0.82$ with a phase of 2.6 days. A Sa(1) host model has been subtracted.}
    \label{fig:Spec06D2bo_1134}
    \end{figure}
    
    \clearpage
    \begin{figure}
    \begin{center}
    \includegraphics[scale=0.45]{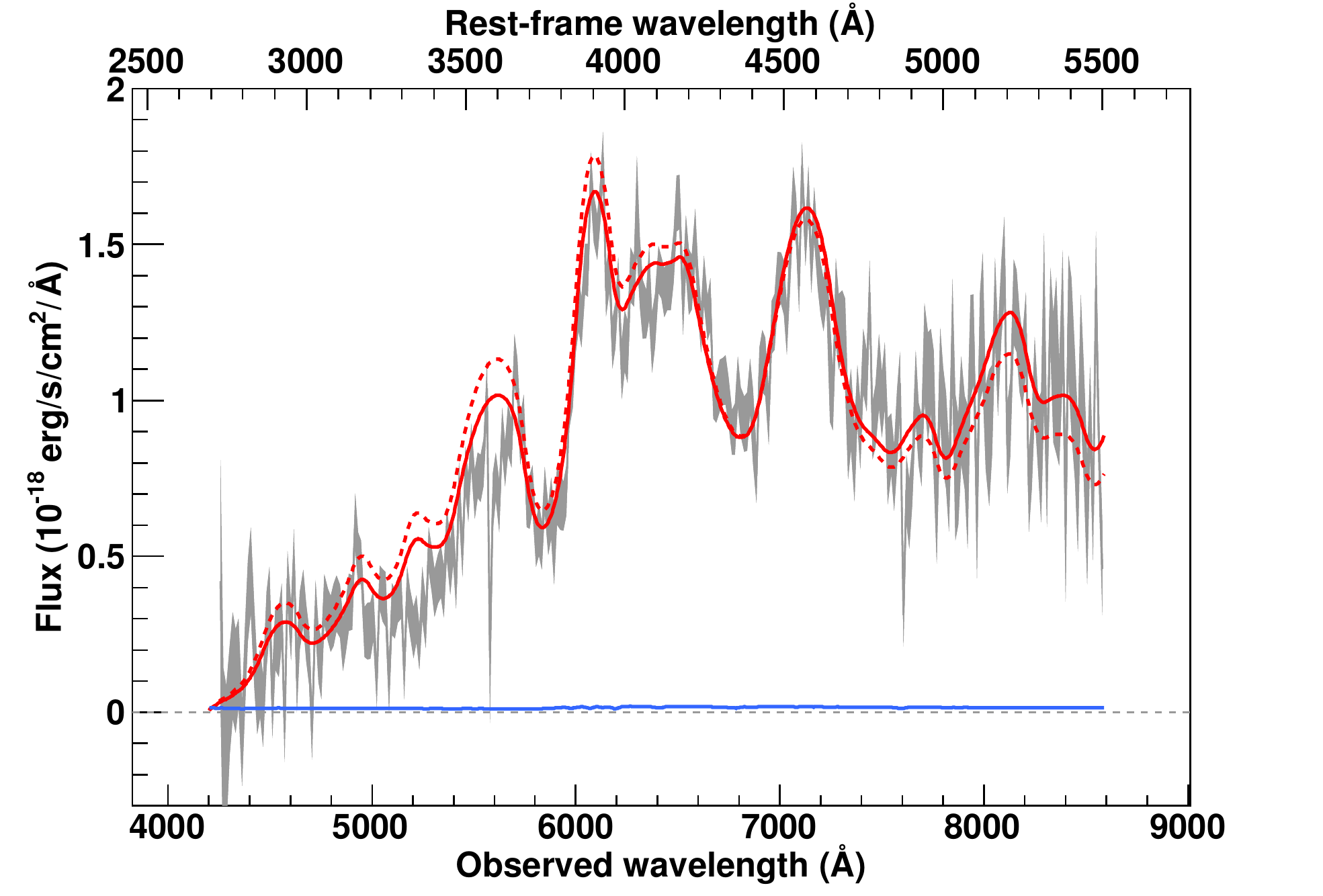}
    \includegraphics[scale=0.45]{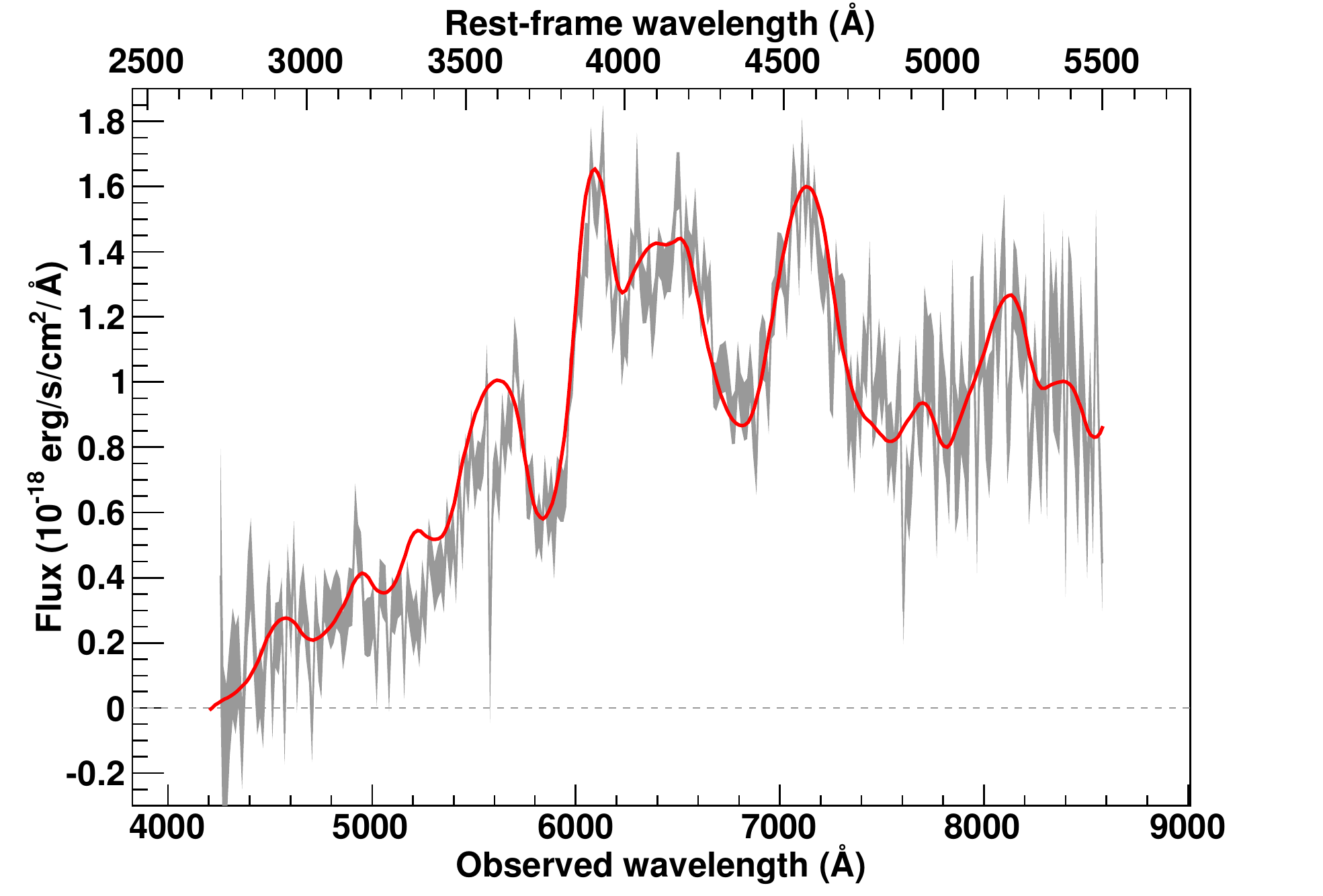}
    \end{center}
    \caption{The SNIa 06D2hm\_1447 spectrum measured at $z=0.56$ with a phase of 7.9 days. A Sa(3) host model has been subtracted.}
    \label{fig:Spec06D2hm_1447}
    \end{figure}
    
    \begin{figure}
    \begin{center}
    \includegraphics[scale=0.45]{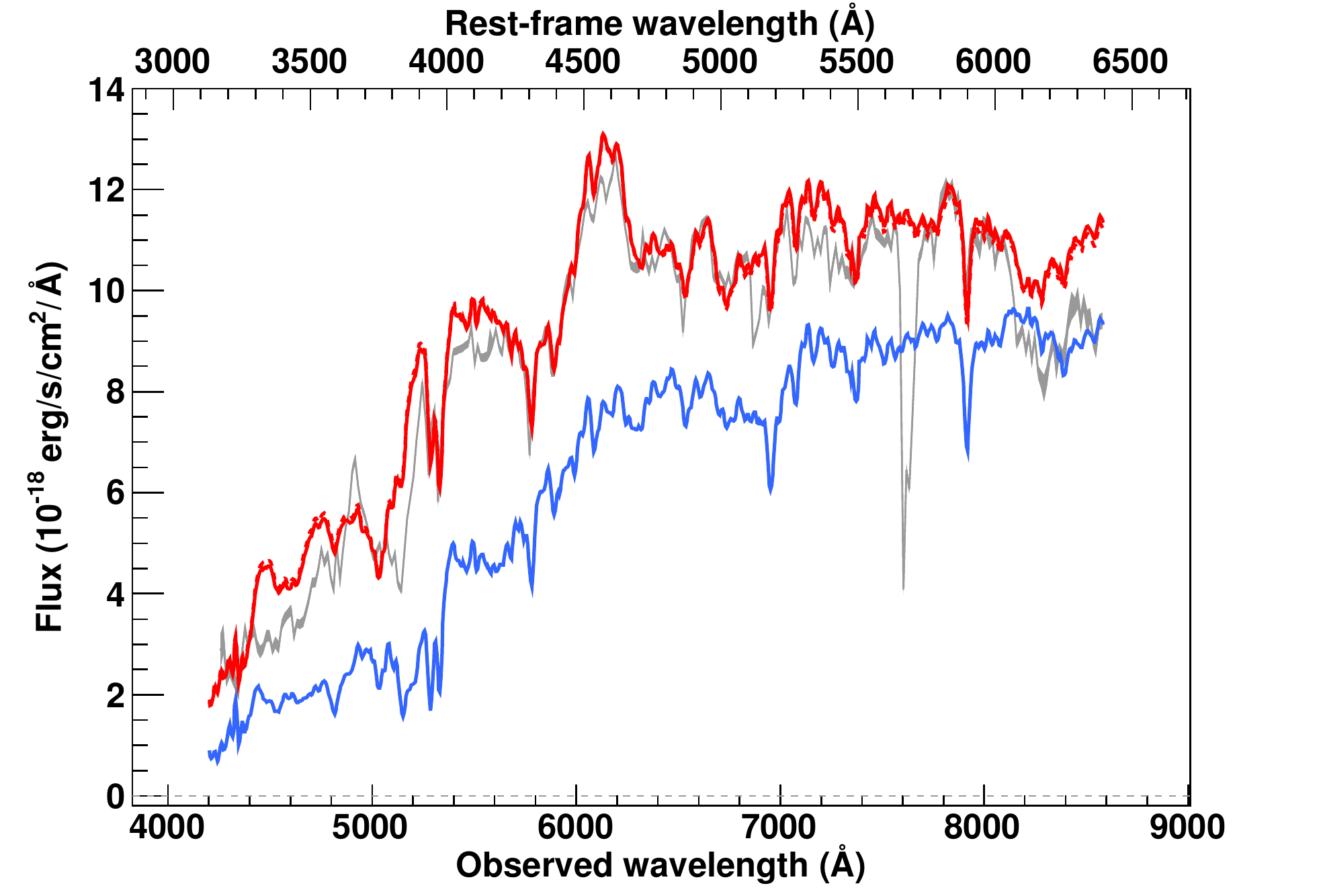}
    \includegraphics[scale=0.45]{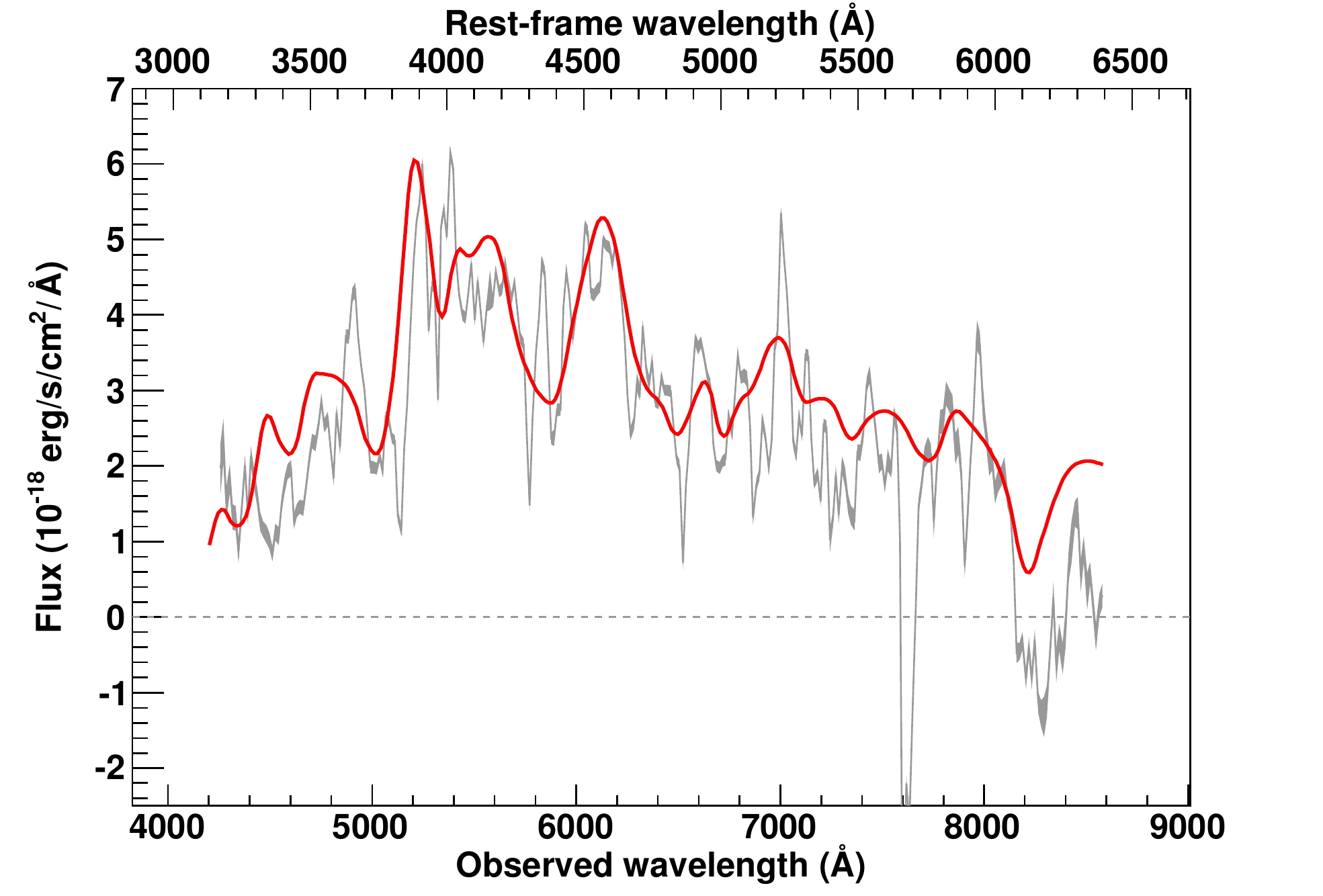}
    \end{center}
    \caption{The SNIa 06D2hu\_1449 spectrum measured at $z=0.342$ with a phase of 7.2 days. A E-S0 host model has been subtracted.}
    \label{fig:Spec06D2hu_1449}
    \end{figure}
    
    \begin{figure}
    \begin{center}
    \includegraphics[scale=0.45]{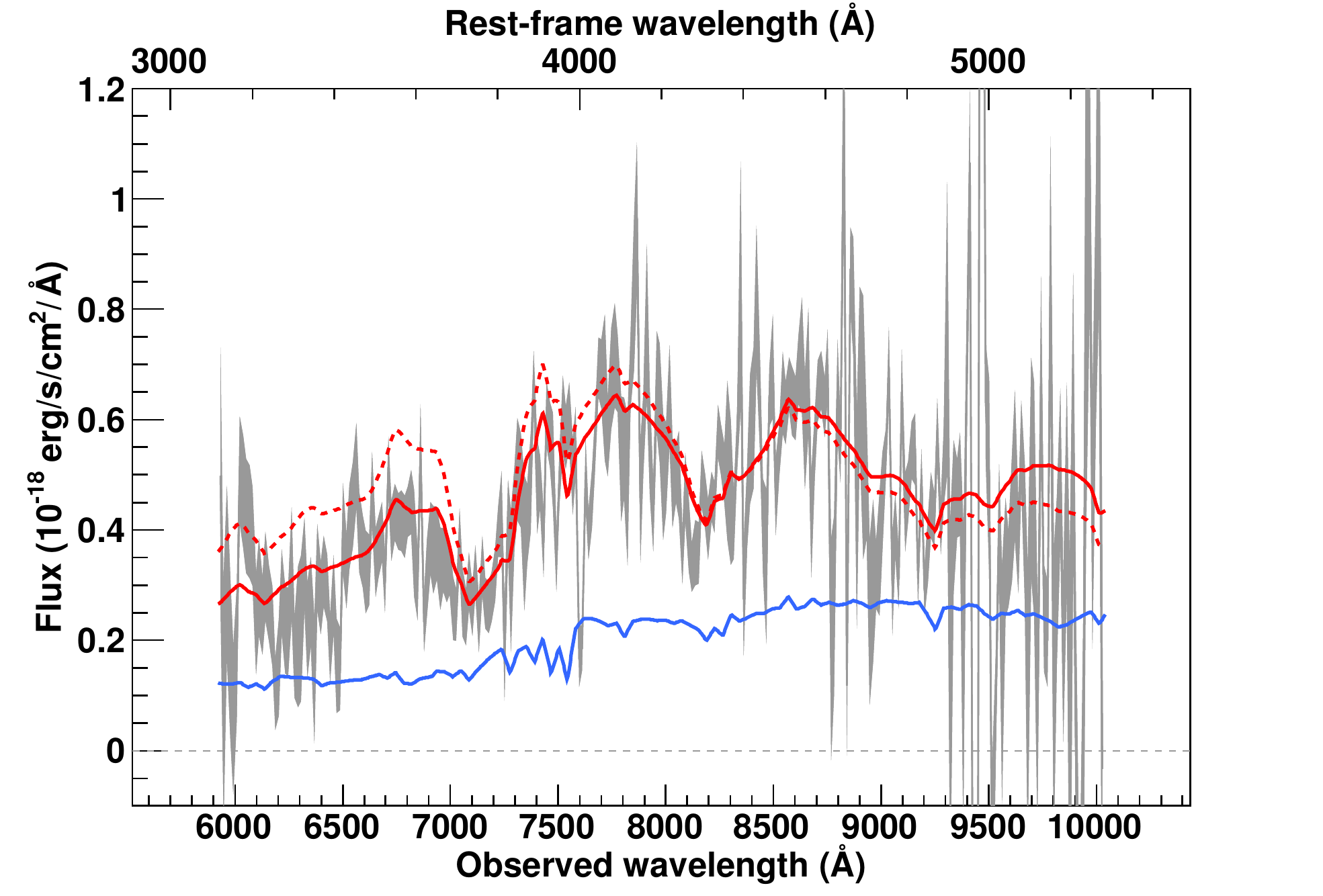}
    \includegraphics[scale=0.45]{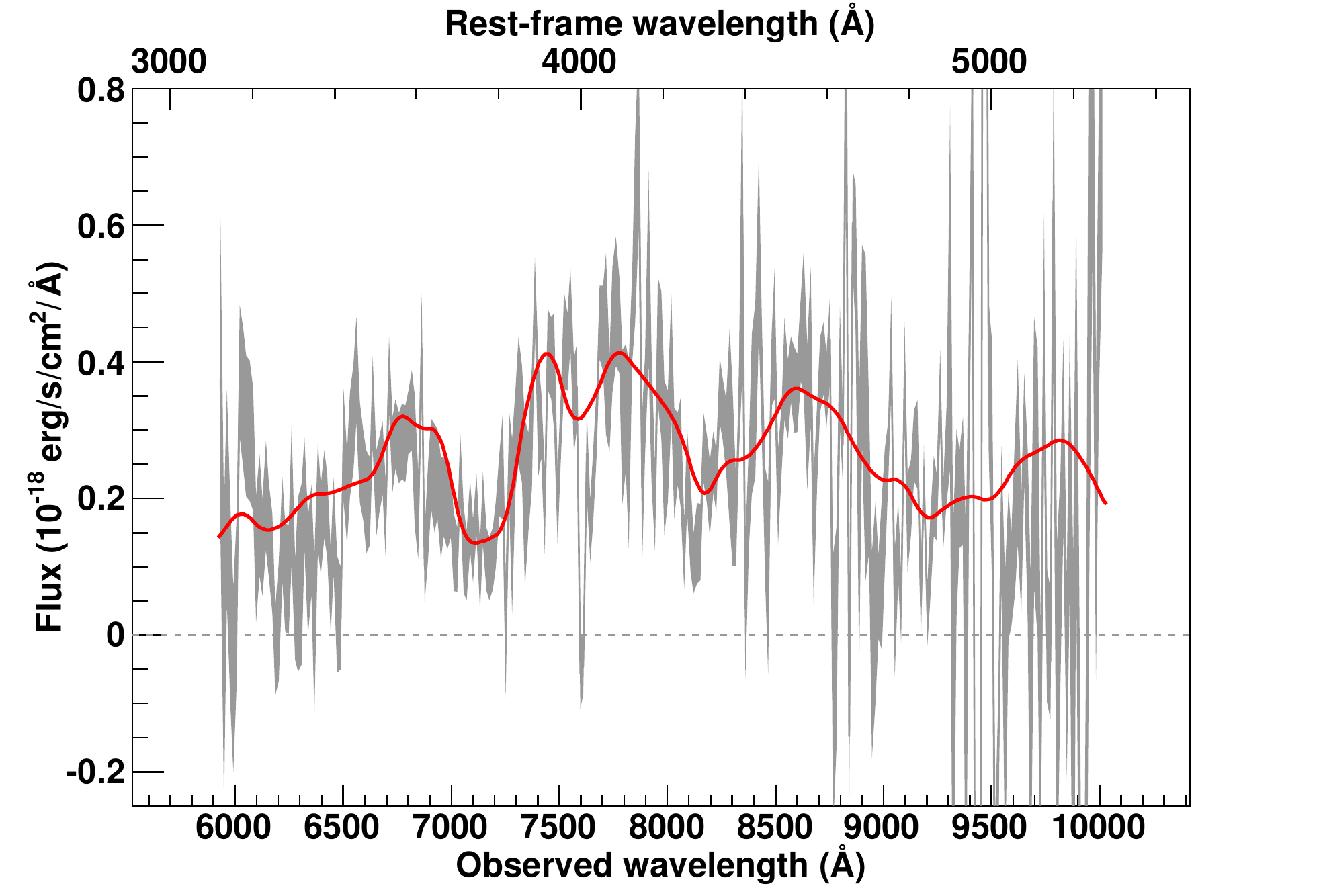}
    \end{center}
    \caption{The SNIa$\star$ 06D2jw\_1456 spectrum measured at $z=0.90$ with a phase of -0.1 days. A E(1) host model has been subtracted.}
    \label{fig:Spec06D2jw_1456}
    \end{figure}
    
    \clearpage
    \begin{figure}
    \begin{center}
    \includegraphics[scale=0.45]{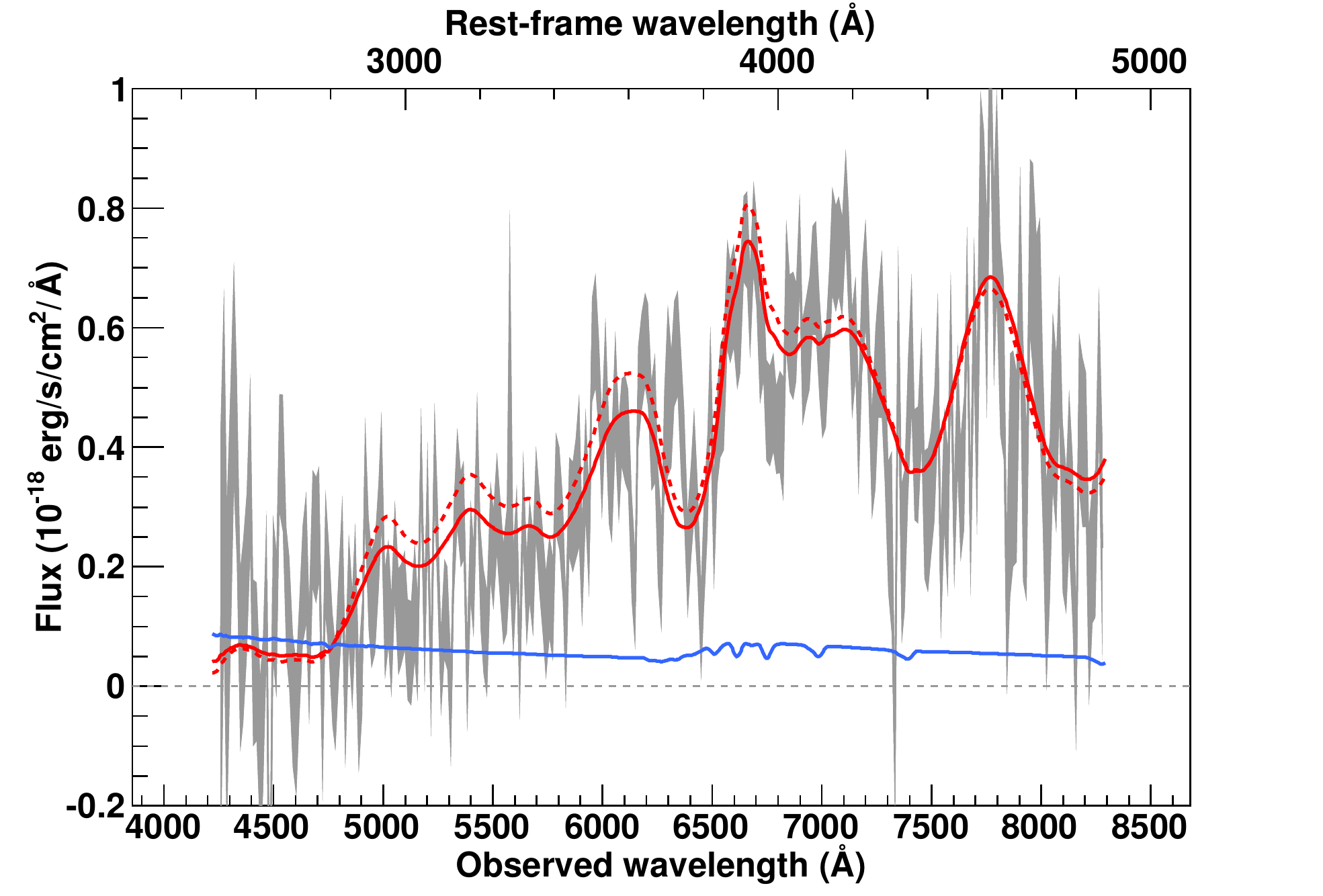}
    \includegraphics[scale=0.45]{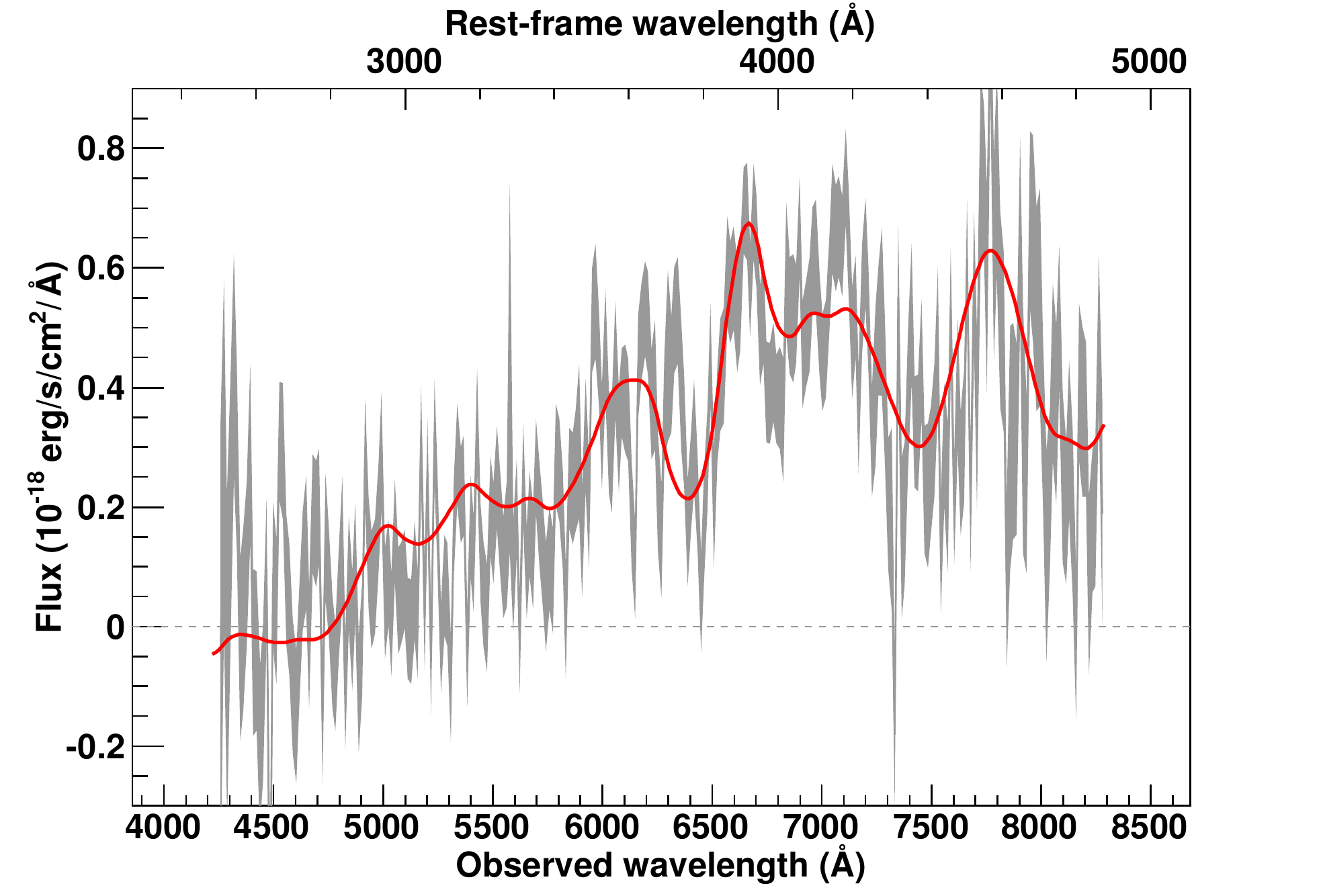}
    \end{center}
    \caption{The SNIa 06D4ba\_1280 spectrum measured at $z=0.70$ with a phase of 9.2 days. A Sd(2) host model has been subtracted.}
    \label{fig:Spec06D4ba_1280}
    \end{figure}
    
    \begin{figure}
    \begin{center}
    \includegraphics[scale=0.45]{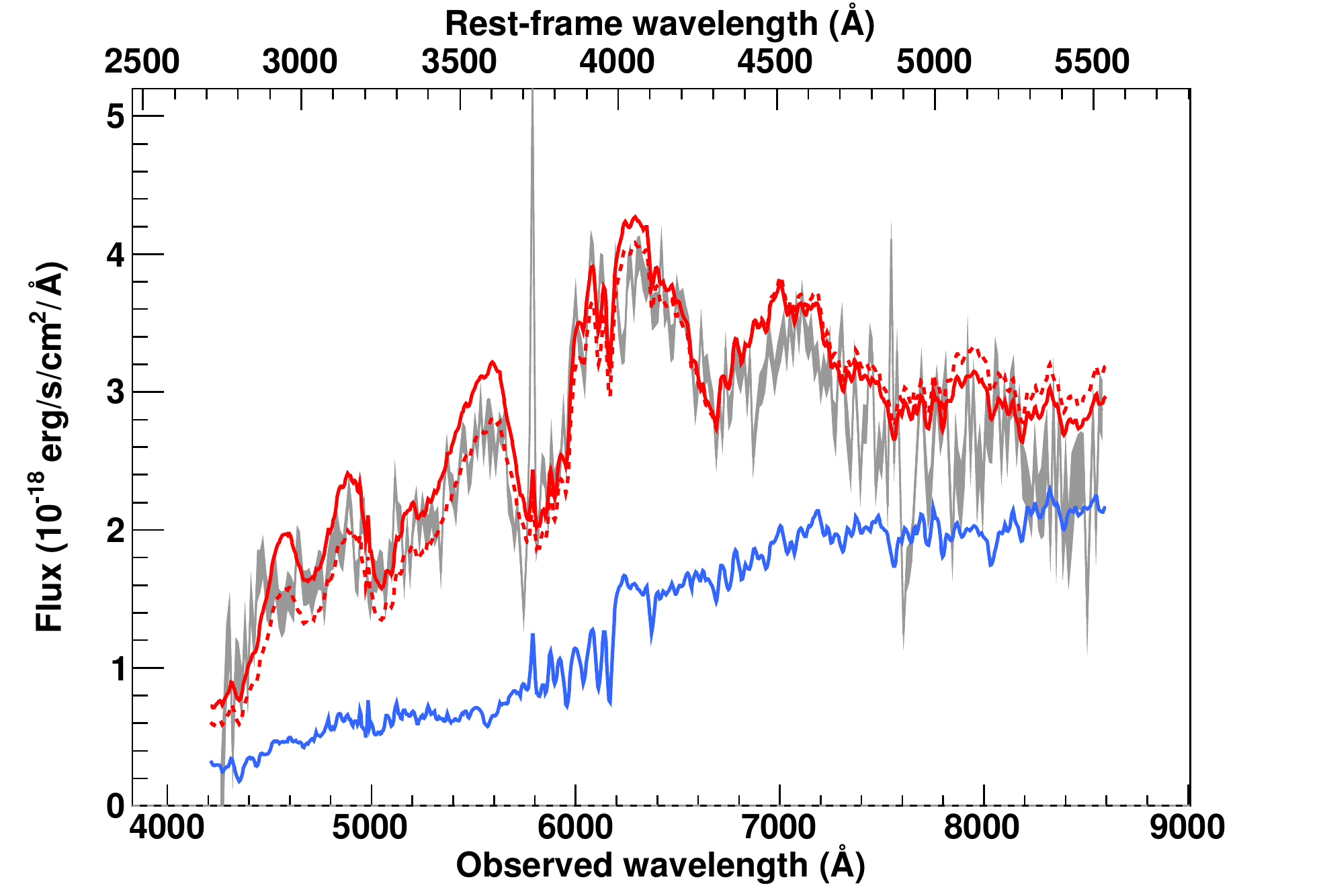}
    \includegraphics[scale=0.45]{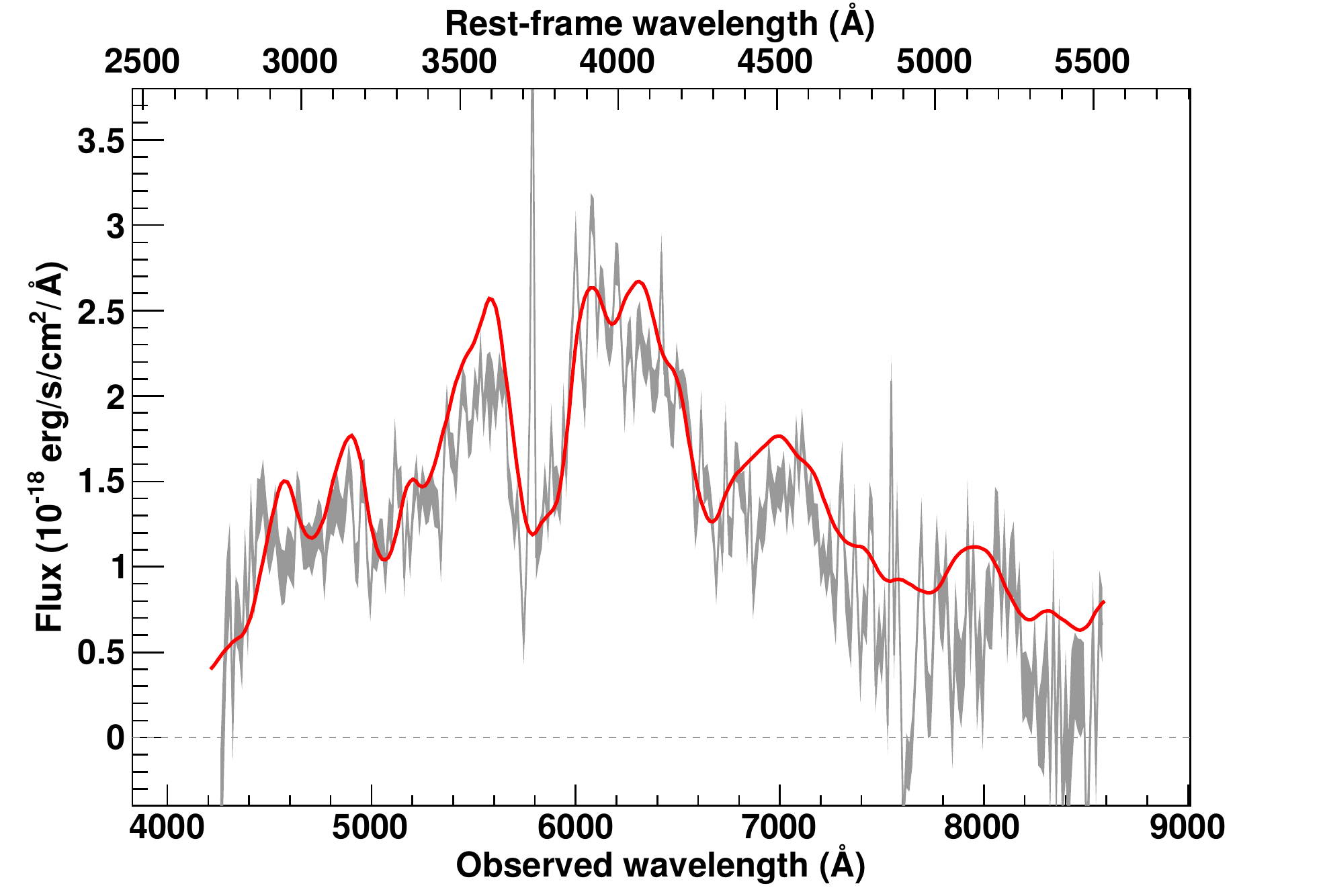}
    \end{center}
    \caption{The SNIa 06D4bo\_1280 spectrum measured at $z=0.552$ with a phase of 1.0 days. A S0-Sb host model has been subtracted.}
    \label{fig:Spec06D4bo_1280}
    \end{figure}
    
    \begin{figure}
    \begin{center}
    \includegraphics[scale=0.45]{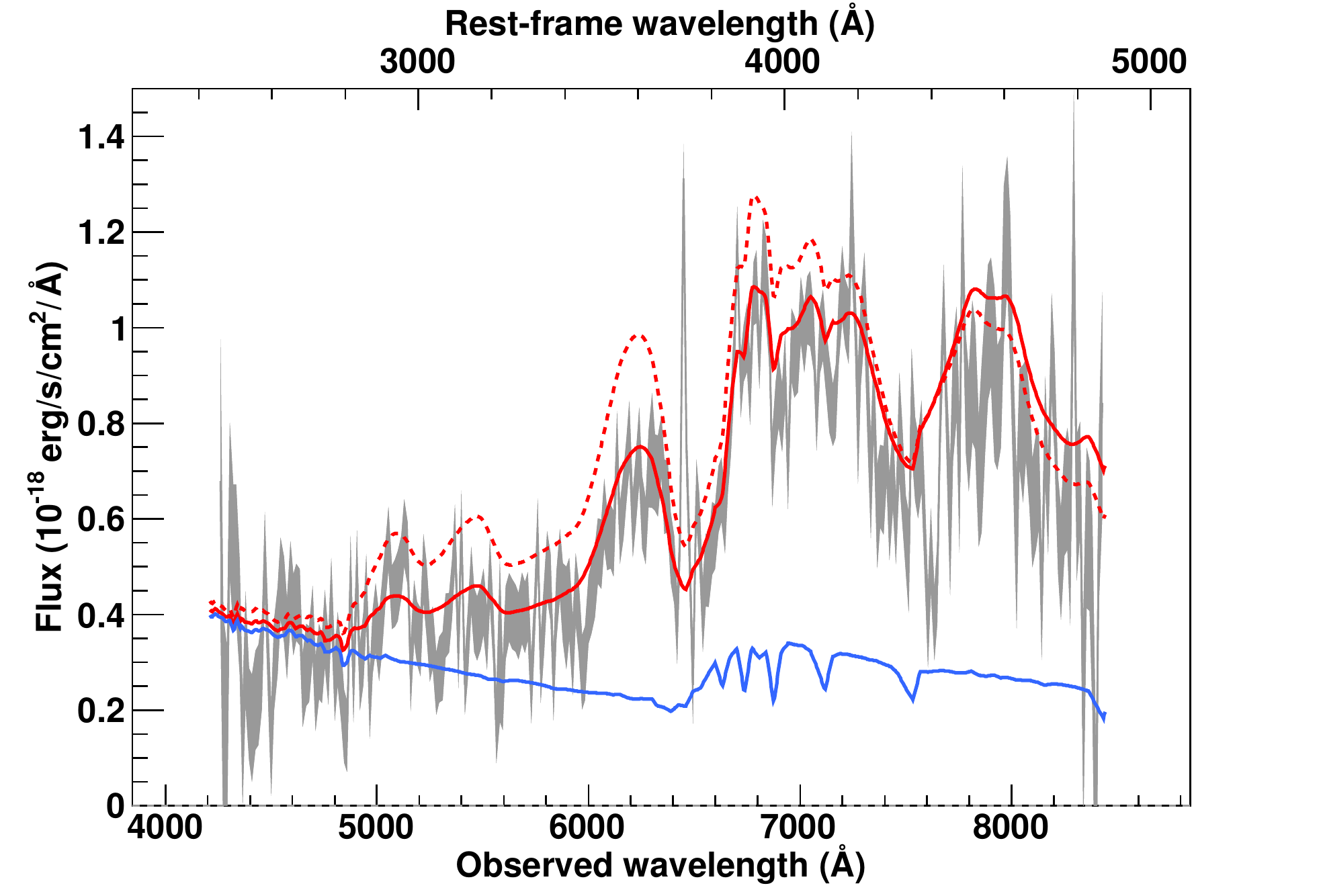}
    \includegraphics[scale=0.45]{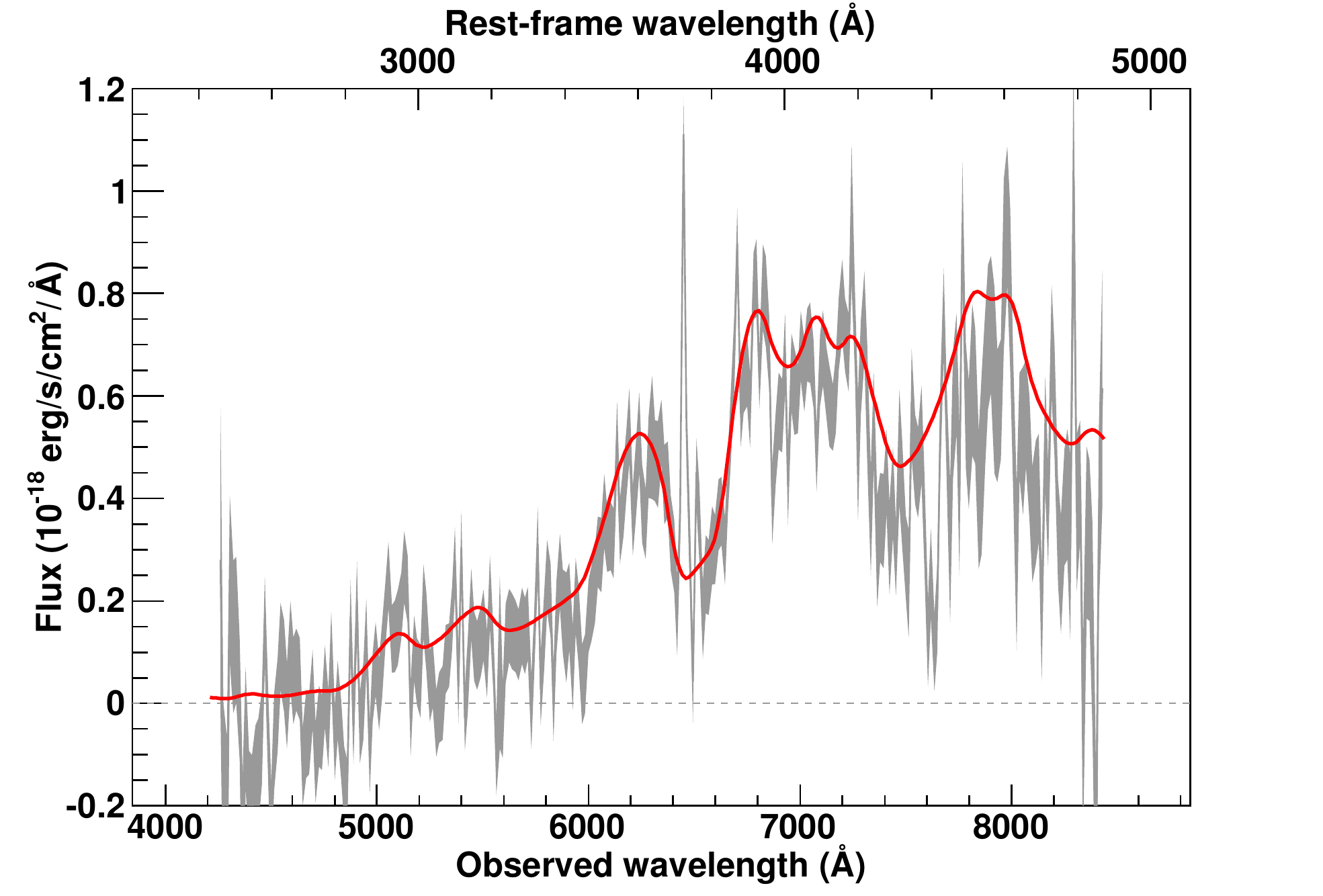}
    \end{center}
    \caption{The SNIa 06D4bw\_1279 spectrum measured at $z=0.732$ with a phase of 5.8 days. A Sa(1) host model has been subtracted.}
    \label{fig:Spec06D4bw_1279}
    \end{figure}
    
    \clearpage
    \begin{figure}
    \begin{center}
    \includegraphics[scale=0.45]{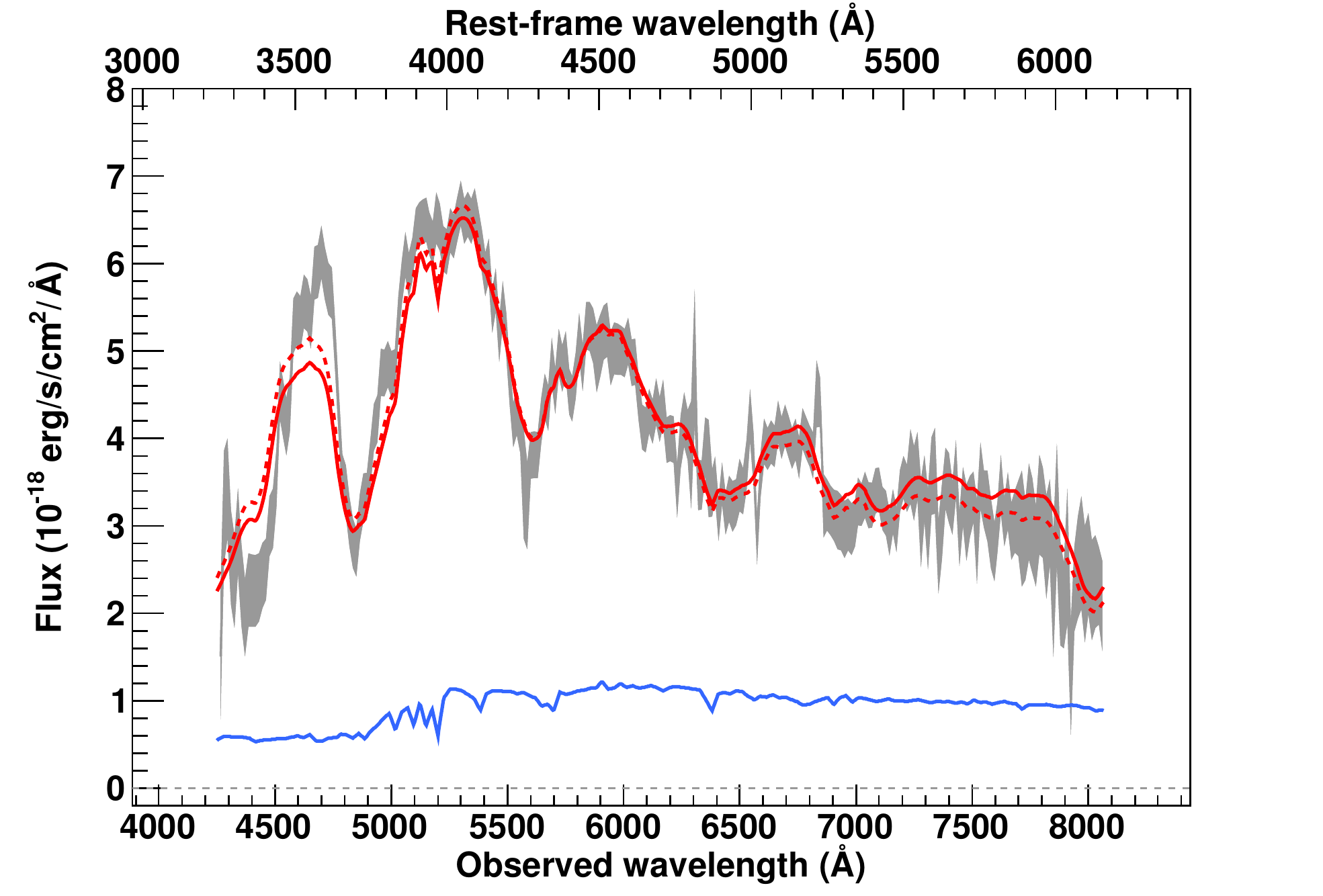}
    \includegraphics[scale=0.45]{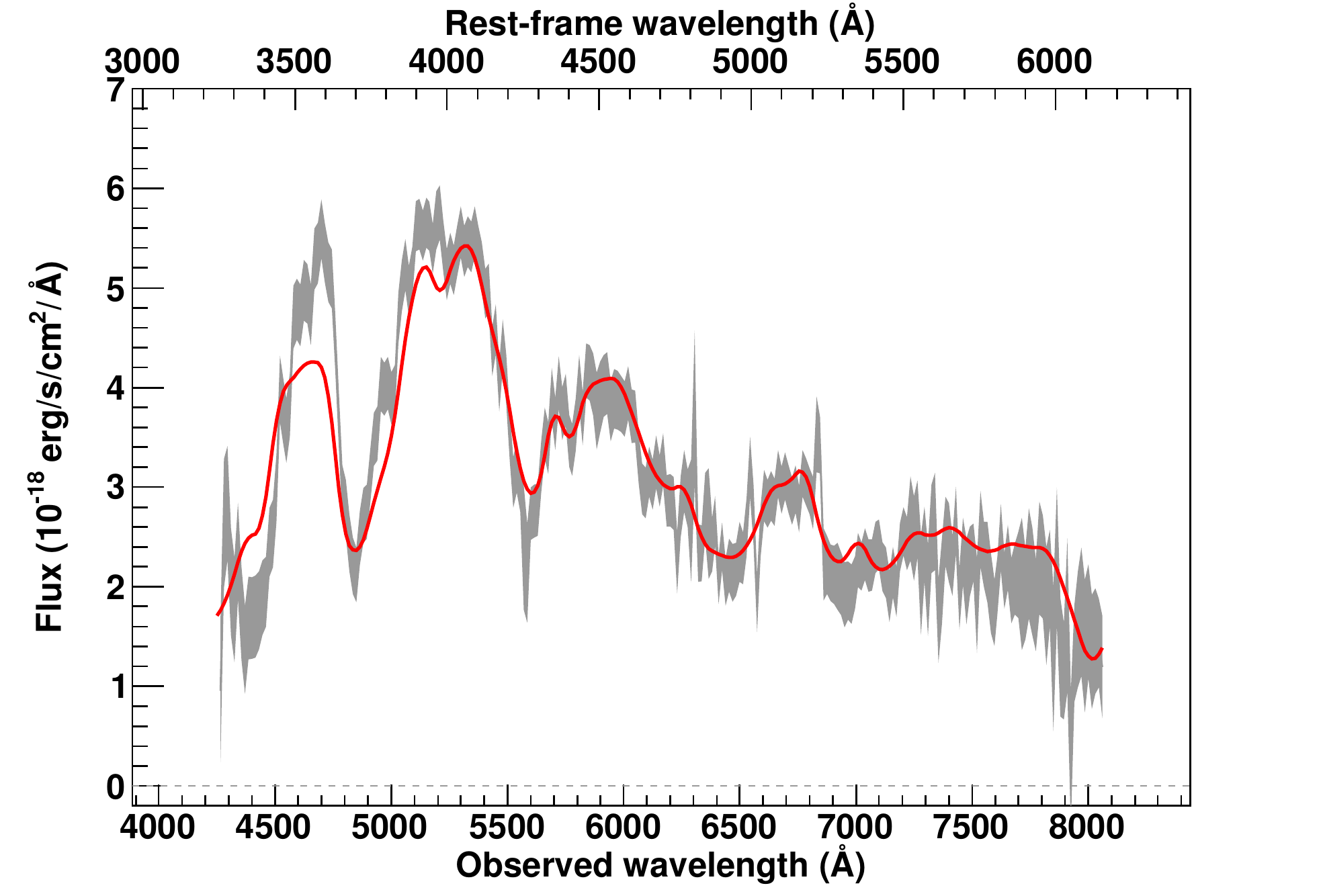}
    \end{center}
    \caption{The SNIa 06D4gs\_1358 spectrum measured at $z=0.31$ with a phase of -4.2 days. A E(1) host model has been subtracted.}
    \label{fig:Spec06D4gs_1358}
    \end{figure}
    
    \begin{figure}
    \begin{center}
    \includegraphics[scale=0.45]{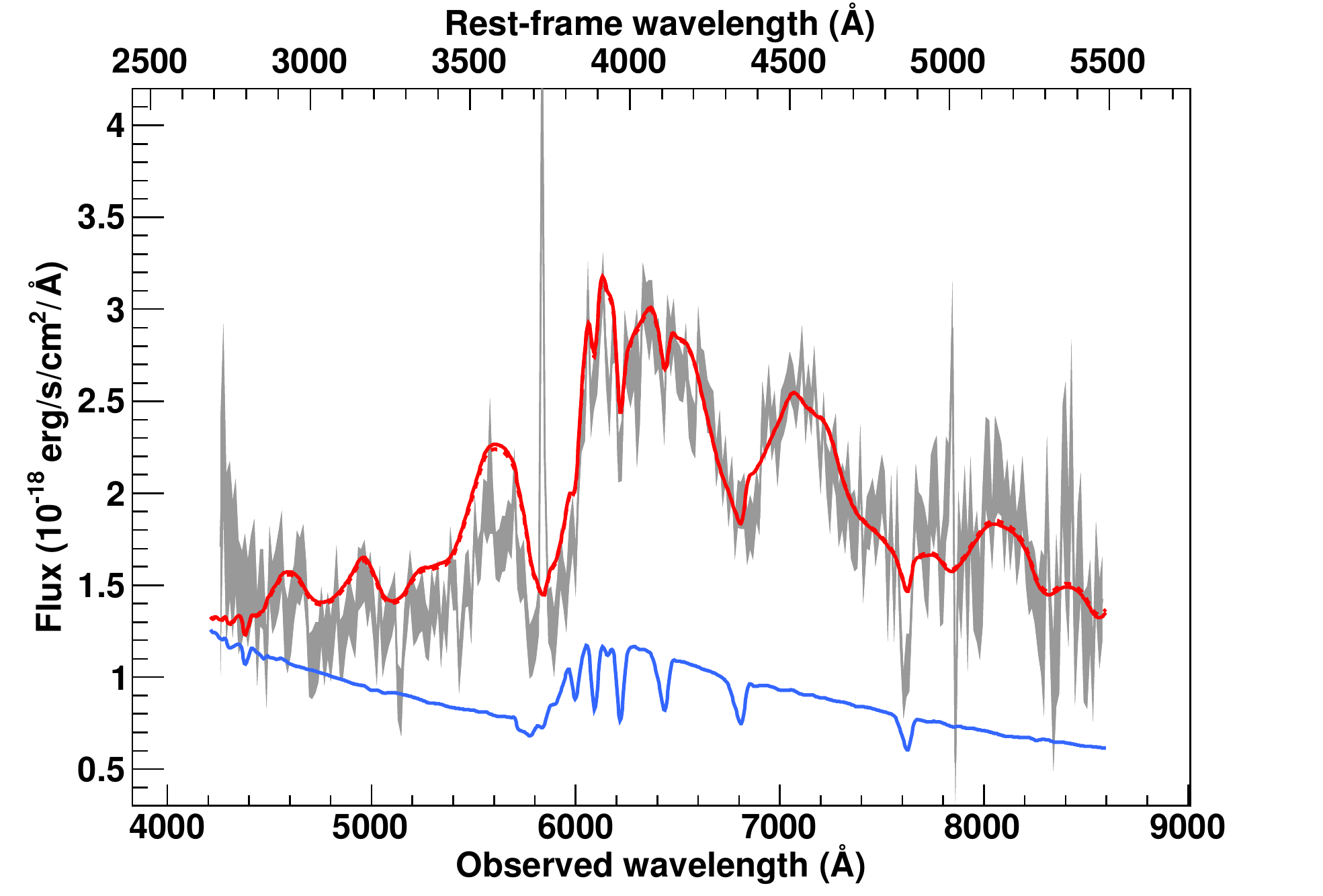}
    \includegraphics[scale=0.45]{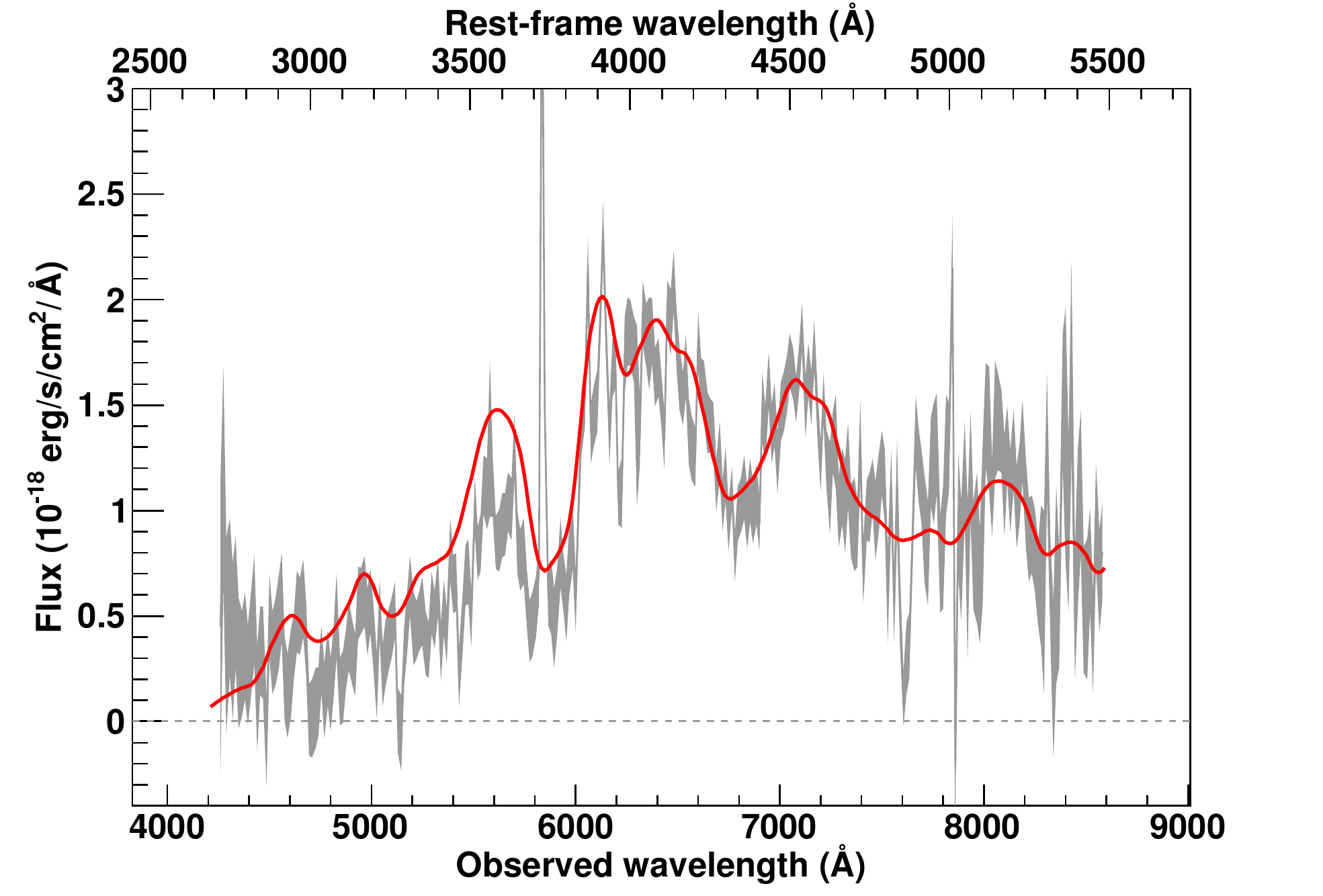}
    \end{center}
    \caption{The SNIa 06D4jh\_1413 spectrum measured at $z=0.566$ with a phase of 3.7 days. A Sd(2) host model has been subtracted.}
    \label{fig:Spec06D4jh_1413}
    \end{figure}
    
    \begin{figure}
    \begin{center}
    \includegraphics[scale=0.45]{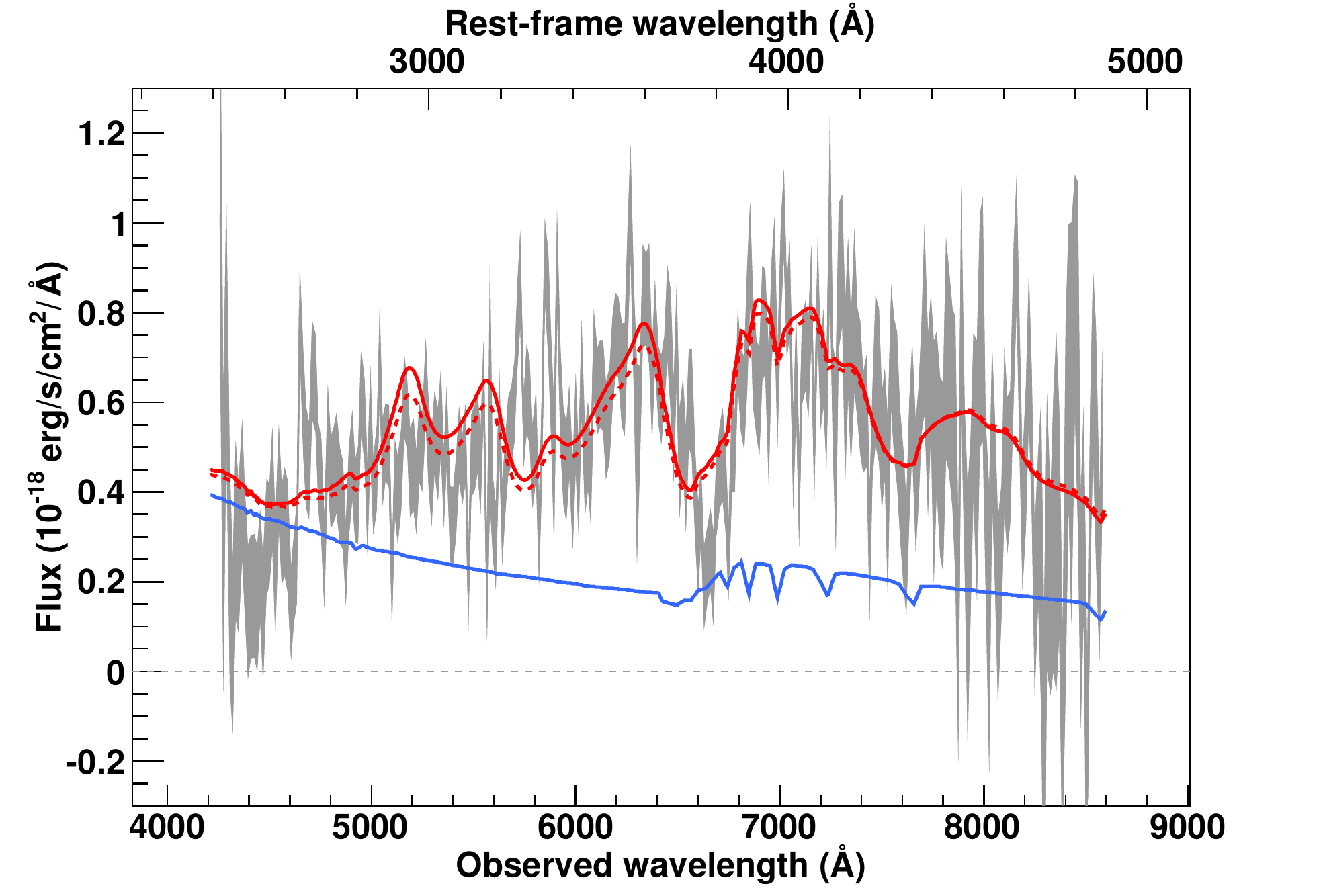}
    \includegraphics[scale=0.45]{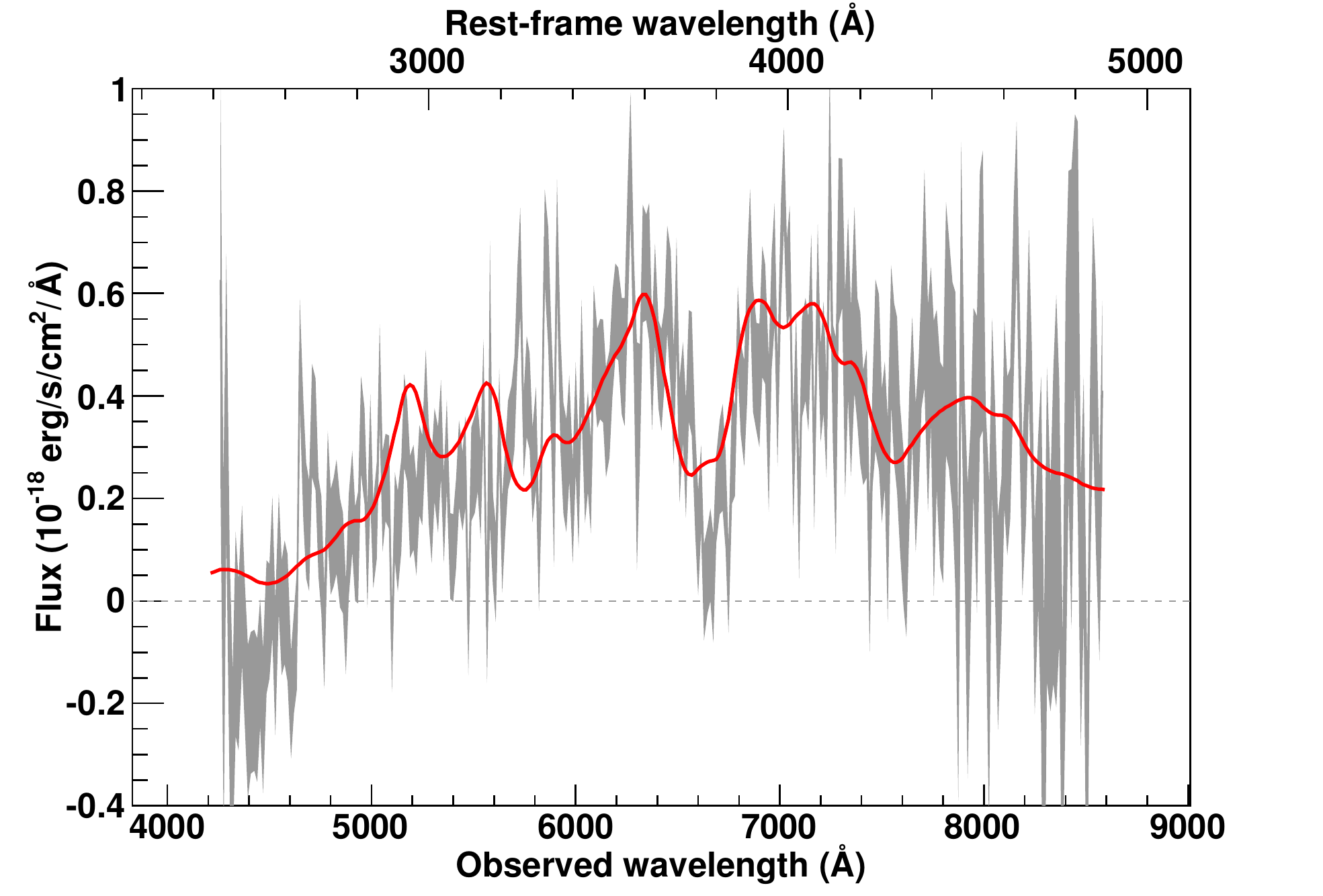}
    \end{center}
    \caption{The SNIa$\star$ 06D4jt\_1418 spectrum measured at $z=0.76$ with a phase of 2.9 days. A Sd(1) host model has been subtracted.}
    \label{fig:Spec06D4jt_1418}
    \end{figure}
    
    \clearpage
    \begin{figure}
    \begin{center}
    \includegraphics[scale=0.45]{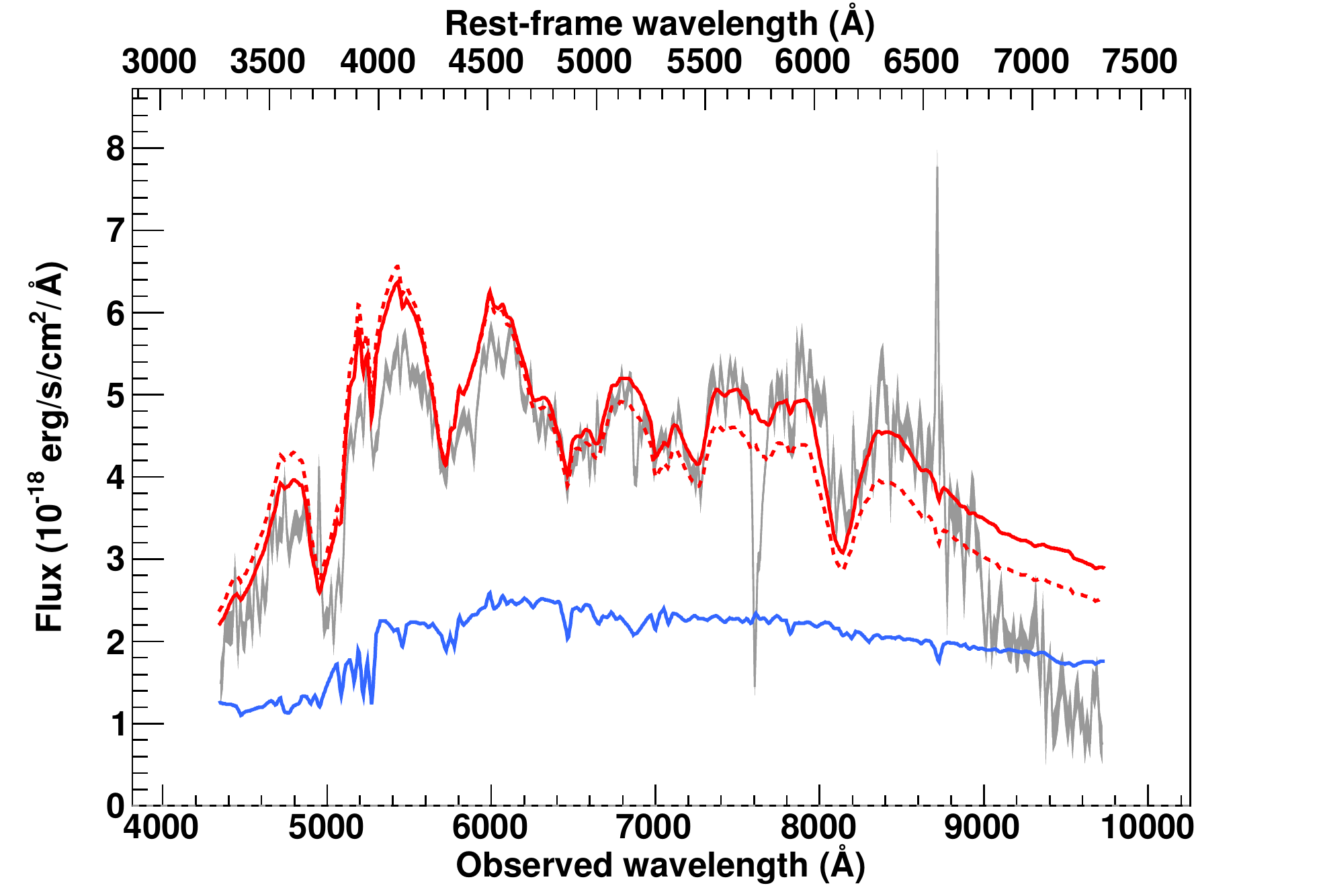}
    \includegraphics[scale=0.45]{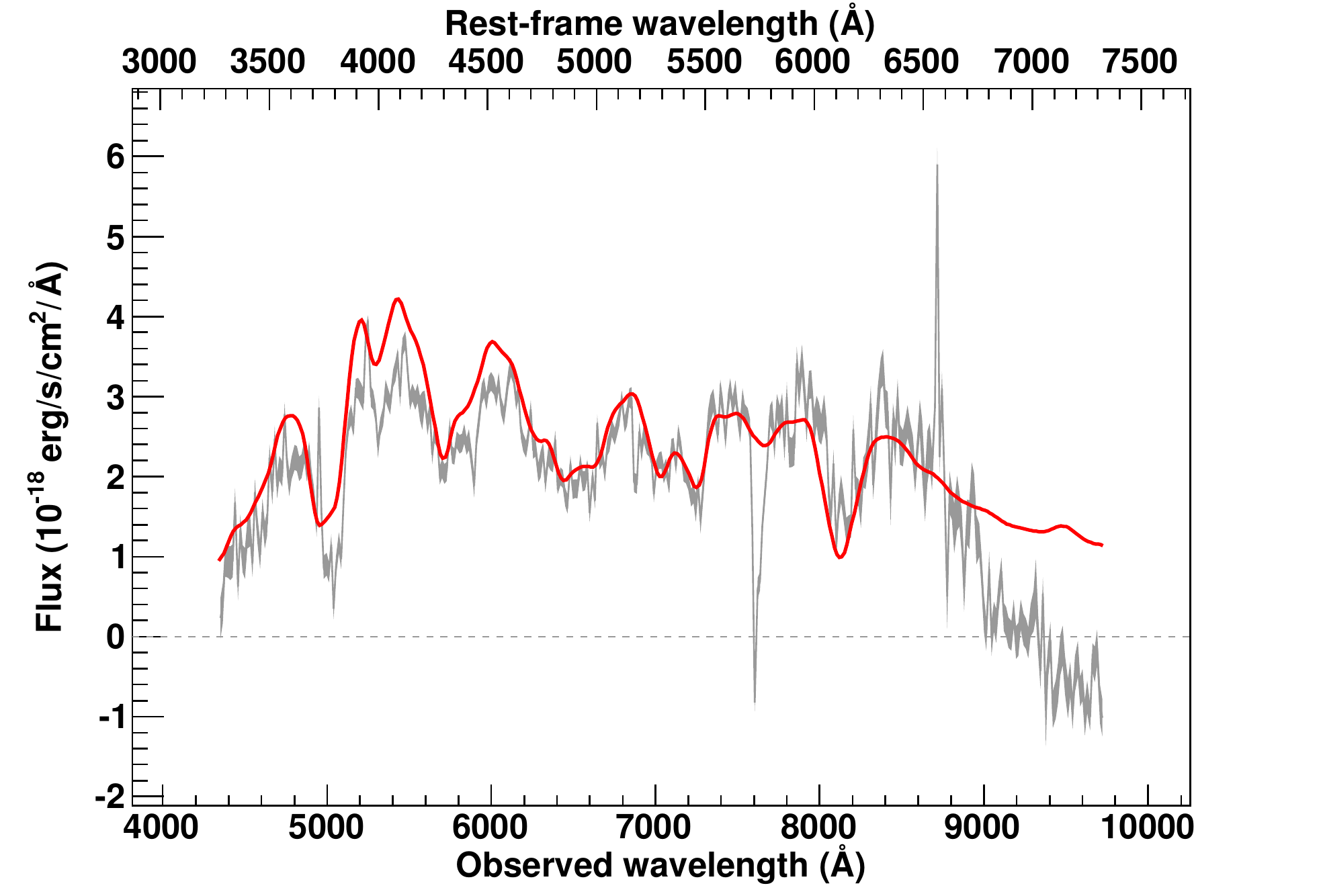}
    \end{center}
    \caption{The SNIa 07D1ab\_1483 spectrum measured at $z=0.328$ with a phase of -0.2 days. A E(1) host model has been subtracted.}
    \label{fig:Spec07D1ab_1483}
    \end{figure}
    
    \begin{figure}
    \begin{center}
    \includegraphics[scale=0.45]{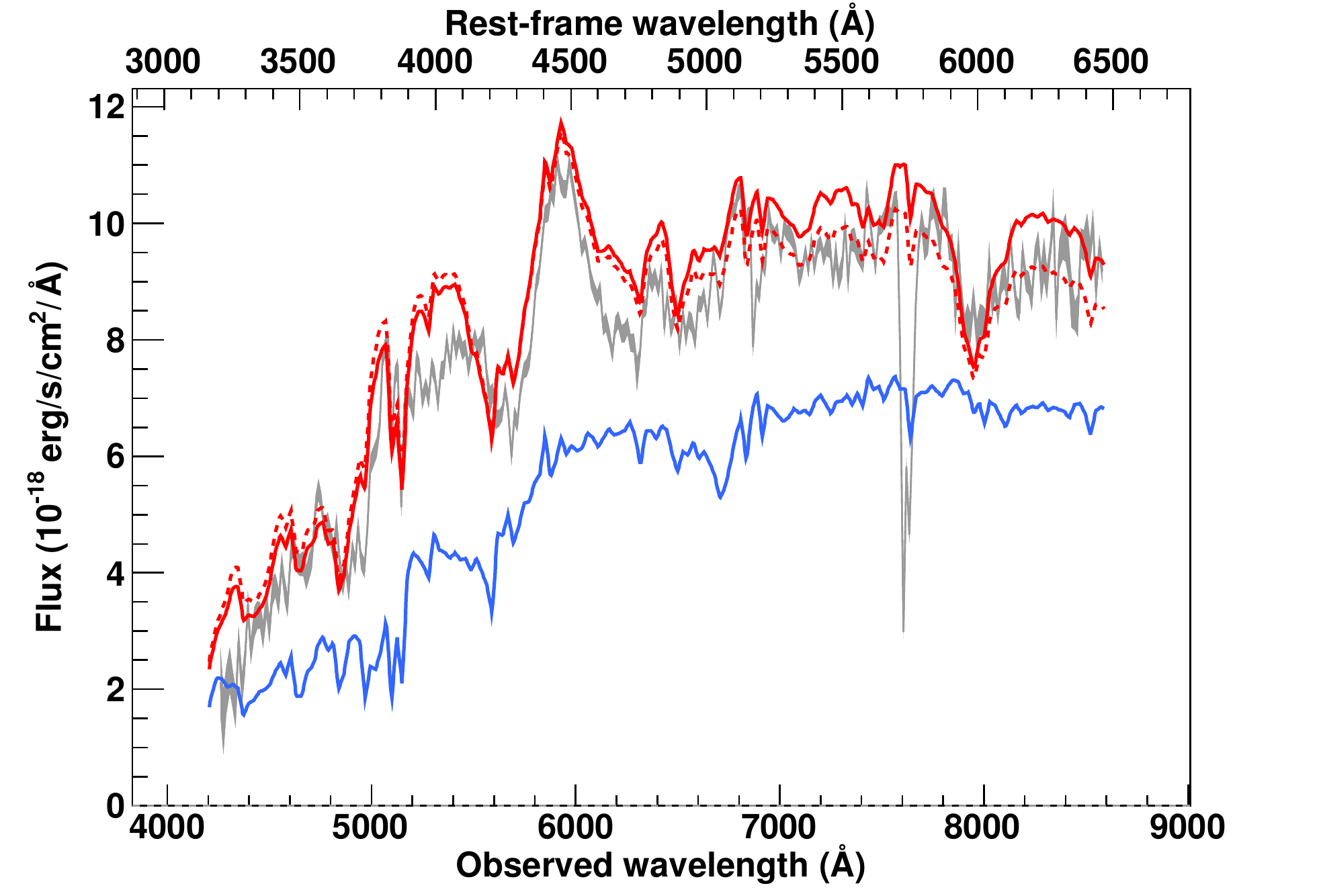}
    \includegraphics[scale=0.45]{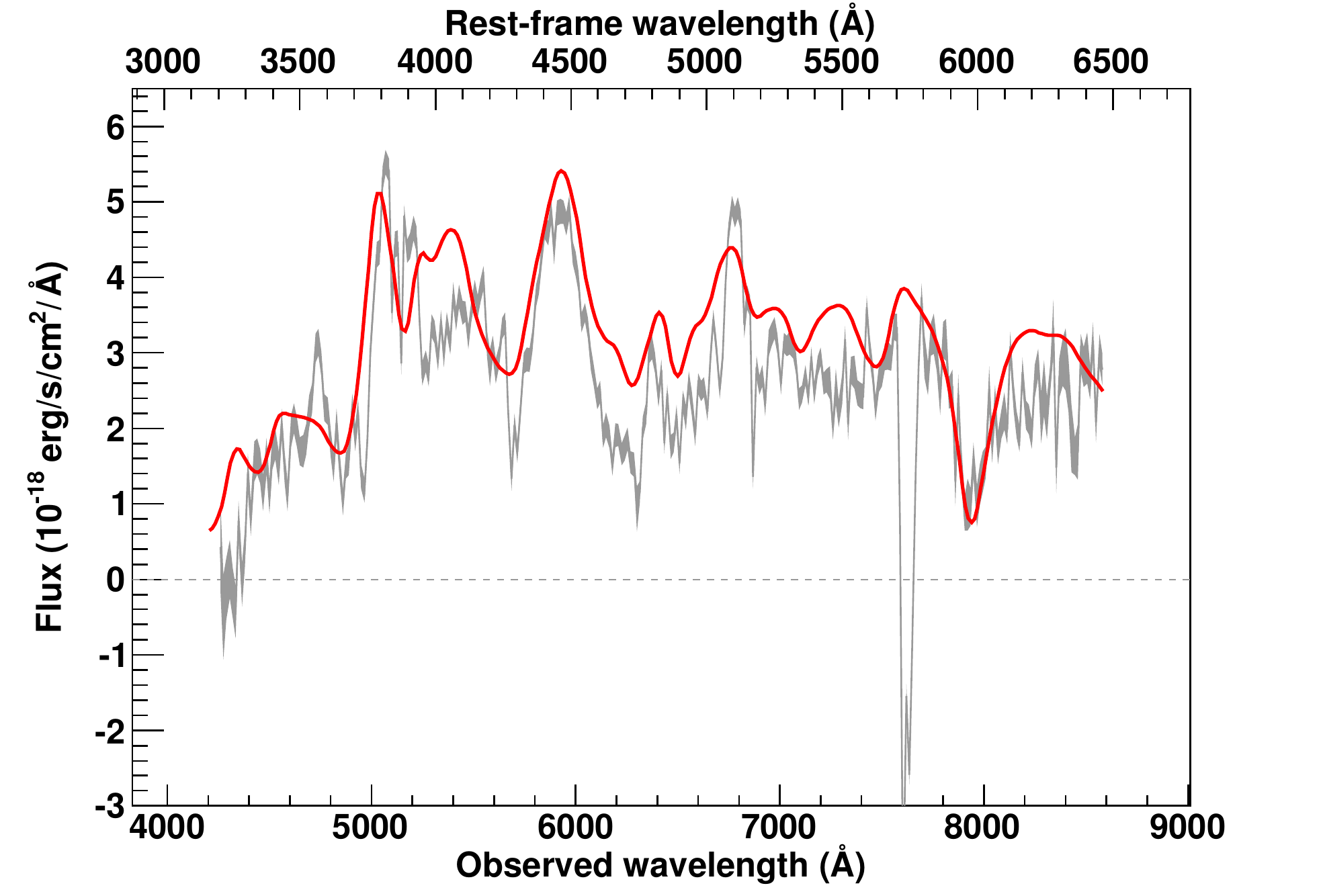}
    \end{center}
    \caption{The SNIa 07D1ad\_1484 spectrum measured at $z=0.297$ with a phase of 6.9 days. A S0(12) host model has been subtracted.}
    \label{fig:Spec07D1ad_1484}
    \end{figure}
    
    \begin{figure}
    \begin{center}
    \includegraphics[scale=0.45]{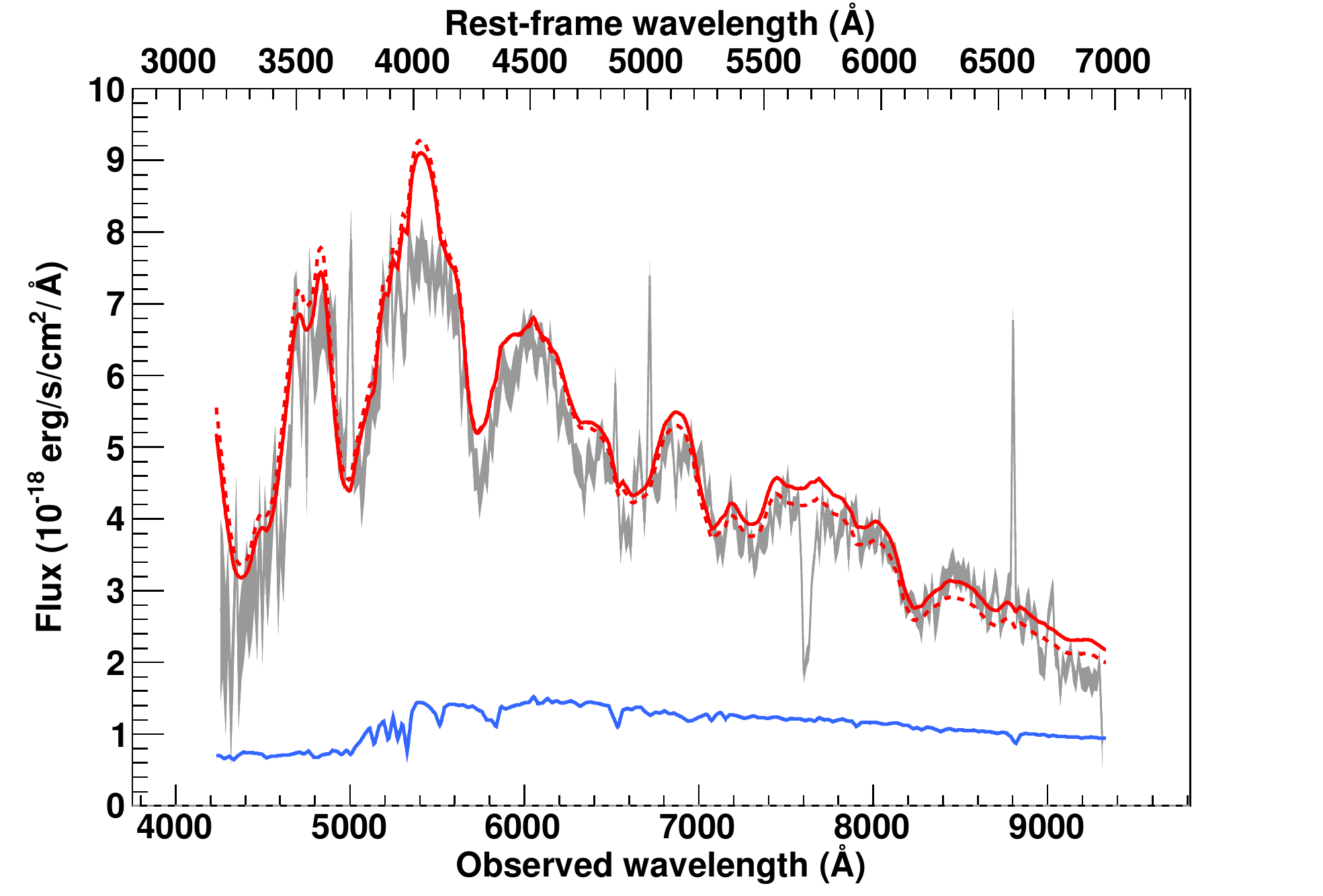}
    \includegraphics[scale=0.45]{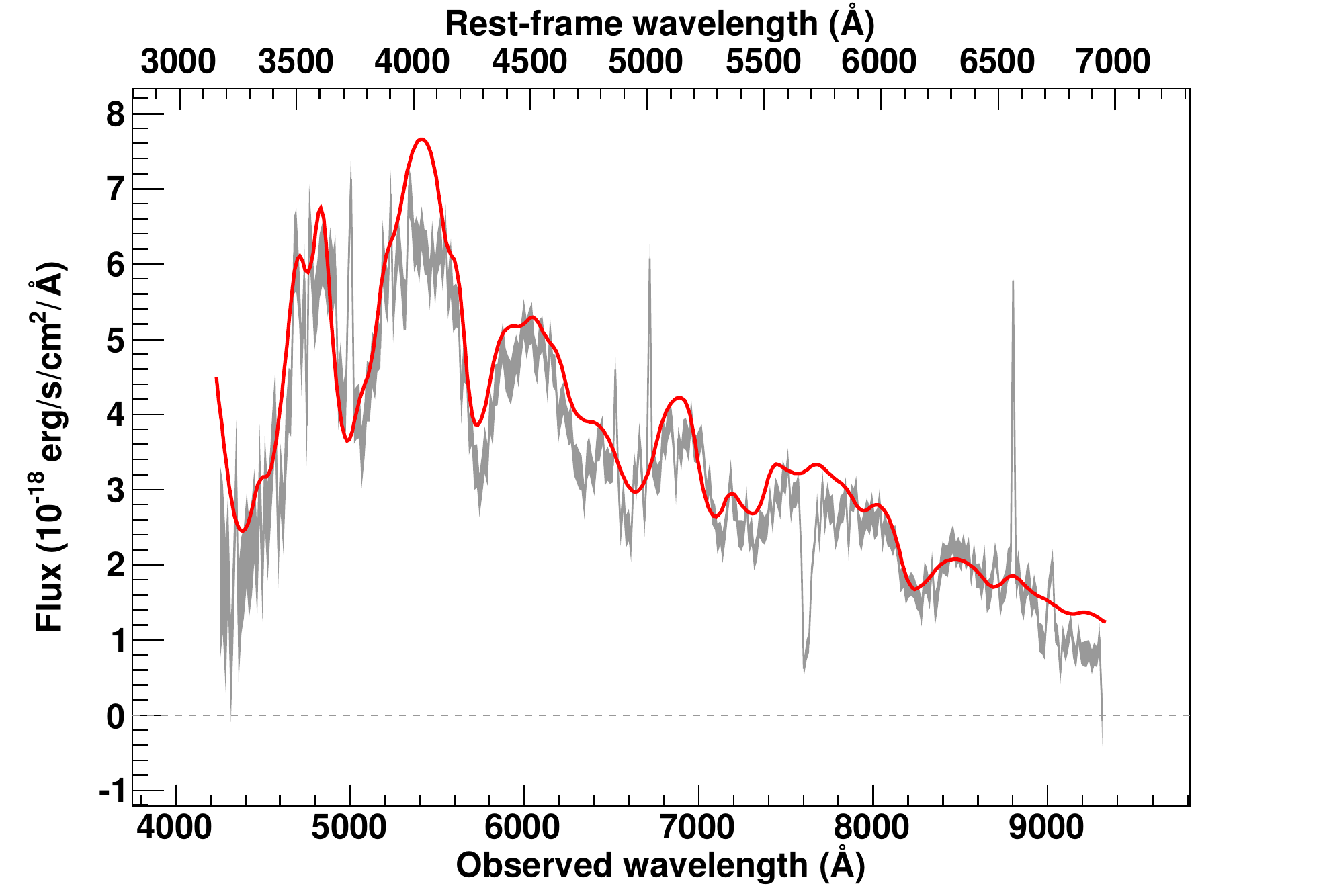}
    \end{center}
    \caption{The SNIa-pec 07D1ah\_1699 spectrum measured at $z=0.342$ with a phase of -0.6 days. A E(1) host model has been subtracted.}
    \label{fig:Spec07D1ah_1699}
    \end{figure}
    
    \clearpage
    \begin{figure}
    \begin{center}
    \includegraphics[scale=0.45]{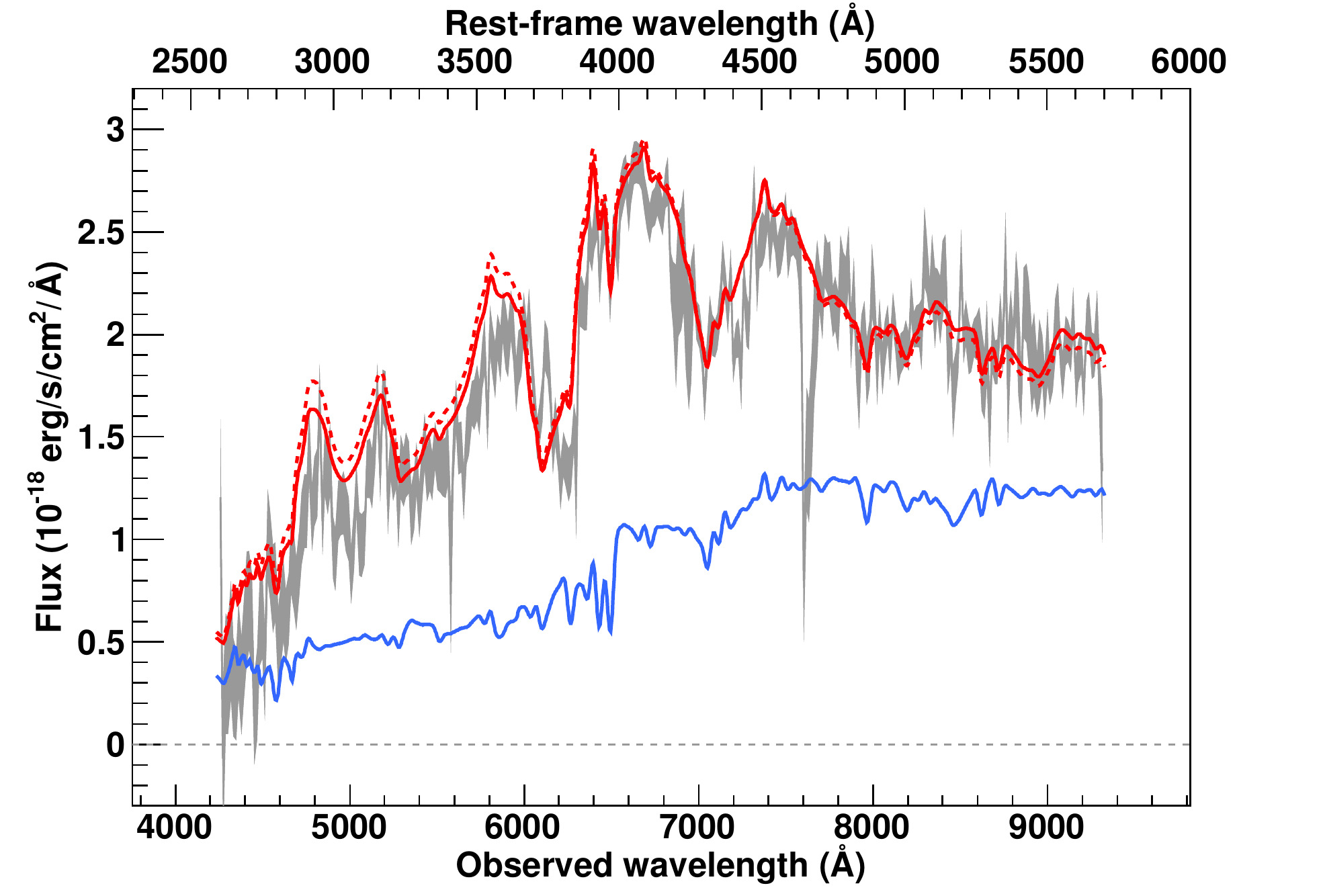}
    \includegraphics[scale=0.45]{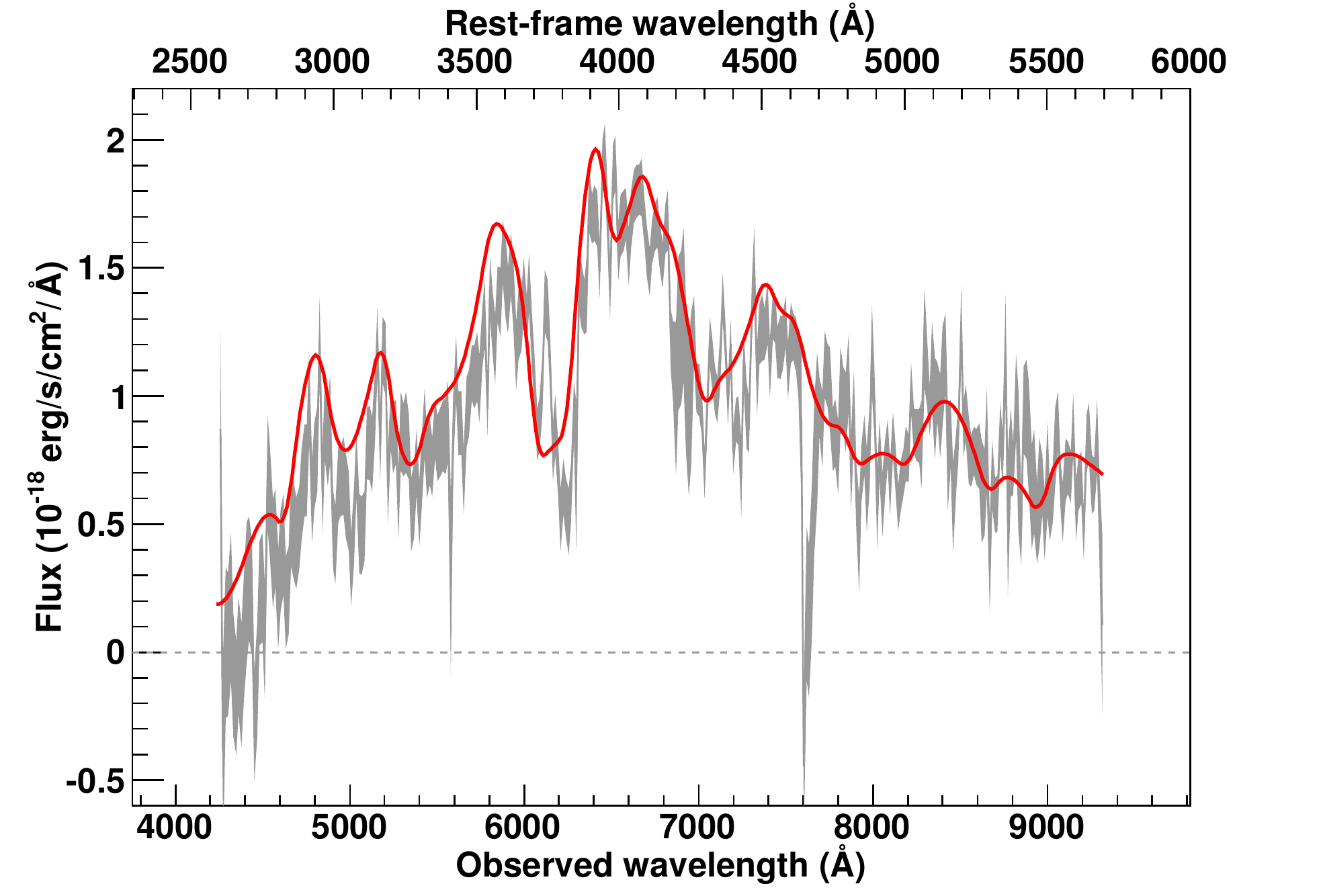}
    \end{center}
    \caption{The SNIa 07D1bl\_1707 spectrum measured at $z=0.636$ with a phase of 2.0 days. A E(2) host model has been subtracted.}
    \label{fig:Spec07D1bl_1707}
    \end{figure}
    
    \begin{figure}
    \begin{center}
    \includegraphics[scale=0.45]{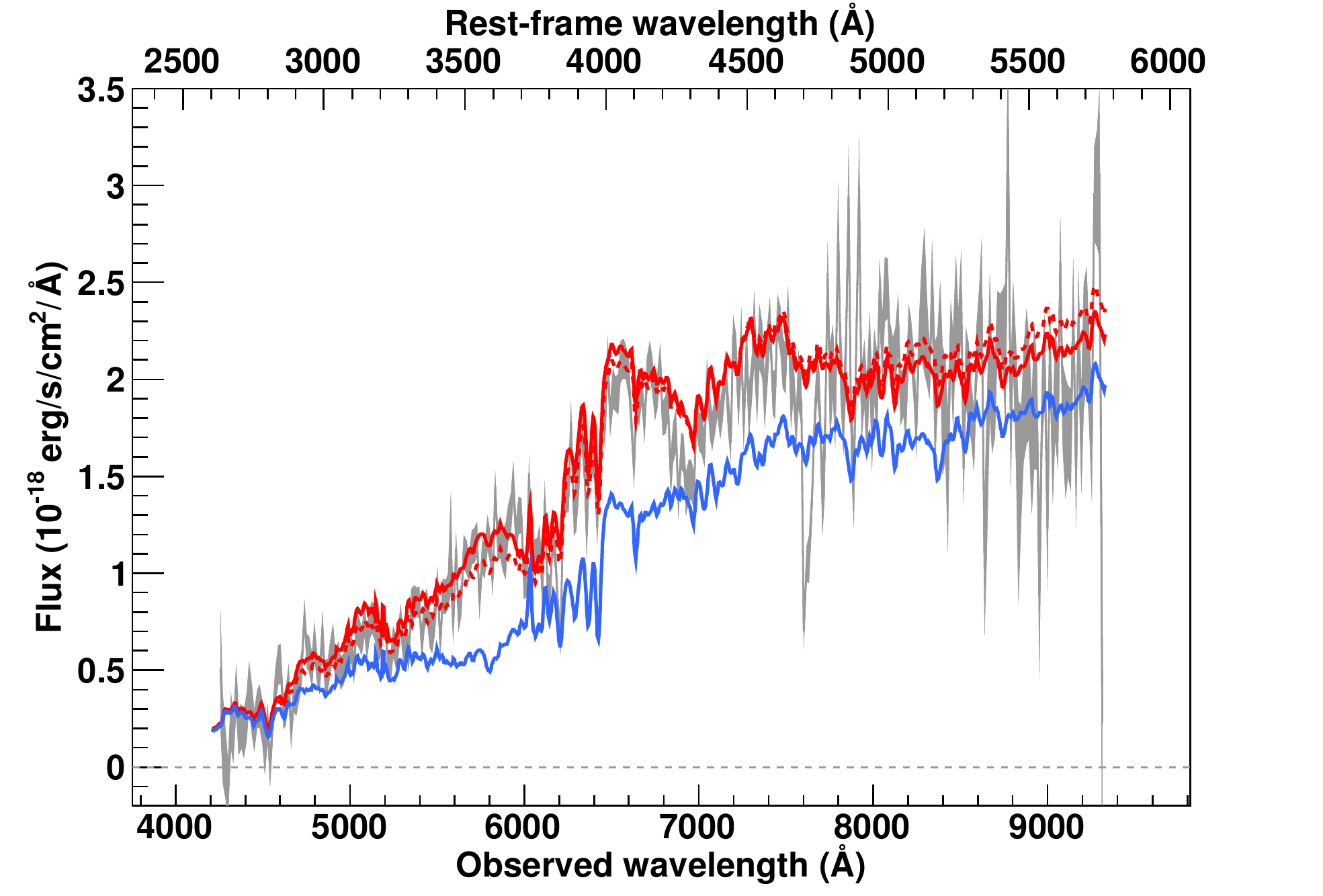}
    \includegraphics[scale=0.45]{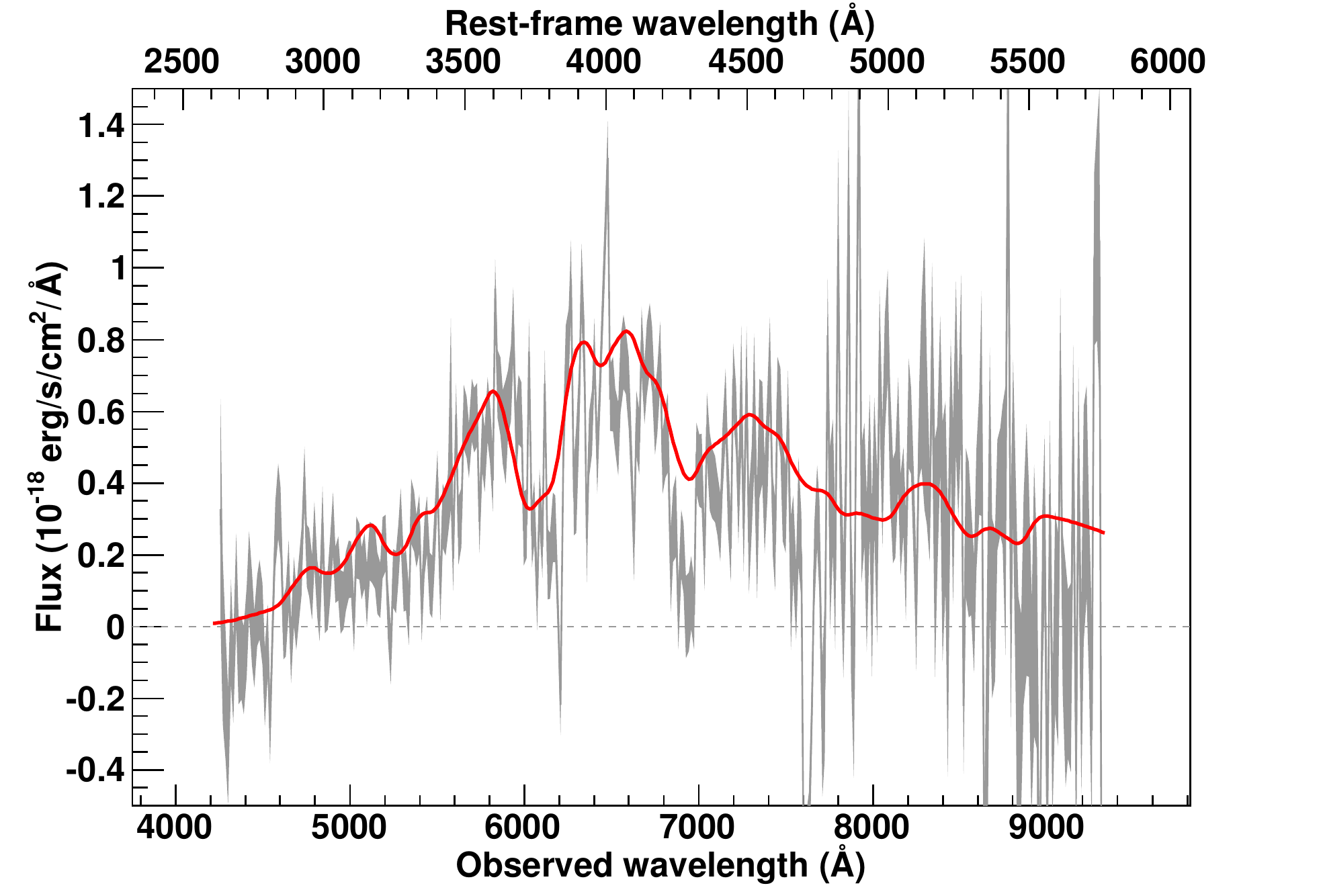}
    \end{center}
    \caption{The SNIa$\star$ 07D1bs\_1711 spectrum measured at $z=0.617$ with a phase of 0.7 days. A Sa-Sb host model has been subtracted.}
    \label{fig:Spec07D1bs_1711}
    \end{figure}
    
    \begin{figure}
    \begin{center}
    \includegraphics[scale=0.45]{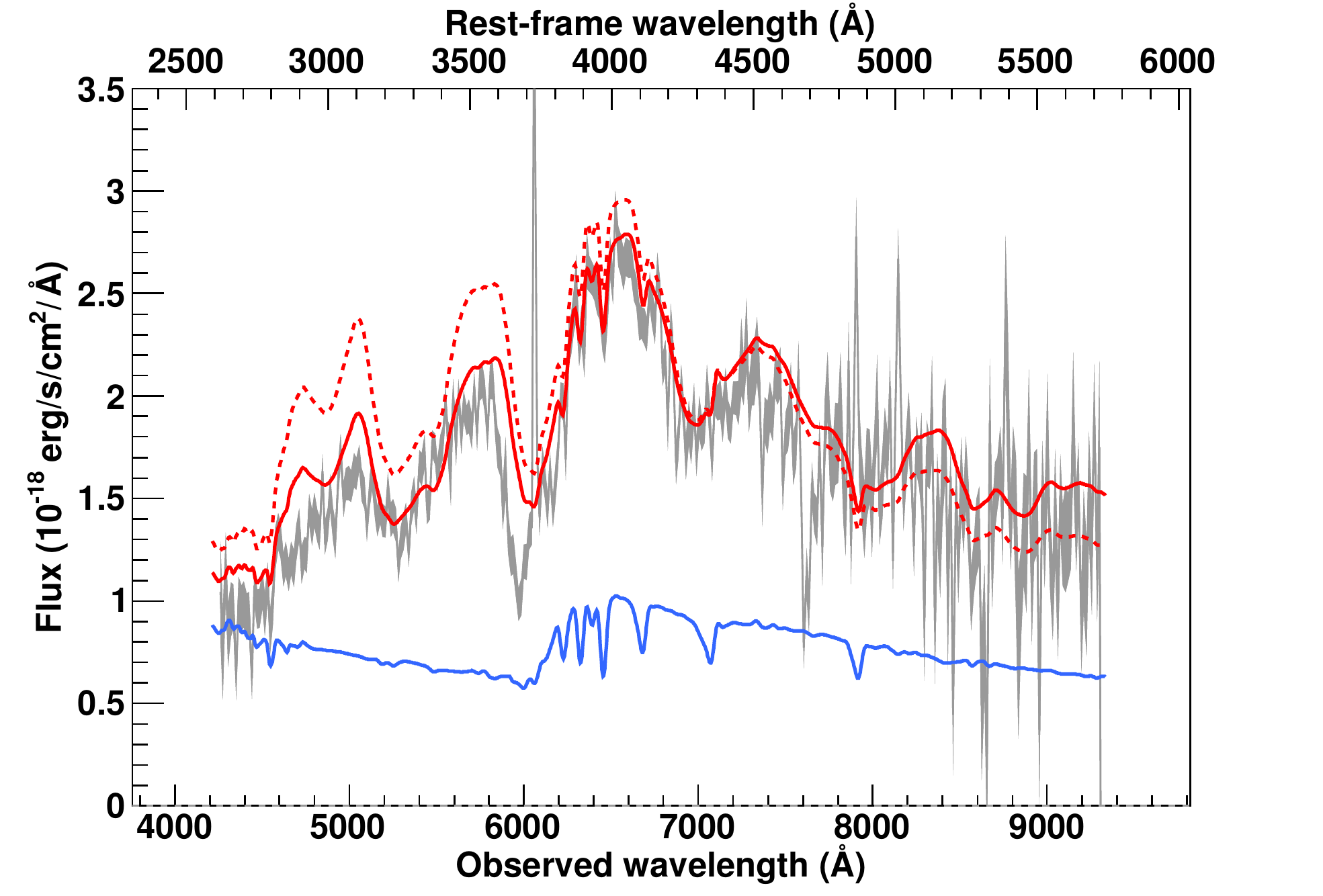}
    \includegraphics[scale=0.45]{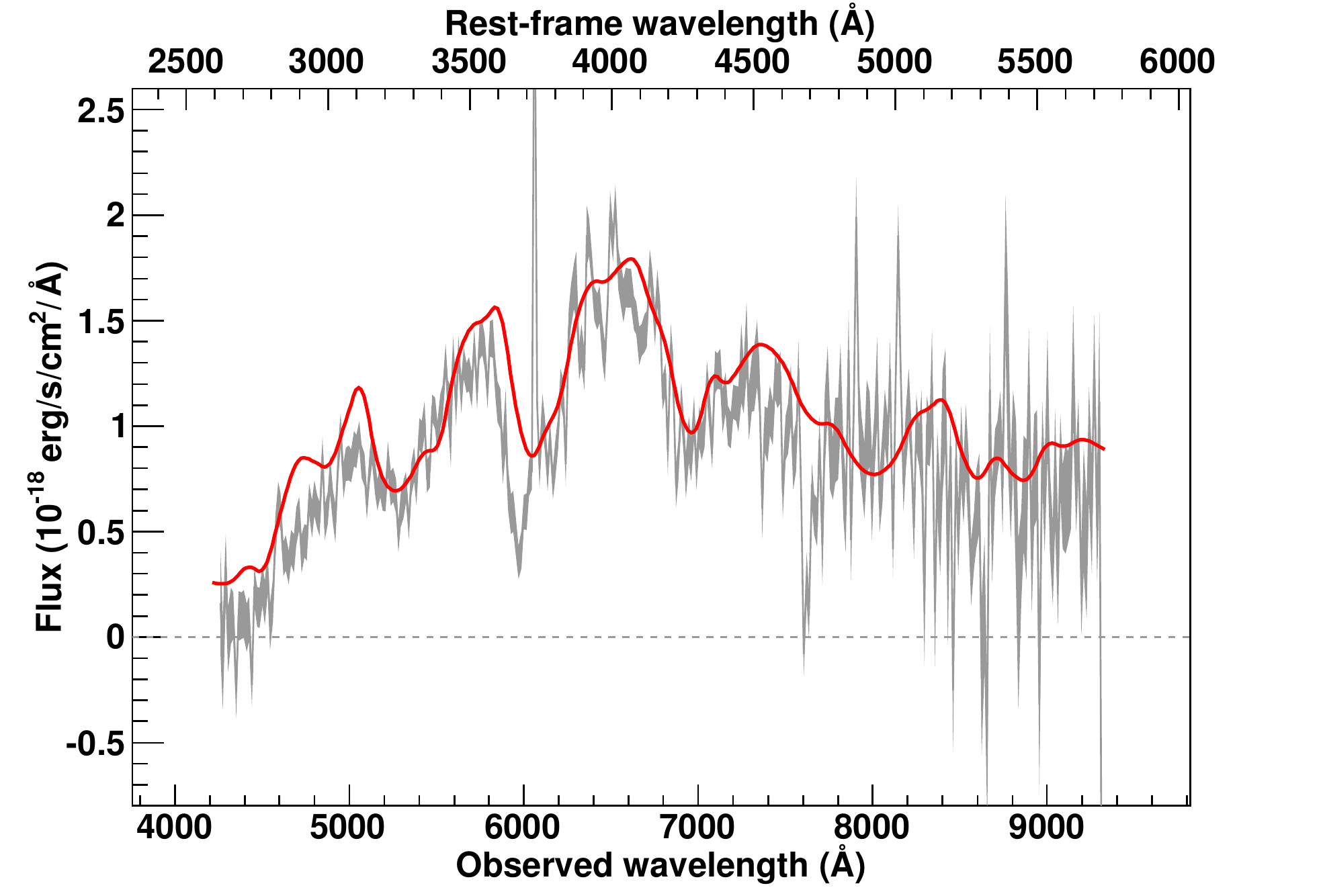}
    \end{center}
    \caption{The SNIa 07D1bu\_1711 spectrum measured at $z=0.626$ with a phase of -2.8 days. A Sd(5) host model has been subtracted.}
    \label{fig:Spec07D1bu_1711}
    \end{figure}
    
    \clearpage
    \begin{figure}
    \begin{center}
    \includegraphics[scale=0.45]{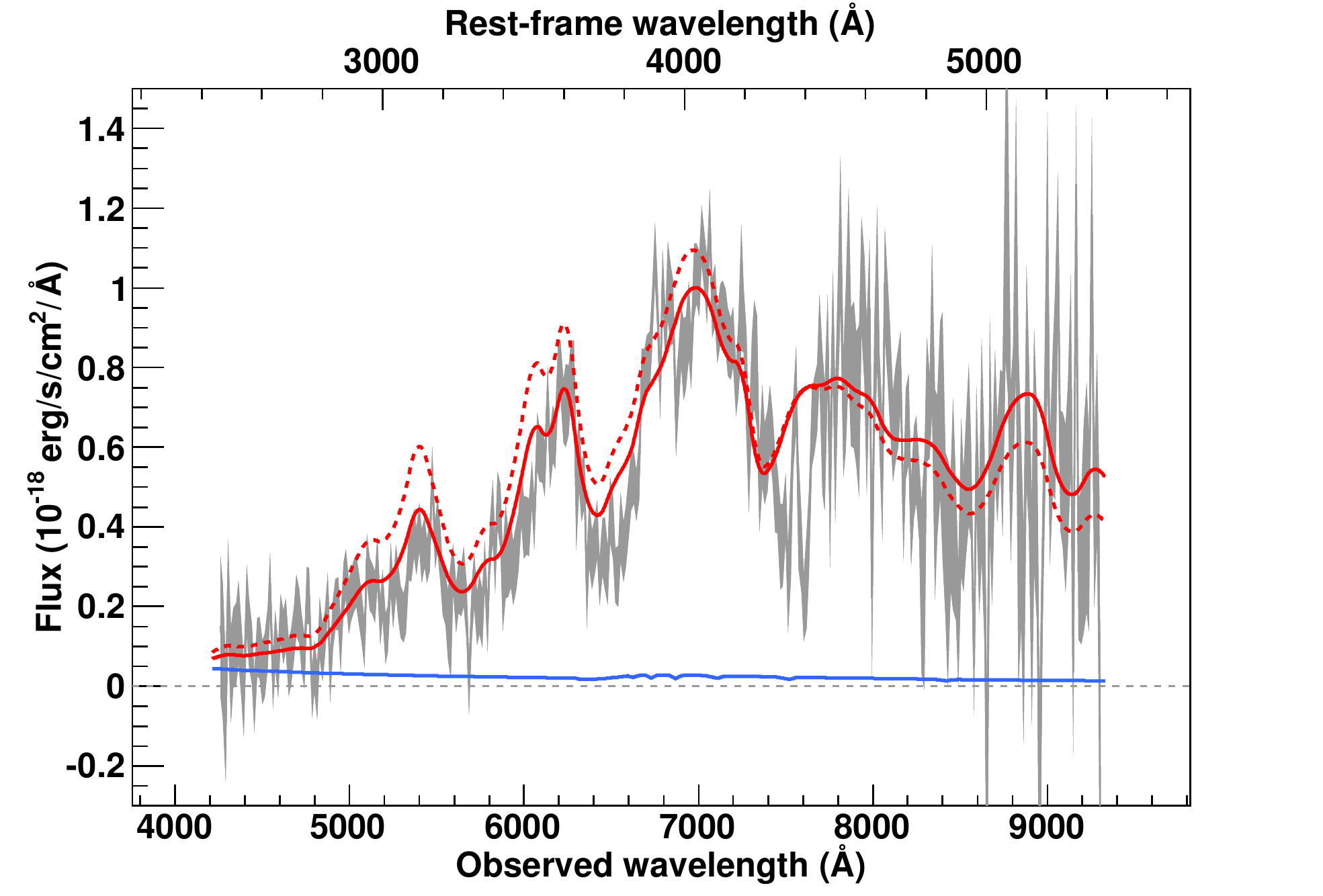}
    \includegraphics[scale=0.45]{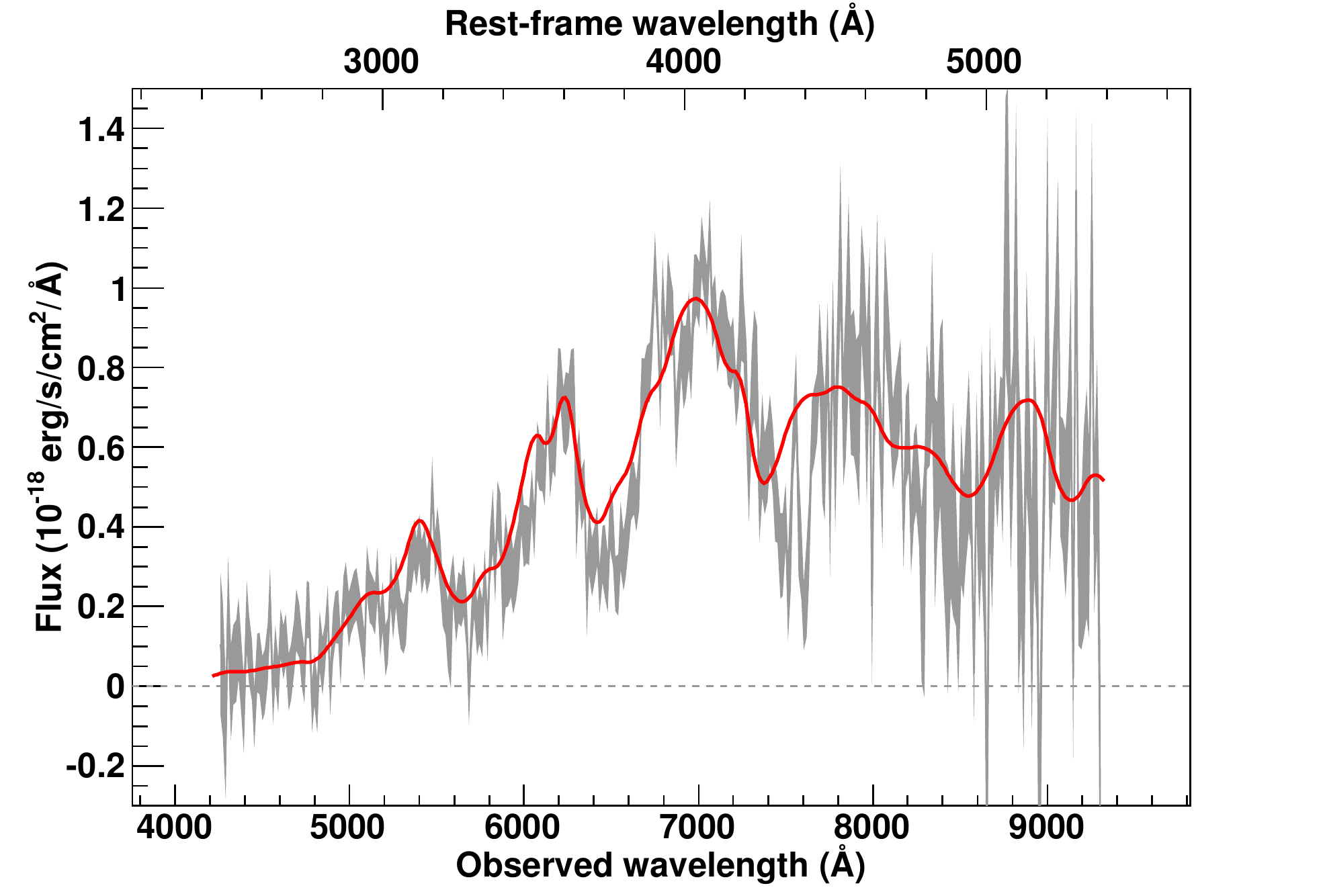}
    \end{center}
    \caption{The SNIa 07D1by\_1715 spectrum measured at $z=0.73$ with a phase of -0.5 days. A Sd(1) host model has been subtracted.}
    \label{fig:Spec07D1by_1715}
    \end{figure}
    
    \begin{figure}
    \begin{center}
    \includegraphics[scale=0.45]{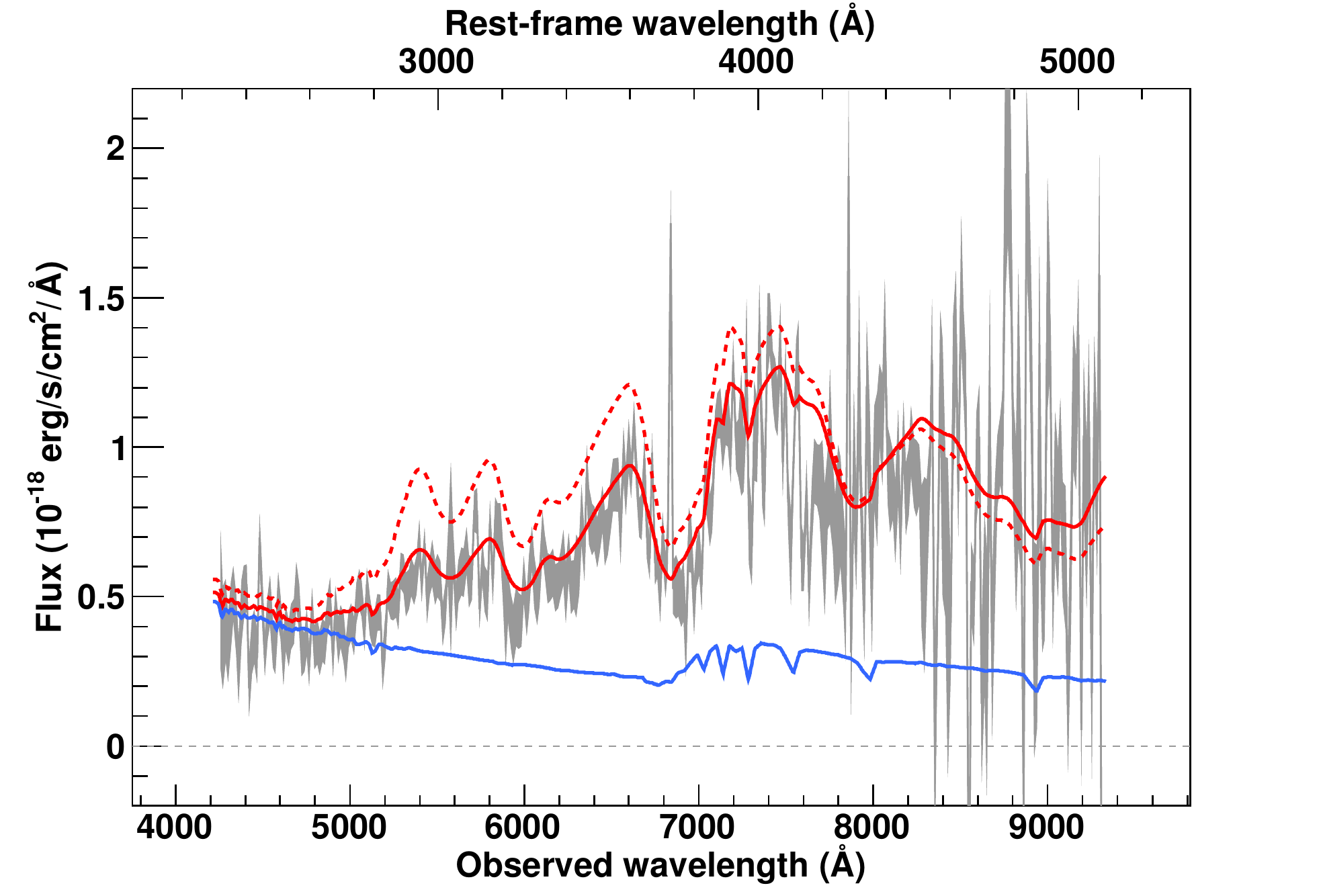}
    \includegraphics[scale=0.45]{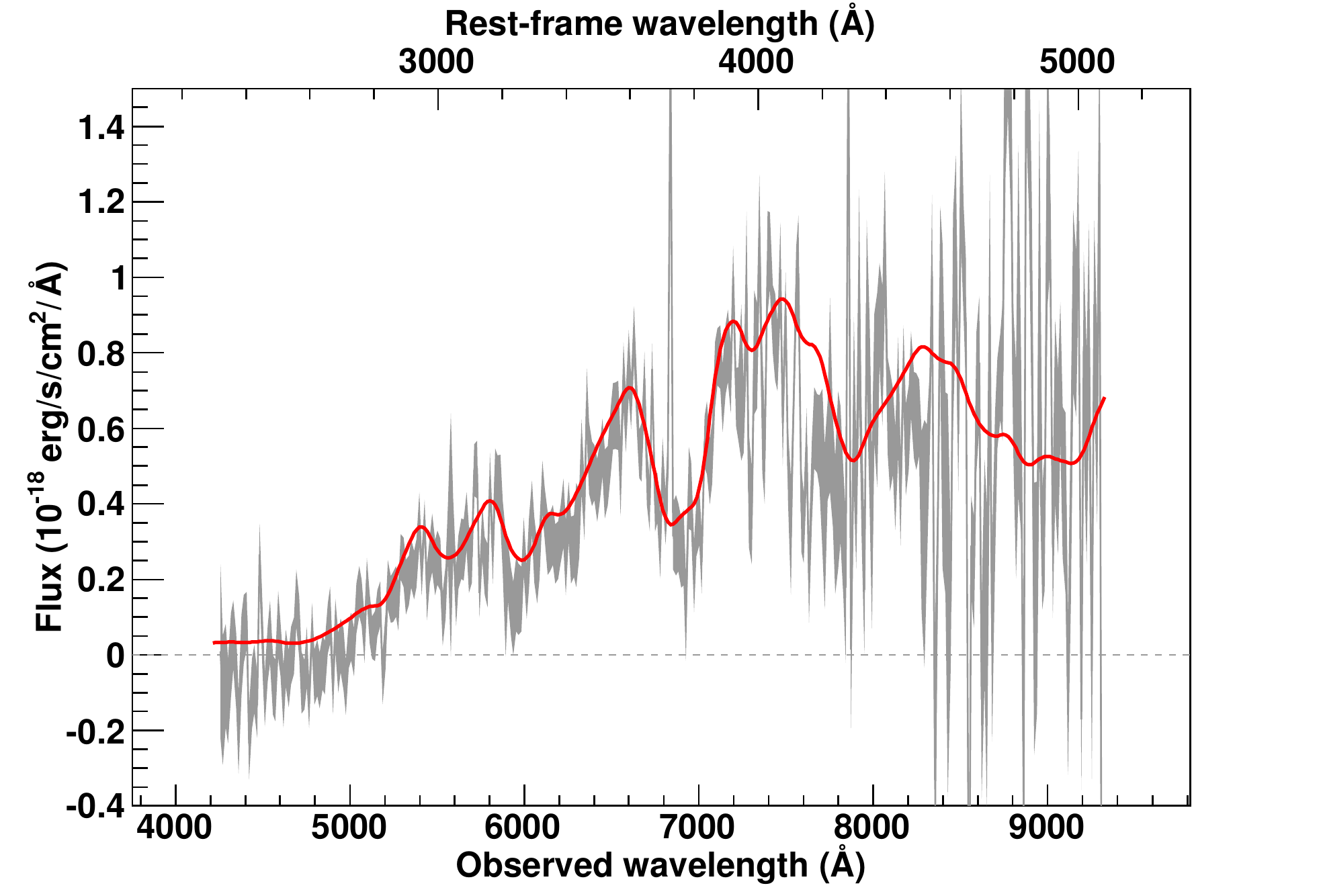}
    \end{center}
    \caption{The SNIa$\star$ 07D1ca\_1719 spectrum measured at $z=0.835$ with a phase of 1.4 days. A Sa(1) host model has been subtracted.}
    \label{fig:Spec07D1ca_1719}
    \end{figure}
    
    \begin{figure}
    \begin{center}
    \includegraphics[scale=0.45]{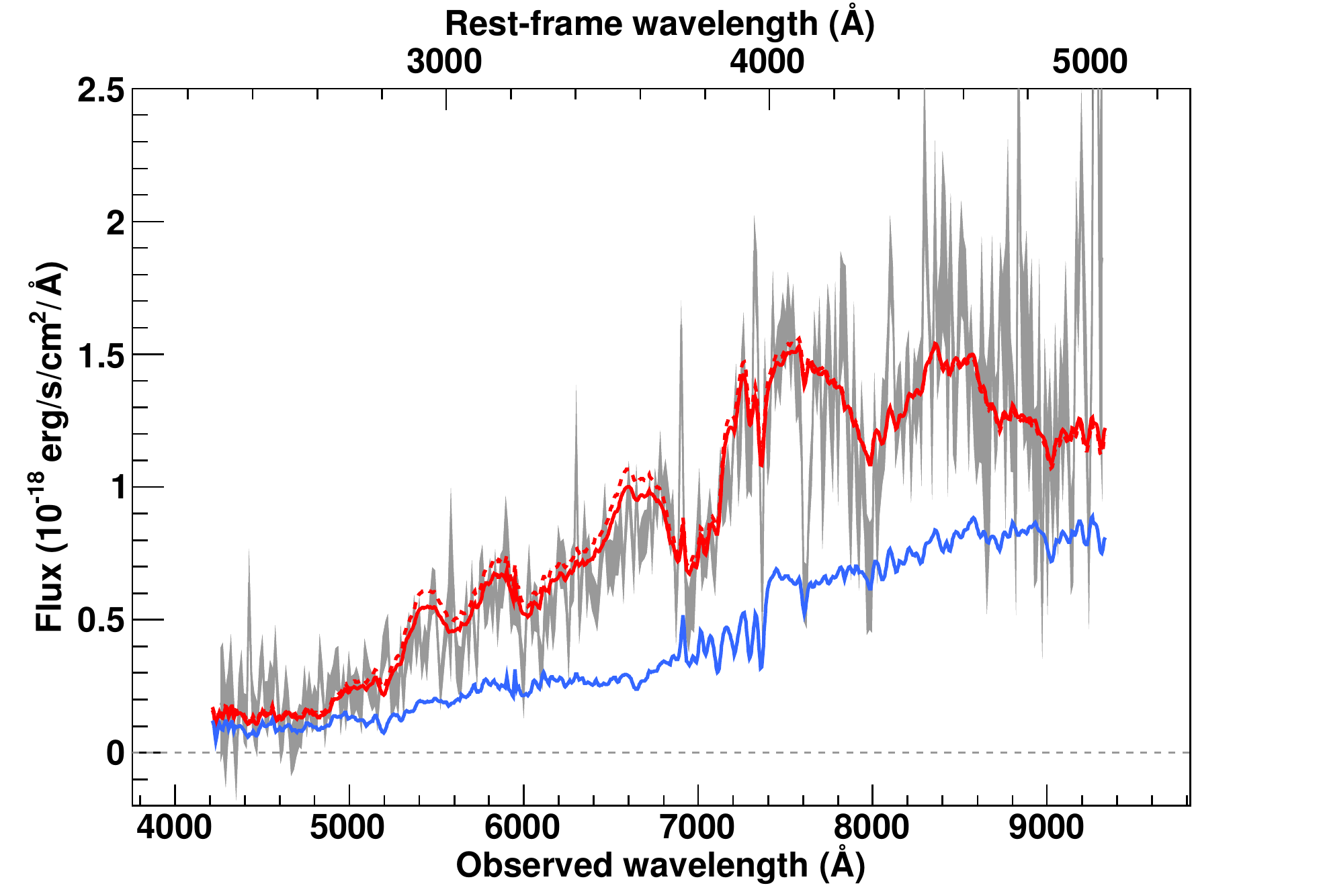}
    \includegraphics[scale=0.45]{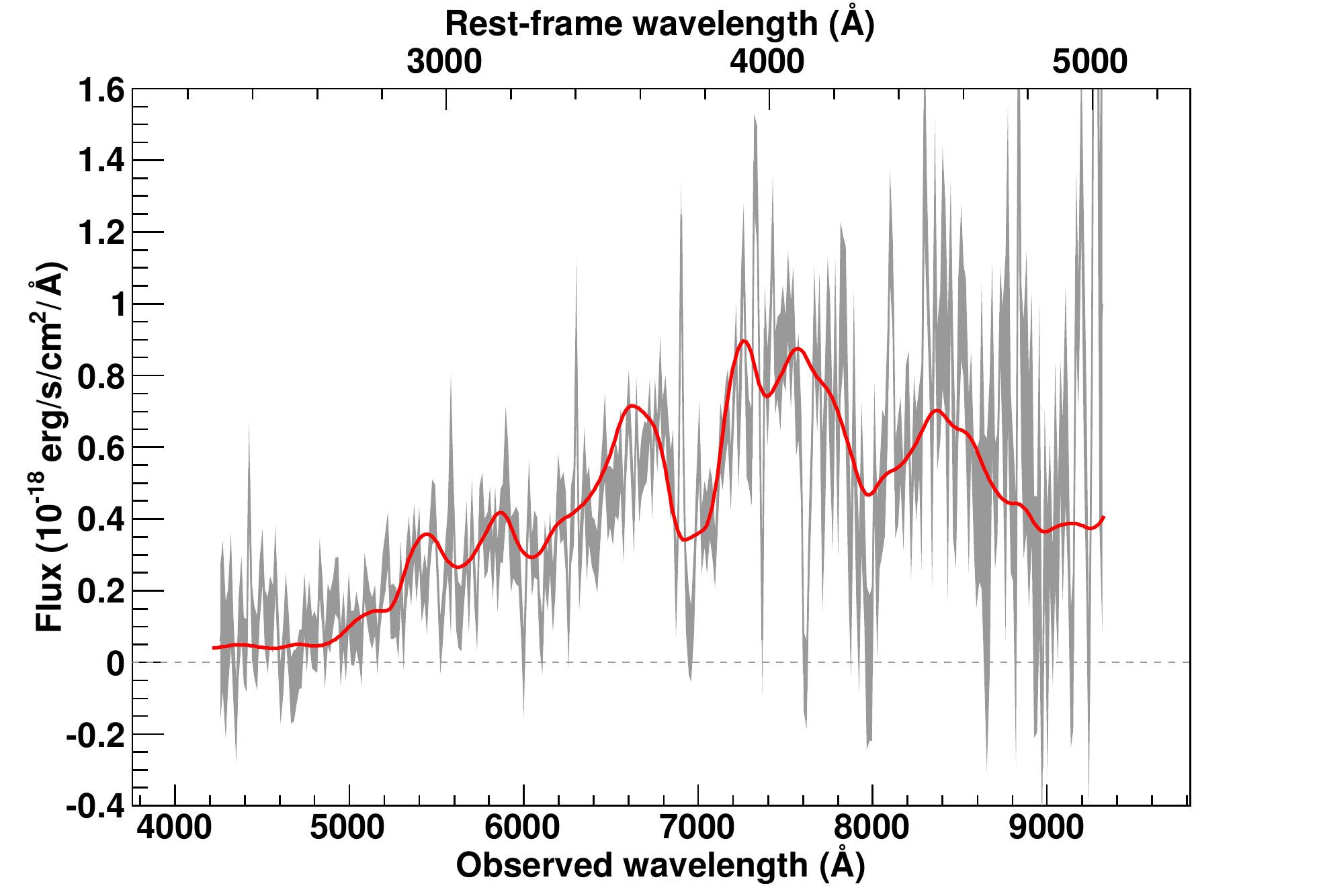}
    \end{center}
    \caption{The SNIa 07D1cc\_1719 spectrum measured at $z=0.853$ with a phase of 1.2 days. A Sa-Sb host model has been subtracted.}
    \label{fig:Spec07D1cc_1719}
    \end{figure}
    
    \clearpage
    \begin{figure}
    \begin{center}
    \includegraphics[scale=0.45]{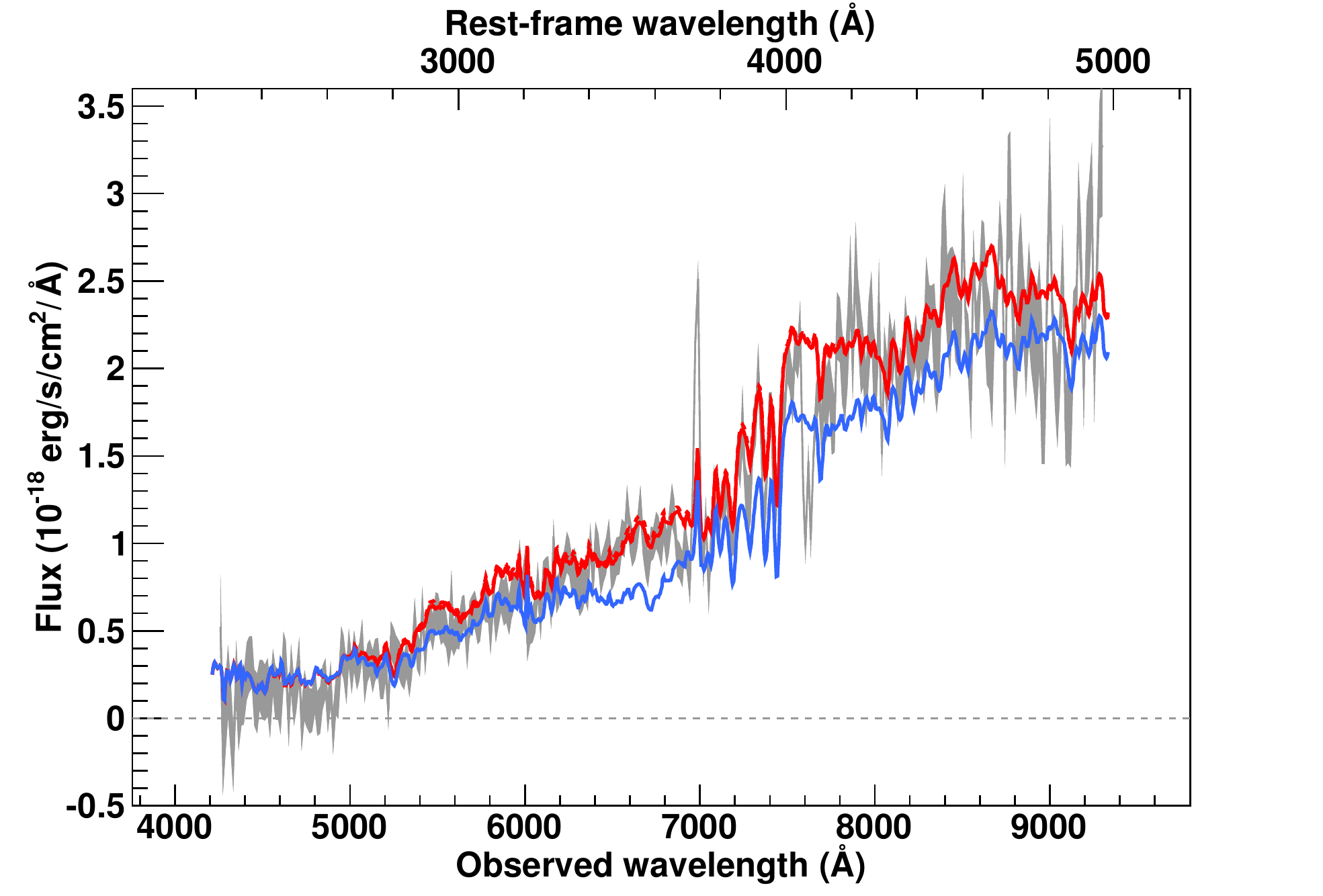}
    \includegraphics[scale=0.45]{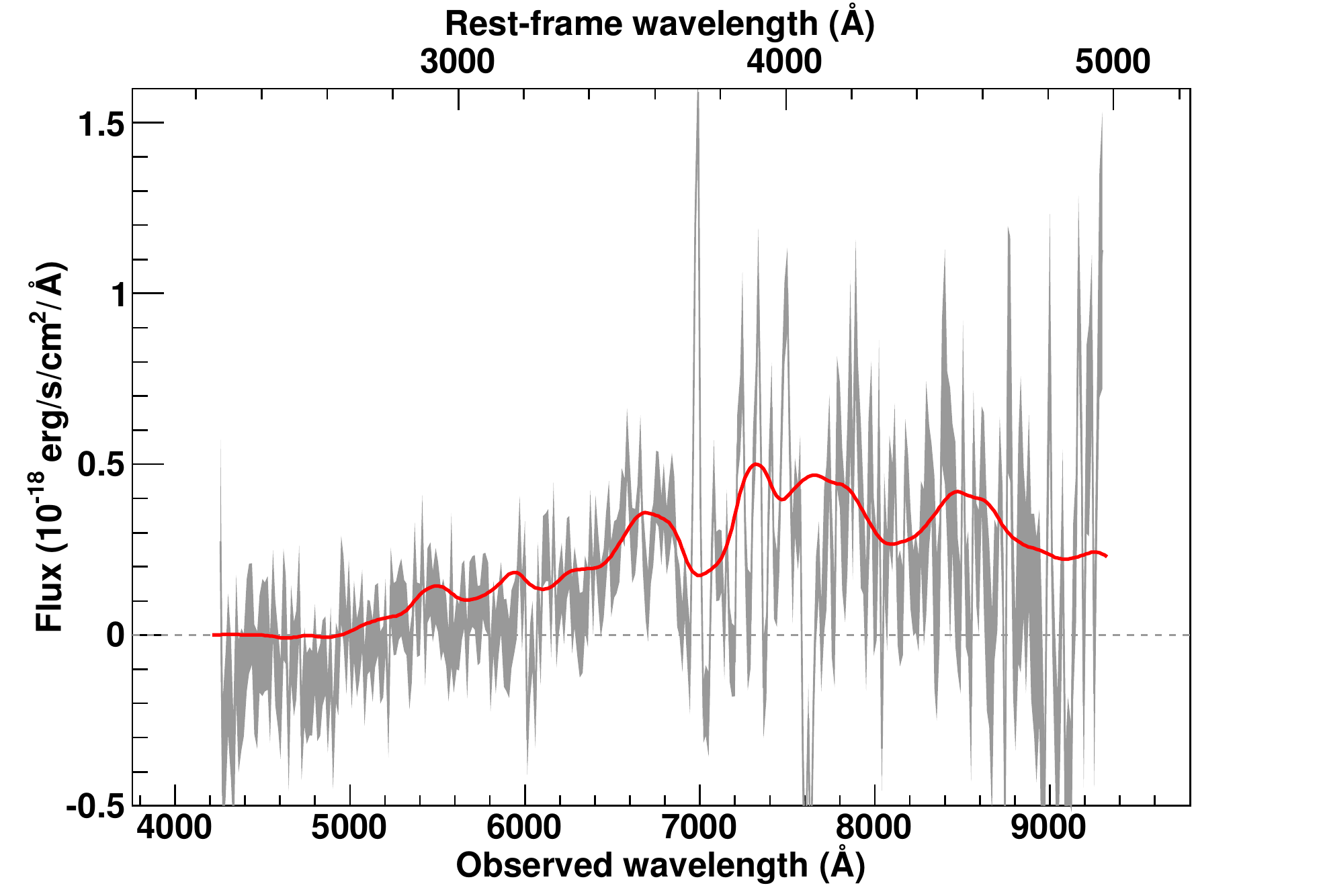}
    \end{center}
    \caption{The SNIa$\star$ 07D1cd\_1724 spectrum measured at $z=0.873$ with a phase of 4.1 days. A S0-Sa host model has been subtracted.}
    \label{fig:Spec07D1cd_1724}
    \end{figure}
    
    \begin{figure}
    \begin{center}
    \includegraphics[scale=0.45]{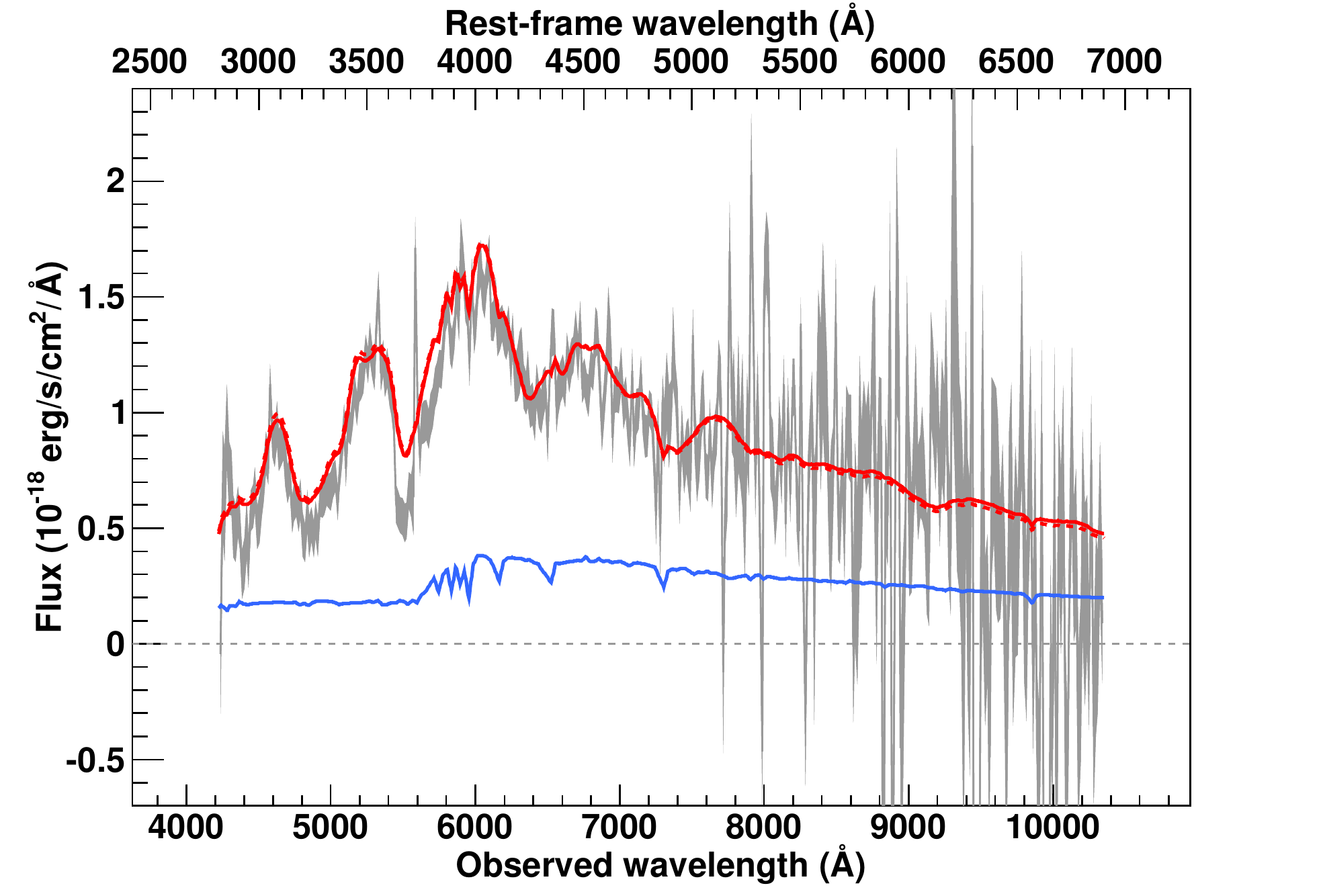}
    \includegraphics[scale=0.45]{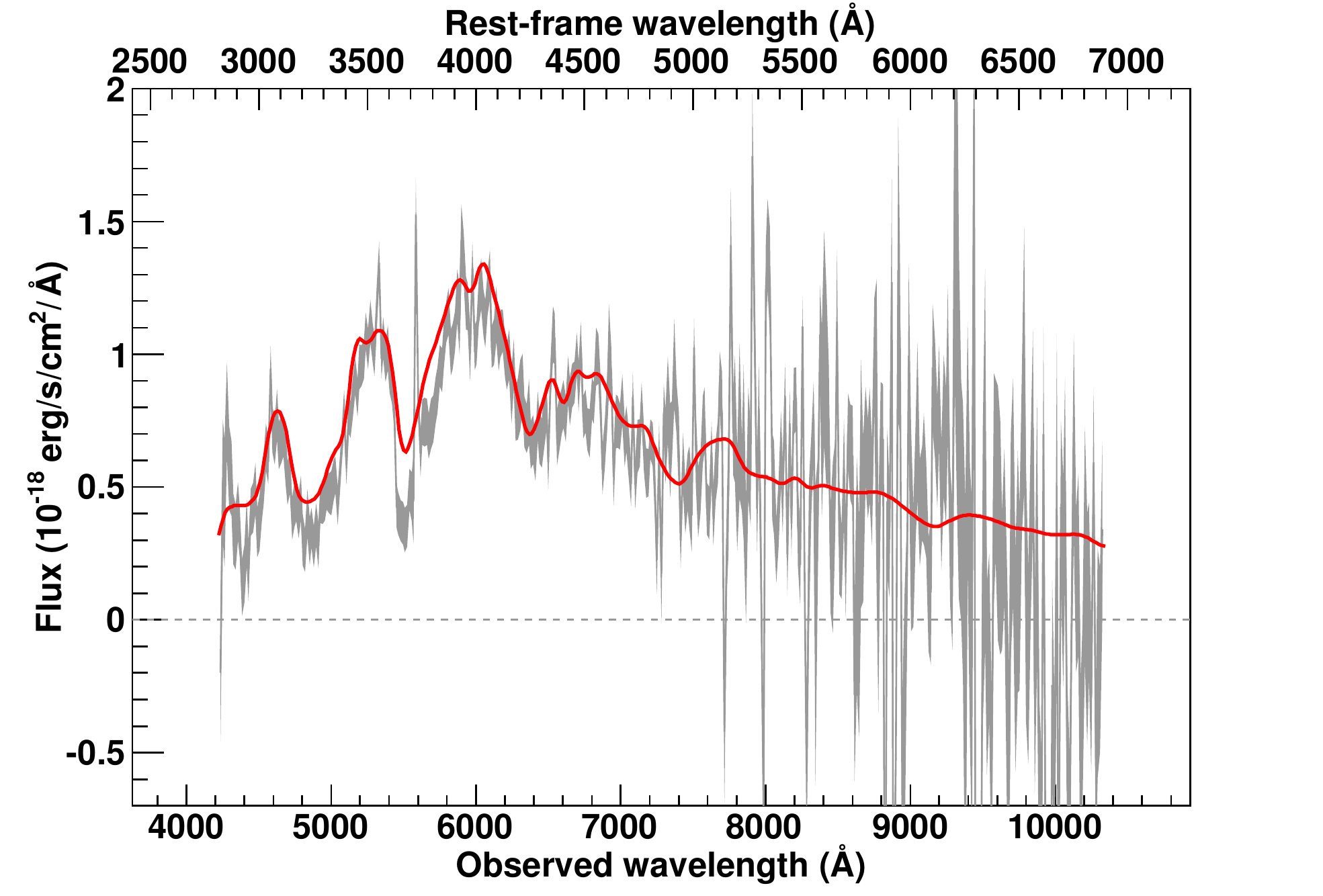}
    \end{center}
    \caption{The SNIa 07D1cf\_1723 spectrum measured at $z=0.500$ with a phase of -8.4 days. A E(1) host model has been subtracted.}
    \label{fig:Spec07D1cf_1723}
    \end{figure}
    
    \begin{figure}
    \begin{center}
    \includegraphics[scale=0.45]{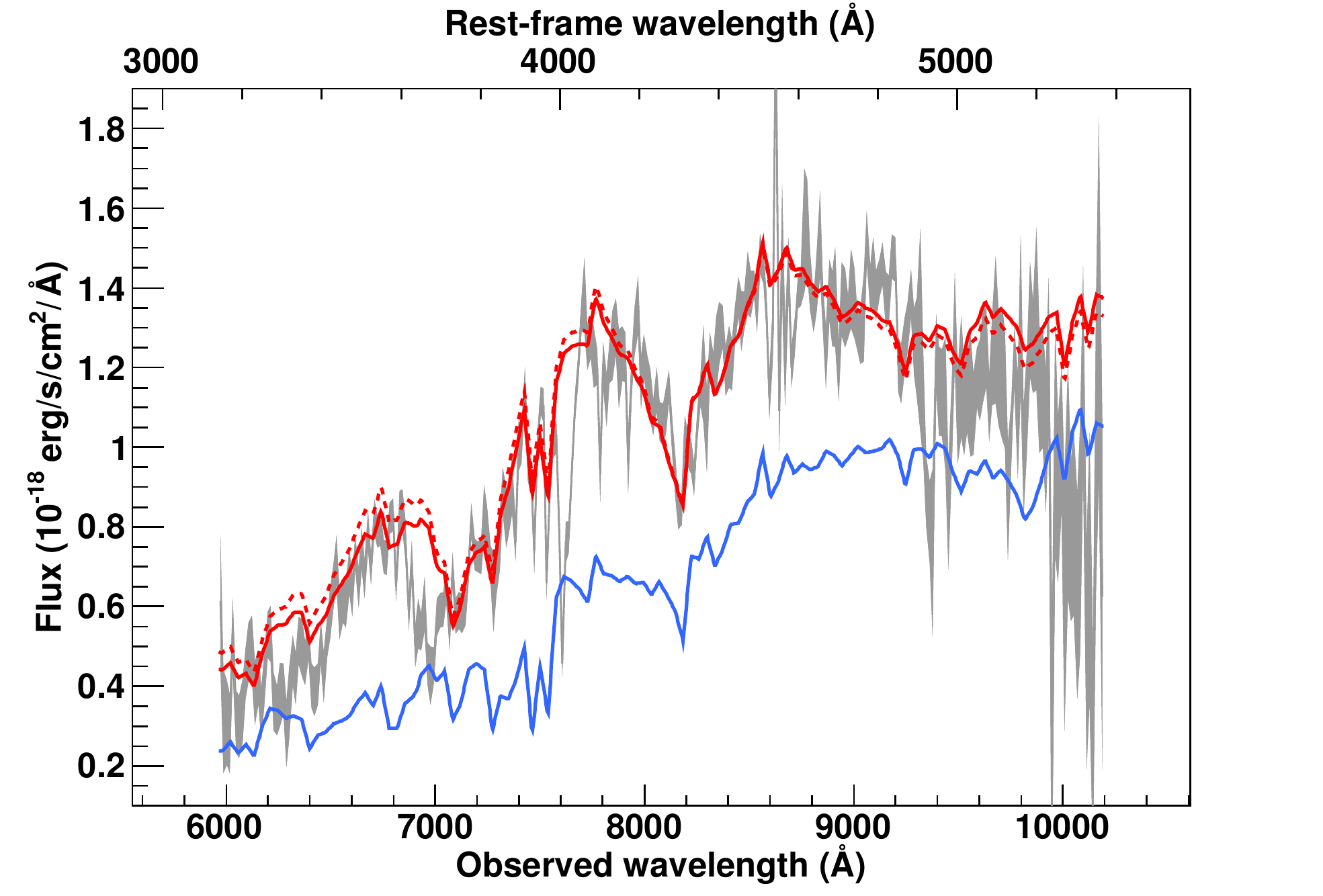}
    \includegraphics[scale=0.45]{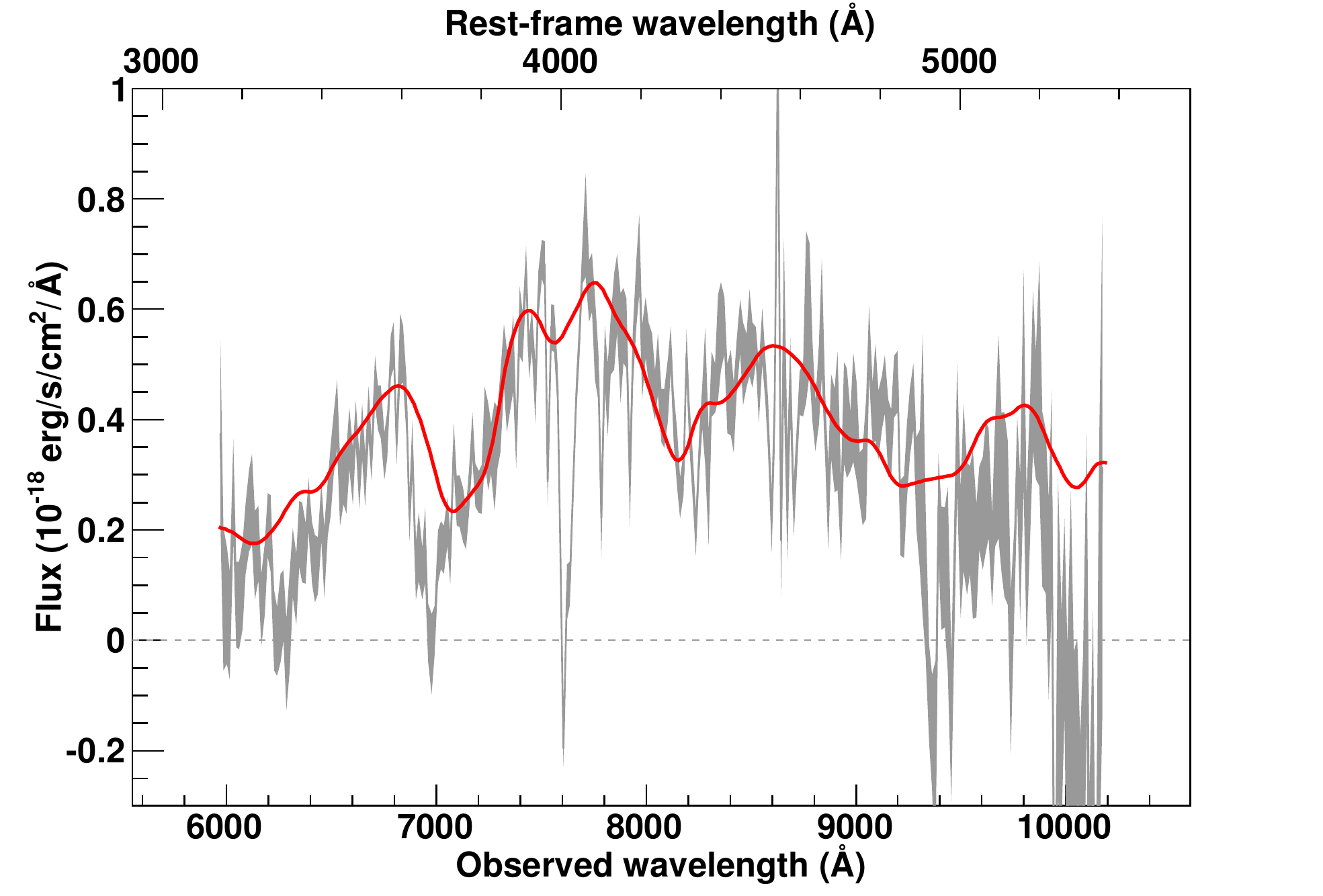}
    \end{center}
    \caption{The SNIa 07D2aa\_1487 spectrum measured at $z=0.899$ with a phase of -1.9 days. A S0(12) host model has been subtracted.}
    \label{fig:Spec07D2aa_1487}
    \end{figure}
    
    \clearpage
    \begin{figure}
    \begin{center}
    \includegraphics[scale=0.45]{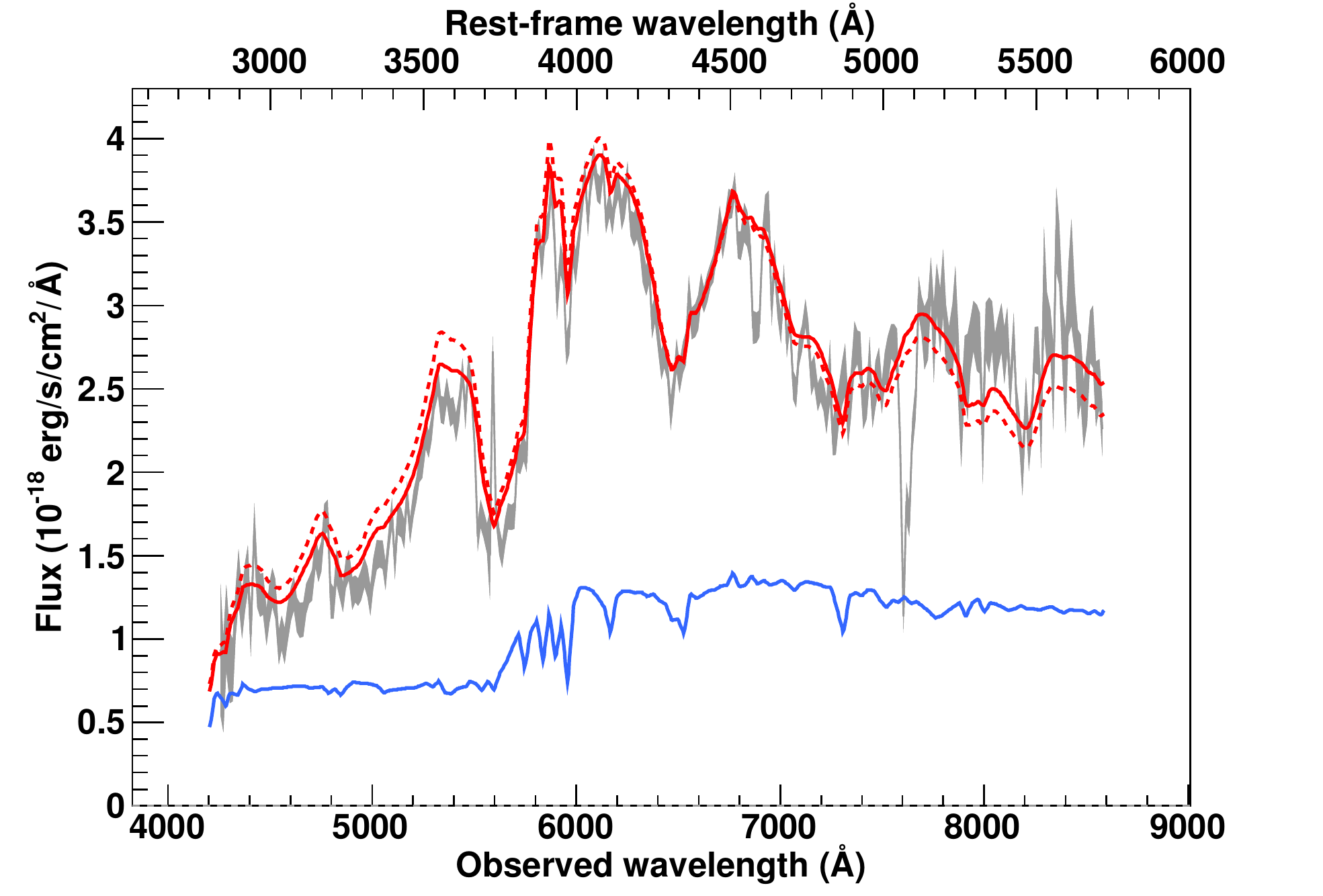}
    \includegraphics[scale=0.45]{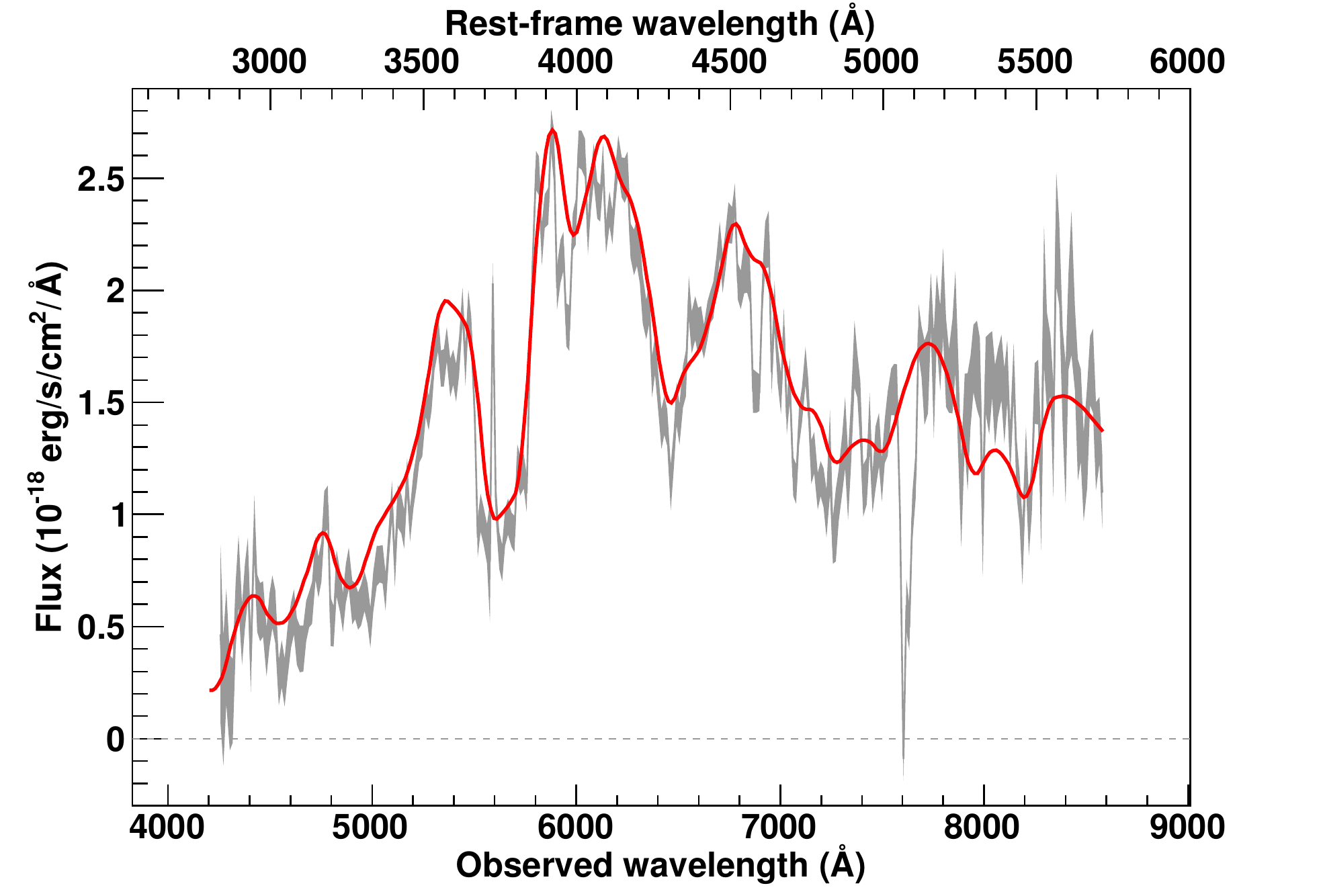}
    \end{center}
    \caption{The SNIa 07D2ae\_1485 spectrum measured at $z=0.501$ with a phase of 1.7 days. A S0(1) host model has been subtracted.}
    \label{fig:Spec07D2ae_1485}
    \end{figure}
    
    \begin{figure}
    \begin{center}
    \includegraphics[scale=0.45]{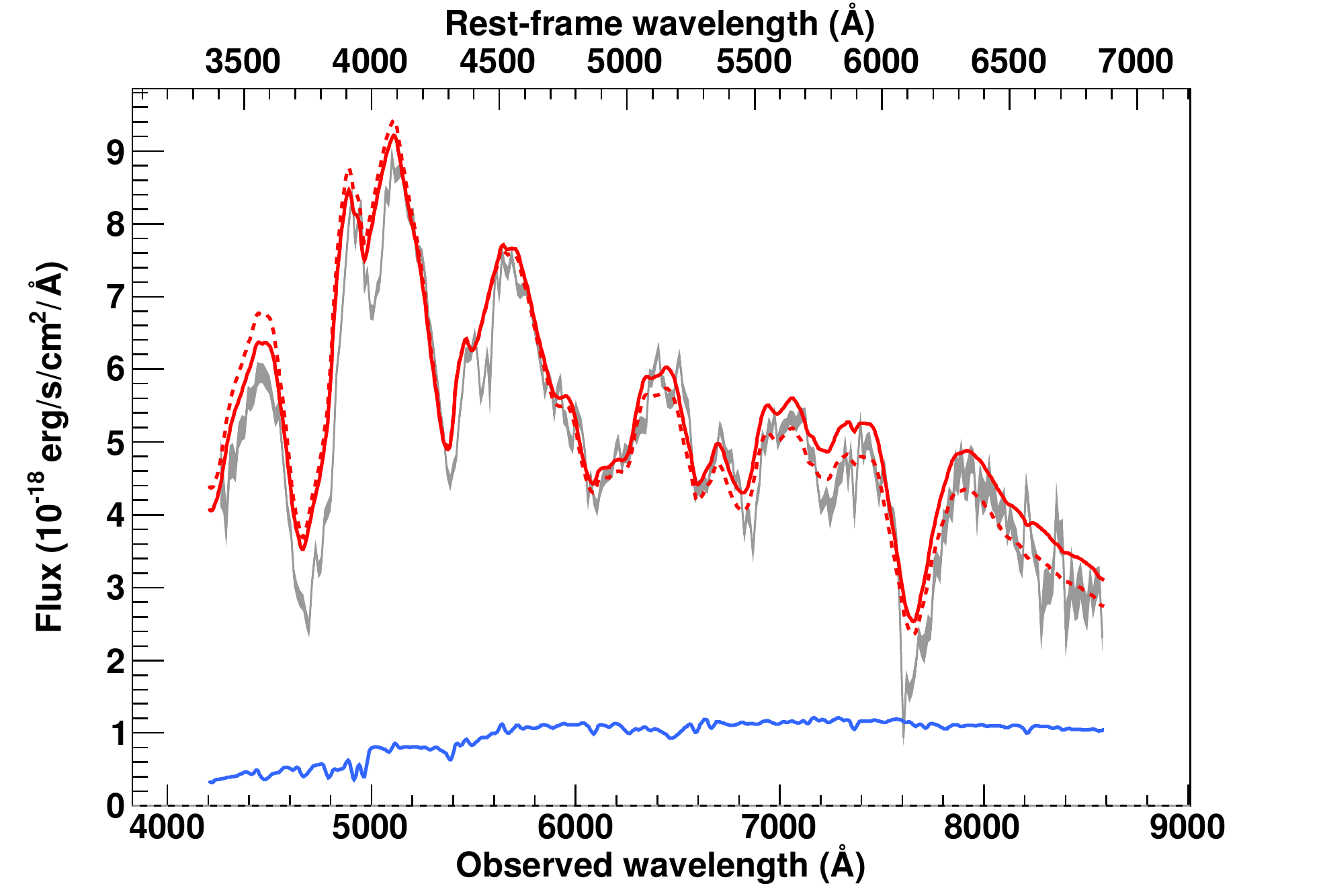}
    \includegraphics[scale=0.45]{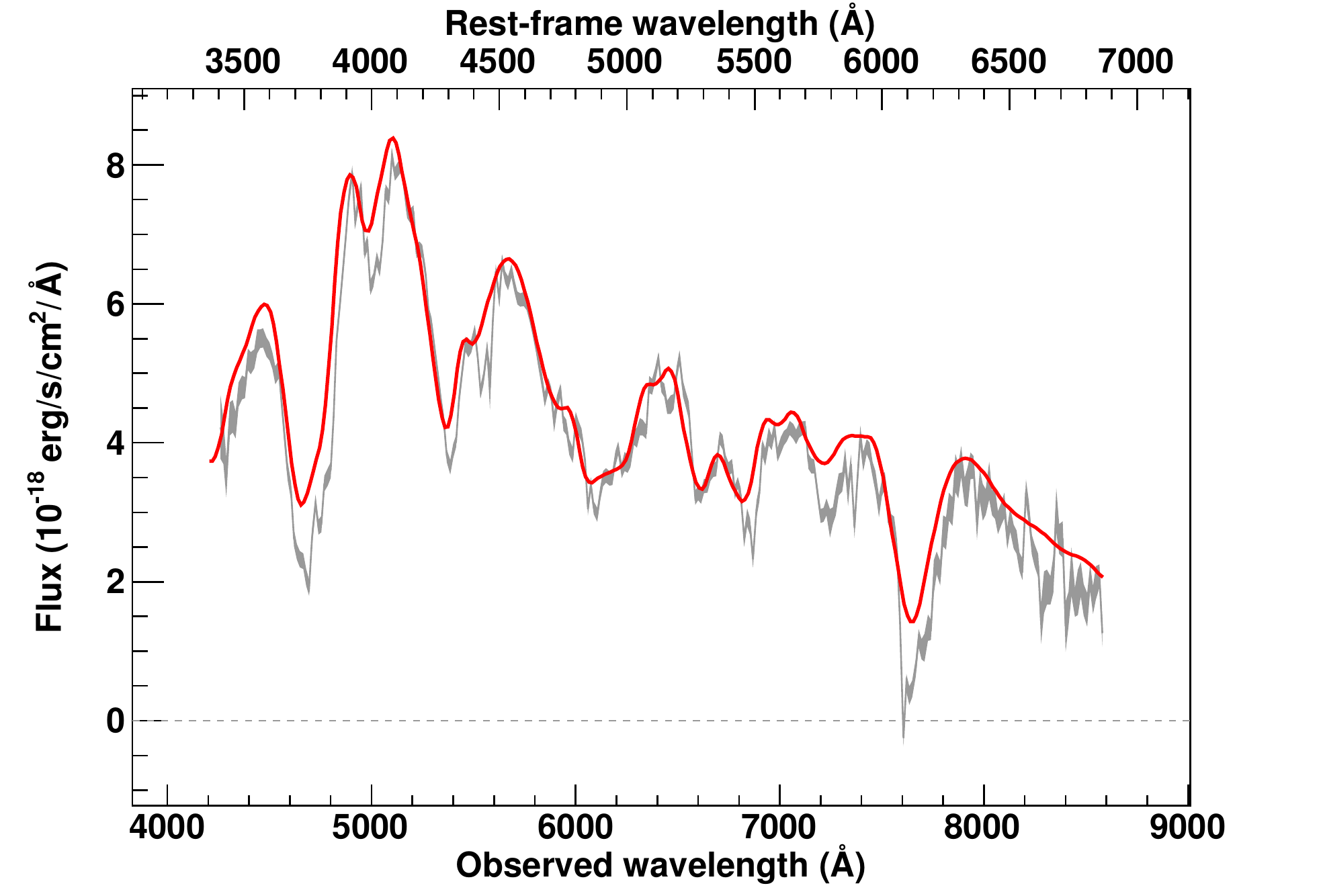}
    \end{center}
    \caption{The SNIa 07D2ag\_1485 spectrum measured at $z=0.25$ with a phase of -2.6 days. A S0(5) host model has been subtracted.}
    \label{fig:Spec07D2ag_1485}
    \end{figure}
    
    \begin{figure}
    \begin{center}
    \includegraphics[scale=0.45]{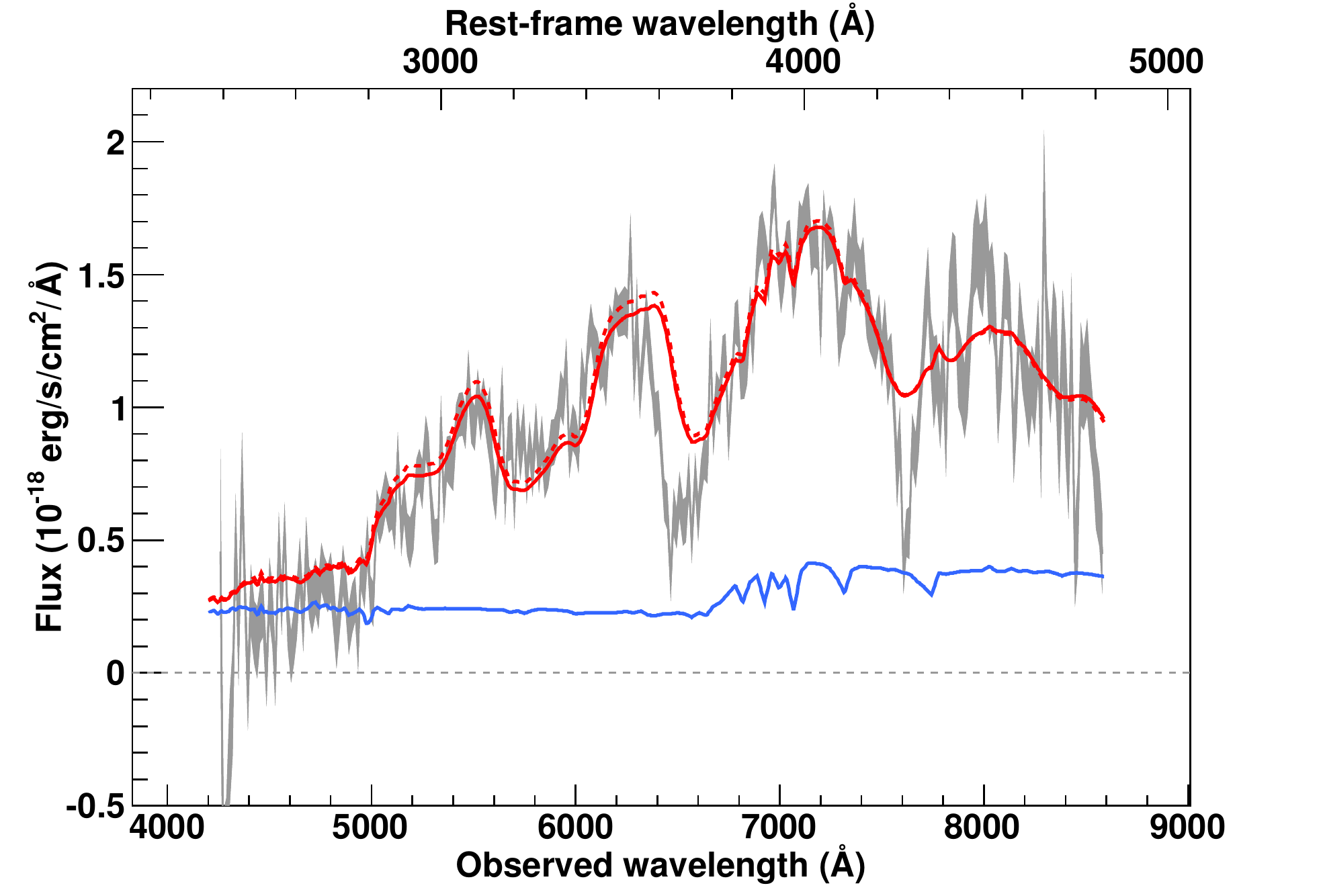}
    \includegraphics[scale=0.45]{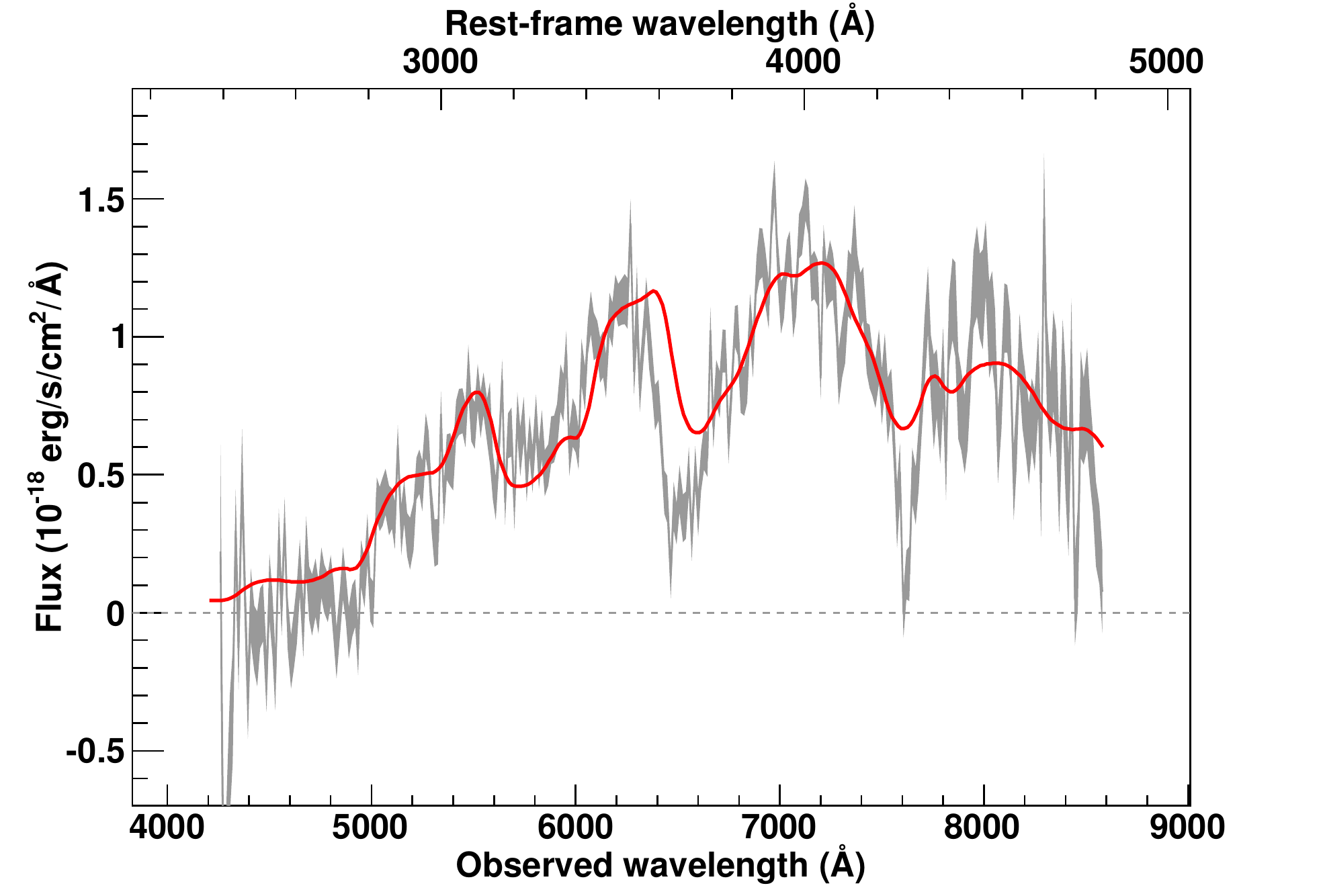}
    \end{center}
    \caption{The SNIa 07D2ah\_1486 spectrum measured at $z=0.780$ with a phase of -3.8 days. A S0(1) host model has been subtracted.}
    \label{fig:Spec07D2ah_1486}
    \end{figure}
    
    \clearpage
    \begin{figure}
    \begin{center}
    \includegraphics[scale=0.45]{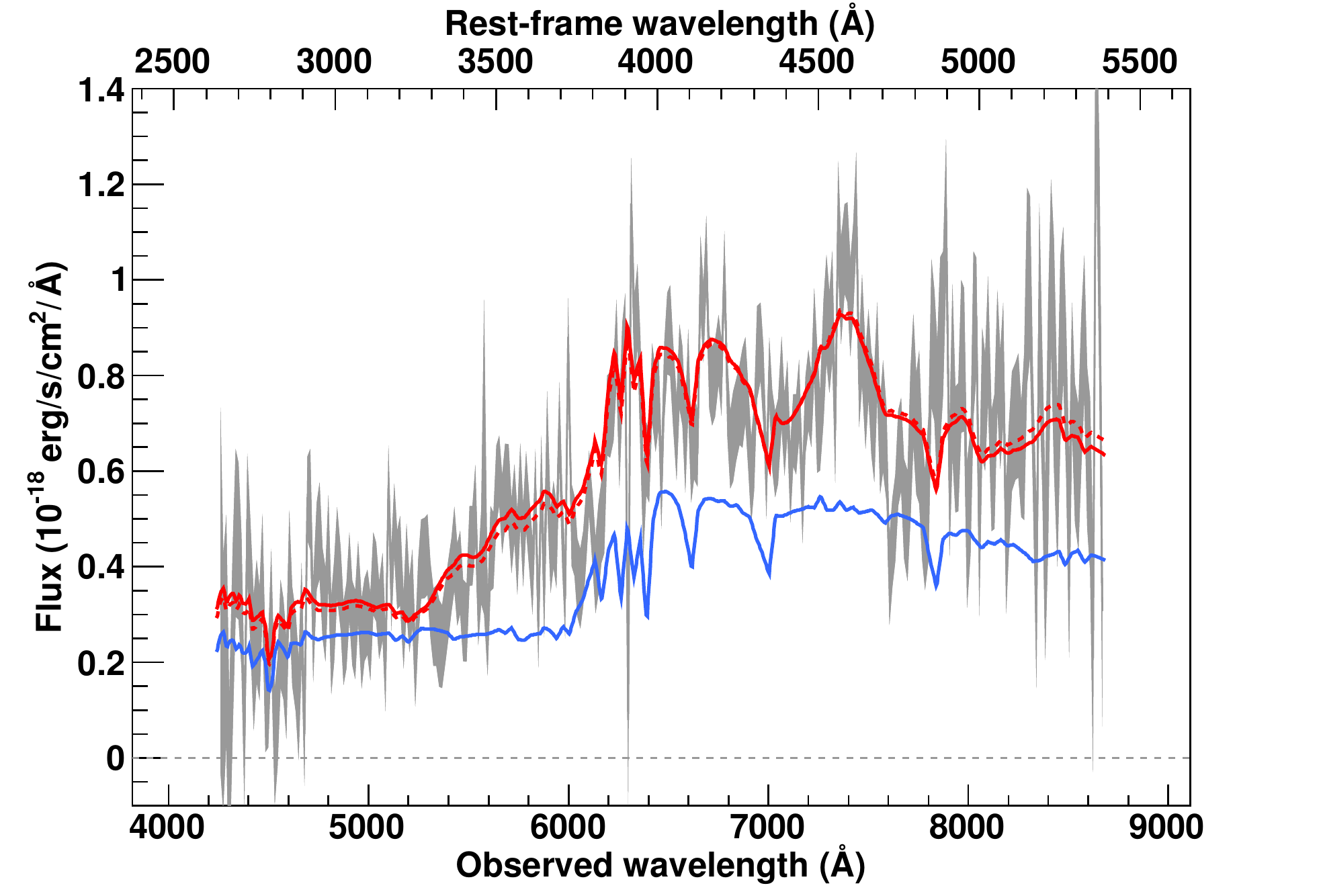}
    \includegraphics[scale=0.45]{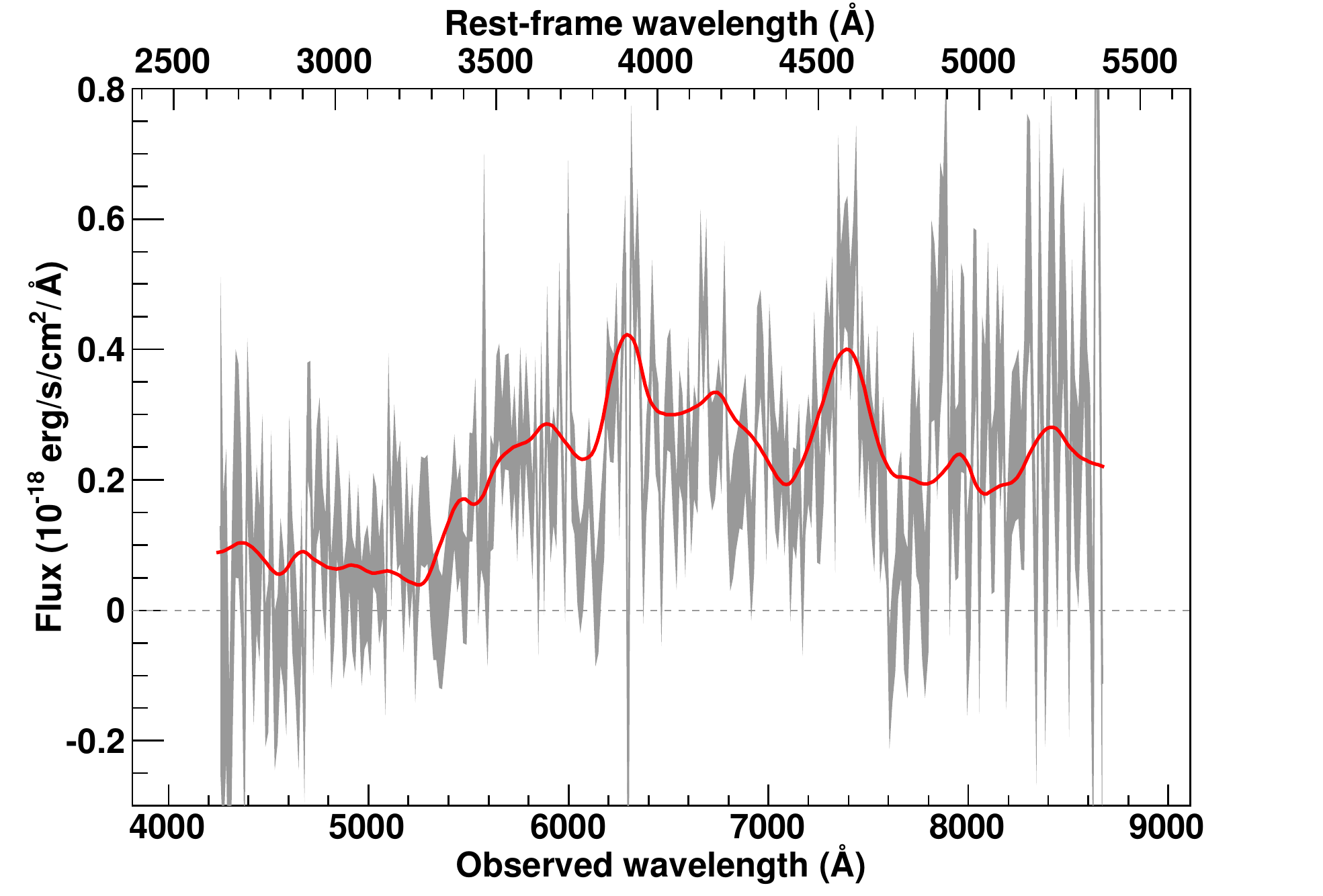}
    \end{center}
    \caption{The SNIa$\star$ 07D2aw\_1515 spectrum measured at $z=0.610$ with a phase of 10.0 days. A E(1) host model has been subtracted.}
    \label{fig:Spec07D2aw_1515}
    \end{figure}
    
    \begin{figure}
    \begin{center}
    \includegraphics[scale=0.45]{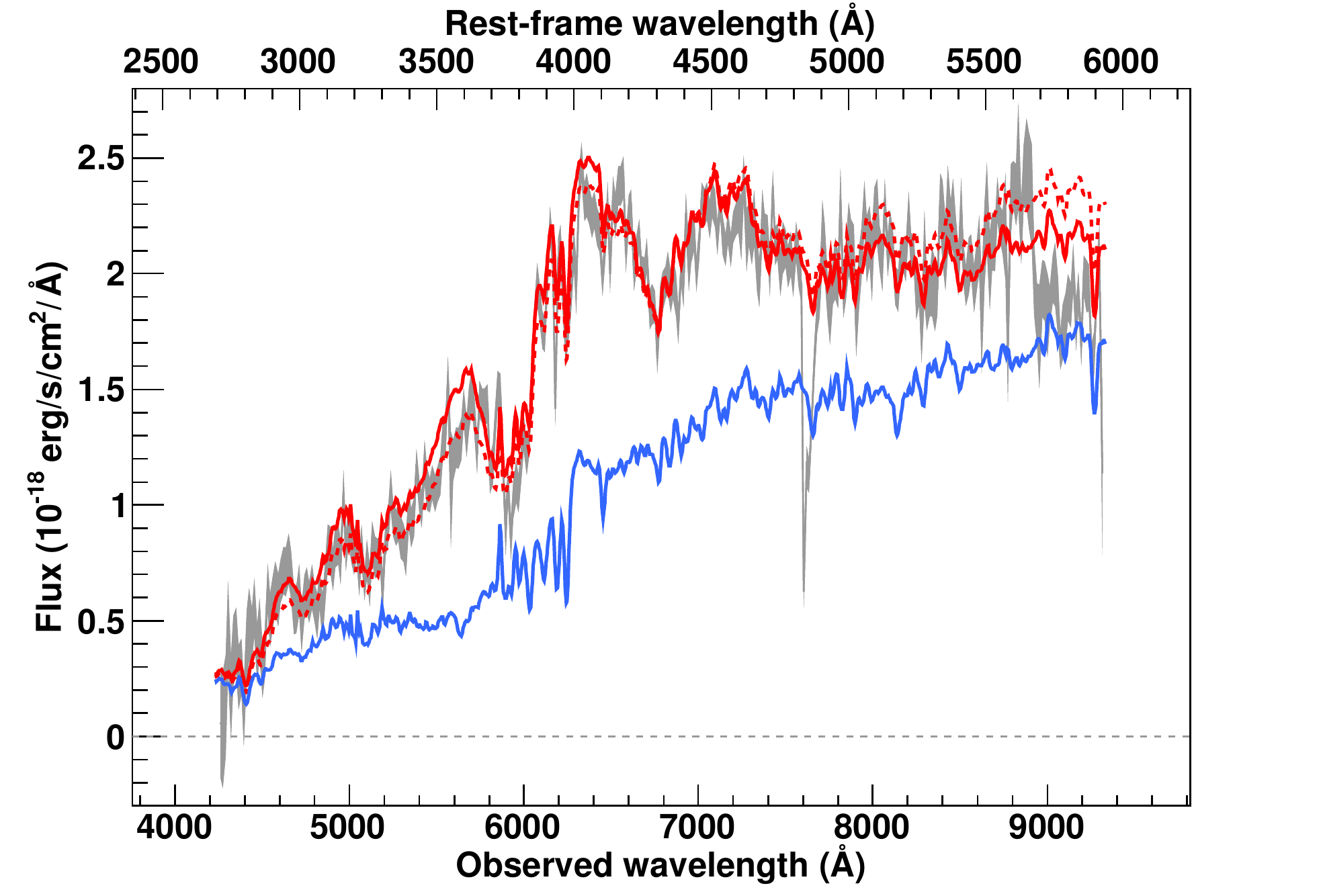}
    \includegraphics[scale=0.45]{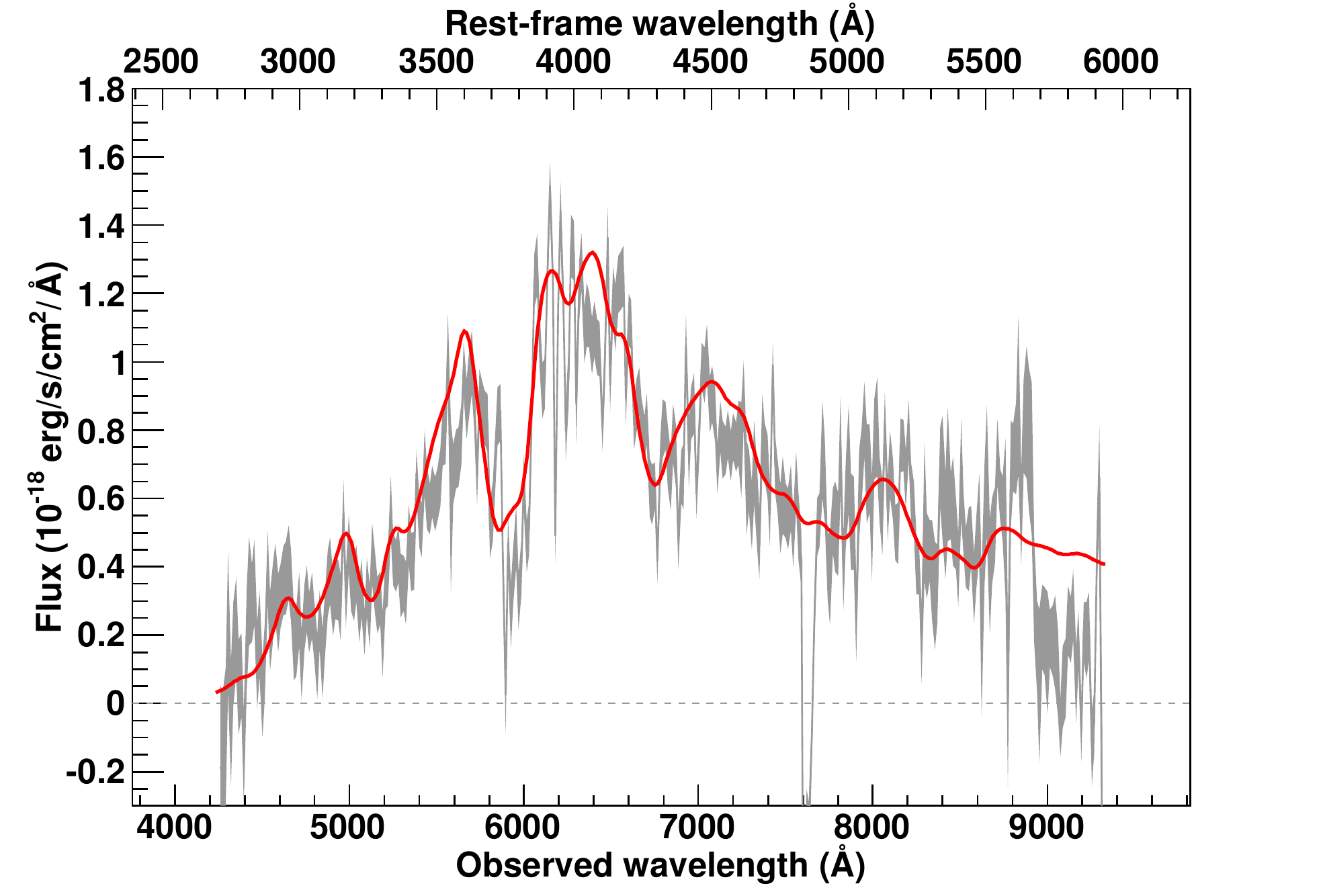}
    \end{center}
    \caption{The SNIa 07D2bd\_1510 spectrum measured at $z=0.572$ with a phase of 2.1 days. A Sa-Sb host model has been subtracted.}
    \label{fig:Spec07D2bd_1510}
    \end{figure}
    
    \begin{figure}
    \begin{center}
    \includegraphics[scale=0.45]{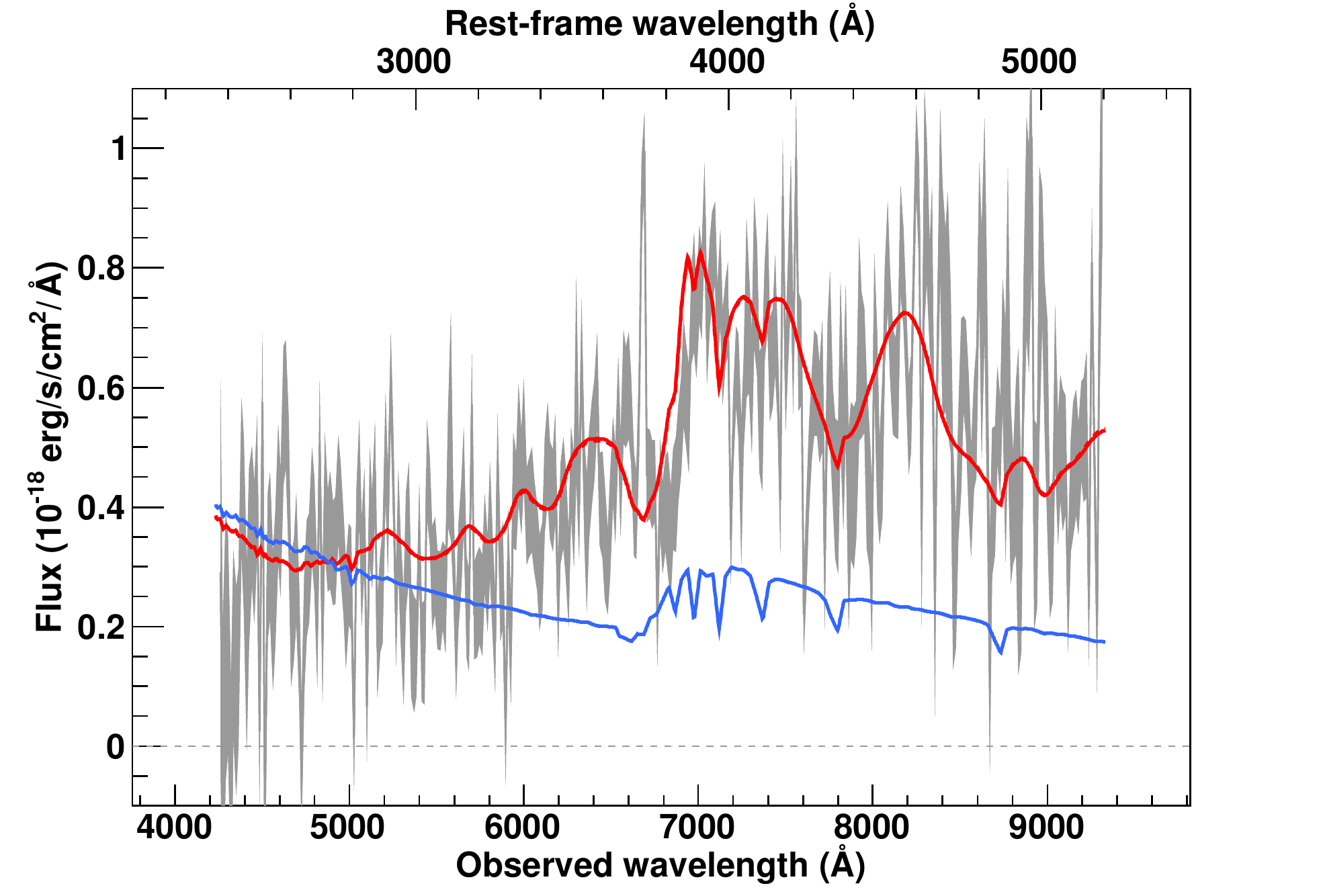}
    \includegraphics[scale=0.45]{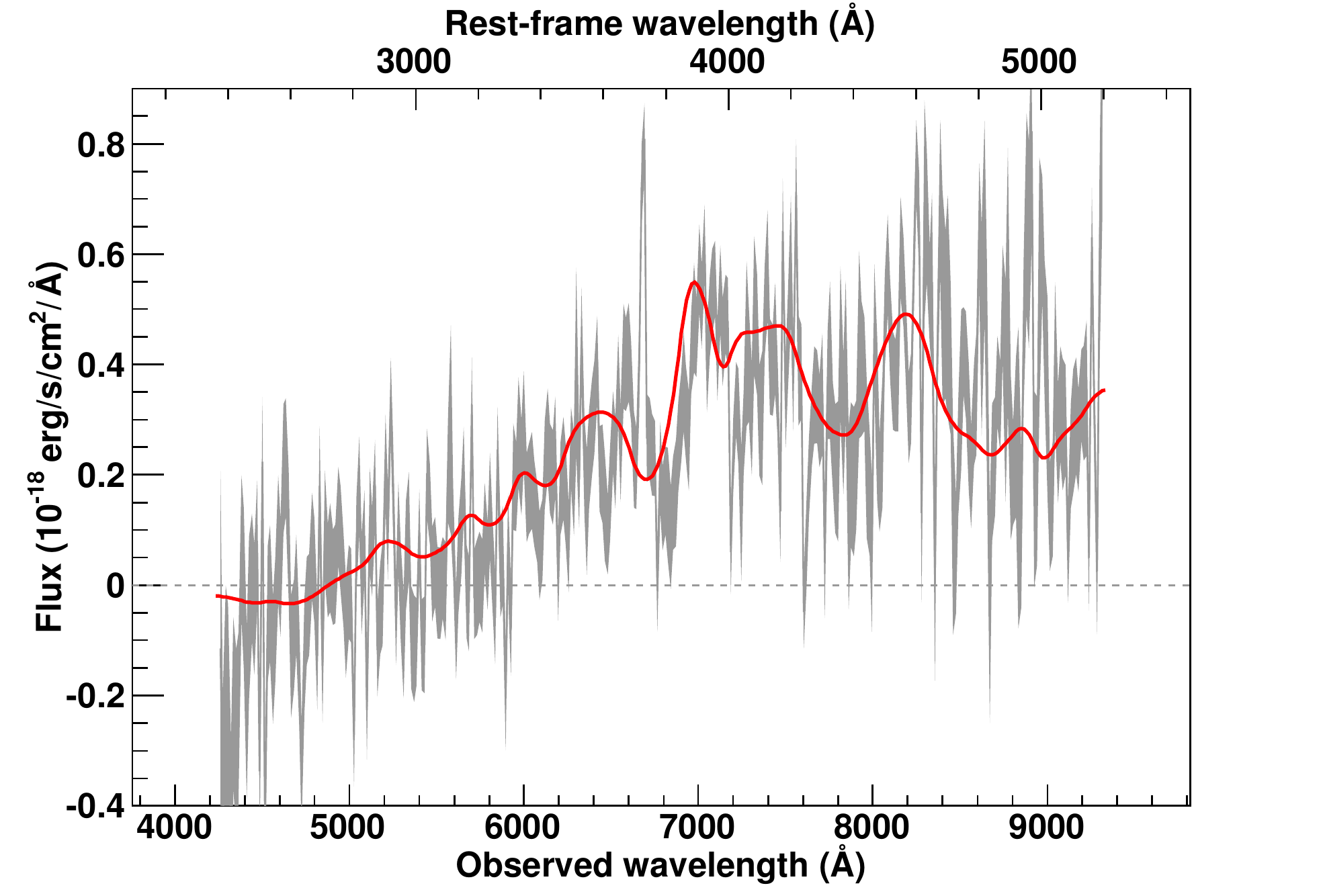}
    \end{center}
    \caption{The SNIa$\star$ 07D2be\_1510 spectrum measured at $z=0.793$ with a phase of 7.0 days. A Sc(1) host model has been subtracted.}
    \label{fig:Spec07D2be_1510}
    \end{figure}
    
    \clearpage
    \begin{figure}
    \begin{center}
    \includegraphics[scale=0.45]{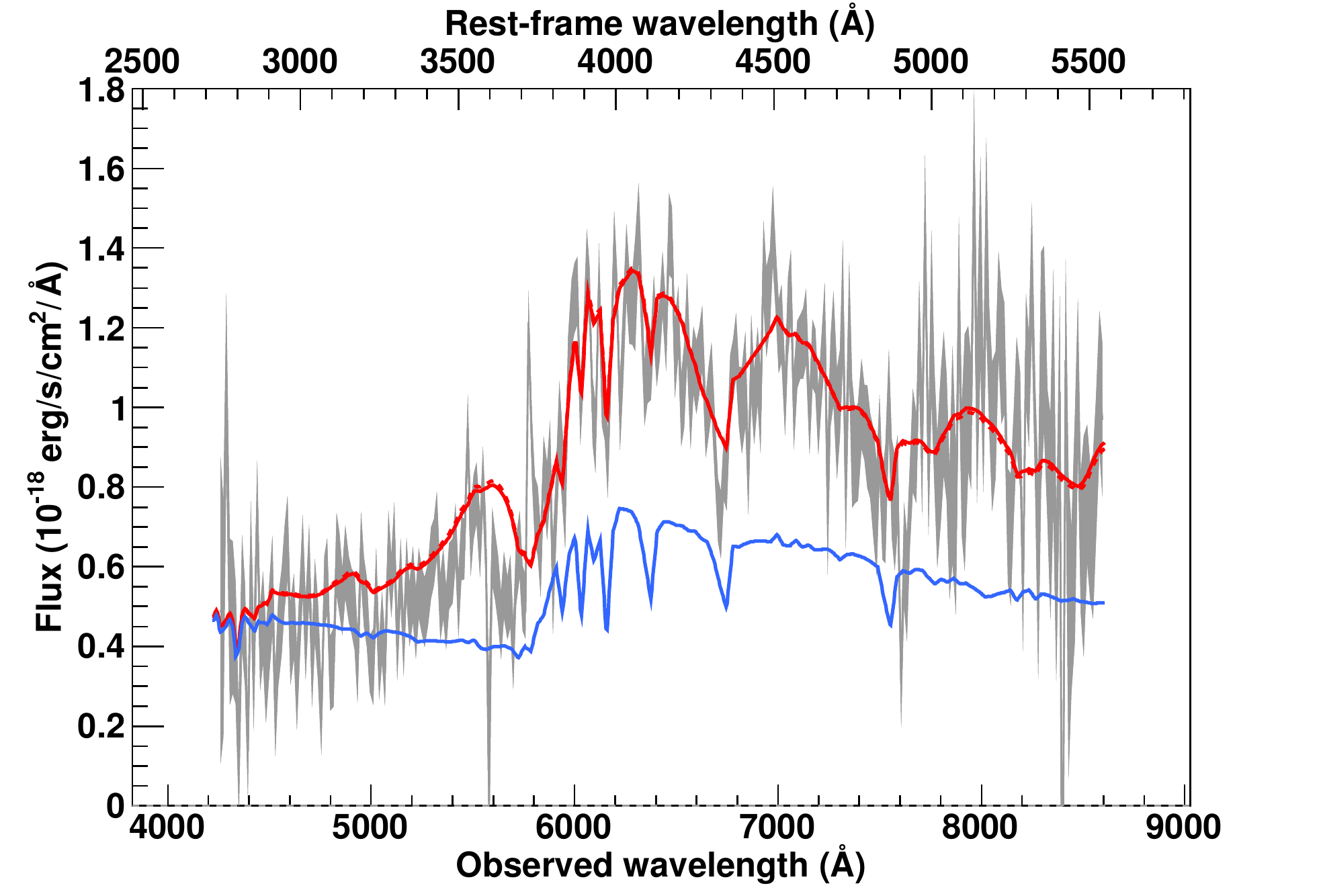}
    \includegraphics[scale=0.45]{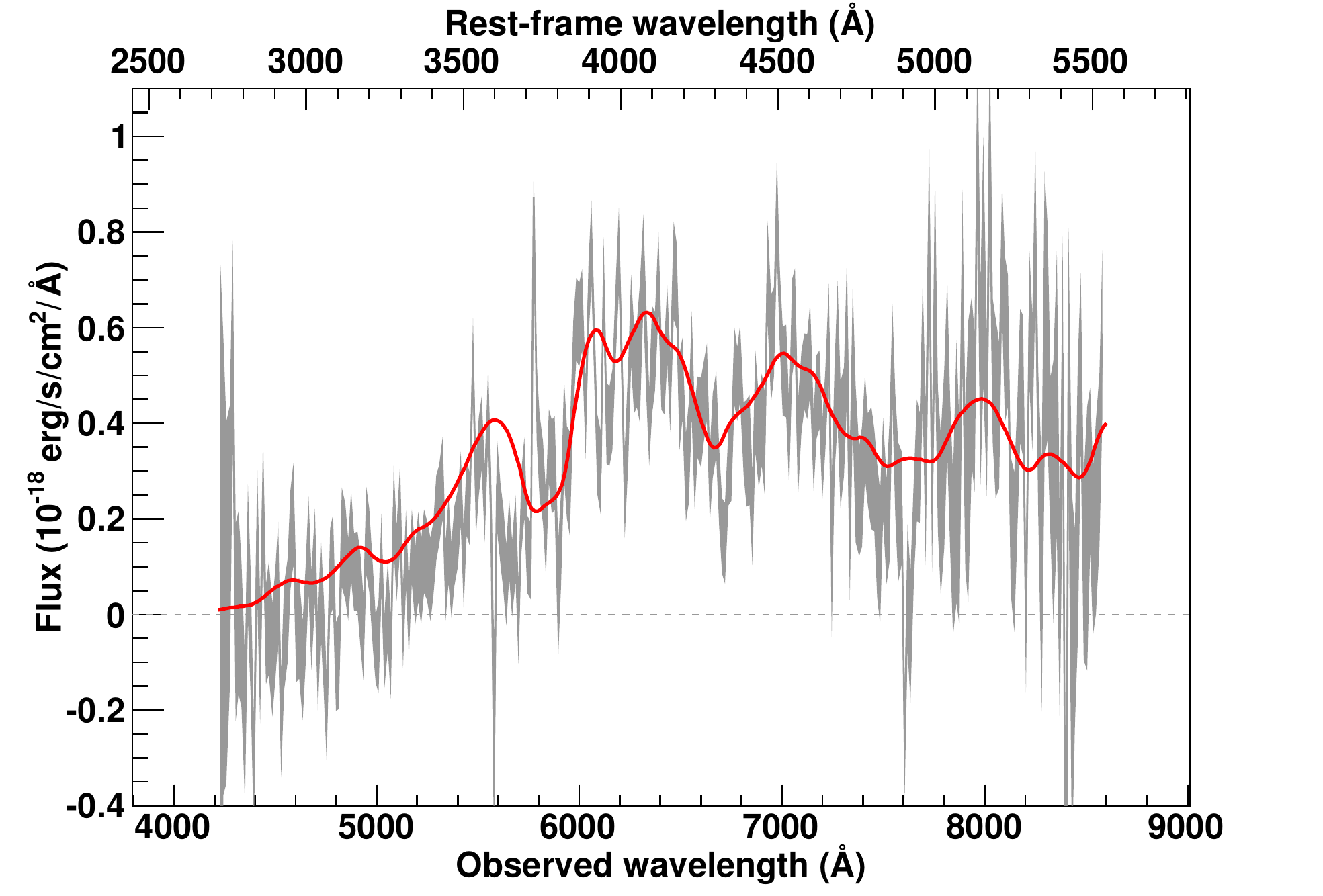}
    \end{center}
    \caption{The SNIa 07D2bi\_1514 spectrum measured at $z=0.551$ with a phase of 0.9 days. A S0(1) host model has been subtracted.}
    \label{fig:Spec07D2bi_1514}
    \end{figure}
    
    \begin{figure}
    \begin{center}
    \includegraphics[scale=0.45]{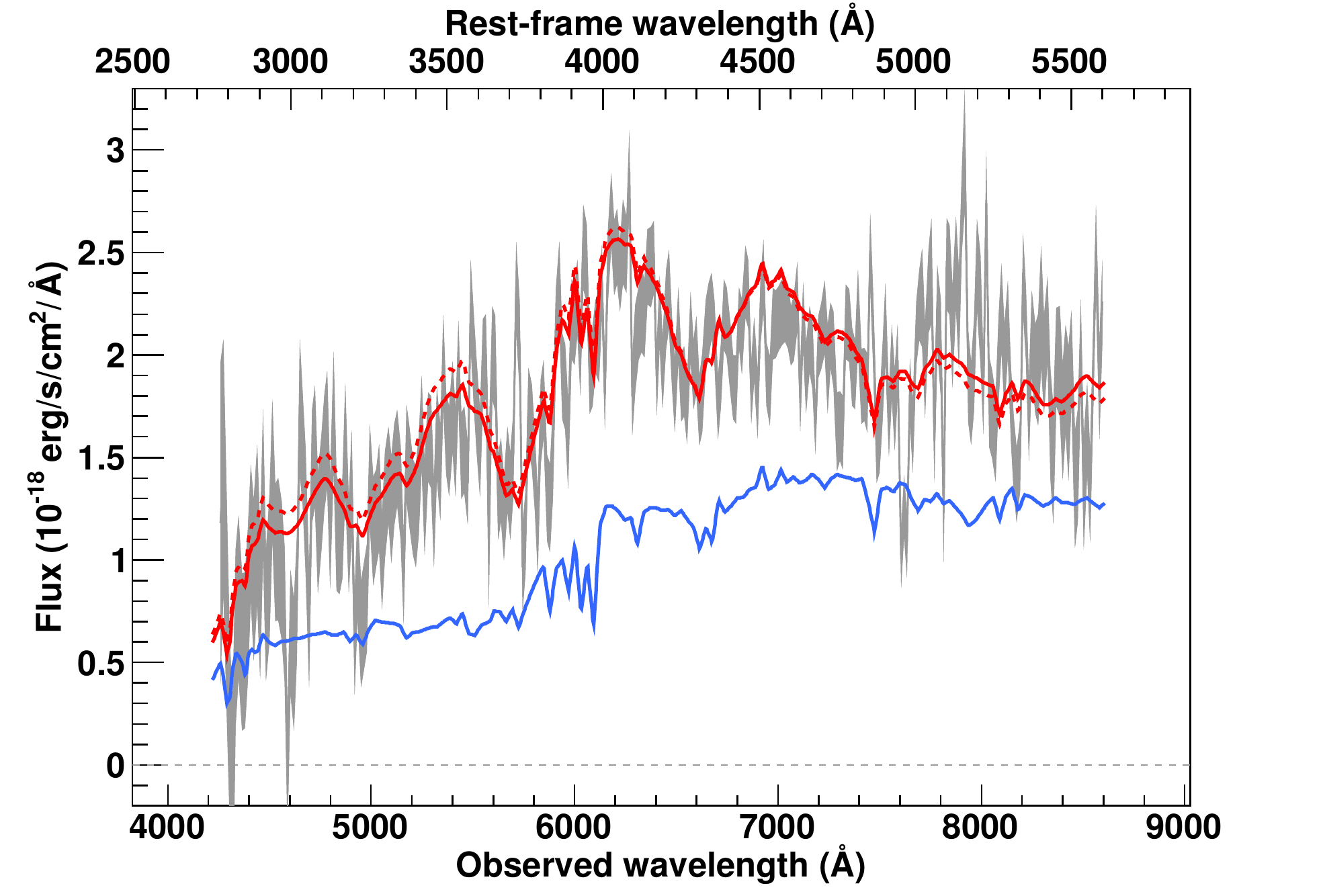}
    \includegraphics[scale=0.45]{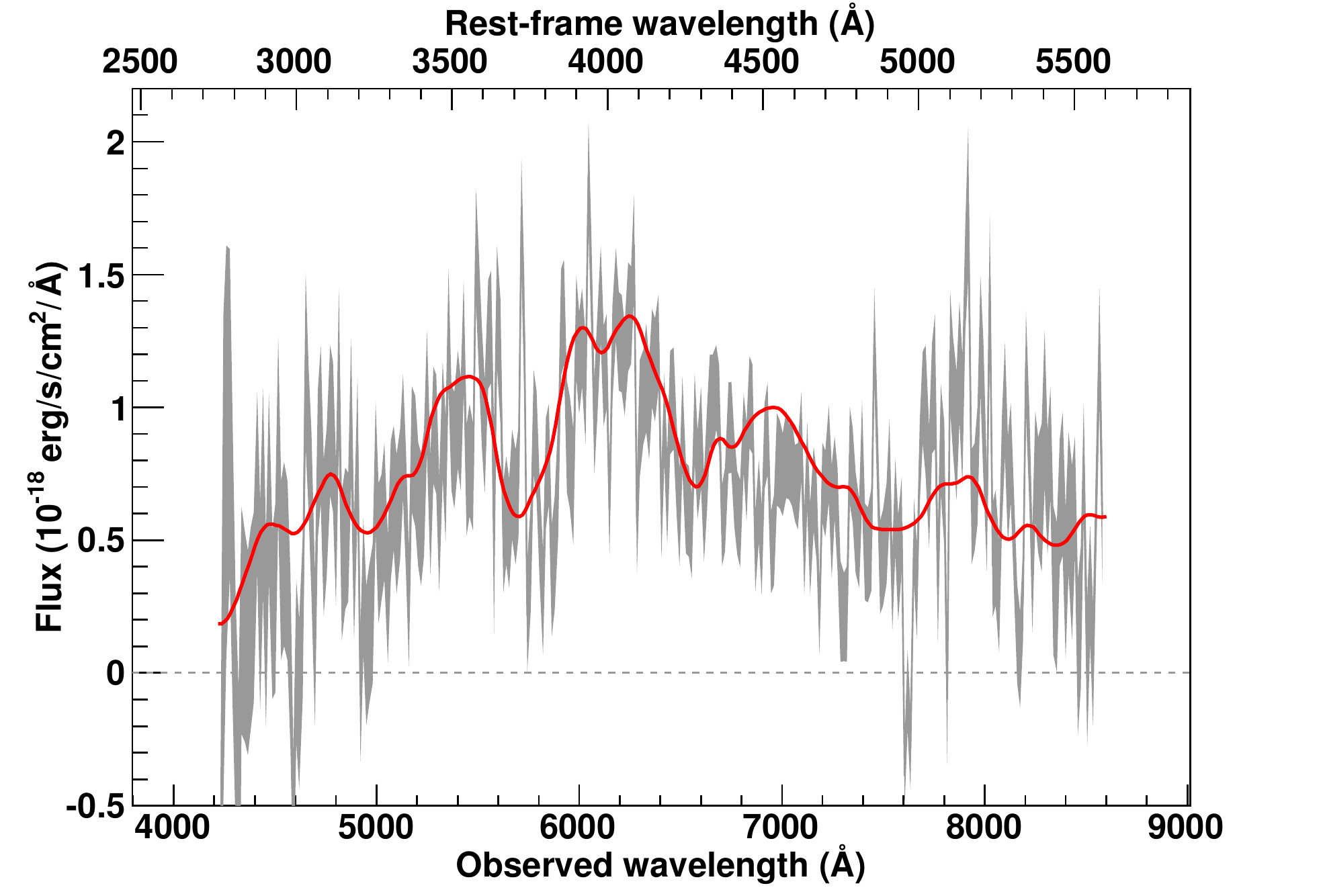}
    \end{center}
    \caption{The SNIa 07D2bq\_1518 spectrum measured at $z=0.535$ with a phase of -3.5 days. A E(1) host model has been subtracted.}
    \label{fig:Spec07D2bq_1518}
    \end{figure}
    
    \begin{figure}
    \begin{center}
    \includegraphics[scale=0.45]{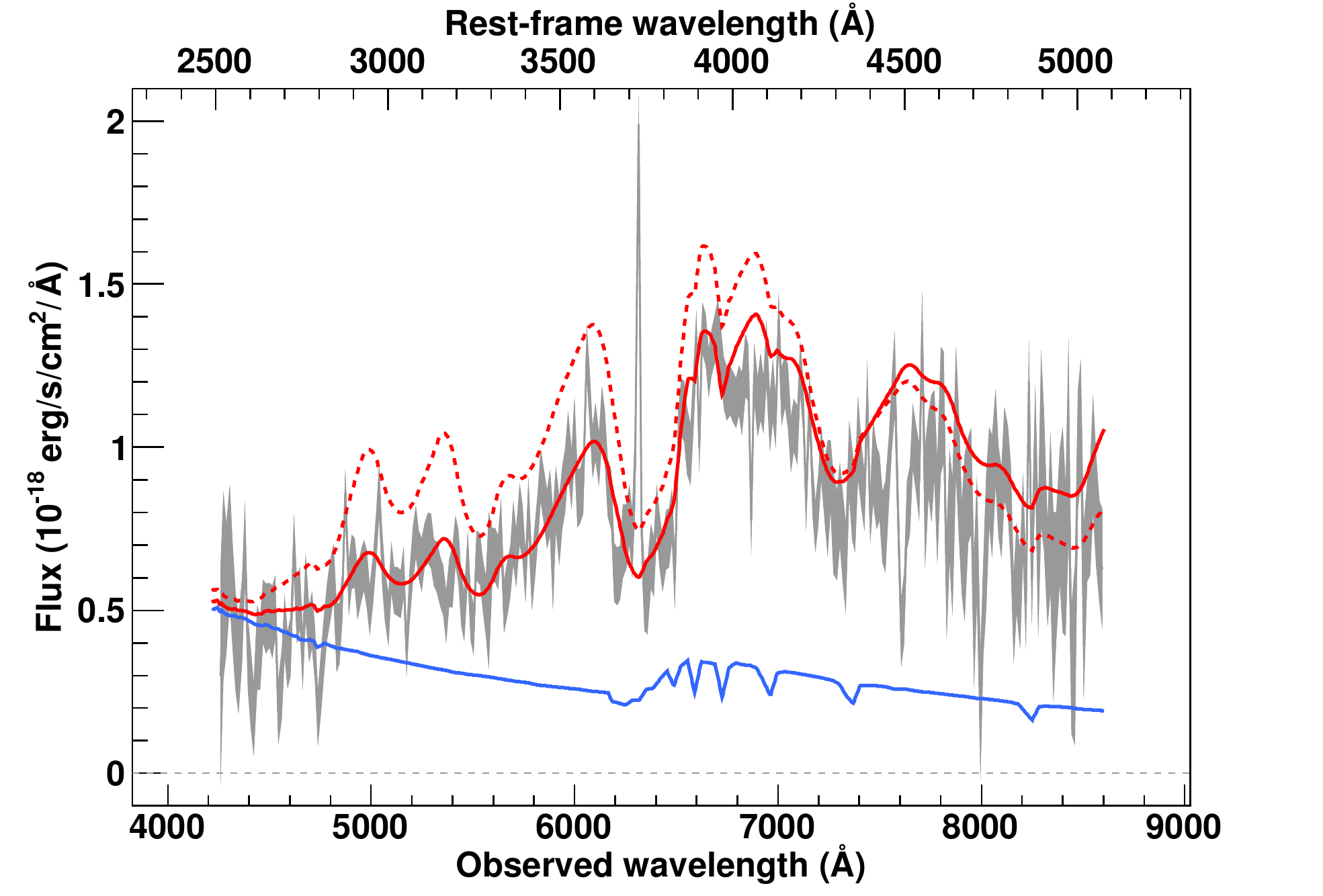}
    \includegraphics[scale=0.45]{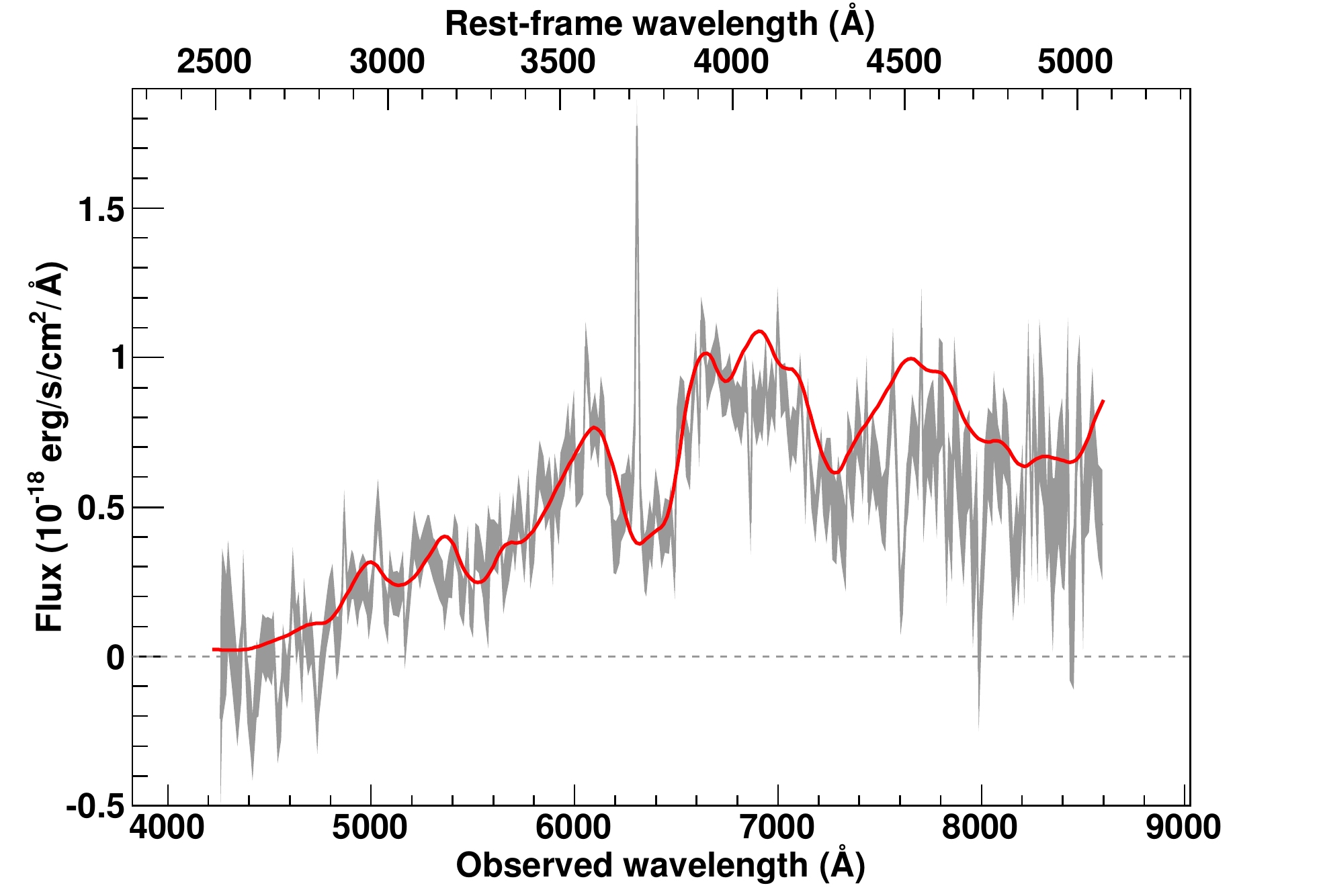}
    \end{center}
    \caption{The SNIa 07D2cb\_1536 spectrum measured at $z=0.694$ with a phase of 1.8 days. A Sd(1) host model has been subtracted.}
    \label{fig:Spec07D2cb_1536}
    \end{figure}
    
    \clearpage
    \begin{figure}
    \begin{center}
    \includegraphics[scale=0.45]{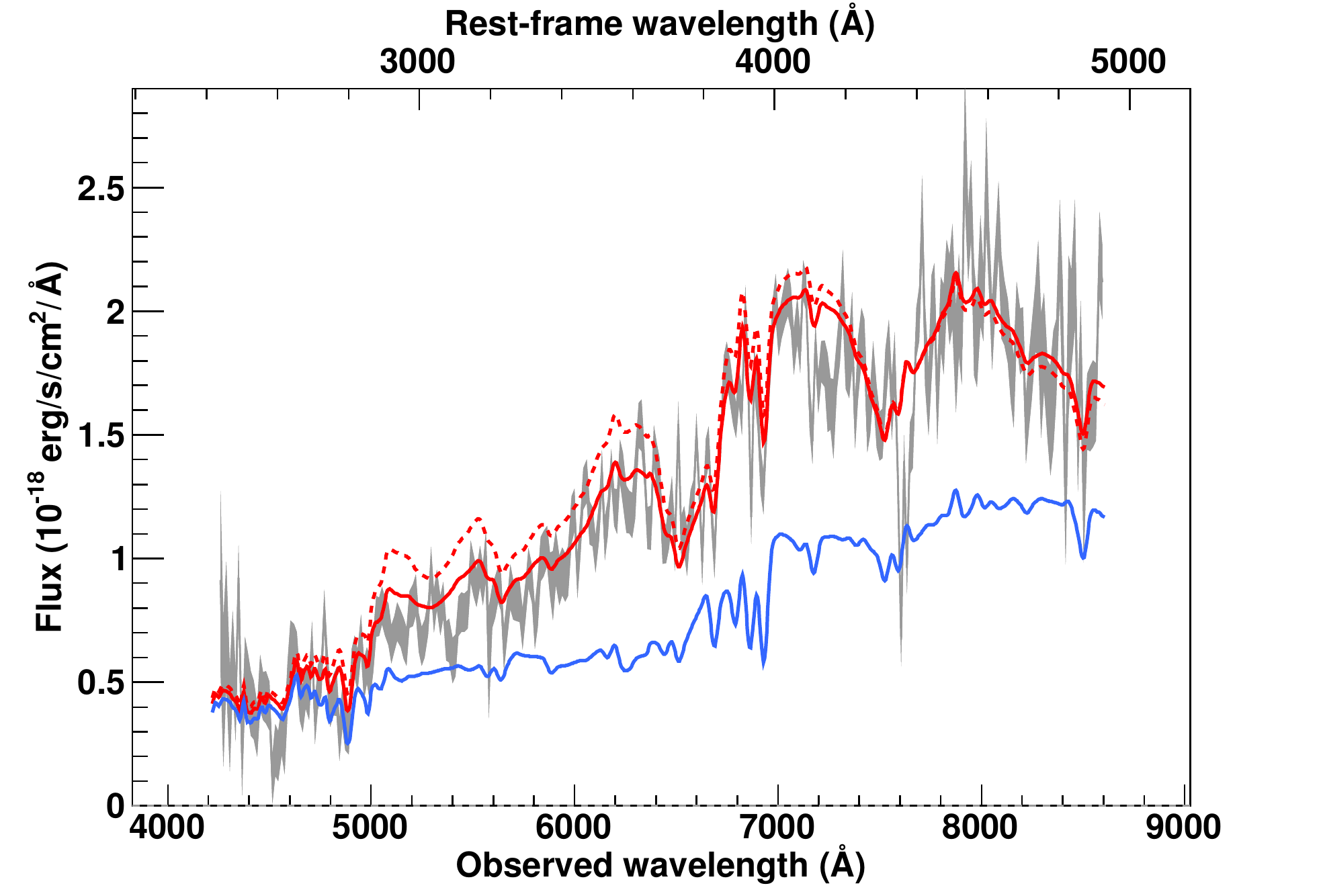}
    \includegraphics[scale=0.45]{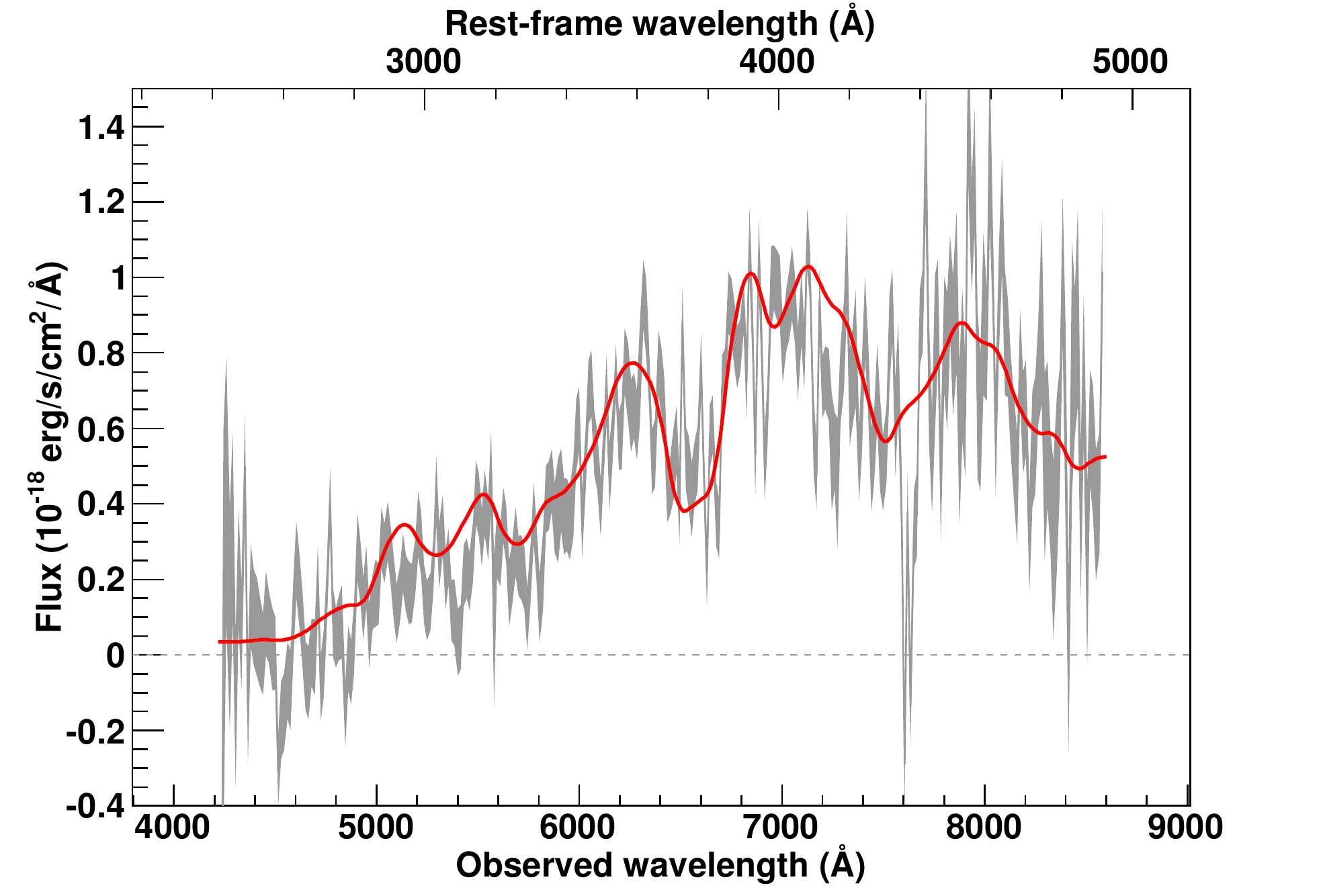}
    \end{center}
    \caption{The SNIa$\star$ 07D2cq\_1539 spectrum measured at $z=0.746$ with a phase of 1.1 days. A E(2) host model has been subtracted.}
    \label{fig:Spec07D2cq_1539}
    \end{figure}
    
    \begin{figure}
    \begin{center}
    \includegraphics[scale=0.45]{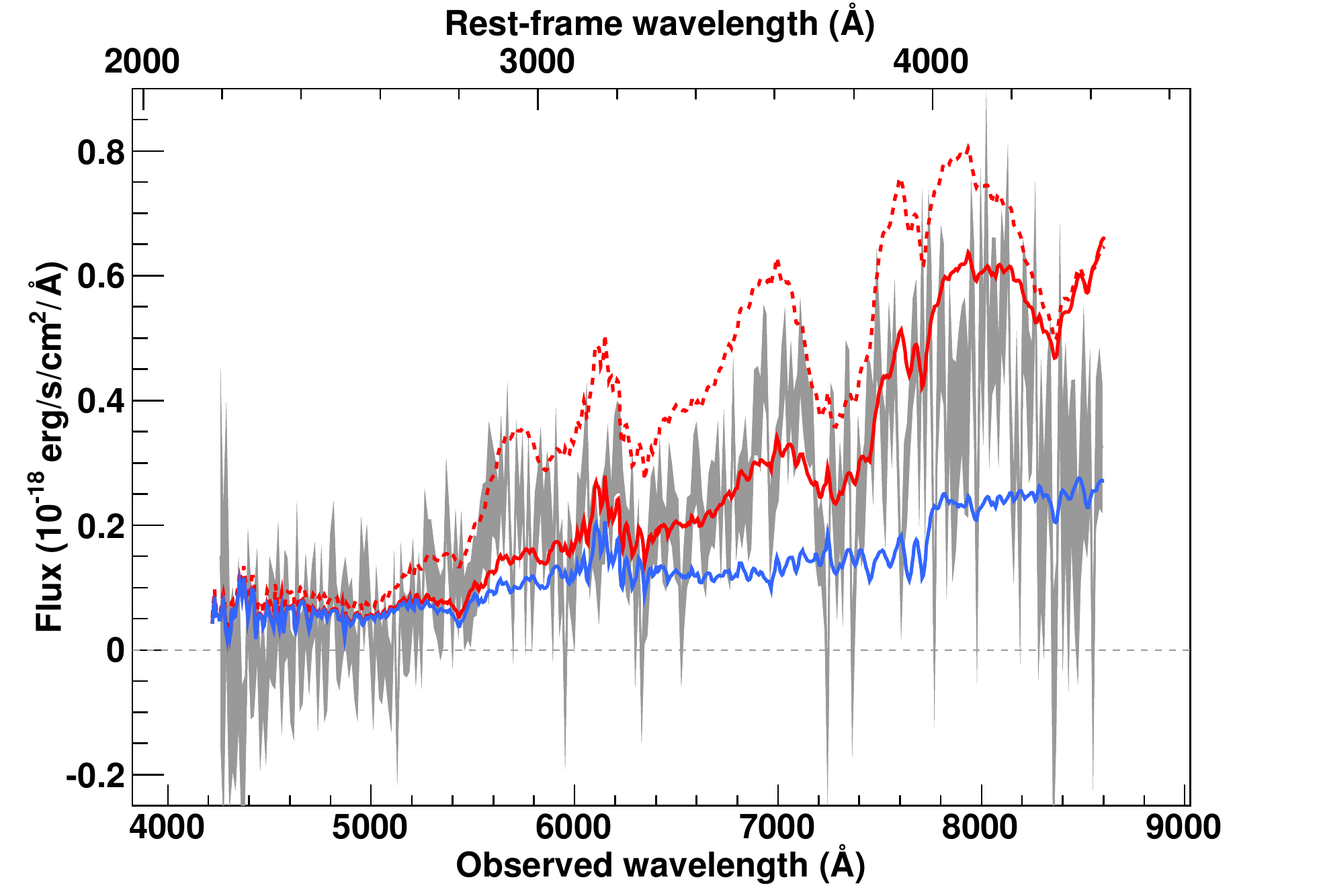}
    \includegraphics[scale=0.45]{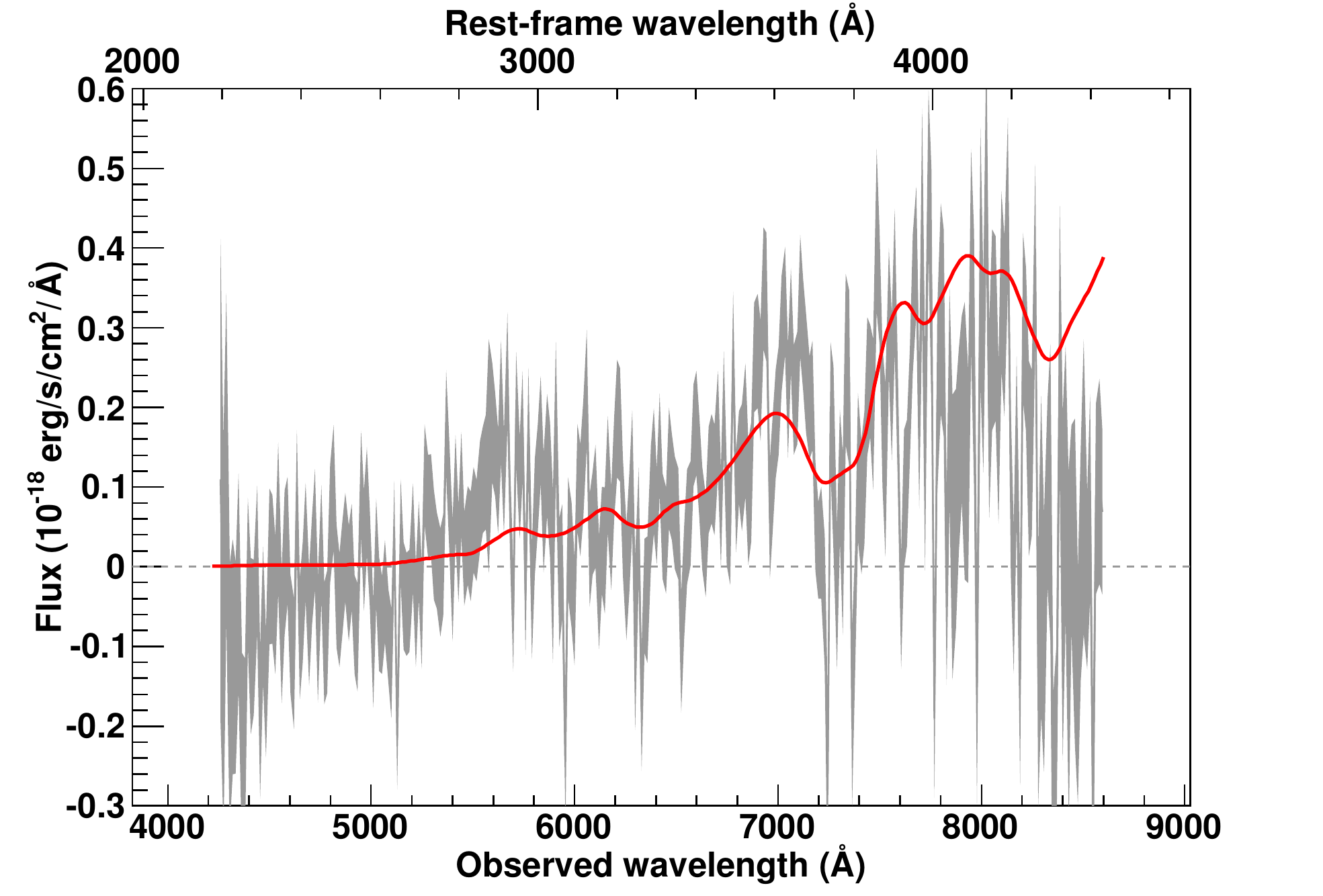}
    \end{center}
    \caption{The SNIa$\star$ 07D2ct\_1540 spectrum measured at $z=0.94$ with a phase of 1.9 days. A Sa-Sb host model has been subtracted.}
    \label{fig:Spec07D2ct_1540}
    \end{figure}
    
    \begin{figure}
    \begin{center}
    \includegraphics[scale=0.45]{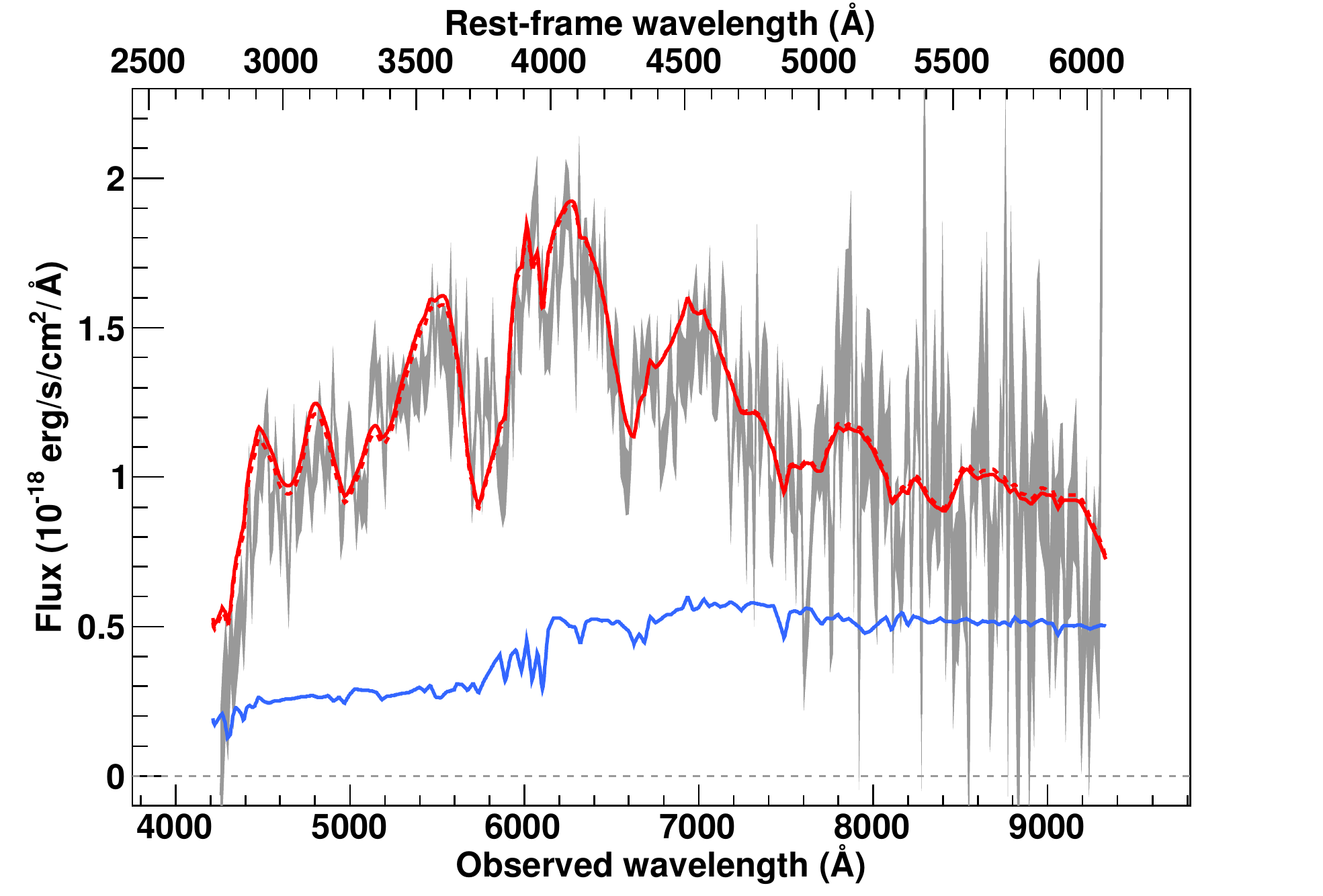}
    \includegraphics[scale=0.45]{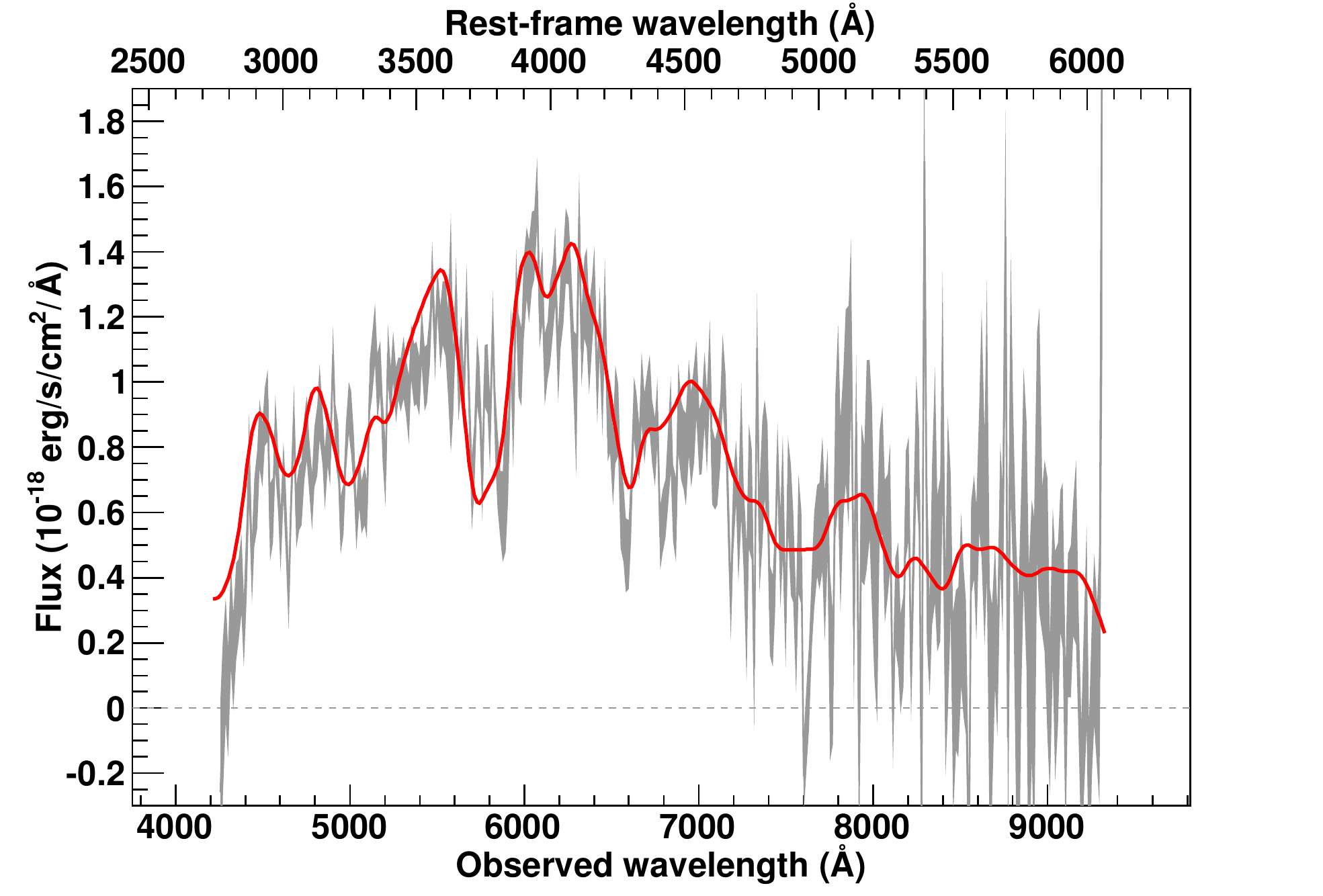}
    \end{center}
    \caption{The SNIa 07D2du\_1570 spectrum measured at $z=0.538$ with a phase of -1.4 days. A E(1) host model has been subtracted.}
    \label{fig:Spec07D2du_1570}
    \end{figure}
    
    \clearpage
    \begin{figure}
    \begin{center}
    \includegraphics[scale=0.45]{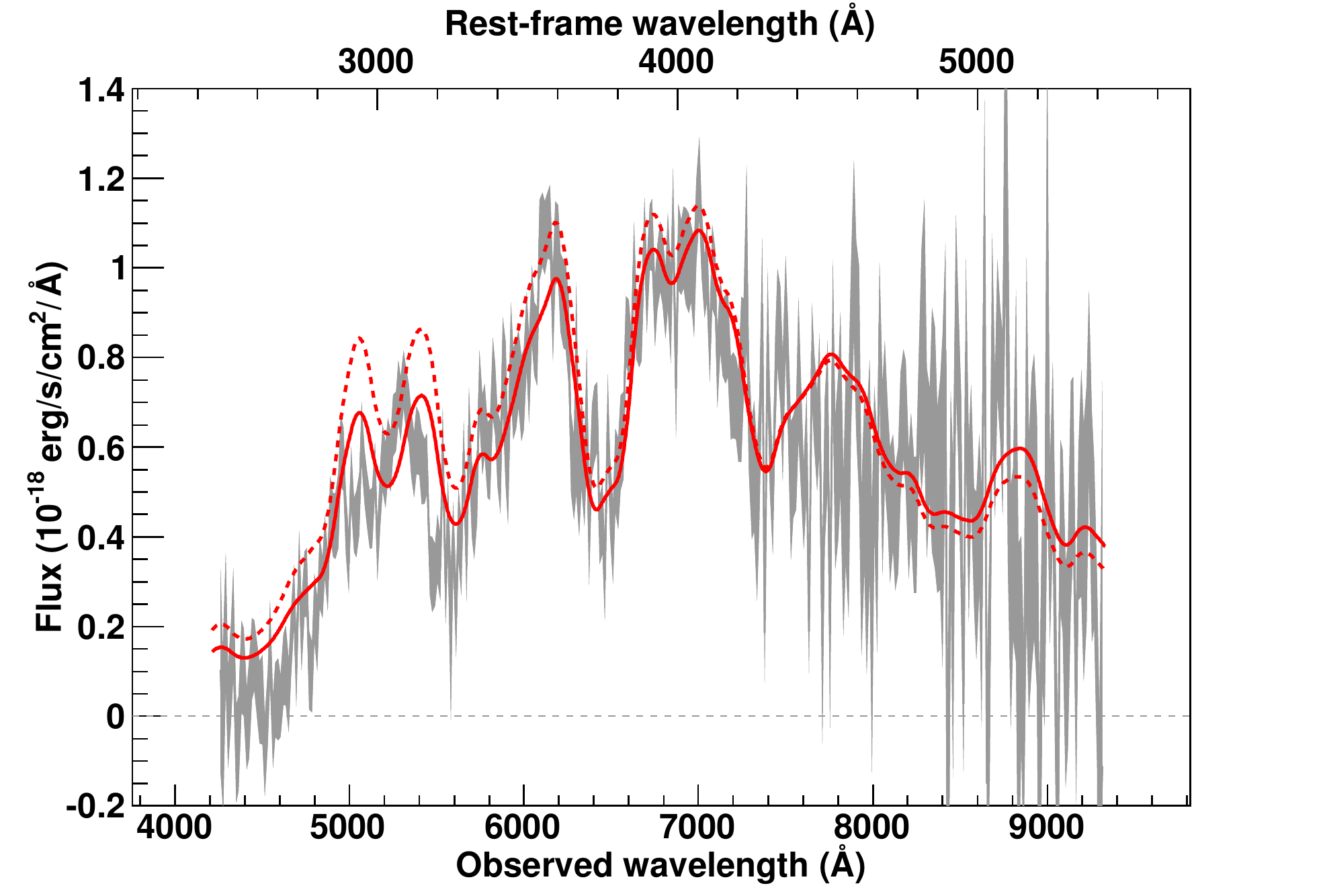}
    \end{center}
    \caption{The SNIa 07D2fy\_1596 spectrum measured at $z=0.72$ with a phase of 0.3 days. Best fit is obtained without galactic component.}
    \label{fig:Spec07D2fy_1596}
    \end{figure}
    
    \begin{figure}
    \begin{center}
    \includegraphics[scale=0.45]{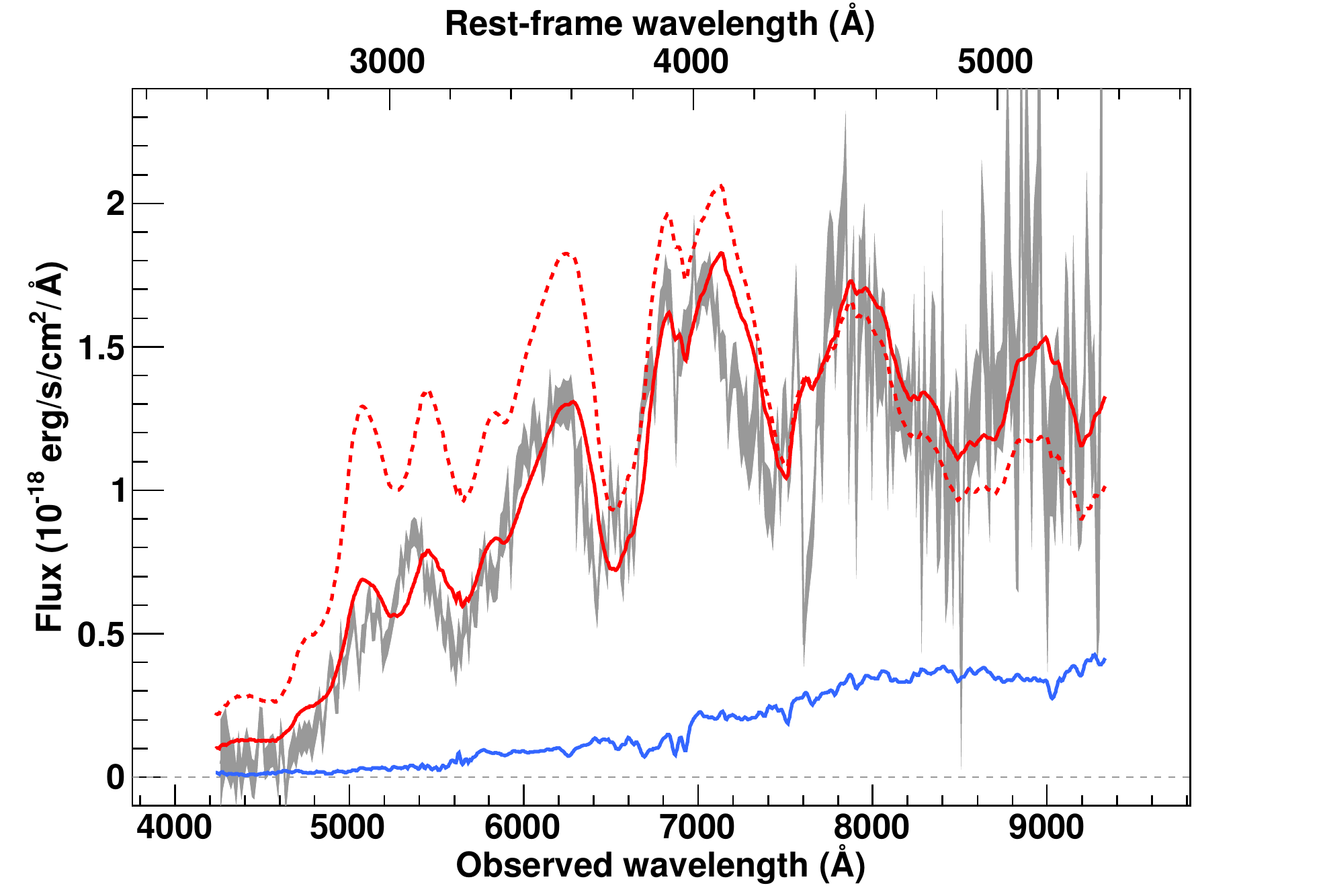}
    \includegraphics[scale=0.45]{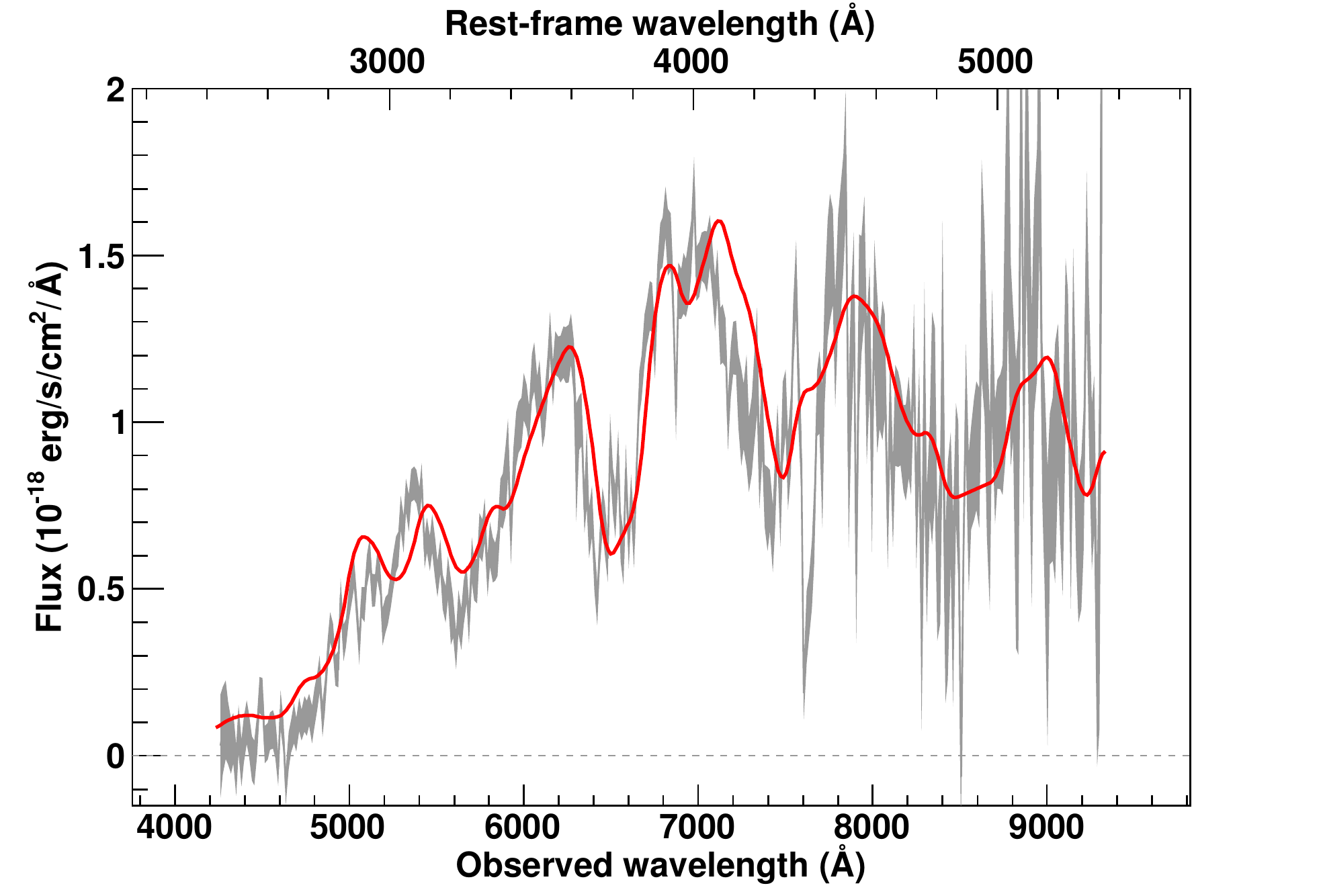}
    \end{center}
    \caption{The SNIa 07D2fz\_1596 spectrum measured at $z=0.743$ with a phase of -1.4 days. A E-S0 host model has been subtracted.}
    \label{fig:Spec07D2fz_1596}
    \end{figure}
    
    \begin{figure}
    \begin{center}
    \includegraphics[scale=0.45]{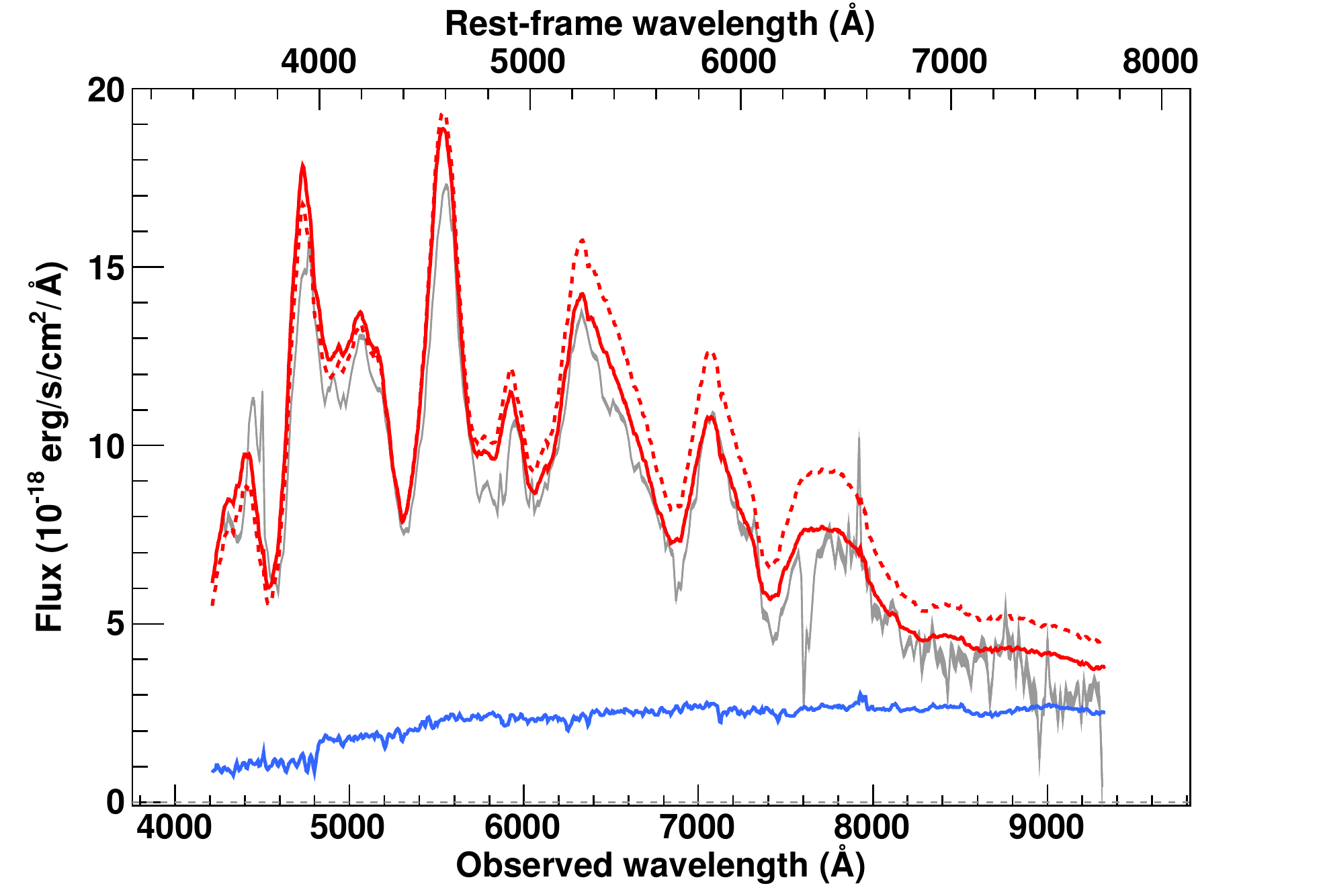}
    \includegraphics[scale=0.45]{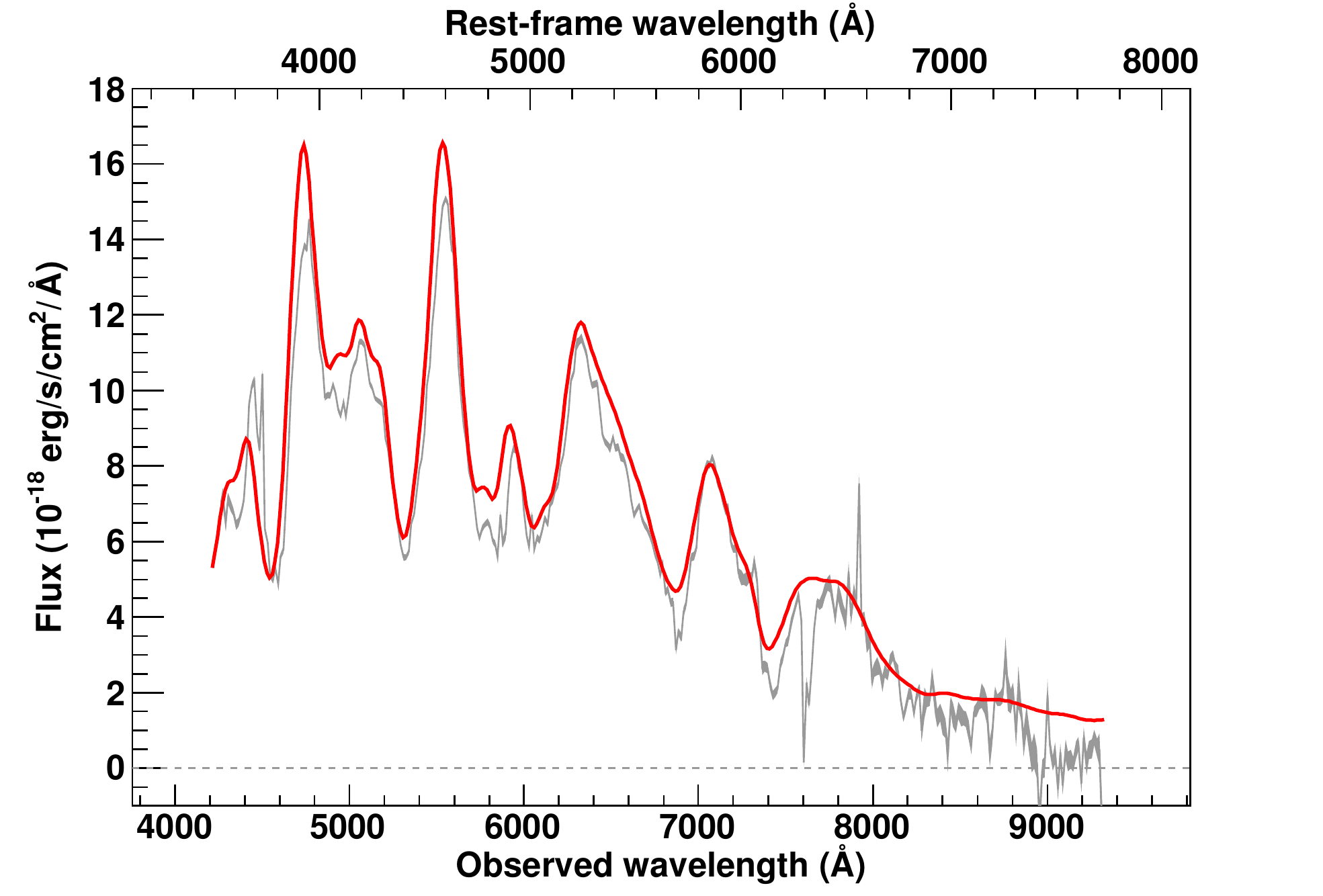}
    \end{center}
    \caption{The SNIa 07D4aa\_1630 spectrum measured at $z=0.207$ with a phase of 13.9 days. A Sb-Sc host model has been subtracted.}
    \label{fig:Spec07D4aa_1630}
    \end{figure}
    
    \clearpage
    \begin{figure}
    \begin{center}
    \includegraphics[scale=0.45]{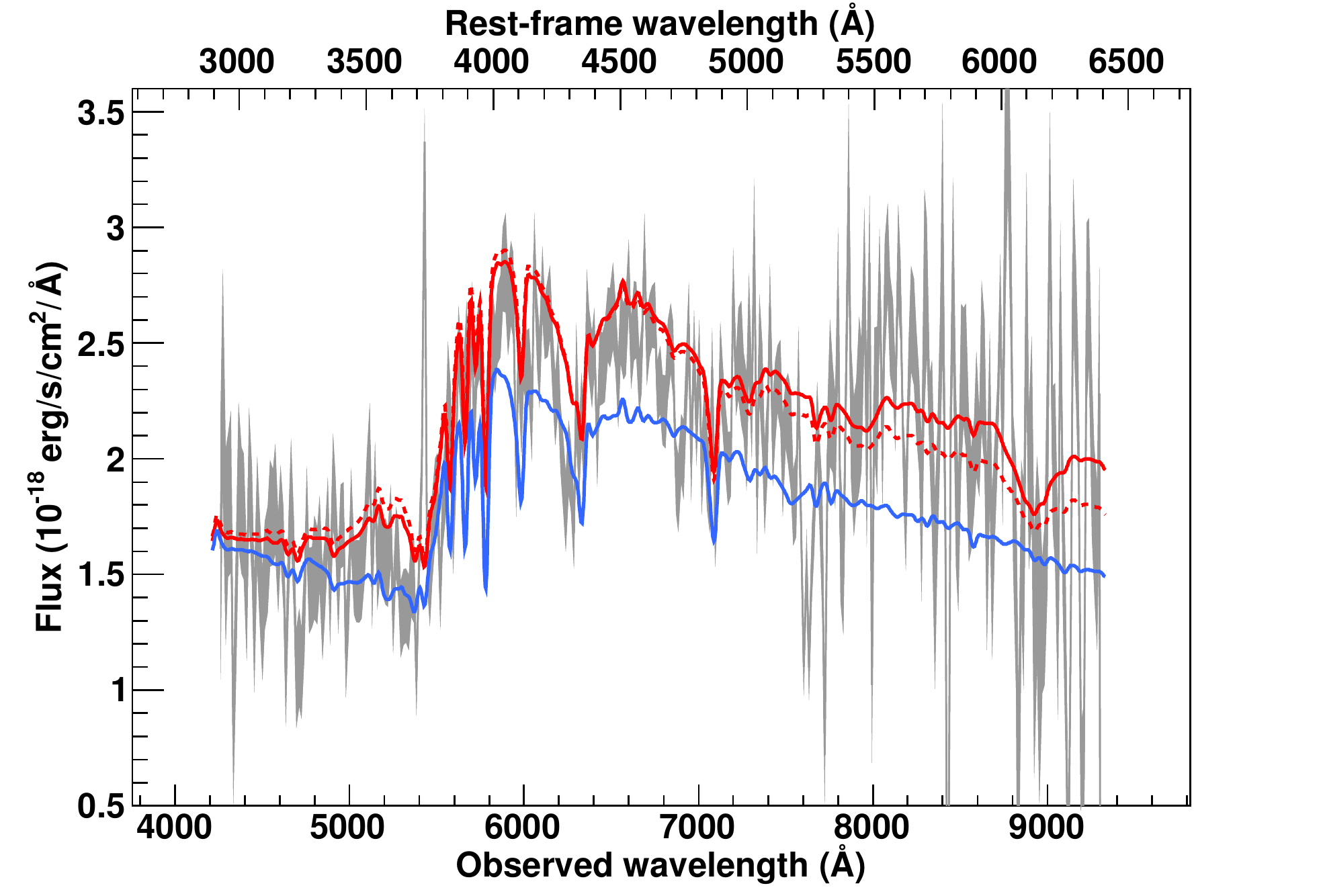}
    \includegraphics[scale=0.45]{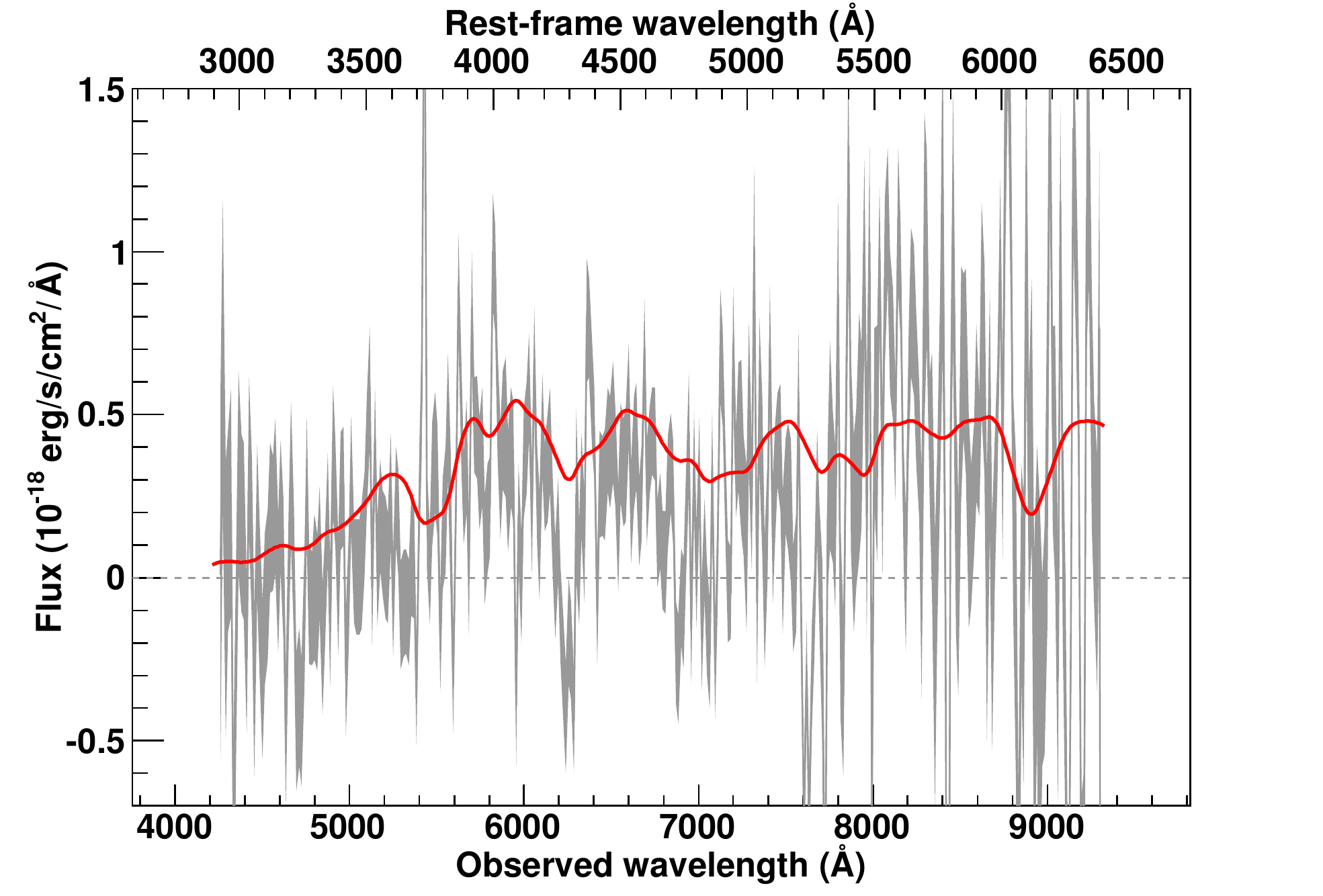}
    \end{center}
    \caption{The SNIa$\star$ 07D4cy\_1694 spectrum measured at $z=0.456$ with a phase of -0.1 days. A Sd(9) host model has been subtracted.}
    \label{fig:Spec07D4cy_1694}
    \end{figure}
    
    \begin{figure}
    \begin{center}
    \includegraphics[scale=0.45]{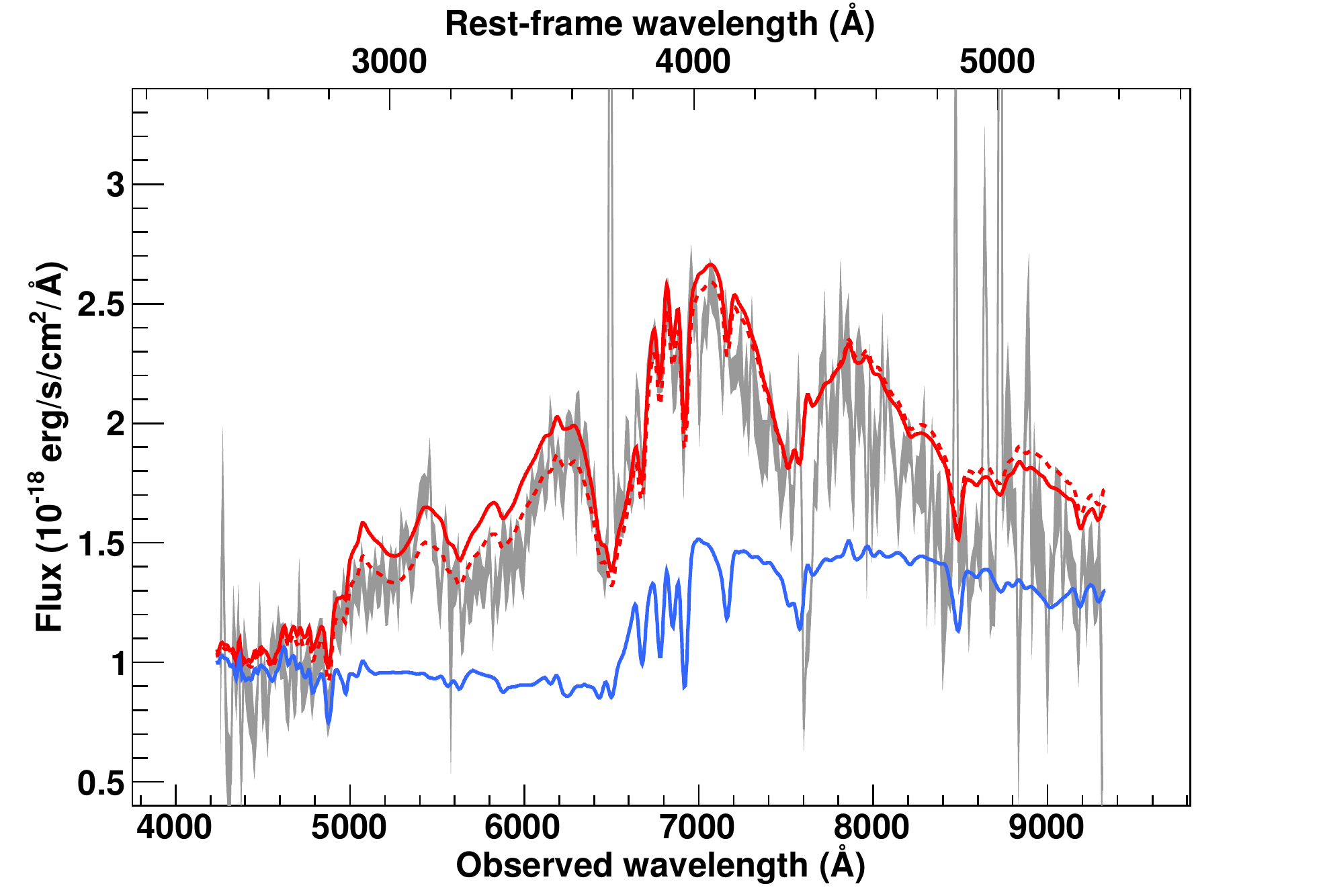}
    \includegraphics[scale=0.45]{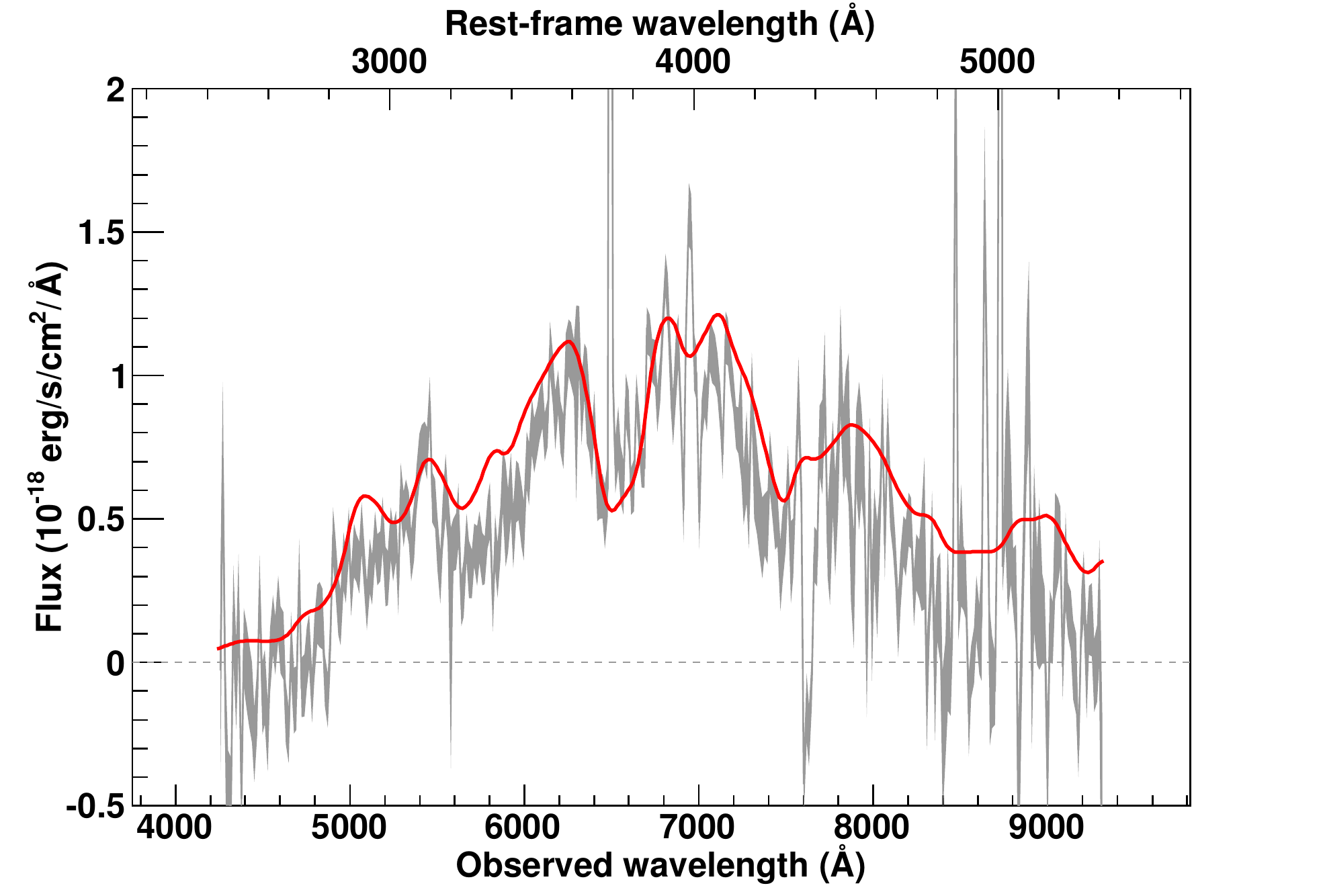}
    \end{center}
    \caption{The SNIa$\star$ 07D4dp\_1713 spectrum measured at $z=0.743$ with a phase of -1.8 days. A Sd(11) host model has been subtracted.}
    \label{fig:Spec07D4dp_1713}
    \end{figure}
    
    \begin{figure}
    \begin{center}
    \includegraphics[scale=0.45]{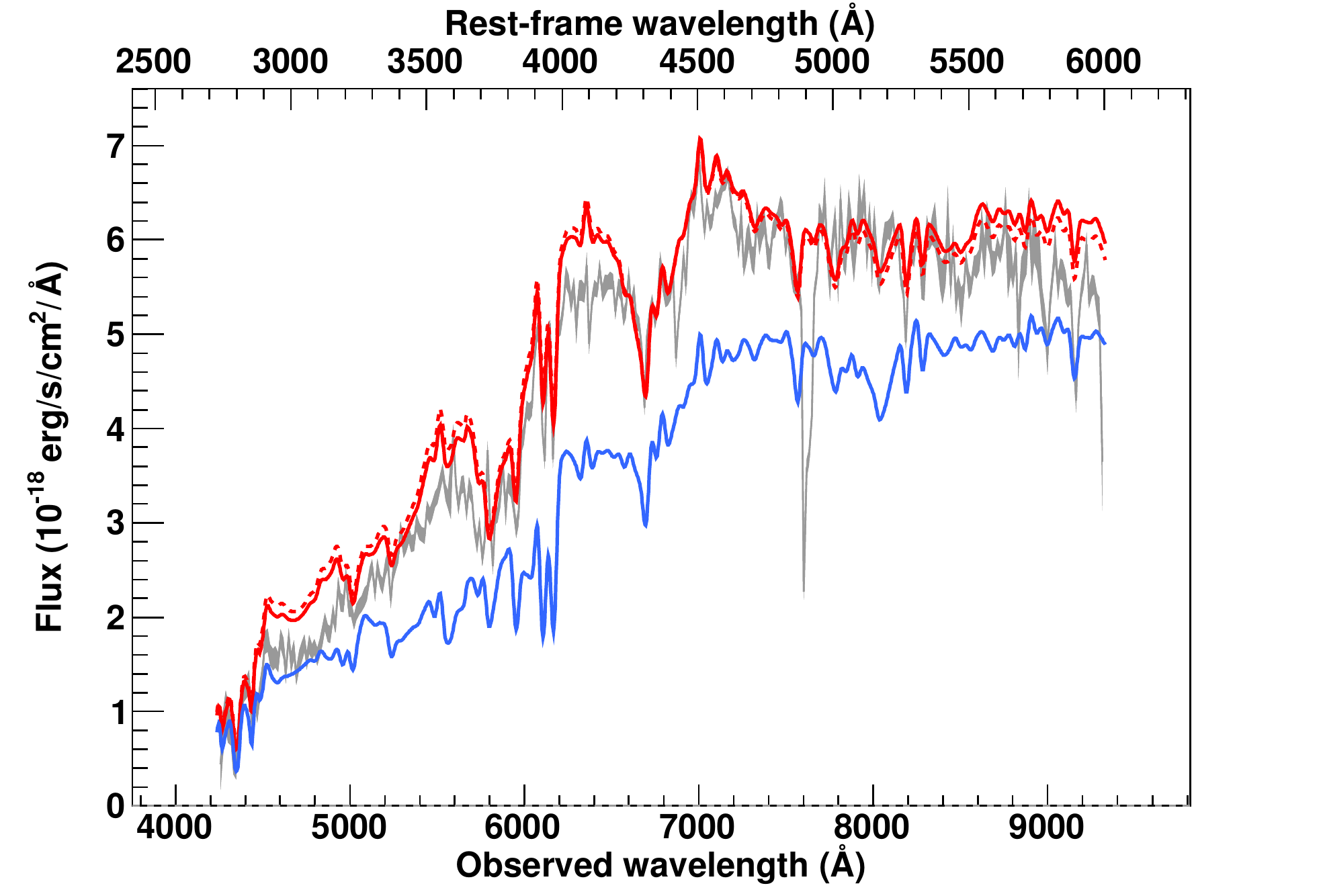}
    \includegraphics[scale=0.45]{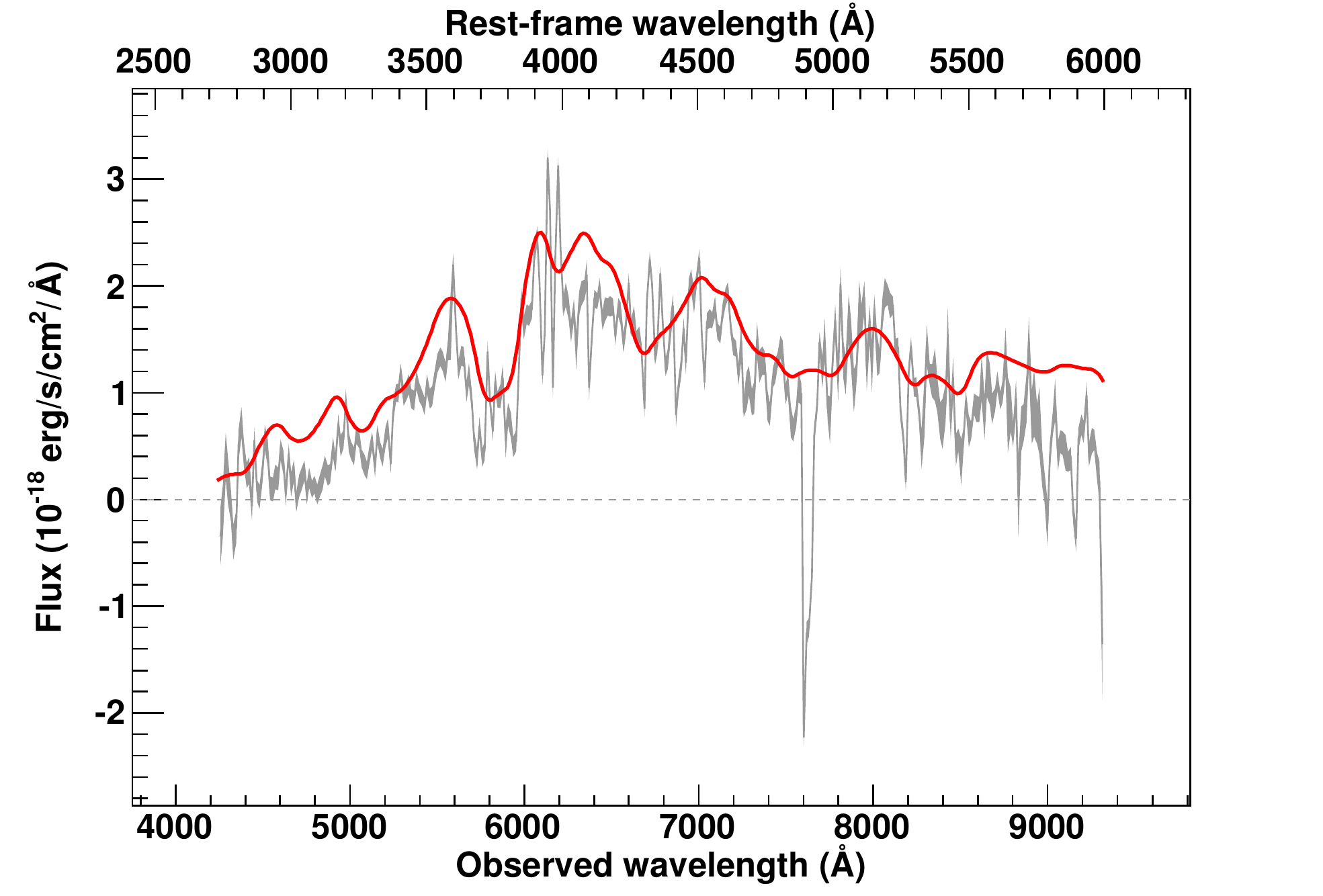}
    \end{center}
    \caption{The SNIa 07D4dq\_1713 spectrum measured at $z=0.554$ with a phase of 1.9 days. A E(3) host model has been subtracted.}
    \label{fig:Spec07D4dq_1713}
    \end{figure}
    
    \clearpage
    \begin{figure}
    \begin{center}
    \includegraphics[scale=0.45]{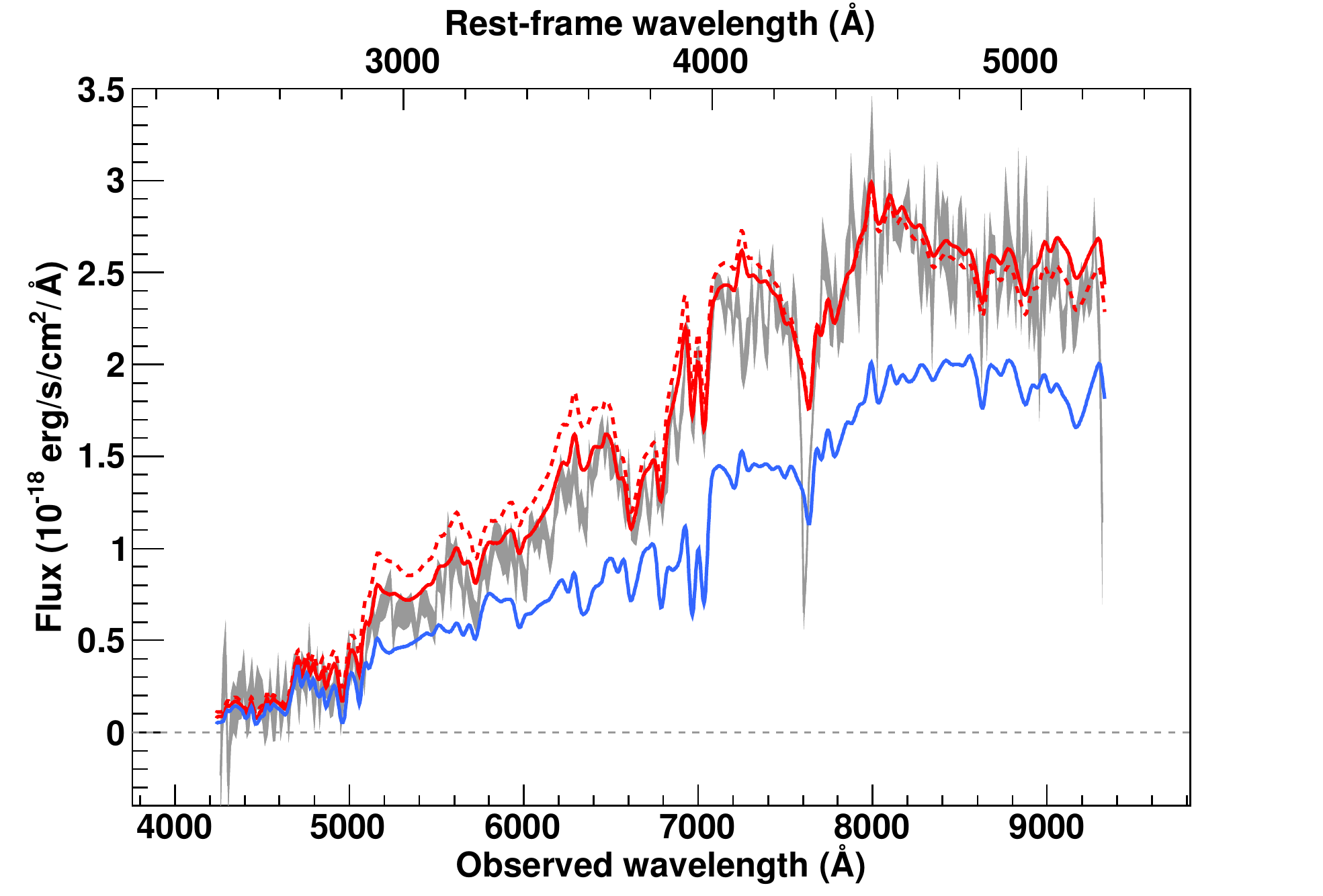}
    \includegraphics[scale=0.45]{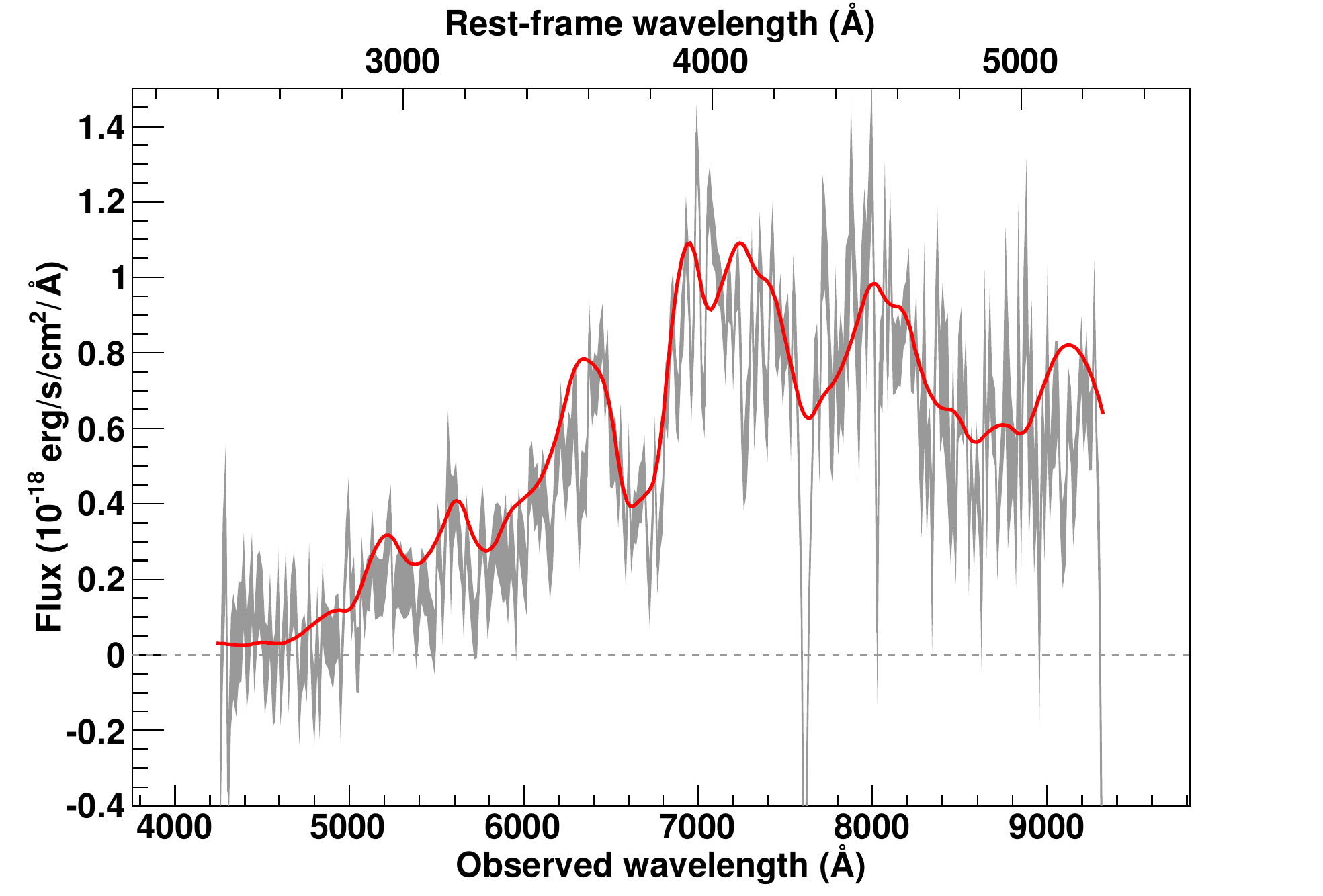}
    \end{center}
    \caption{The SNIa 07D4dr\_1713 spectrum measured at $z=0.772$ with a phase of 2.2 days. A E(4) host model has been subtracted.}
    \label{fig:Spec07D4dr_1713}
    \end{figure}
    
    \begin{figure}
    \begin{center}
    \includegraphics[scale=0.45]{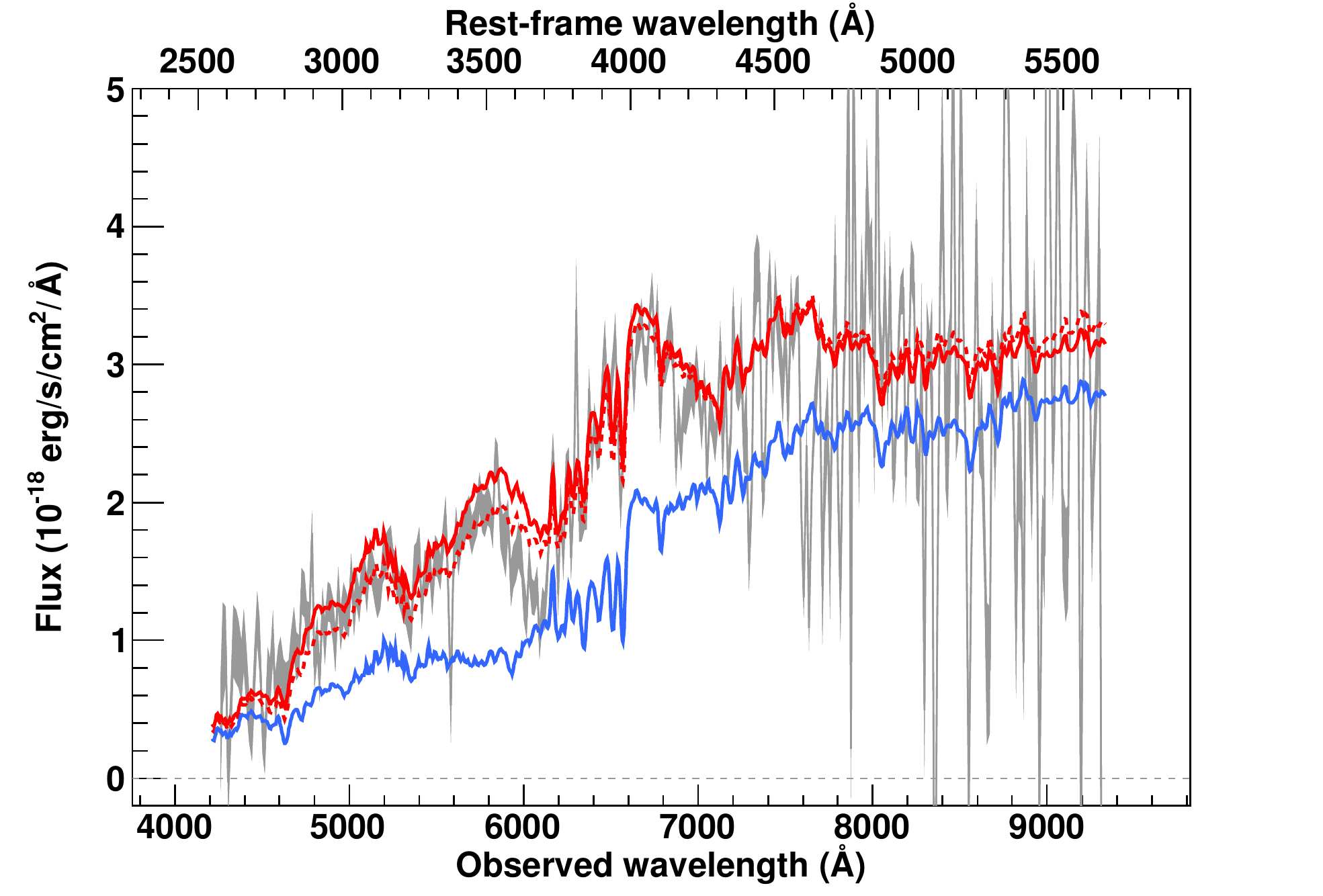}
    \includegraphics[scale=0.45]{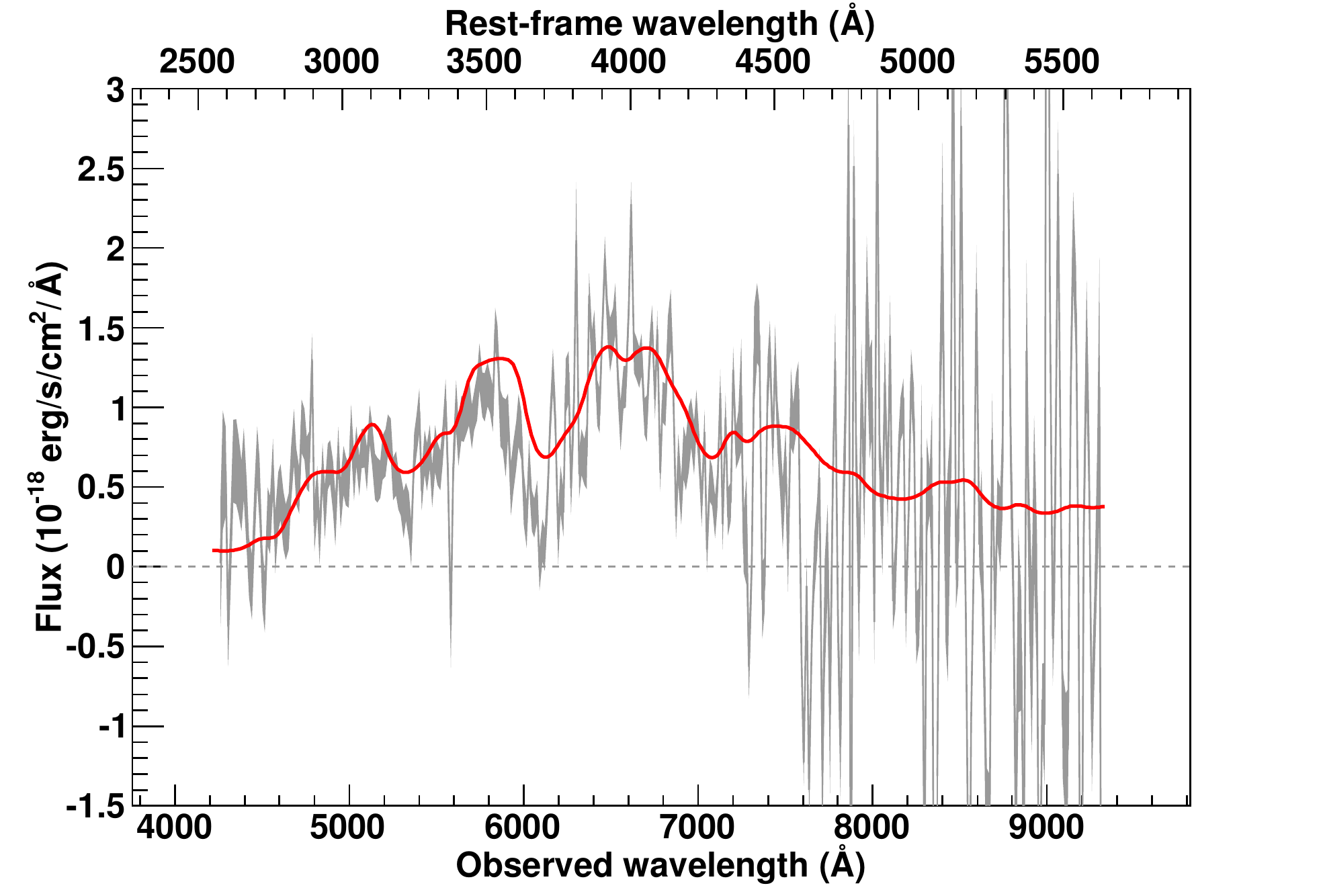}
    \end{center}
    \caption{The SNIa 07D4ec\_1722 spectrum measured at $z=0.653$ with a phase of -4.0 days. A Sa-Sb host model has been subtracted.}
    \label{fig:Spec07D4ec_1722}
    \end{figure}
    
    \begin{figure}
    \begin{center}
    \includegraphics[scale=0.45]{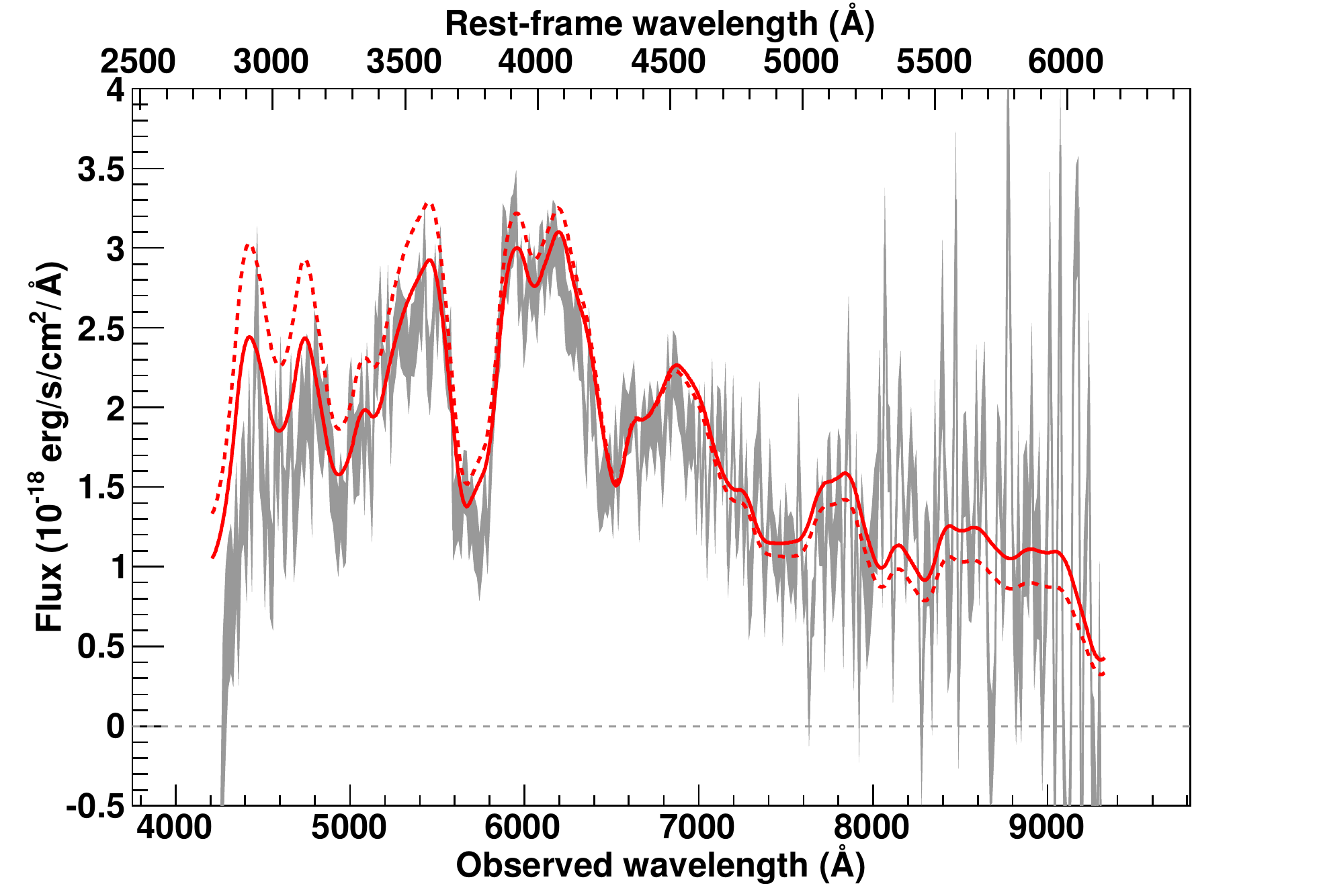}
    \end{center}
    \caption{The SNIa 07D4ed\_1731 spectrum measured at $z=0.52$ with a phase of -1.5 days. A Best fit is obtained without galactic component.}
    \label{fig:Spec07D4ed_1731}
    \end{figure}
    
    \clearpage
    \begin{figure}
    \begin{center}
    \includegraphics[scale=0.45]{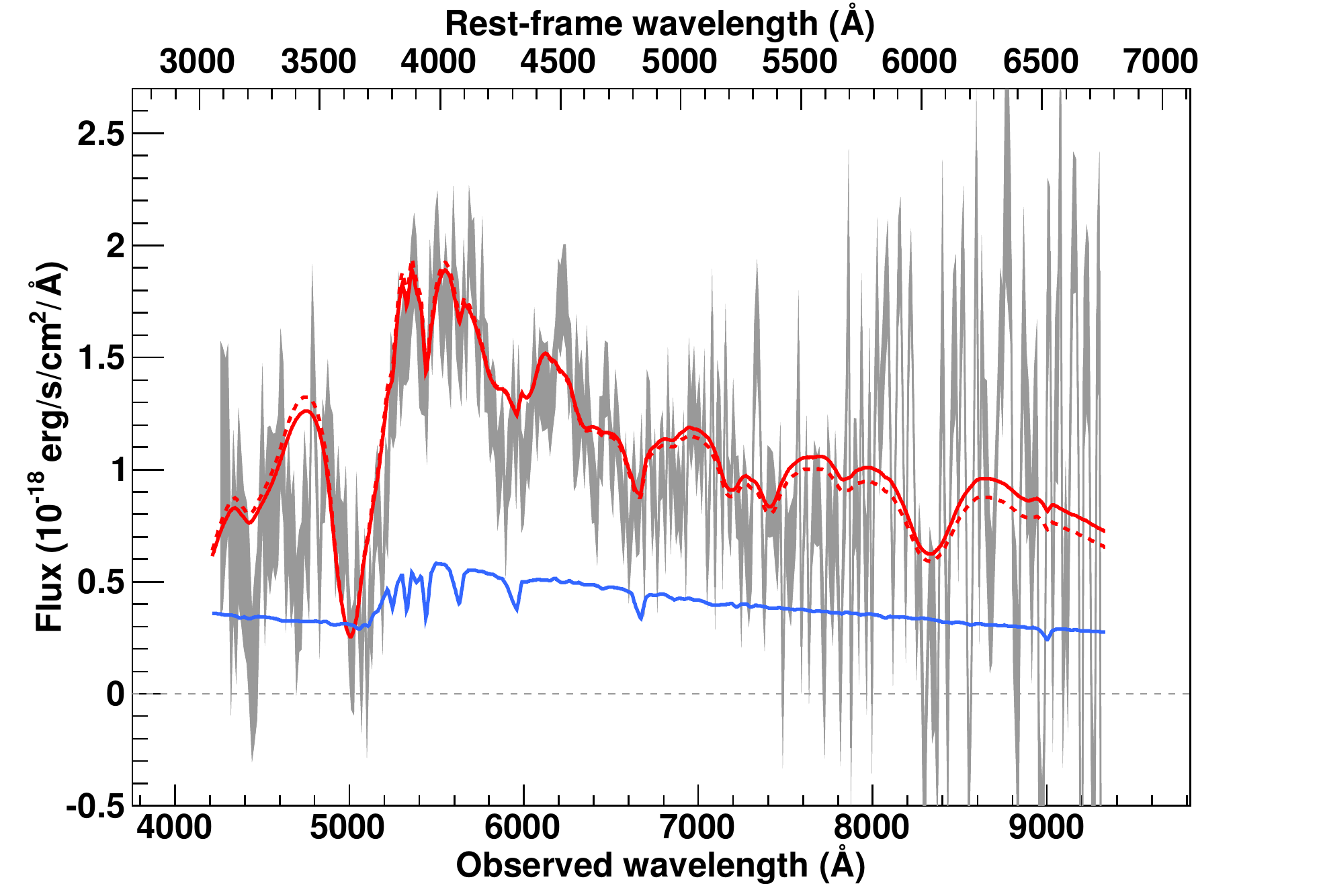}
    \includegraphics[scale=0.45]{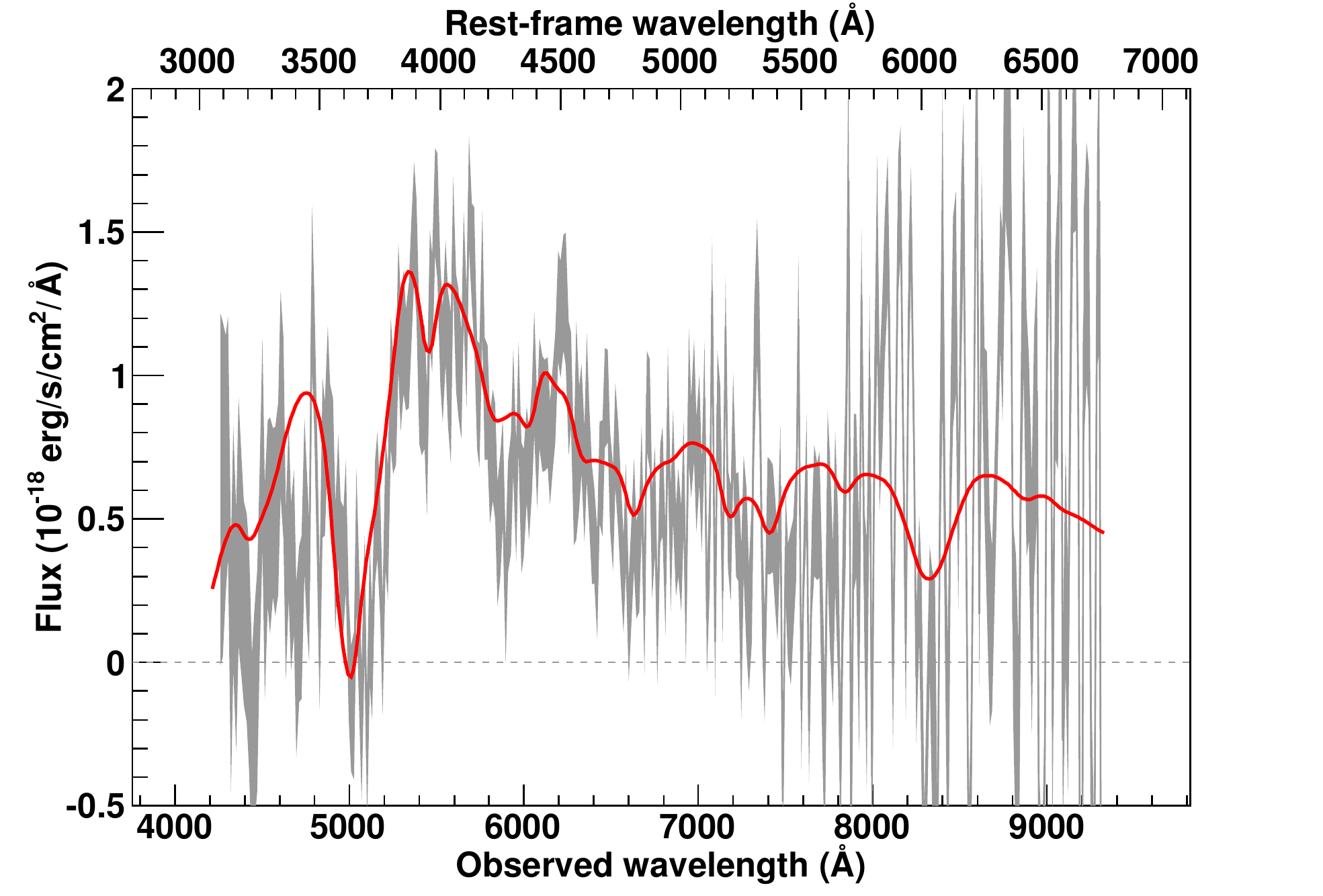}
    \end{center}
    \caption{The SNIa 07D4ei\_1725 spectrum measured at $z=0.37$ with a phase of -6.7 days. A S0(1) host model has been subtracted.}
    \label{fig:Spec07D4ei_1725}
    \end{figure}